\documentclass[12pt,a4paper,reqno]%
{amsart}
\usepackage{amssymb, amsthm, amscd}

\newcommand{\No}{n.~}

\include{diagrams}
\diagramstyle[small,scriptlabels,
midhshaft,nohug]
\newarrow{TeXto}----{->}
\newcommand{\rto}{\rTeXto}
\newcommand{\lto}{\lTeXto}
\newcommand{\uto}{\uTeXto}

\newcommand{\ruto}{\ruTeXto}
\newcommand{\rdto}{\rdTeXto}
\newcommand{\luto}{\luTeXto}
\newcommand{\ldto}{\ldTeXto}

\newtheorem{theor}{Theorem}
\newtheorem*{theorNo}{Theorem}
\newtheorem{state}[theor]{Proposition}
\newtheorem{prop}[theor]{Proposition}
\newtheorem{lemma}[theor]{Lemma}
\theoremstyle{definition}
\newtheorem{define}{Definition}

\theoremstyle{remark}
\newtheorem{cor}[theor]{Corollary}
\newtheorem{rem}{Remark}
\newtheorem{example}{Example}
\newtheorem{case}{Case}

\newcommand{\Id}{{\mathrm d}}
\newcommand{\ID}{{\mathrm D}}

\newcommand{\IL}{{\mathrm L}}

\newcommand{\pinner}{\mathbin{\mathchoice
   {\hbox{\vrule width0.6em depth0pt height0.4pt
   \vrule width0.4pt depth0pt height0.8ex}}
   {\hbox{\vrule width0.6em depth0pt height0.4pt
   \vrule width0.4pt depth0pt height0.8ex}}
   {\hbox{\kern0.14em
   \vrule width0.48em depth0pt height0.4pt
   \vrule width0.4pt depth0pt height0.6ex\kern0.14em}}
   {\hbox{\kern0.1em
   \vrule width0.39em depth0pt height0.4pt
   \vrule width0.4pt depth0pt height0.5ex\kern0.1em}}}}
\newcommand{\inner}{\pinner\,}

\DeclareFontFamily{OML}{cyr}{}
\DeclareFontShape{OML}{cyr}{m}{n}{
   <5> <6> <7> <8> <9> gen * wncyr
   <10> <10.95> <12> <14.4> <17.28> <20.74> <24.88> wncyr10
  }{}
\DeclareSymbolFont{rusletters}{OML}{cyr}{m}{n}
\DeclareSymbolFontAlphabet{\rusmath}{rusletters}
\DeclareMathSymbol\re{\rusmath}{rusletters}{"03}
\newcommand{\cEv}{\re}

\newcommand{\BBC}{{\mathbb{C}}}
\newcommand{\BBE}{{\mathbb{E}}}
\newcommand{\BBR}{{\mathbb{R}}}
\newcommand{\BBN}{{\mathbb{N}}}
\newcommand{\BBZ}{{\mathbb{Z}}}

\newcommand{\id}{\mathop{\rm id}\nolimits}

\newcommand{\Sym}{\mathop{\rm Sym}\nolimits}
\newcommand{\sym}{\mathop{\rm sym}\nolimits}

\newcommand{\scal}{\mathop{\rm scal}\nolimits}
\newcommand{\const}{\mathop{\rm const}\nolimits}
\newcommand{\grad}{\mathop{\rm grad}\nolimits}

\newcommand{\gothg}{\mathfrak{g}}

\newcommand{\gm}{\mathfrak{m}}
\newcommand{\gt}{\mathfrak{t}}
\newcommand{\bi}{\text{\textit{\textbf{i}}}}
\newcommand{\bu}{\text{\textit{\textbf{u}}}}
\newcommand{\bx}{\text{\textit{\textbf{x}}}}
\newcommand{\bE}{\mathbf{E}}
\newcommand{\bT}{\mathbf{T}}
\newcommand{\gA}{\mathfrak{A}}

\newcommand{\gB}{\mathfrak{B}}
\newcommand{\gC}{\mathfrak{C}}
\newcommand{\gD}{\mathfrak{D}}
\newcommand{\gv}{{\mathfrak{v}}}

\newcommand{\cB}{\mathcal{B}}
\newcommand{\cC}{\mathcal{C}}
\newcommand{\cD}{\mathcal{D}}
\newcommand{\cE}{\mathcal{E}}
\newcommand{\cF}{\mathcal{F}}
\newcommand{\cH}{\mathcal{H}}
\newcommand{\cI}{\mathcal{I}}
\newcommand{\cL}{\mathcal{L}}

\newcommand{\cP}{\mathcal{P}}

\newcommand{\cT}{\mathcal{T}}

\newcommand{\cU}{\mathcal{U}}
\newcommand{\cV}{\mathcal{V}}
\newcommand{\cX}{\mathcal{X}}
\newcommand{\cY}{\mathcal{Y}}
\newcommand{\dd}{\partial}
\newcommand{\zz}{\bar{z}}
\newcommand{\bun}{\mathbf{1}}
\newcommand{\vph}{\varphi}
\newcommand{\vvph}{{\vec{\vph}}}
\newcommand{\vpsi}{{\vec{\psi}}}
\newcommand{\vth}{\vartheta}

\newcommand{\vu}{{\vec{u}}}
\newcommand{\rme}{{\mathrm{e}}}
\newcommand{\rSi}{\stackrel{\leftrightarrow}{\varSigma}}

\newcommand{\pKdV}{{\text{\textup{pKdV}}}}
\newcommand{\pmKdV}{{\text{\textup{pmKdV}}}}
\newcommand{\KdV}{{\text{\textup{KdV}}}}
\newcommand{\mKdV}{{\text{\textup{mKdV}}}}
\newcommand{\Liou}{{\text{\textup{Liou}}}}
\newcommand{\Toda}{{\text{\textup{Toda}}}}
\newcommand{\heav}{{\text{\textup{heav}}}}
\newcommand{\NLS}{{\text{\textup{NLS}}}}

\DeclareMathOperator{\CDiff}{\mathcal{C}Diff}

\DeclareMathOperator{\proj}{proj}

\DeclareMathOperator{\Hom}{Hom}

\DeclareMathOperator{\tr}{tr}

\DeclareMathOperator{\wt}{wt}

\newcommand{\pp}{\phantom{+}}

\newcommand{\fn}[2]{{[\![{#1},{#2}]\!]}^{\mathrm{FN}}}
\newcommand{\fnh}[2]{{[\![{#1},{#2}]\!]}^{\mathrm{FN}}}
\newcommand{\dy}[2]{{#1}_{\overline{{#2}}}}

\newcommand{\omx}{\Id y\wedge\Id z}
\newcommand{\omy}{\Id z\wedge\Id x}
\newcommand{\omz}{\Id x\wedge\Id y}

\newcommand{\ttbox}[1]{\mbox{\ttfamily#1}}
\newcommand{\by}[1]{\textit{{#1}}}
\newcommand{\jour}[1]{\textrm{{#1}}}
\newcommand{\vol}[1]{\textbf{{#1}}}
\newcommand{\book}[1]{\textrm{{#1}}}

\title[Symmetries and conservation laws in integrable models]%
{Methods of geometry of differential equations
in analysis of the integrable field theory models}
\author{Arthemy~V.~Kiselev}
\thanks{Partially supported by the scholarship
of the Government of the Russian Federation, the INTAS grant
YS~2001/2-33, and the Lecce University grant n.~650~CP/D}

\address{153003 Russia, Ivanovo, Rabfakovskaya str.\ 34,
Ivanovo State Power University, Chair of Mathematics.}

\curraddr{Dipartimento di Matematica `Ennio De Giorgi',
Universit\`a degli Studi di Lecce,
Via per Arnesano, 73100 Lecce (LE), Italy.}

\email{arthemy@poincare.unile.it}

\date{June 1, 2004.}

\subjclass[2000]%
{35Q53, 37K05}


\begin{document}

\begin{abstract}
In this paper, we investigate the algebraic and geometric properties of
the hyperbolic Toda equations $u_{xy}=\exp(Ku)$ associated with
nondegenerate symmetrizable matrices~$K$.
A hierarchy of analogs to the potential modified Korteweg\/--\/de Vries
equation $u_t=u_{xxx}+u_x^3$ is constructed, and its relation with the
hierarchy for the Korteweg\/--\/de Vries equation $T_t=T_{xxx}+TT_x$ is
established. Group\/--\/theoretic structures for the dispersionless
$(2+1)$-dimensional Toda equation $u_{xy}=\exp(-u_{zz})$ are obtained.
Geometric properties of the multi\/--\/component
nonlinear Schr\"odinger equation type systems
$\Psi_t=\bi\Psi_{xx}+\bi f(|\Psi|)\,\Psi$ (multi\/--\/soliton complexes)
are described.
\end{abstract}

\maketitle

\noindent{UDC 517.957 + 514.763.85}

\tableofcontents

\begin{verse}
`You boil it in sawdust: you salt it in glue:\\
You condense it with locusts and tape:\\
Still keeping one principal object in view ---\\
To preserve its symmetrical shape.'
\end{verse}
\rightline{\textit{The Hunting of the Snark}, Lewis Carroll.}

\bigskip
\noindent%
The Toda equation (\cite{MToda})
and the Toda type equations associated with the semisimple
Lie algebras (\cite{LeznovSaveliev}) play an important role in
the models of modern conformal field theory.
We recall that the Toda equations appear in gravity
theory (\cite{Alfinito, BoyerFinley}),
in the Yang\/--\/Mills field theory (\cite{LSYangMills}),
in differential geometry (\cite{GervaisENS, Kazdan}),
and in the classification problems for nonlinear partial differential
equations (\cite{SokolovUMN}).
Also, the Toda equations are related with the integrable dynamical
systems (\cite{DSViniti84}),
the Frobenius manifolds, and the associative algebra
structures (\cite{Dubrovin3}).
The detailed study of the following systems that appear in the
above\/--\/mentioned theories is reduced to the study of the Toda
equations: these systems are the antiselfdual vacuum Einstein
equations, the Yang\/--\/Mills equations, the
Gauss\/--\/Mainardi\/--\/Codazzi equations for complex curves in the
K\"ahler manifolds, the dynamical equations for the Laplace invariants
of differential equations, the Korteweg\/--\/de Vries equation, and the
WDVV (Witten\/--\/Dijkraaf\/--\/H.~Verlinde\/--\/E.~Verlinde) equation,
respectively.

The algebraic approach towards the description of properties
of the two\/--\/dimensional hyperbolic Toda equations
\[
\bu_{xy}=\exp(K\bu)
\]
was developed in the papers by
A.~N.~Leznov and M.~V.~Saveliev (\cite{LeznovSaveliev}),
V.~G.~Drinfel'd and V.~V.~Sokolov (\cite{DSViniti84}),
B.~A.~Dubrovin (\cite{Dubrovin3}) \textit{et al.}
In these papers, the Toda equations are interpreted as the flat
connection equations on semisimple complex Lie algebras (or
the Kac\/--\/Moody algebras)
such that $K$ is the corresponding Cartan matrix.
Then the Toda equations are said to be associated with the Lie
algebras (respectively, the Kac\/--\/Moody algebras). These equations
are exactly integrable~(\cite{LeznovSaveliev}).
In the fundamental paper~\cite{DSViniti84}, the integrable
Drinfel'd\/--\/Sokolov hierarchies were assigned to the Toda equations
associated with the Kac\/--\/Moody algebras.
The Drinfel'd\/--\/Sokolov equations are the
bi\/--\/Hamiltonian Korteweg\/--\/de Vries type equations.
Still, the Lie\/--\/algebraic approach does not provide the exhaustive
description of geometric properties of the Toda equations themselves.
Indeed, the structure of generators of the Noether symmetries Lie
subalgebra for the Toda equation, the existence of recursion operators
for its symmetry algebra, or the relation between conservation laws for
the Toda equation and the Hamiltonian structures for the
Korteweg\/--\/de Vries equation was not revealed.
In particular, until recently it remained unnoticed that all these
properties of the Toda equation are preserved for the general case of
the Toda equations $\bu_{xy}=\exp(K\bu)$ associated with nondegenerate
symmetrizable matrices~$K$, which are not assumed to be the Cartan
matrices.

The homological methods developed by I.~S.Krasil'shchik,
V.~V.~Lychagin, A.~M.~Vinogradov (\cite{ClassSym, JKKersten, Opava,
Vin84}), and their scientific school are a powerful tool in the
analysis of algebraic and geometric structures for differential
equations. Owing to essential achievements of this theory, it
was quite natural to apply the homological methods to the study of the
Toda equation and related systems.

In this article, we carry out the detailed investigation of
the geometric properties of the hyperbolic Toda equations.
Also, we construct new Hamiltonian evolution systems associated with
the Toda equations and establish their nontrivial links with other
important classes of the mathematical physics equations such as the
Korteweg\/--\/de Vries equation.

\newpage
\specialsection*{\textbf{Introduction}}
The discrete one\/--\/dimensional Toda lattice~(\cite{MToda})
\begin{equation}\label{IeqMToda}
\ddot q_n=\exp(q_{n-1}-q_n)-\exp(q_n-q_{n+1}),\qquad n\in\BBZ,\quad
q_n=q_n(\tau)
\end{equation}
is a model nonlinear integrable
equation in mathematical physics.
The equation
\begin{equation}\label{IeqDubrovinStart}
\varepsilon^2\cdot q_{\tau\tau}=
\exp\bigl(q(z-\varepsilon)-q(z)\bigr)-
\exp\bigl(q(z)-q(z+\varepsilon)\bigr)
\end{equation}
is a continuous analog of Eq.~\eqref{IeqMToda}.
Here $\varepsilon\geq0$ and $q=q(\tau,z)$.
Equation~\eqref{IeqDubrovinStart} is then obtained
by using the relation
$q_n(\tau)\mathrel{{=}}q(\tau,n\varepsilon)$.
The equation
\[       
u_{\tau\tau}=\pm D_z^2\circ\exp(u),\qquad u=u(\tau,z),
\]       
is the limit for the family of equations~\eqref{IeqDubrovinStart}
as $\varepsilon\to+0$.
Here $D_z$ is the total derivative with respect to $z$ and
$u_n(\tau)\equiv q_{n-1}-q_n\mathrel{{=}}u(\tau,n\varepsilon)$.
In the paper~\cite{Dubrovin3} Dubrovin proved that
Eq.~\eqref{IeqDubrovinStart} is reduced to the nonlinear Schr\"odinger
equation if $\varepsilon=\bi$.
In this paper, we consider the multi\/--\/component analogs
(multi\/--\/soliton complexes, see~\cite{Chaos})
\begin{equation}\label{INLS}
\Psi_t=\bi\Psi_{xx}+\bi f(|\Psi|)\,\Psi,
\end{equation}
of the nonlinear Schr\"odinger equation.
Here $\Psi$ is the $m$-component vector, $\bi=\sqrt{-1}$,
and $f\in C^\infty(\BBR)$.

The passage from one\/--\/dimensional equation~\eqref{IeqMToda}
to the two\/--\/di\-men\-si\-o\-nal Toda equations
\begin{equation}\label{IGeneqToda}
u_{xy}=\exp(Ku)
\end{equation}
was described in the
paper~\cite{LeznovSaveliev}. By~\cite{LeznovSaveliev},
the acceleration $\dd^2/\dd\tau^2$ with respect to the time
$t\in\BBR$ is replaced by the d'Alambertian~$\dd^2/\dd x\dd y$;
we also assume~$(x$, $y)\in\BBC^2$.
The Toda fields $u^j(x$, $y)$ are counted by the superscript
$j\in[1$, $r]$.
Suppose $\langle\vec\alpha_i\rangle$ is the rood system of a semisimple
complex Lie algebra $\gothg$ of rank $r$ and
\[
K=\left\|k_{ij}=\frac{2(\vec\alpha_i,\vec\alpha_j)}{|\vec\alpha_j|^2}\right\|
\]
is its Cartan matrix. In~\cite{LeznovSaveliev},
equations~\eqref{IGeneqToda} were supposed to be associated with the
Cartan matrices by default.
In the sequel, we extend the existing picture and
analyse the properties of the Toda equations~\eqref{IGeneqToda}
such that $K$ is a nondegenerate symmetrizable matrix (see precise
definitions below).

Let the matrix $K$ in Eq.~\eqref{IGeneqToda} be the Cartan matrix
of the Lie algebra of type~$A_r$. Then we set
\[
u^j(x, y)=u(x, y, {z)\bigr|}_{z=j\varepsilon}
\]
for each $1\leq j\leq r$.
The continuous limit of Eq.~~\eqref{IGeneqToda} as
$r\to\infty$ and $\varepsilon\to+0$ is the dispersionless Toda equation
\begin{equation}\label{IOurHeavenly}
u_{xy}=\exp(-u_{zz}).
\end{equation}
This equation appeares in many problems of mathematical physics, for
example, in the gravity theory (\cite{BoyerFinley}, see also the
paper~\cite{SavelievTMPh} and references therein).

In this paper, we investigate algebraic and geometric properties of the
Toda equations~(\ref{IGeneqToda}--\ref{IOurHeavenly}) and the nonlinear
Schr\"odinger equation~\eqref{INLS}. Also, we describe the relation
between the Toda equation~\eqref{IGeneqToda} and the hierarchies of the
Korteweg\/--\/de Vries equations~\eqref{IKdV} and~\eqref{IpKdV} (see
below).
We apply modern cohomological methods and algorights in our study of
the geometric properties of all these equations.
By using the algebraic approach, we reject the idea of a
decisive break\/--\/through and the inevitably
immense calculations in coordinates.
We operate with the notions of homological algebra in
the category of infinite prolongations of differential equations instead.

The results exposed in this article
are also found in the sequence of publications
\cite{Thesis}--\cite{KisOvchTodaHam}.

The paper is organized as follows.
In the introduction, we fix notation, formulate necessary definitions,
and give a brief list of the final results.

In Chapter~1, we describe important properties of the symmetry algebra as
well as conservation laws for the Toda equations.
In Sec.~\ref{SecTodaEq}, we review some known properties of the
hyperbolic Toda equations associated with the semisimple Lie algebras.
Also, we define the Toda equations associated with nondegenerate
symmetrizable matrices. In what follows, we study these equations that
require less restrictions upon the matrix~$K$.
In Sec.~\ref{SecTodaNoeth} within Chapter~1, we obtain the Noether
symmetries of the Lagrangian for the Toda equations.
In Sec.~\ref{SecTodaRec}, we construct a continuum of recursion
operators for the Toda equation.

In Chapter~2, we construct the commutative hierarchy $\gA$ of analogs for
the potential modified Kortweg\/--\/de Vries equation. This hierarchy
is identified with a commutative subalgebra of the Noether symmetries
algebra for the Toda equations.
Also, some aspects of the Hamiltonian formalism for the hyperbolic Toda
equations themselves are discussed. Then we relate the symmetry
hierarchy $\gA$ with the higher Korteweg\/--\/de Vries equations.

In Chapter~3, we illustrate the methods of geometry of partial differential
equations. Namely, we investigate the properties of the dispersionless
Toda equation and the multi\/--\/component analog of the nonlinear
Schr\"odinger equation, which is related with the former equation.

In Chapter~4, we consider B\"acklund transformations for
the Toda equation associated with the algebra
$\mathfrak{sl}_2(\BBC)$ (the Liouville equation) and related systems.
Also, we obtain one\/--\/parametric families of these transformations.
Examples of integrating B\"acklund transformations in nonlocal
variables are given in Chapter~4, zero\/--\/curvature representations are
pointed out, and the relations between all these structures are
analysed.

In Appendix~\ref{SecAppendix}, several geometrical methods of solving the
boundary problems for equations of the mathematical physics are listed.

\subsection*{Basic definitions and notation}
First let us introduce some notions from the geometry
of partial differential equations. We follow~\cite{Arnold, ArnoldGeom,
ClassSym, JKKersten, Opava, Vin84} and the
articles~\cite{Vestnik2000, DeformLiou}.

\subsubsection*{Differential equations and their symmetries}
Let $\cE=\{F^\alpha($\textbf{\textit{x}},
\textbf{\textit{u}}, $\mathbf{p})=0$,
$1\leq\alpha\leq r\}$ be a partial
differential equation of order $k$
imposed by the smooth functions $F^\alpha$
on $n$ independent variables
$\vec{x}={}^t(x^1$,
$\ldots$, $x^n)$, $r$ dependent variables
$\bu={}^t(u^1$, $\ldots$, $u^r)$, and the derivatives
\begin{multline*}
{\bf p}=\{p^j_\sigma\mid p^j_\sigma=\dd^{|\sigma|}u^j/\dd
(x^1)^{i_1}\dots\dd(x^n)^{i_n},\\
\sigma=\{i_1,\dots,i_n\},\quad
|\sigma|=i_1+\dots+i_n\leq k\}.
\end{multline*}

Consider the trivial $r$-dimensional smooth fibre bundle
\[\pi\colon\BBR^r\times\BBR^n\to \BBR^n.\]
The space $J^k(\pi)$ of $k$th jets of the bundle $\pi$
is the union $\bigcup_xJ^k_x$, where
$J^k_x$ is the set of all equivalence classes
$[s]^k_x$ of sections $s$ in the bundle
$\pi$ such that the sections $s$ are tangent with order
$k$ at the point $x\in\BBR^n$.
Define the sequence of smooth fibre bundles
\[\pi_{k+1,k}\colon J^{k+1}(\pi)\to J^k(\pi)\]
by the formula
\[\pi_{k+1,k}\bigl({[s]}^{k+1}_x\bigr)={[s]}_x^k.\]
We denote by $\pi_k$ the smooth fibre bundle
$\pi_k\colon J^k(\pi)\to\BBR^n$ defined by
$\pi_k\bigl({[s]}^{k}_x\bigr)=x$.
Also, by $\cF_k(\pi)$ we denote the ring of smooth ($C^\infty$)
functions on $J^k(\pi)$. Finally, we set
$\cF_{-\infty}(\pi)=C^\infty(\BBR^n)$.

We treat the variables \textbf{\textit{x}},
\textbf{\textit{u}}, and ${\bf p}$ as coordinates in the jet space
$J^k(\pi)$. By the Hadamars lemma~(\cite{JetNestruev}),
two sections $s$, $s'\in\Gamma(\pi)$
are equivalent in $J_x^k(\pi)$ iff their partial derivatives at the
point $x$ coincide up to order~$k$.

A submanifold \[\cE=\{F^\alpha=0\}\subset J^k(\pi)\] is a\label{DefPDE}
\emph{differential equation} of order not greater than $k$
imposed on $r$ functions of $n$ independent variables.
An equation $\cE\subset J^k(\pi)$ is \emph{regular}\label{DefRegular}
if the mapping
\[{\pi_k\bigr|}_{\cE}\colon\cE\to M=\BBR^n\] is a surjection.

\begin{define}\label{DefCartanDistribution}
Consider the sections $s$ of the fiber bundle $\pi$ such that
$[s]_x^k=\theta$ for some point $\theta\in J^k(\pi)$. Consider
the tangent planes to the graphs $\Gamma_s^k$ of the
$k$th jets of~$s$.
The linear span of all these tangent planes is the
\emph{Cartan plane} $C_\theta=C_\theta^k$ at the point~$\theta$.
The union $\cC$ of the mappings $\theta\mapsto C_\theta$
by all $\theta\in J^k(\pi)$ is the \emph{Cartan distribution}.
\end{define}

\begin{define}
The subset
\begin{multline*}
\cE^{(l)}=\{\theta_{k+l}=[s]_x^{k+l}, s\in\Gamma(\pi)\mid{}\\
 {}\mid j_k(s)(x)
\mbox{ is tangent to }\cE\mbox{ at the point }
\theta_k=[s]_x^k\mbox{ with order }
\geq l\}
\end{multline*}
in $J^{k+l}(\pi)$ is the $l$th \emph{prolongation} of an equation
$\cE\subset J^k(\pi)$. The inverse limit
\[\cE^\infty=\proj\,\lim_{l\to\infty}\cE^{(l)}\] with respect to the
projections \[\pi_{l+1,l}\colon J^{l+1}(\pi)\to J^l(\pi)\]
is called the \emph{infinite prolongation}
of the equation $\cE$.
In what follows, we omit the superscript $\infty$ if it is clear from the
context that the infinite prolongation $\cE^\infty$ but not the
differential equation $\cE$ is considered.

The infinite prolongation of the empty equation
$\{0=0\}\simeq J^0(\pi)$ is called
the \emph{infinite jets space}~$J^\infty(\pi)$.
The projections
\begin{align*}
\pi_{\infty,k}&\colon J^\infty(\pi)\to J^k(\pi)\\
\intertext{and}
\pi_\infty&\colon J^\infty(\pi)\to\BBR^n
\end{align*}
are defined by the formulas
\begin{align*}
\pi_{\infty,k}(\theta_\infty)&=\theta_k\\
\intertext{and}
\pi_\infty(\theta_\infty)&=x,
\end{align*}
respectively,
where
\[
\theta_\infty=\{x, \theta_k\in
J^k(\pi)\mid k\in\BBN\}\in J^\infty(\pi).
\]
The algebra $\cF(\pi)$ of smooth functions on $J^\infty(\pi)$
is defined by using the following procedure. We set
\[
\cF(\pi)=\bigcup_k\cF_k(\pi), \quad k\in\{-\infty\}\cup\BBN.
\]
The module $\Lambda^i(\pi)$ of differential $i$-forms
on $J^\infty(\pi)$ is defined by the relation
\[
\Lambda^i(\pi)=\bigoplus_{k\geq0}\Lambda^i(J^k(\pi)).
\]
\end{define}

The restriction $\cC_\cE$ of the Cartan distribution $\cC$ onto
the equation $\cE^\infty$ is an $n$-dimensional Frobenius distribution.
This distribution defines the decomposition of the tangent space to
$\cE^\infty$ into the horizontal and vertial subspaces.
The horizontal component is generated by the restrictions
of the total derivatives
\[
D_i=\widehat{\frac{\dd}{\dd x^i}}\]
onto $\cE^\infty$.
These restrictions are denoted by~$\bar D_i$.
The dual description of the Cartan distribution $\cC_\cE$ in the language
of differential forms is also useful. The de~Rham differential on
$\cE^\infty$ is representable as the restriction onto $\cE^\infty$ of the
sum composed by the horizontal differential
\[
\Id_h=\sum_{i=1}^n \Id x^i\otimes D_i
\]
(this is the lifting of the differential on the base of the fibre
bundle~$\pi$) and the Cartan differential
\[\Id_\cC=\Id-\Id_h. \]
Therefore the space $\Lambda^l(\cE)$
of differential $l$-forms on the equation $\cE^\infty$
is the direct sum
\[
\Lambda^l(\cE)=\bigoplus_{i+j=l}
\bar\Lambda^i(\cE)\otimes\cC^j\Lambda(\cE)
\]
of the horizontal $i$-forms $\bar\Lambda^i(\cE)$ and
the Cartan $j$-forms~$\cC^j\Lambda(\cE)$.
The Cartan forms $\omega^j_\sigma\equiv\Id_\cC(u^j_\sigma)$
constitute a basis in~$\cC^1\Lambda(\cE)$. Here
$u^j_\sigma$ are coordinates on~$\cE^\infty$.

The horizontal differential $\bar\Id_h$ generates the
horizontal de~Rham complex
\[
0\to\cF(\pi)\xrightarrow{\Id_h}\bar\Lambda^1(\pi)
\xrightarrow{\Id_h}\ldots\xrightarrow{\Id_h}
\bar\Lambda^n(\pi)\to0
\]
of the space $J^\infty(\pi)$. The cohomologies of this complex are called
the horizontal cohomologies and are denoted by~$\bar H^i(\pi)$.
We denote by~$\bar H^i(\cE)$ the cohomologies of
the restriction of this complex onto the equation~$\cE^\infty$.
From the definition it follows that the equivalence classes
$[\eta]\in\bar H^{n-1}(\cE)$ are conservation laws for the
equation~$\cE$.

\begin{define}
\begin{enumerate}
\item The \emph{evolutionary derivation} is the operator
\[
\cEv_\vph=\sum_{j,\sigma}D_\sigma(\vph^j)\,{\dd}/{\dd
p^j_\sigma},
\]
where $\vph^j\in C^\infty(J^k(\pi))$ for some $k$ and
$D_\sigma$ is the composition of total derivatives
$D_i$ that corresponds to the multiindex~$\sigma$.
We denote by $\varkappa$\label{DefKappa}
the $\cF(\pi)$-module
$\Gamma(\pi)\otimes_{\cF_{-\infty}}\cF(\pi)$.
Also, we denote
\[\hat\varkappa=\Hom_{\cF(\pi)}(\varkappa, \bar\Lambda^n(\pi)).\]
\item \label{UnivLinear}%
Suppose $\Delta_\psi$ is the nonlinear differential operator
determined by a function $\psi\in C^\infty(J^k(\pi))$.
The operator $\ell_\psi$ that acts by the rule
\[
\ell_\psi(\vph)=\cEv_\vph(\psi)
\]
is called the \emph{universal
linearization} operator of $\Delta_\psi$.
In coordinates, we have
\[
\ell_\psi=\left\|\sum_\sigma \frac{\dd \psi^i}{\dd u^j_\sigma}\cdot D_\sigma
\cdot\bun_{ij}\right\|.
\]
\end{enumerate}
\end{define}

Any Lie field $X$ that preserves the Cartan distribution $\cC$
can be decomposed to the sum
$X=\cEv_\vph+Y$, where $Y\in\cC$ and $\cEv_\vph$
is the evolutionary vector field.
Any infinitesimal transformation of the space
$J^0(\pi)$ can be extended up to the Lie field.
In coordinates, the lifting rules are the following.
The field
\[
\hat X=\sum_ia_i\,D_i+\sum_{i,j}\cEv_{b_j-a_ip_i^j}
\]
is assigned to the field
\[
X^0=\sum_ia_i\,\frac{\dd}{\dd x^i}+\sum_jb_j\,\frac{\dd}{\dd u^j}.
\]
An example is found in Eq.~\eqref{IShadowLift} on
page~\pageref{IShadowLift}.

\begin{define}\label{DefSymmetry}
A \emph{symmetry} of an equation $\cE^\infty$ is a
$\pi$\/-\/vertical vector
field $X$ such that $X$ preserves the Cartan distribution
$\cC_\cE=\bigcup_{\theta\in\cE^\infty}C_\theta$:
\[ [X,\cC_\cE]\subset\cC_\cE.\]
\end{define}

\begin{theor}[\textup{\cite{ClassSym}}]\label{ThLieSym}
Suppose $\cE\subset J^k(\pi)$ is an equation
\[
\{F^1=0, \dots, F^r=0\},
\]
such that
\[
\pi_{\infty,0}(\cE^\infty)=J^0(\pi).
\]
Then the symmetry Lie
algebra $\sym\cE^\infty$ is isomorphic to the Lie algebra
of solutions to the system of the determining equations
\[\cEv_\vph(F)=0\quad\text{on $\cE$}\]
or, equivalently,
of the equations
\[\ell_F(\vph)=0\quad\text{on $\cE$}\]
owing to the definition of the linearization operator~$\ell$.
Here $\vph\in{\varkappa\bigr|}_{\cE^\infty}$\textup{.}
The bracket
$$\{\vph,\psi\}_{\cE^\infty}
=\left(\cEv_\vph(\psi)-\cEv_\psi(\vph)\right)\bigr
|_{\cE^\infty}.
$$
endowes the algebra $\sym\cE^\infty$ of solutions $\vph$
with the Lie algebra structure.
\end{theor}

In what follows, we identify the generating sections
\[
\vph={}^t(\vph^1, \ldots, \vph^r)
\]
of symmetries
\[
\cEv_\vph=\sum_{i,\sigma}\bar D_\sigma(\vph^i)\cdot\frac{\dd}{\dd
u^i_\sigma}
\]
(we denote $u^i_\sigma\equiv D_\sigma(u^i)$)
of differential equations with the symmetries $\cEv_\vph$ themselves
(see~\cite{DSViniti84, Dirichlet}).
Roughly speaking, the component $\vph^i$
of a generating section $\vph$
measure the velocity $u^i_t$ of the dependent variable $u^i$
along the "integral trajectories'' of the field~$\cEv_\vph$.

\begin{define}\label{DefInvarSolution}
Suppose $\varphi(u$, $\ldots$, $u_\sigma)$ is a symmetry
of a differential equation ${\cE}$,
where $\sigma$ is a multiindex and
\[u_\sigma=\partial^{|\sigma|}u/\partial{(x^1)}^{\sigma_1}
\dots\partial {(x^n)}^{\sigma_n}\]
is the derivative of the dependent variable $u$.
Assume that the symmetry $\varphi$ has
the flow \[A_\tau\colon u(x, 0)\mapsto u(x, \tau).\]
This flow is defined on solutions of the equation $u_\tau=\varphi$
and maps the solutions $s(x)={u(x,\tau)|}_{\tau=0}$ of equation ${\cE}$ to
solutions of the same equation at points~$\tau>0$.
A solution $s(x)$ of the equation ${\cE}$ is called
$\varphi$-\emph{invariant} if $s(x)$ is a stationary solution of the
evolution equation \[u_\tau=\varphi(u, \ldots, u_\sigma).\]
%
Therefore, the search of the $\vph$-invariant solutions for a given
equation $\cE=\{F=0\}$ is reduced to solving the system
$\{F=0$, $\vph=0\}$.
\end{define}

%
No we specify an important class of \emph{Hamiltonian} equations
within the set of evolution equations
$\bu_t=f(t$, $\bx$, $\bu$, $\mathbf{p})$.
Here we formulate the definitions of the Poisson bracket and the
Hamiltonian operators class (see~\cite{ClassSym, GelfandDorfman,
DIPS6-2002} for further detailes).

\begin{define}\label{DefPoissonBracket}
Let $A\in\CDiff(\hat\varkappa(\pi),\varkappa(\pi))$ be an $(r\times
r)$-matrix operator in total derivatives:
\[A=\|A^{ij}\|,\qquad
A^{ij}=A^{ij}_\sigma\cdot D_\sigma,
\]
and let $\cL_1$, $\cL_2\in\bar
H^n(\pi)$ be two Lagrangians. By definition, set the Poisson bracket
(the variational bracket) on $\bar H^n(\pi)$ by the formula
\begin{equation}\label{PBracket}
\{\cL_1,\cL_2\}_A=\bigl\langle \bE(\cL_1),A(\bE(\cL_2))\bigr\rangle =
\Bigl[\sum_{i,j}\frac{\delta\cL_1}{\delta u^i}\cdot
A^{ij}\left(\frac{\delta\cL_2}{\delta u^j}\right)
\Id\text{\textbf{\textit{x}}}\Bigr],
\end{equation}
where $\langle\cdot,\cdot\rangle$ is the natural coupling
$\varkappa(\pi)\times\hat\varkappa(\pi)\to\bar H^n(\pi)$
and $[{\cdot}]$ denotes the equivalence class of differential forms.
\end{define}

\begin{define}\label{DefHamOperator}
Suppose the operator $A$ is subject to the assumptions
of the previous definition. Then the
operator $A$ is called \emph{Hamiltonian} if the Poisson bracket
defined in Eq.~\eqref{PBracket} endows $\bar H^n(\pi)$ with the Lie
algebra structure over $\BBR$ such that the followin relations hold
\begin{subequations}
\begin{gather}
\{\cL_1,\cL_2\}_A+\{\cL_2,\cL_1\}_A=0,\label{SkewLie}\\
\{\{\cL_1,\cL_2\}_A,\cL_3\}_A+\{\{\cL_2,\cL_3\}_A,\cL_1\}_A+
\{\{\cL_3,\cL_1\}_A,\cL_2\}_A=0.\label{Jacobi}
\end{gather}
\end{subequations}
The bracket $\{\cdot,\cdot\}_A$ is the \emph{Hamiltonian structure}.

Two Hamiltionian operators $A_1$, $A_2$ are \emph{compatible} if
$\lambda A_1+\mu A_2$ is a Hamiltonian operator again for any
$\lambda$, $\mu\in\BBR$.
\end{define}

Condition~\eqref{SkewLie} is  satisfied iff $A+A^*=0$. Refer
\cite{ClassSym, Opava, JKKersten, DIPS6-2002}
for the suitable criteria%
\footnote{The\label{FootSchouten}
of notion of the \emph{Poisson bi\/-\/vectors}
is a useful intrument in the Hamiltonian operators theory
(see \cite{GelfandDorfman, DIPS6-2002}).   
For example, the first Hamiltonian structure for the Korteweg\/--\/de
Vries equation~\eqref{IKdV} is $\bun\wedge D_x$ and the second structure
is $\bun\wedge\hat B_2$, see Eq.~\eqref{IKdVHam}.
A bi\/-\/vector $A$ is Poisson if and only if $A$
satisfies the equation $[\![A, A]\!]=0$, where
$[\![{\cdot}, {\cdot}]\!]$ is the Schouten bracket.
A pait of the Poisson bi\/-\/vectors $A_1$ and $A_2$ is compatible
if $[\![{A_1}, {A_2}]\!]=0$. In the papers~\cite{ForKac,
NoteDiMatematica}, the equation $[\![A, A]\!]=0$ was considered in a more
general situation such that $A$ is not necessarily a bi\/-\/vector.} %
which check whether
Jacobi's identity \eqref{Jacobi} holds for a given operator $A$.
For example, any
skew\/--\/symmetric $\cC$-differential operator $A$ with
constant coefficients is Hamiltonian.

\begin{define}\label{HamEquation}
The evolution equation
\begin{equation}\label{HamEv}
\bu_t=A(\bE_{\bu}(\cH))
\end{equation}
is a \emph{Hamiltonian evolution equation} assigned to the Hamiltonian
$\cH\in\bar H^n(\pi)$ by the Hamiltonian operator~$A$.
\end{define}

\begin{example}[\textup{\cite{Magri}}]
The Korteweg\/--\/de Vries equation
\begin{equation}\label{IKdV}
T_t=-\beta\,T_{xxx}+3T\,T_x,\qquad\beta=\const,
\end{equation}
is hamiltonian with respect to\ the pair
\begin{equation}\label{IKdVHam}
\hat B_1=D_x,\qquad \hat B_2=-\beta\,D_x^3+2T\cdot D_x+T_x
\end{equation}
of the compatible Hamiltonian operators.
Indeed, we have
\[
T_t=\hat B_1\circ\bE_T\bigl((\tfrac{1}{2}\beta
T_x^2+\tfrac{1}{2}T^3)\,\Id x\bigr)=
\hat B_2\circ\bE_T\bigl(\tfrac{1}{2}T^2\,\Id x\bigr).
\]
\end{example}

\subsubsection*{Conservation laws}\label{SecCL}
In this subsection, we formulate important definitions and statements
concerning conservation laws for differential equations. Also, we recall
the relation between symmetries and conservation laws for the Euler
equations.

\begin{define}
A \emph{conservation law}
\begin{multline*}
[\eta]\in\bar H^{n-1}(\cE)\equiv
\{\omega\in\bar\Lambda^{n-1}(\cE) \mid \bar\Id_h(\omega)=0\}/{}\\
{}/\,\{\omega\in\bar\Lambda^{n-1}(\cE)$\,$\mid$\,$\omega=\bar\Id_h
\gamma,\ \gamma\in\bar\Lambda^{n-2}(\cE)\}
\end{multline*}
for an equation $\cE$ is an equivalence class of horizontal
$(n-1)$-forms $\eta\in\bar\Lambda^{n-1}(\cE)$ closed on~$\cE$,
\[\bar\Id_h\eta=\nabla(F)\,\Id\bx.\] Here
\[\bar\Id_h=\sum_{i=1}^n \Id x^i\otimes\bar D_i\] is the restriction
of the horizontal differential $\Id_h$ on $\cE$,
$D_i$ is the total derivative with respect to\ $x^i$,
$\nabla$ is an operator in total derivatives,
and the equivalence is the factorization by exact forms
$\bar \Id_h\gamma$, $\gamma\in\bar\Lambda^{n-2}(\cE)$.
Representatives $\eta$ of the equivalence classes $[\eta]\in\bar
H^{n-1}(\cE)$ are called \emph{conserved currents} for
the equation~$\cE$.
\end{define}

\begin{example}[\textup{\cite{HeavLaws}}]
The horizontal $2$-form
\[
\eta=u_{xz}\exp(-u_{zz})\,\Id x\wedge\Id y
+ \bigl(\tfrac{1}{2}u_{xz}^2-u_{xx}\bigr)\,\Id x\wedge\Id z
\]
is a conserved current for the dispersionless Toda equation
\[
u_{xy}=\exp(-u_{zz}).
\]
Indeed, the relation
\begin{equation}\label{EMTHeav}
\bar D_y\bigl(\tfrac{1}{2}u_{xz}^2-u_{xx}\bigr) =
\bar D_z\bigl(u_{xz}\,\exp(-u_{zz})\bigr)
\end{equation}
holds on that equation.
\end{example}

\begin{define}
A regular equation $\cE=\{F=0\}$ is called
$\ell$-\emph{normal} if the
condition
\[\nabla\circ\bar\ell_F=0\] implies~$\nabla=0$.
\end{define}

\begin{state}[\cite{ClassSym}]
Let $n$ be the number of independent variables $x^1$, $\ldots$, $x^n$
and let $\cE$ be a regular equation. Consider a coordinate neighbouhood
$\Omega(\theta^\infty)\subset\cE^\infty$ of a point
$\theta^\infty\in\cE^\infty$ and suppose that there is a set
$\{v\}$of the internal coordinates $v$ on $\Omega$ such that the total
derivatives $D_i(v)$ can be expressed in terms of these coordinates
$\{v\}$ for any $i$, $1\leq i<n$. Then the equation $\cE$ is
$\ell$-normal.
\end{state}

If $\cE$ is an $\ell$-normal equation, then the compatibility complex
(\cite{ClassSym, Opava}) for the equation $\cE$ has length $2$, and
the equation satisfies the assumptions of the '$2$-line' theorem.

\begin{rem}
The Maxwell, the Yang\/--\/Mills, and the Einstein equations are
\emph{not} $\ell$-normal since there is a nontrivial dependence
between the equations that is provided by the gauge symmetry
pseudogroup.
\end{rem}

Now we need a test that checks whether a given equation is
$\ell$-normal. We have

\begin{example}[\textup{\cite{ClassSym, Opava}}]\label{EvolutAreNormal}
The evolution equations are $\ell$-normal.
\end{example}

Conservation laws $[\eta]$ for the $\ell$-normal equations
are described by their \emph{generating sections}
$\vpsi\equiv\nabla^*(1)\in\hat\varkappa$.
We see that
\[
\Id_h\eta=\langle\nabla(F), 1\rangle=\langle F,
\nabla^*(1)\rangle+\Id_h\gamma.
\]
Then the coupling
\begin{equation}\label{CouplingIsExact}
\langle\nabla^*(1), F\rangle=\Id_h(\eta-\gamma)
\end{equation}
is an exact horizontal $n$-form. Suppose $\eta$ is trivial and
therefore $\nabla=0$. Then, obviously, then $\vpsi=0$.

\begin{lemma}[\textup{\cite{Vin84}}]\label{PsiZeroEtaZero}
Let $\cE=\{F=0\}$ be an $\ell$-normal equation and assume that
$H^n(\cE)\subset\bar H^n(\cE)$
\textup{(}\textit{e.g.}, $H^n(\cE)=0$\textup{).}
Suppose that the generating section $\vpsi$ of a current $\eta$
is zero. Then the conserved current $\eta$ is trivial.
\end{lemma}

In what follows, we shall use the following remarkable theorem
many times.

\begin{theor}[\cite{Vin84}]
Let $\cE=\{F=0\}$ be an $\ell$-normal equation in the fibre bundle
$\pi\colon\BBR^m\times\BBR^n\to\BBR^n$.
Then the generating sections $\vpsi$ of conservation laws $[\eta]$
satisfy the equation
\begin{equation}\label{DetEqOnPsi}
\bar\ell_{F}^*(\vpsi_\eta)=0,
\end{equation}
where $\ell_F^*$ is the operator formally adjoint to $\ell_F$
and $\bar\ell_F^*$ is the restriction of $\ell_F^*$
onto the equation~$\cE$.
\end{theor}

\begin{proof}
Apply the Euler operator $\bE$ to both sides of
Eq.~\eqref{CouplingIsExact}. Thence we obtain
\begin{equation}\label{PreDetEq}
0=\bE(\langle\psi,F\rangle)=\ell_{\langle\psi,F\rangle}^*(1)=
\ell^*_F(\psi)+\ell^*_\psi(F)=0
\end{equation}
by the Leibnitz rule. Now restrict Eq.~\eqref{PreDetEq} onto the
prolongation of $F=0$ and obtain the determining equation
\eqref{DetEqOnPsi} imposed on the generating sections and satisfied on
the initial equation~$\cE$.
\end{proof}

The generating sections $\psi_\eta$ are often named \emph{gradients}
of the conservation laws $[\eta]$ (see~\cite{DSViniti84}) by the
following reason. The generating sections for evolution equations lie in
the image of the Euler operator (that is, the
"gradient" $\delta/\delta u$) applied to the corresponding conserved
density.

\begin{lemma}[\cite{Opava, JPW}]\label{OpavaGenf}
Let $\cE=\{u_t=f(t,x,u,u_1,\ldots)\}$ be an evolution equation.
Assume that
\[
\eta=\eta_0\,d\text{\textbf{\textit{x}}}
+\sum_{i=1}^n(-1)^{i-1} \eta_i\,dt\wedge dx^1
\wedge\dots\wedge\widehat{dx^i}\wedge\dots\wedge dx^n
\]
is a conserved current for $\cE$\textup{:}
$\bar d_h(\eta)=0$, that is,
\[
\bar D_t(\eta_0)+\sum_i\bar D_{x^i}(\eta_i)=0.
\]
Then its generating function is
\[
\psi_\eta=\bE(\eta_0)\equiv\ell^\ast_{\eta_0}(1).
\]
\end{lemma}

The proof of Lemma~\ref{OpavaGenf} is straightforward.

%
The Euler--Lagrange equation $\cE_{E-L}$ assigned to a Lagrangian
$\cL\in\bar H^n(\pi)$ is
\begin{equation}\label{ELE}
\cE_{E-L}=\{{G}\equiv\bE_{\bu}(\cL)=0\}.
\end{equation}
We recall that the Helmholz condition $\ell_{\mathstrut G}=\ell_G^*$ is
valid for the image $G$ of the Euler operator~$\bE$.

The correlation between conservation laws $[\eta]$, their generating
sections $\psi_\eta$, and the Noether symmetries $\vph_\cL\in\sym\cL$
of an Euler--Lagrange equation $\cE=\{\bE(\cL=0\}$ is guided by
the following version of the Noether theorem.

\begin{theor}[\textup{\cite{Mystique}}]\label{InverseNoether}
Let $\cE=\{\bE(\cL)=0\}$ be the
Euler-Lagrange equation for a Lagrangian
$\cL$. Then the
evolutionary derivation $\cEv_\vph$ is a Noether symmetry of the
Lagrangian $\cL$\textup{:}
\[\cEv_\vph(\cL)=0\] if and only if
$\vph$ is the generating section of a conservation law
$[\eta]$\textup{:} $\Id_h\eta=0$ on~$\cE$.
\end{theor}

\begin{proof}
Suppose $\vph$ is the generating section of a conservation laws
$[\eta]$ for the equation~$\cE$:
$$
\Id_h(\eta)=
\langle1,\square(F)\rangle=\langle\square^*(1),F\rangle+
\Id_h(\gamma),
$$
where the coupling $\langle{\cdot},{\cdot\rangle}$
takes values in the space $\bar\Lambda^n(\pi)$ of horizontal
$n$-forms. Thence,
$\langle\vph,F\rangle=\Id_h(\eta-\gamma)$
is an exact horizontal $n$-form.
We have $F=\bE(\cL)=\ell_L^*(1)$ by the initial assumption,
whence we easily obtain
\[ \langle
\vph,F\rangle=\langle\vph,\ell_L^*(1)\rangle=
\langle\ell_L(\vph),1\rangle+\Id_h\beta=
\langle\cEv_\vph(L),1\rangle+\Id_h\beta.
\]
Therefore, $\vph$ is a Noether symmetry.
The second implication in the theorem's statement,
the sufficiency, is proved by inverting the reasonings.
\end{proof}

\subsubsection*{Coverings}
Let $\cE$ be a differential equation in the fibre bundle
$\BBR^{n}\times\BBR^{m}\to\BBR^n$
and $\cE^\infty$ be the infinite prolongation of $\cE$.
The $n$\/-\/dimensional
Cartan plane $\cC_\theta\subset T_\theta(\cE^\infty)$
is defined at each point $\theta\in\cE^\infty$.
The Cartan distribution $\cC_\cE$ on $\cE^\infty$
is Frobenius:
\[[\cC_\cE, \cC_\cE]\subset\cC_\cE.\]
In local coordinates, $\cC_\cE$ is defined by $n$ vector fields
$\bar D_1$, $\dots$, $\bar D_n$, where $\bar D_i$ is the
restriction of the total derivative with respect to\ the $i$th independent
variable $x^i$ onto $\cE^\infty$.

\begin{define}[\textup{\cite{ClassSym}}]
The equation $\tilde\cE^\infty$ with
the $n$-dimensional Cartan distribution
$\tilde\cC$ and the regular map $\tau$ are called a
\emph{covering} over the equation $\cE^\infty$ if
at any point $\theta\in\tilde\cE^\infty$ the tangent map
$\tau_{\ast,\theta}$ is an isomorphim of the plane
$\tilde\cC_\theta$ onto the Cartan plane
$\cC_{\tau(\theta)}$ on the equation $\cE^\infty$ at the point
$\tau(\theta)$.
The equation $\tilde\cE$ is called the \emph{covering equation}.
The dimension of the fibre in the bundle $\tau$ is the
\emph{dimension of the covering}.
\end{define}

In coordinates, the structure of a covering is realized in the
following way. Locally, we can realize the manifold
$\tilde\cE$ and the mapping $\tau\colon\tilde\cE\to\cE^\infty$
as the direct product $\cE^\infty\times W$
($W\subseteq\BBR^N$ is an open subset, $0<N\leq\infty$) and
the natural projection $\cE^\infty\times W\to\cE^\infty$, respectively.
Then we describe the distribution
$\tilde\cC$ on $\tilde\cE$ by the vector fields
$$
\tilde D_i=\bar D_i+\sum\nolimits_{j=1}^N X_{ij}\,\frac{\dd}{\dd s^j},
\qquad i=1,\dots,n,
$$
where $X_{ij}\in{C^\infty}(\tilde\cE)$ are the coefficients of the
$\tau$-vertical fields on $\tilde\cE$ and
$s_1$, $\dots$, and $s_N$ are the Cartesian coordinates in
$\BBR^N$. Then, the Frobenius condition
$[\tilde\cC,\tilde\cC]\subset\tilde\cC$ of
integrability for the distribution $\tilde\cC$
is equivalent to the system of relations
\[
[\tilde D_i,\tilde D_j]=0, \qquad i, j=1,\dots,n,
\]
which are equivalent to the equations
\[
\tilde D_i(X_{jk})=\tilde D_j(X_{ik})
\]
that hold for all $i,j=1,\dots,n$, $0\leq k\leq N$.

The coordinates $s_i$ are called nonlocal variables.
In coordinates $x^i$, $u^j_\sigma$, and $s_j$ the rules
$\tilde D_i(s_j)=X_{ij}$ to differentiate the nonlocal variables
$s_j$ as well as the initial equation $\cE^\infty$
define the covering equation~$\tilde\cE$.

\begin{example}
Again, consider the Korteweg\/--\/de Vries equation~\eqref{IKdV}.
Extend the set of variables $t$, $x$, $T_j\equiv D_x^j(T)$
with the "nonlocality" $s=\int T\,\Id x$:
\begin{equation}\label{IpKdV}
s_x=T,\qquad s_t=-\beta\,s_{xxx}+\tfrac{3}{2}s_x^2.
\end{equation}
We see that the equation that covers~\eqref{IKdV} is the potential
Korteweg\/--\/de Vries equation.
\end{example}

A symmetry of the covering equation $\tilde\cE$ is a
\emph{nonlocal symmetry} of the equation $\cE^\infty$.
Suppose that the field $\hat X$ is a symmetry of an equation
$\cE^\infty$ and $\tau\colon\tilde\cE\to\cE^\infty$ is a covering.
Two radically different cases are possible:
\begin{enumerate}
\item the symmetry $\hat X$ of the equation $\cE^\infty$
can be extended up to a symmetry $\tilde X$ of the covering
equation $\tilde\cE$, and
\item the converse: any lifting
of the symmetry $\hat X$ is not a symmetry of the
covering equation.
\end{enumerate}
In the second case, the field $\hat X$
generates the one\/-\/parametric family of equations $\tilde\cE_t$
that cover the equation $\cE^\infty$ for any $t$.

\subsubsection*{Recursion operators}\label{ISecRec}
In this subsection, we describe an explicit
method (\cite{JKKersten}) of constructing recursion operators for
differential equations. The Cartan generating forms approach developed
by I.~S.~Krasil'shchik is the key-point of our reasonings.

Let $\cE=\{F=0\}$ be a determined system of differential equations,
that is, the number $m$ of dependent variables in this equation is
equal to the number $r$ of equations.
Suppose $\vph\in\sym\cE^\infty$ is a symmetry of this equation
and consider an $r$-component column
\[
\omega={}^t(\omega^1, \ldots, \omega^r)
\]
such that its components
\[\omega^i\in
C^\infty(\cE^\infty)\otimes\cC^1\Lambda(\cE)\]
are the Cartan $1$-forms that vanish on the Cartan
distribution $\cC_\cE$ on $\cE^\infty$ and the coefficients at these
forms are functions on~$\cE^\infty$.
Obviously, the component\/-\/wise action
$\cEv_\vph\inner\omega\equiv\vph'$
is again an element of the module $\varkappa$
of evolutionary derivations.
The condition for $\vph'$ to be a symmetry of the initial equation
$\cE$ is
\begin{equation}\label{DetEqGFRec}
\bar\ell_F^{{}\,[1]}(\omega)=0,
\end{equation}
where $\bar\ell_F^{{}\,[1]}$ is the restriction of the
linearization operator onto~$\cC^1\Lambda(\cE)$.

Now we formulate the rule that assigns differential operators $R$ in
total derivatives to the columns (the \emph{generating forms})
$\omega$. Suppose the components $\omega^i$ of a generating form
$\omega={}^t(\omega^1$, $\ldots$, $\omega^r)$ are
\[
\omega^i=\sum_{j,\sigma} a^{ij}_\sigma\,\omega^j_\sigma.
\]
Then we obtain
\[
\cEv_\vph\inner\omega={}^t\bigl(\ldots,
\sum\nolimits_{j,\sigma} a^{ij}_\sigma\,D_\sigma(\vph^j),\ldots\bigr).
\]
We see that the component\/-\/wise action of the Cartan forms $\omega^i$
is equivalent to the action of the operators
\[
\sum_{j,\sigma} a^{ij}_\sigma\,\ell_{u^j_\sigma}
\]
on the components of the initial generating section
$\vph={}^t(\vph^1$, $\ldots$, $\vph^r)$.
Hence we deduce that the formal differential recursion operator
$R$ is the $(r\times r)$-matrix,
\[
R=\left\|\sum\nolimits_\sigma a^{ij}_\sigma\,D_\sigma\right\|.
\]

Still, we observe that ``usually'' equation~\eqref{DetEqGFRec}
admits only the trivial solution
\[
\omega_\varnothing={}^t(\omega^1_\varnothing, \ldots,
\omega^r_\varnothing)
\]
that corresponds to the identity recursion operator
\[\id\colon\vph\mapsto\vph'=\vph.\]
The explanation is that the recursion operators for known equations of
mathematical physics usually contain summands that involve the negative
powers $D_{x^\alpha}^{-1}$ of the total derivatives $D_\sigma$.

\begin{example}
The recursion operator $R_\KdV=\hat B_2\circ\hat B_1^{-1}$
for the Korteweg\/--\/de Vries equation~\eqref{IKdV}
is equal to
\begin{subequations}\label{IKdVRec}
\begin{align}
R_\KdV&=-\beta\,D_x^2+2T+T_x\cdot D_x^{-1}.\label{IRKdV}\\
\intertext{The recursion operator $R_\pKdV$ for the potential
Korteweg\/--\/de Vries equation~\eqref{IpKdV} is}
R_\pKdV&=-\beta\,D_x^2+2s_x-D_x^{-1}\circ s_{xx}.\label{IRpKdV}
\end{align}
\end{subequations}
\end{example}

From the geometric viewpoint, the situation is like follows.
We obtain nontrivial solutions of Eq.~\eqref{DetEqGFRec}
by extending the set of local coordinates
$\langle x^1$, $\ldots$, $x^n$, $u^j_\sigma\rangle$
by the ``nonlocal'' variables~$s^i$. These nonlocalities are defined by
the compatible rules of their derivation with respect to\ the independent
variables. Also, to each $s^i$ we assign the Cartan form
\[
\Id_\cC(s^i)=\Id s^i-\sum_{j=1}^n s^i_{x^j}\,\Id x^j,
\]
which now belongs to some larger
distribution~$\cC^1\Lambda(\tilde\cE)$.

Consider the restriction $\tilde\ell_F^{{}\,[1]}$
of the linearization operator $\ell_F$ onto the extended
set of the Cartan forms.
Substitute the new linearization operator in
equation~\eqref{DetEqGFRec} and assume that a nontrivial solution
$\omega$ appeared.
Now we see that the generating form~$\omega$ is coupled with some
prolongation
\[
\tilde\cEv_{\vph,A(\vph)}=\cEv_\vph+\sum_i A_i(\vph)\cdot
\frac{\dd}{\dd s^i}+\cdots
\]
of the initial evolutionary field~$\cEv_\vph$.
Possibly, this prolongation involves a larger
(ingeneral, an infinite) set of nonlocal variables,
see the paper~\cite{NGHMatemZametki}.
The result
\[\vph'=\tilde\cEv_{\vph,A(\vph)}\inner\omega\]
of the action of $\omega$ satisfies the equation
\[\tilde\ell_F(\vph)=0.\]
These sections $\vph'(x$, $u^j_\sigma$, $s^i$, $\ldots)$
are called \emph{shadows} of nonlocal symmetries.
In general, there are shadows that cannot be extended up to a true
nonlocal symmetry by using a prescribed set of the nonlocal
variables~$s^i$. An example is found in
Proposition~\ref{CannotLiftState} on page~\pageref{CannotLiftState}.

\begin{example}
Consider the following one\/--\/dimensional covering over
the potential Korteweg\/--\/de Vries equation~\eqref{IpKdV}.
Introduce the nonlocal variable $\zeta$ such that
\begin{equation}\label{IKNpKdV}
\zeta_x=-\tfrac{1}{2}s_1^2,\qquad
\zeta_t=\beta s_1s_3-\tfrac{1}{2}\beta s_2^2-s_1^3,
\end{equation}
where $s_j\equiv D_x^j(s)$ and we similarly denote
$\zeta_j\equiv\tilde D_x^j(\zeta)$.
One easily checks that the covering equation is the
Krichever\/--\/Novikov type equation
\begin{equation}\label{IKN}
\zeta_t=-\beta\,\zeta_3+\tfrac{3}{4}\beta\zeta_2^2\zeta_1^{-1}-
2\sqrt{2}{(-\zeta_1)}^{3/2}.
\end{equation}
Then the equation \[\tilde\ell_\pKdV^{\,[1]}(\omega)=0\] admits a
nontrivial solution in the extended set of variables
$t$, $x$, $s_j$, and $\zeta$. Indeed, the Cartan generating form of the
recursion operator for Eq.~\eqref{IpKdV} is
\begin{equation}\label{IgformpKdV}
\omega_\pKdV=-\beta\,\Id_\cC(s_2)+s_1\,\Id_\cC(s)-\Id_\cC(\zeta).
\end{equation}
\end{example}

Now we describe the relation between recursion operators~$R$
in total derivatives, which are convenient in applications,
and solutions $\omega$ solutions of the determining equation
\[\tilde\ell_F^{{}\,[1]}(\omega)=0.\] It is sufficient to solve this
problem for the Cartan forms~$\Id_\cC(s^i)$. Now it is clear that
the problem is reduced to calculation of the linearization
$\ell_{s^i}$ with respect to\ the nonlocal variable~$s^i$.
To do this, we use the following lemma.

\begin{lemma}\label{LinWRTDLemma}
Let $s\in\cF(\pi)$ be a function\textup{;}
fix a superscript $i\in[1$, $n]$ of the independent variable~$x^i$.
Then the relation
\begin{equation}\label{LinearizeNonlocal}
\ell_{D_i(s)}= D_i\circ\ell_{s}
\end{equation}
holds.
\end{lemma}

\begin{proof}
Suppose $\vph\in\varkappa$, then we obtain
\[
\ell_{D_i(s)}(\vph)=\cEv_\vph\circ D_i(s)=
D_i\circ\cEv_\vph(s)=D_i\circ\ell_s(\vph),
\]
whence follows relation~\eqref{LinearizeNonlocal}.
\end{proof}

Lemma~\ref{LinWRTDLemma} defines the rule
\begin{equation}\label{EqLinNonlocal}
\tilde\ell_{s^j}=\tilde D_i^{-1}\circ\tilde\ell_{\tilde D_i(s^j)}
\end{equation}
of calculating the linearization of a nonlocality~$s^j$.
Here $i\in[1$, $n]$ is arbitrary.
Therefore, to each generating form
$\omega={}^t(\omega^1$, $\ldots$, $\omega^r)$,
where
\[
\omega^i=\sum_{j,\sigma}a^{ij}_\sigma\,\omega^j_\sigma+
\sum_j a^{ij}\,\Id_\cC(s^j),
\]
we assign the matrix operator
\[
R=\Bigl\|\sum\limits_{j,\sigma}
   a^{ij}_\sigma\cdot D_\sigma\cdot\ell_{u^j}+
\sum\limits_j
   a^{ij}\cdot D_{x^k}^{-1}\circ\ell_{D_{x^k}(s^j)}\Bigr\|,
\]
where the index $i$ enumerates the rows and
$1\leq k\leq n$ is arbitrary.

\begin{example}\label{IExampleGenForms}
Recursion operator~\eqref{IRpKdV} corresponds to the
generating Cartan form~\eqref{IgformpKdV}.
Operator~\eqref{IRKdV} corresponds to the Cartan form
\[
\omega_\KdV=-\beta\,\Id_\cC(T_2)+2T\,\Id_\cC(T)+T_1\,\Id_\cC(s).
\]
\end{example}

\subsubsection*{B\"acklund transformations}
The notion of a covering is very useful in description of
B\"acklund transformations between differential equations.

\begin{define}[\textup{\cite{ClassSym}}]
Let $\cE_i\subset J^{k_i}(\pi_i)$, $i=1,2$, be two differential
equations and $\tau_i\colon \tilde\cE\to\cE_i^\infty$
be coverings with the
same total space $\tilde\cE$. Then the diagram
\begin{equation}\label{BacklundDiag}
\cB(\tilde\cE,\tau_i,\cE_i)=\bigl\{
\cE_1\xleftarrow{\phantom{Ml}\tau_1\phantom{Ml}}
\tilde\cE\xrightarrow{\phantom{Ml}\tau_2\phantom{Ml}}\cE_2\bigr\}
\end{equation}
is called a \emph{B\"acklund transformation} $\cB(\tilde\cE,\tau_i,\cE_i)$
between the equations $\cE_i$. Diagram (\ref{BacklundDiag})
is called a \emph{B\"acklund autotransformation} if
$\cE_1^\infty=\cE_2^\infty=\cE^\infty$.
\end{define}

In what follows, any B\"acklund transformation between equations
$\cE$ and $\cE'$ is a system of differential relations imposed
on the unknown
functions $u$ and $u'$ such that the following property holds.
Suppose the function $u$ is a solution to the equation
$\cE$ and both functions $u$ and $u'$ satisfy these relations, then
the function $u'$ is a solution to the equation  $\cE'$,
and \textit{vice versa}.

\begin{example}
B\"acklund autotransformation for the two\/--\/dimensional
La\-p\-l\-ace equation $\Delta_2 v=0$ is defined by the relations
\[
v_y'=v_x'',\ v_x'=-v_y''.
\]
\end{example}

\begin{rem}
Let $\tau_j\colon\tilde\cE_j\to\cE_j^\infty$,
 $j=1$, $2$, be two coverings and
$\mu\colon \tilde\cE_1\to\tilde\cE_2$ be a diffeomorphism that maps the
Cartan distribution $\cC_{\tau_1}\ID(\tilde\cE_1)$ into
$\cC_{\tau_2}\ID(\tilde\cE_2)$. Then the diagram
$\cB(\tilde\cE_1$, $\tau_1$, $\tau_2\circ\mu$, $\cE_j)$
is also a B\"acklund
transformation between the equations $\cE_j$ and these coverings
$\tau_1$ and $\tau_2\circ\mu$ are called \emph{equivalent}.
\end{rem}

\begin{rem}\label{BacklundBySymmetryRem}
Let $\tau\colon\tilde\cE\to\cE^\infty$ be a covering and $\mu\colon
\tilde\cE\to\tilde\cE$ be a nontrivial diffeomorphism of manifolds
that preserves the Cartan distribution,
\textit{e.g.}, a discrete symmetry that cannot be
restricted onto~$\cE^\infty$. Then the diagram
\begin{equation}\label{AlternativeAuto}
\cE\xleftarrow{\phantom{M}\tau\phantom{M}}
\tilde\cE\xrightarrow{\phantom{M}\mu\phantom{M}}\tilde\cE
\xrightarrow{\phantom{M}\tau\phantom{M}}\cE
\end{equation}
is also B\"acklund autotransformation for $\cE$.
In Chapter~4, we apply this contruction to the study
of B\"acklund autotransformation for the Liouville
equation~$u_{xy}=\exp(2u)$.
\end{rem}

\bigskip
Now we give a brief exposition of the mail results
obtained in the present article.

\subsection*{Main results}
In Chapter~1, we consider a  problem standard in the geometry of
differential equations. Namely, we describe relations between
symmetries, conservation laws, the Noether symmetries, and recursion
operators for the Toda equations. The structures obtained in this
chapter are essentially used in what follows.

Let $K=\|k_{ij}$, $1\leq i$, $j\leq r\|$ be a nondegenerate
$(r\times r)$-matrix and $K^{-1}=\|k^{ij}\|$ be its inverse.
Suppose there is the set $\vec{a}$ of real numbers
$\{a_i\not=0$, $1\leq i\leq r\}$ such that the
symmetry condition $\kappa_{ij}=\kappa_{ji}$ holds fot the
matrix $\kappa=\|\kappa_{ij}\|$ whose elements are
\[\kappa_{ij}=a_i\cdot k_{ij}.\] Then we say that the matrix
$K$ is \emph{symmetrizable}.

The hyperbolic Toda equations associated with a nondegenerate
symmetrizable $(r\times r)$-matrix $K$ are
\begin{equation}\label{IeqToda}
\cE_\Toda=\Bigl\{F_i\equiv u^i_{xy}-\exp\bigl(\sum\limits_{j=1}^r
k_{ij}u^j\bigr)=0,
\quad 1\leq i\leq r\Bigr\}.
\end{equation}
In particular, suppose $\gothg$ is a semisimple Lie algebra of rank
$r$, let $\{\alpha_i$, $1\leq i\leq r\}$ be the system of its
simple roots, and
$K=\|k_{ij}=2(\alpha_i,\alpha_j)\cdot|\alpha_j|^{-2}$, $1\leq
i,j\leq r\|$ be the Cartan matrix of the algebra~$\gothg$.
Tne we set $a_i=|\alpha_i|^{-2}$. The Toda equation~\eqref{IeqToda}
associated with this matrix $K$ is said to be~(\cite{LeznovSaveliev})
\emph{associated with the Lie algebra}~$\gothg$.

The Toda equations~\eqref{IeqToda} are the Euler\/--\/Lagrange
equations with respect to the action functional
\[\cL_\Toda=\int L_\Toda\,\Id x\wedge\Id y\]
with the density
$$
L_\Toda=-\tfrac{1}{2}\sum\limits_{i,j=1}^r
\kappa_{ij}u^i_{x}u^j_{y} - \sum\limits_{i=1}^r a_i \cdot
\exp\Bigl(\sum\limits_{j=1}^r k_{ij}u^j\Bigr).
$$
The canonical Hamiltonian structure is known for the Toda equations
(see~\cite{Ovch}).

Suppose $K$ is a nondegenerate symmetrizable matrix.
Then the Toda equation~\eqref{IeqToda} associated with $K$
admits at least one
{integral}~(\cite{SokolovUMN}), that is, an expession dependent of
$u^j_\sigma$ whose total derivative $\bar D_y$ vanishes on this
equation. Namely, we have
\begin{equation}\label{IEMT}
T=\tfrac{1}{2}\sum\limits_{i,j=1}^r \kappa_{ij}u^i_xu^j_x
-\sum\limits_{i=1}^ra_i\cdot{u^i_{xx}}\in\ker\bar D_y.
\end{equation}
By definition, put \[T_j\equiv \bar D_x^j(T).\]
The differential consequences $T_j$ to the functional $T$ generate
the subspace $\bT\subset\ker\bar D_y$ within the kernel of the total
derivative $\bar D_y$. Indeed, any smooth function $Q$ supplies the
functional
$$Q(x,\bT)\equiv Q(x,T,T_1,\ldots,T_\mu)\in\ker\bar D_y.$$
We say that the nondegenerate symmetrizable matrix $K$ is
\emph{generic} if the integral $T$ in Eq.~\eqref{IEMT} is a unique
solution to the equation $\bar D_y(T)=0$ on the corresponding Toda
equation~\eqref{IeqToda}.

Still, by a special choice of the matrix $K$ one can achieve
the situation such that the functional $T$ will \emph{not} be a unique
integral for the Toda equation~\eqref{IeqToda}.
By~\cite{ShabatYamilov},
a necessary and sufficient condition for $r$ nontrivial independent
solutions $\Omega^i$ of the equation \[\bar D_y(\Omega^i)=0\] to exist
is that $K$ be the Cartan matrix of a semisimple Lie algebra~$\gothg$.
The Toda equations associated with $\gothg$ are exactly
integrable~(\cite{LeznovIntegrate, Ovch}).
From now on, by $\mathbf{\Omega}\subseteq\ker\bar D_y$ we denote the
subspace in $\ker\bar D_y$, which is differentially generated by
\emph{all} solutions $\Omega^i$ to the equation $\bar D_y(\Omega^i)=0$.
The total number of these solutions is denoted by $q$: $1\leq
i\leq q\leq r$. We always assume $\Omega^1\equiv T$.

By $\vec\Delta=|\Delta^i|$, where
${\Delta^i=\sum\limits_{j=1}^r k^{ij}}$,
we denote the vector
of the conformal dimensions of the Toda fields
$\exp(u)\equiv {{}^t(}\exp(u^1)$, $\ldots$, $\exp(u^r))$.
The transformation
\begin{equation}\label{IFiniteSymDiffeo}
\begin{aligned}
x&\mapsto\cX(x),\\
y&\mapsto\cY(y),\\
u^i(x,y)&\mapsto\tilde u^i=u^i(\cX,\cY)+\Delta^i\log\cX'(x)\cY'(y)
\end{aligned}
\end{equation}
is a finite conformal symmetry of the Toda
equation~$\cE_\Toda$\textup{.}
The generating section $\vph$ of the infinitesimal form
of conformal transformation~\eqref{IFiniteSymDiffeo} is
\[\vph=\square(\phi(x)),\] where the vector\/--\/valued, first\/--\/order
differential operator $\square$~is
\begin{equation} \label{ISquare}
\square = u_x + \Delta\cdot\bar D_x
\end{equation}
and $\phi$ is an arbitrary smooth function.
The structure of the Lie algebra $\sym\cE_\Toda$ generators
is described in the paper~\cite{MeshkovTMPh}. We have

\begin{enumerate}
\item
Suppose $K$ is a generic nondegenerate $(r\times r)$-matrix
and the operator $\square$ is assigned to this matrix
by using formula~\eqref{ISquare}\textup{.}
Then any symmetry of the Toda equation~\eqref{IeqToda} is
\begin{equation} \label{IsymToda}
\vph=\square(\phi(x,\mathbf{\Omega})),
\end{equation}
where $\phi$ is an arbitrary smooth function
that depends on a set of the integrals
$\mathbf{\Omega}\subset\ker\bar D_y$.
\item
Suppose the matrix $K$ is subject to additional constraints
such that the Toda equation~\eqref{IeqToda} admits $q$
independent integrals $\Omega^i\in\ker\bar D_y$,
where $1\leq i\leq q\leq r$ and
\[
\mathbf{\Omega}=\{\Omega^i_j\equiv\bar
D_x^j(\Omega^i)\}\supset\bT.
\]
We also assume that the first\/--\/order operator
$\square$ is the $(r\times q)$-matrix and satisfies
some additional restrictions \textup{(see~ \cite{MeshkovTMPh})}.
Then symmetries $\vph$ of the Toda equation
are of the form~\eqref{IsymToda} with respect to the new notation.
\end{enumerate}

The Jacobi bracket on the symmetries $\vph\in\sym\cE_\Toda$
induces a bracket on the arguments of the operator~$\square$.
Suppose
\begin{align*}
\vph_1&=\square\,(\phi_1(x, \bT))\\
\intertext{and}
\vph_2&=\square\,(\phi_2(x, \bT)).\\
\intertext{Then we have}
\{\vph_1, \vph_2\}&=\square\,(\phi_{\{1,2\}}),
\end{align*}
where
\[
\phi_{\{1,2\}}=\cEv_{\vph_1}(\phi_2)-\cEv_{\vph_2}(\phi_1)+\bar
D_x(\phi_1)\,\phi_2-\phi_1\,\bar D_x(\phi_2).
\]
Here
$\phi_{\{1,2\}}=\phi_{\{1,2\}}(x$, $\bT)$,
since the evolution $\dot T_\phi$ of integral~\eqref{IEMT}
along any symmetry $\vph=\square\,(\phi)$ is equal to
\begin{equation} \label{IEvolEMT}
\dot T_\phi=-\beta\,\bar D_x^3(\phi)+T\,\bar D_x(\phi)
+\bar D_x(T\cdot\phi).
\end{equation}
Here and below we use the notation
\[\beta\equiv\sum_{i=1}^r a_i\cdot\Delta^i.\]
Consider the operator applied to the function $\phi$ in the
right\/-\/hand side of Eq.~\eqref{IEvolEMT}. We see that this operator
is the second Hamiltonian structure $\hat B_2$ for the Korteweg\/--\/de
Vries equation~\eqref{IKdV}.

In this paper, we prove that the generating sections $\psi_\eta$ of
conservation laws $[\eta]$ and the Noether symmetries $\vph_\cL$ for
the Toda equation~\eqref{IeqToda} are related by the equation
$\psi_\eta=\kappa\,\vph_\cL$. Hence we obtain

\begin{theorNo}[\textup{theorem \ref{ThNoetherStructure}
on page~\pageref{ThNoetherStructure}}]
\begin{enumerate}\item
Suppose $\Omega^i$ is an integral for the Toda equation~\eqref{eqToda}
such that $D_y(\Omega^i)=\tilde\nabla_i(F)$.
Then there is the operator $\nabla_i$ such that
\[
\tilde\nabla_i=\nabla_i\circ\square^*\circ\kappa.
\]
In particular\textup{,} consider the integral
$\Omega^1=T$ defined in Eq.~\eqref{EMT}. Then
\[\nabla_1=\mathbf{1}.\]
\item
The Noether symmetries of the Toda equation are
\[
\vph_\cL=\square\circ\sum\limits_i
   \nabla_i^*\circ\bE_{\Omega^i}(Q(x,\bar{\mathbf{\Omega}})),
\]
where $\Omega^i\in\ker\bar D_y$ are the integrals for the
equation~ $\cE_\Toda$\textup{,}
\[
\bE_{\Omega^i}=\sum_{j\geq0}(-1)^jD_x^j\cdot\frac{\dd}{\dd\Omega^i_j}
\]
is the Euler operator with respect to\ $\Omega^i$\textup{,}
$\bar{\mathbf{\Omega}}$ is an arbitrary set of differential consequences
to $\Omega^i$\textup{,} and $Q$ is a smooth function.
\end{enumerate}
\end{theorNo}

\begin{example}[\textup{\cite{TodaLawsActa}}]
The Noether symmetries $\vph_\cL$ of the Toda
equation \eqref{eqToda} associated with a generic nondegenerate
symmetrizable matrix $K$ are
\[
\vph_\cL=\square\circ\bE_T(Q(x,\bT))
\]
up to the transformation~$x \leftrightarrow y$.
\end{example}

In Sec.~\ref{SecTodaRec}, we
construct a continuum of recursion operators
for the symmetry algebra of the Toda equations.
These operators are parameterized by
arbitrary smooth functions.
Although the structure of the symmetry algebra itself is
known, see Eq.~\eqref{IsymToda}, presence of the recursion operators
gives us additional information about the Toda equation and permits to
establish the relation between $\cE_\Toda$ and other mathematical
physics equations.

\begin{theorNo}[\textup{theorem~\ref{ThContinuumTodaRec}
on page~\pageref{ThContinuumTodaRec}}]
\begin{enumerate}\item
Equation~\eqref{eqToda} admits continuum of
\emph{local} recursion operators
\[R\colon\sym\cE_\Toda\to\sym\cE_\Toda,\]
which are
\[
R=\square\circ\sum\limits_{i,j}f_{ij}(x,\mathbf{\Omega})\cdot
\bar D_x^j\circ \ell_{\Omega^i}.
\]
Here $f_{ij}$ are arbitrary smooth functions and the linearizations
$\ell_{\Omega^i}$ with respect to\ the integrals $\Omega^i$ for the Toda
equation are
\begin{equation}\label{ILinIntegral}
\ell_{\Omega^i}=\Bigl(\ldots,\underbrace{\sum\nolimits_\sigma
{\dd\Omega^i}/{\dd u^k_\sigma}\cdot
\bar D_\sigma}_{\text{$k$\textup{th}
component}},\ldots\Bigr).
\end{equation}
\item
There is a continuum of \emph{nonlocal} recursion operators
for Eq.~\eqref{IeqToda}, which are constructed in the following way.
Assign the nonlocal variables $s^i$ to the integrals $\Omega^i$ by the
compatible differentiation rules
\[s^i_x=\Omega^i,\quad
s^i_y=0.\] The linearizations $\ell_{s^i}$ are defined by the
formulas \[\ell_{s^i}=\bar D_x^{-1}\circ\ell_{\Omega^i},\]
where $\ell_{\Omega^i}$ are calculated by Eq.~\eqref{ILinIntegral}.
Then the required recursion operators are
\[
R=\square\circ\sum\limits_i f_i(x,\mathbf{s},\mathbf{\Omega})\cdot\bar
D_x^{-1} \circ\ell_{\Omega^i},
\]
where the functions $f_i$ are arbitrary.
In general, these operators do not preserve the locality of
elements~\eqref{IsymToda} of the symmetry algebra~$\sym\cE_\Toda$.
\end{enumerate}
\end{theorNo}

\medskip
In Chapter~2, we construct the commutative Hamiltonian hierarchy
$\gA$ of $r$-component analogs for the potential modified
Korteweg\/--\/de Vries equation. This hierarchy is
related with the Korteweg\/--\/de Vries hierarchy for
equation~\eqref{IKdV} that describes the dynamics of the
integral~\eqref{IEMT}.

First, we consider an example (\cite{YS2003}) with $r=1$.
Namely, we consider the scalar Liouville equation
\[u_{xy}=\exp(2u)\]
and construct the symmetry sequence for this equation. This sequence is
identified with the hierarchy $\gA$ of the potential modified
Korteweg\/--\/de Vries equation
\[u_t=-\tfrac{1}{2}u_{xxx}+u_x^3\]
Also, we assign the hierarchy $\gB$ of the Korteweg\/--\/de Vries
equation
\[T_t=-\tfrac{1}{2}T_{xxx}+3TT_x\]
to the initial Liouville equation.

Now suppose $r\geq1$. The hierarchy $\gA$ is constructed in
Sec.~\ref{SecAandB} by using the following procedure.
Introduce the nonlocal variable $s$ such that $s_x=T$ and $s_y=1$.
Choose the initial function $\phi_{-1}=1$.
Then construct two sequences,
\begin{align*}
\phi_i&=D_x^{-1}(\dot T_{\phi_{i-1}}),\\
\intertext{see Eq.~\eqref{IEvolEMT}, and}
\vph_i&=\square(\phi_{i-1}).
\end{align*}
We denote the first sequence $\phi_i$ by~$\gB$. This is the commutative
bi\/--\/Hamiltonian hierarchy of local higher symmetries for the
potential Korteweg\/--\/de Vries equation
\[
 s_{t}=-\beta\,s_{xxx}+\tfrac{3}{2}s_x^2,\qquad
 \beta=\sum_i a_i\Delta^i.
\]
The second sequence $\vph_i$ is denoted by~$\gA$. This is the required
hierarchy of the higher analogs for the potential modified
Korteweg\/--\/de Vries type equation
\[
u_{t}=\vph_1.
\]
The following theorem is valid.

\begin{theorNo}[\textup{theorem~\ref{AIsCommutative} on
page~\pageref{AIsCommutative}}]
The elements $\vph_i$ of the sequence
$\gA\subset\sym\cE_\Toda$ compose a commutative Lie
algebra\textup{:} \[ [\gA, \gA]=0.\]
\end{theorNo}

Then we describe the elements $\vph_k\in\gA$ such that $k<0$.
We prove that the symmetry $\vph_{-1}$
is defined by the Toda equation itself represented in the Hamiltonian
form
\[
u_y=A_1\circ\bE_u\bigl( (\vec a\cdot\exp(Ku))\,\Id x\bigr),
\]
where \[A_1=\kappa^{-1}\cdot\bar D_x^{-1}\] is the first Hamiltonian
structure for the equation within the hierarchy $\gA$ and
$\vec{a}={}^t(a_1,\ldots,a_r)$,~$\kappa=\|a_i k_{ij}\|$.


\begin{theorNo}
\textup{1 (theorem \ref{NoetherSubalgebra} on
page~\pageref{NoetherSubalgebra}).}
The generators $\vph_k$ of the commutative Lie algebra $\gA$
are the Noether symmetries of the Toda equation\textup{:}
$\vph_k\in\sym\cL_\Toda$.

\textup{2 (proposition~\ref{CylinderLaws} on
page~\pageref{CylinderLaws}).}
Suppose $k\geq0$ is arbitrary.
Then the $k$\textup{th} term $\phi_k=\bE_T(h_k\,\Id x)$
of the hierarchy $\gB$ is a conserved density for the
$k$\textup{th} higher potential modified
Korteweg\/--\/de Vries equation.
\end{theorNo}


Now we formulate the most remarkable relation between
the hierarchy $\gA$ for the potential modified Korteweg\/--\/de Vries
equation~\eqref{pmKdV} and the hierarchy~$\gB$ for scalar
equation~\eqref{KdV}.

\begin{theorNo}\label{ICylinder}
\textup{1 (proposition \ref{NoethAreHamTh} on
page~\pageref{NoethAreHamTh}).}
Each Noether symmetry
\[\vph_\cL=\square\,\circ\,\bE_T(Q(x,\bT))\in\sym\cL_\Toda\]
of the Toda equation
associated with a nondegenerate symmetrizable matrix $K$
is Hamiltonian with respect to\ the Hamiltonian structure
$A_1=\kappa^{-1}\cdot D_x^{-1}$ and the Hamiltonian
$\cH=[Q(x,\bT)]$\textup{:}
\[\vph_\cL=A_1\circ\bE_u(\cH).\]

\textup{2 (theorem \ref{CylinderTh} on page~\pageref{CylinderTh}).}
The Hamiltonian $[h_k\,\Id x]$ for the $k$\textup{th}
higher Korteweg\/--\/de Vries equation is the Hamiltonian for the
Noether symmetry $\vph_k\in\gA$ for any integer $k\geq0$.
\end{theorNo}

%

\medskip
In Chapter~3, we apply the methods of geometry of partial differential
equations to the study of the properties of
equations~\eqref{IOurHeavenly} and~\eqref{INLS}.

In Sec.~\ref{SecNLS} of Chapter~3
we obtain the geometric structures for the $m$-component analog
\[\Psi_t=\bi\Psi_{xx}+\bi f(|\Psi|)\,\Psi\]
of the nonlinear Schr\"odinger equation~\eqref{INLS}.
By~\cite{NwavesNLS}, this equation admits a commutative
bi\/--\/Hamiltonian hierarchy of higher symmetries and an infinite set
of the conserved densities in involution if $f=\id$. Suppose $f$ is
arbitrary, then this is not true. We compute the symmetry algebra for
this equation if $f$ is the homogeneous function of weight~$\Delta$;
this case is actual for applications in physics.
Also, we point out $m^2$ conserved currents
\[
\eta_{ij}=\Psi^i\bar\Psi^j\,\Id x +
  \bi\left(\Psi^i_x\,\bar\Psi^j - \Psi^i\,\bar\Psi^j_x\right)\,\Id t
\]
for this equation. Here $f$ is arbitrary.
These currents generalize the known conservation laws of energy for  an
$i$th component $\Psi^i$ in the multi\/--\/component nonliear
Schr\"odinger equation.

In Sec.~\ref{SecHeav} of the same chapter we consider the
dispersionless Toda equation~\eqref{IOurHeavenly},
\[u_{xy}=\exp(-u_{zz}),\]
and realize the following scheme for it. First, we find the Lie algebra
of its classical symmetries and assign classes of exact solutions to
these symmetries; then we construct conserved currents for this
equation. The results of our calculations are rather cumbersome. They
are found on pages~\pageref{SymPointHeav}--\pageref{SecNLS} and also in
the paper~\cite{HeavLaws}.

\medskip
In Chapter~4, recent geometric concepts in the theory of B\"acklund
transformations and zero\/--\/curvature representations are illustrated
for the scalar Liouville equation
\begin{equation} \label{Ieqhyp2}
\cE_{\mathrm{Liou}}=\{u_{xy}=\exp(2u)\},
\end{equation}
which is the Toda equation associated with the Lie algebra
$\gothg=\mathfrak{sl}_2(\BBC)$.

In Sec.~\ref{SecDeform} we construct one\/--\/parametric families of
B\"acklund (auto)\-trans\-for\-ma\-ti\-ons~$\tilde\cE_t$ for
equation~\eqref{Ieqhyp2}. These transformations are
\begin{equation} \label{IBacklundAuto}
\tilde\cE_t=\left\{\begin{aligned}
(\tilde{u}-u)_{x}&=\rme^{-t}\,\exp(\tilde{u}+u),\\
(\tilde{u}+u)_{y}&=2\rme^t\,\sinh(\tilde{u}-u)
\end{aligned}\right\}.
\end{equation}
By definition, put
\[
u_k\equiv\frac{\dd^ku}{\dd x^k},\quad
\dy{u}{k}\equiv\frac{\dd^ku}{\dd y^k},
\qquad k\in\BBN.
\]
Consider the scaling symmetry
\begin{equation} \label{IShadowLift}
\hat X=-{x}\,\frac{\dd}{\dd{x}}+{y}\,\frac{\dd}{\dd{y}}
+\sum\limits_{k\geq1}ku_k\,\frac{\dd}{\dd u_k}
-\sum\limits_{k\geq1}k\dy{u}{k}\,\frac{\dd}{\dd\dy{u}{k}}
\end{equation}
of the Liouville equation~\eqref{Ieqhyp2}.
Our reasonings are based on the following property of $\hat X$.
This symmetry cannot be extended up to a symmetry of the covering
equation~$\tilde\cE_t$.
Recently I.~S.~Krasil'shchik described~(\cite{JKIgonin}) the scheme
of generating one\/--\/parametric families of coverings over
differential equations. We have

\begin{theorNo}[\textup{\cite{JKIgonin}}]
Let $\tau\colon \tilde\cE\to\cE$ be a covering and $A_t\colon
\tilde\cE\to \tilde\cE$ be a smooth family of diffeomorphisms such
that $A_0=\id$ and \[\tau_t=\tau\circ A_t\colon \tilde\cE\to \cE\] is the
covering for any $t\in\BBR$.  Then the evolution of the Cartan
connection form $U_{\tau_t}$ is
\begin{equation}\label{IDeformStructElem}
\frac{dU_{\tau_t}}{dt}=\fn{\hat X_t}{U_{\tau_t}},
\end{equation}
where $\hat X_t$ is a $\tau_t$-shadow for any $t\in\BBR$
and $\fn{{\cdot}}{{\cdot}}$ is the Fr\"olicher\/--\/Nijenhuis
bracket, see Eq.~\eqref{DefFNBracket} on page~\pageref{DefFNBracket}.
\end{theorNo}

We prove that scaling symmetry~\eqref{IShadowLift}
is the $\tau_t$-shadow such that the evolution of the
connection form $U_t$ in the covering
$\tau_t\colon\tilde\cE_t\to\cE_\Liou$, see
Eq.~\eqref{IBacklundAuto}, is given by Eq.~(\ref{IDeformStructElem}).
The proof is straighforward. It is based on a useful identity in total
derivatives.

\begin{theorNo}[\textup{lemma~\ref{StrangeRelation} on
page~\pageref{StrangeRelation}}]\label{IStrangeRelation}
Let $u(x)$ and $f(u)$ be smooth functions and
$D_x$ be the total derivative
with respect to~$x$. Denote $u_k\equiv D^k_x(u(x))$, $k\geq0$,
$u_0\equiv u$. Then the relation
\[  
n\cdot D_x^n(f(u))=\sum\limits_{m=1}^nm\,u_m\,\frac{\dd}{\dd u_m}
  D_x^n(f(u))
\]  
holds for any integer $n\geq1$.
\end{theorNo}

In Sec.~\ref{SecIntegrating}, we consider the problem of constructing
the pairs of solutions to the hyperbolic Liouville
equation~\eqref{Ieqhyp2} and the wave equation
\[s_{xy}=0\]
such that these solutions are related by the B\"acklund transformation.
Namely, we obtain the set of the nonlocal variables such that
B\"acklund transformations are successfully integrated and each
of the solutions related by B\"acklund transformation is obtained
explicitly in terms of these variables.

In Sec.~\ref{SecBuildZCR}, we analyse the correspondence
between B\"acklund transformations and zero\/--\/curvature
representations. We use two distinct representations of the Lie algebra
$\gothg=\mathfrak{sl}_2(\BBC)$ generated by the elements
$\langle e$, $h$,~$f\rangle$. The first representation is
the representation of $\gothg$ in the traceless matrices:
\begin{align*}
\varrho(e)&=\begin{pmatrix}0&1\\0&0\end{pmatrix},&
\varrho(h)&=\begin{pmatrix}1&\pp0\\0&-1\end{pmatrix},&
\varrho(f)&=\begin{pmatrix}0&0\\1&0\end{pmatrix}.\\
\intertext{%
The second representation involves vector firlds on the straight line:}
\rho(e)&=1\cdot\frac{\dd}{\dd\Xi},&
\rho(h)&=-2\Xi\cdot\frac{\dd}{\dd\Xi},&
\rho(f)&=-\Xi^2\cdot\frac{\dd}{\dd\Xi}.
\end{align*}
We point out B\"acklund transformations assigned to the known
zero\/--\/curvature representations, and \textit{vice versa}.

In Appendix~\ref{SecAppendix}, we list
several geometrical methods of solving the
boundary problems for equations of the mathematical physics.
Constructing solutions invariant w.r.t.\ symmetries of
the problem is discussed. A method based on the respesentation of the
equation at hand in the evolutionary form is pointed out.
The methods based on the deformation of the boundary problem along
discrete or continuous parameters are described. Among these methods,
we note the direct iterations, the boundary conditions homotopy, the
relaxation method, and the deformation of the initial equation.
Then, comparative analysis of the results of
computer experiments in applying these
methods is carried out.

\newpage
\specialsection*{{
\textbf{Part~I.}}
The Korteweg\/--\/de Vries type equations associated with the
Toda systems}
In Part~I, we analyse the properties of the symmetry algebra
$\sym\cE_\Toda$ for the hyperbolic Toda equations $\cE_\Toda$
associated with nondegenerate symmetrizable matrices~$K$.
By using the canonical nonlocal recursion operator for the algebra
$\sym\cE_\Toda$, we construct its commutative Lie subalgebra $\gA$
generated by the local higher Noether symmetries of the latter equation.
We identify this subalgebra with the Hamiltonian hierarchy
of the higher analogs for the $r$-component potential modified
Korteweg\/--\/de Vries equation associated with~$K$.
Then we relate the hierarchy $\gA$ with the commutative
bi\/-\/Hamiltonian hierarchy $\gB$ of the scalar Korteweg\/--\/de Vries
equation. The hierarchy $\gB$ itself is a commutative Lie subalgebra of
the Noether symmetries of the scalar wave equation.
By using all these relations between the hyperbolic equations and the
evolution hierarchies of their symmetries, we prove that the
hierarchies $\gA$ and $\gB$ share the same sequence of the Hamiltonians
and demonstrate that the Toda equation itself is the first nonlocal
term in the symmetry algebra~$\gA$.

\specialsection*{\textbf{Chapter~1.}
Conservation laws and the Noether symmetries
of the Toda equations}
In this chapter, we study the geometric properties of the
two\/--\/di\-men\-si\-o\-nal Toda equations $u_{xy}=\exp(Ku)$ (in particular,
of the Toda equations associated with the complex semisimple Lie
algebras, see~\cite{LeznovSaveliev, Razumov}).
Namely, we analyse the relations between their conservation laws
(\cite{LeznovSmirnovShabat, ShabatYamilov}), Noether's symmetries of
the Lagrangian functional (\cite{SakovichZCR}), and the
recursion operators for the symmetry algebra of the latter equations.

In Sec.~\ref{SecTodaEq}, we pass from the scalar Liouville equation to
the hyperbolic Toda equations associated with generic nondegenerate
symmetrizable matrices~$K$. Further, we consider
properties of the Lagrangian for
these equations. Then we point out the minimal integral
$T\in\ker {\bar D_y}$ and its differential span
$\bT\subset {\bar D_y}$. Finally, we assign generators of the Lie
algebra $\sym\cE_\Toda$ to elements $\mathbf{\Omega}$ of the kernel
$\ker\bar D_y$.

In Sec.~\ref{SecTodaNoeth},
we construct a one\/-\/to\/-\/one correspondence between the generating
sections of conservation laws and the Noether symmetries of the
Lagrangian~$\cL_\Toda$ for the Toda equation.
The description of the Lie algebra
$\sym\cL_\Toda\subset\sym\cE_\Toda$ of the Noether symmetries for the
Toda equation associated with a generic matrix $K$
is based on this correspondence. Also, we
establish some properties which are common to all integrals
$\Omega^i\in\ker {\bar D_y}$ of the Toda equation associated
with the Cartan matrix $K$ of a semisimple Lie algebra.

Finally, in Sec.~\ref{SecTodaRec} we construct a continuum of the recursion
operators for the symmetry algebra of the Toda equation.
There are two types of the
recursion operators: they are either local or nonlocal with respect to\ the total
dervatives.

The exposition follows the papers~\cite{VestnikNoether,
Acta, TodaLawsActa}.

\section{The Toda equation}\label{SecTodaEq}
\subsubsection*{The Liouville equation and its generalizations}
The Liouville equation
\begin{equation}\label{eq}
\cE_\Liou=\{u_{\xi\xi}+u_{\eta\eta}=\exp(2u)\}
\end{equation}
is a model exactly integrable nonlinear differential equation
that appeares in many parts of mathematics and mathematical
physics.
%
The study of this equation was initiated by
Liouville~(\cite{Liouville}) and H.~Po\-i\-n\-ca\-r\'e~(\cite{Poincare}).
One of the problems that they posed, the uniformization problem for
algebraic curves (compact Riemannian surfaces), was investigated later by
Kazdan and Warner~(\cite{Kazdan}).
Suppose the Lagrangian
%
$$\cL_\Liou=
-{\tfrac{1}{2}[\bigl(u_\xi^2+u_\eta^2}+\exp(2u)\bigr)\,\Id\xi\wedge
\Id\eta]$$
is regularized as described in~(\cite{Zograph}) and then
calculated on a classical solution of the Liouville equation. In this case,
$\cL_\Liou$ is known~(\cite{Zograph})
to be the generating function for
the accessor parameters that characterize the uniformization of a
Riemannian surface of genus $0$.
Also, the Lagrangian $\cL_\Liou$ for Eq.~(\ref{eq})
is the potential for the Weil\/--\/Peterson metric on the Teichm\"uller
space of marked Riemannian surfaces (\textit{ibid}).

Equation~(\ref{eq}) plays an important role in modern field theory. In
particular, the quantum Liouville field appeares as a quantum anomaly
in the string theory~(\cite{Strings}).
Further, consider the free self\/--\/dual
Yang\/--\/Mills equation
\[F_{\mu\nu}=\ast F_{\mu\nu},\]
where $F_{\mu\nu}$ is the stress tensor. Then, finding
the $N$-instanton solutions, which minimize the action for the latter equation,
also leads (\cite{Witten}) to Eq.~(\ref{eq}).

In Riemannian geometry, the Liouville equation (\ref{eq})
is the Gauss equation for a conformal metric on the Lobachevsky
plane (see~\cite{DNF} and also~\cite{Kazdan}).

\begin{example}\label{ConfEquivExample}
Consider two pointwise equivalent Riemannian metrics
$$\Id s_j^2=f_j(x,y)(\Id x^2+\Id y^2),$$
where $f_j>0$ and $j$ is either 1 or 2,
on open twofolds of constant Gaussian curvature
$$K_j=- {(2f_j)^{-1}}\Delta\log f_j=\const_j.$$
Here $\Delta$ is the Laplace operator.
Namely, suppose $f_2=f_1\cdot\exp(2u)$,
then $u$ satisfies the equation
\begin{equation}\label{confequiv}
\Delta u=-K_2f_1\exp(2u)-K_1f_1.
\end{equation}
Obviously, the two\/--\/dimensional elliptic Liouville equation
corresponds to the case of the flat metric such that
$f_1\equiv 1$ (\textit{i.e.}, $K_1\equiv 0$) and the Lobachevsky plane metric
($K_2\equiv -1$).
This interpretation of Eq.~(\ref{eq}) is treated as a limit case
($K_1<0$) in the superconductivity theory in description of the
Abrikosov curls in the 2D model~(\cite{Kazdan}).
The Liouville equation appeares in the study of the
Kadomtsev\/--\/Pogutse equations. The latter equation is a reduction of
the general magneto\/--\/hydrodynamic system, where some detailes that
are inessential in the stability problem for high\/--\/temperature
plasma in TOKAMAK are omitted~(\cite{Pogutse}).

For any solution of the elliptic Liouville equation
(\ref{eq}) there is a Lobachevsky plane model that is conformally
equivalent to the Euclidean plane with the diagonal metric
$g_{ij}=\delta_{ij}$. Consider the standard Poincar\'e model in the upper
half\/-\/plane $y>0$ such that $f_2=1/{y^2}$. Then we get a particular
solution $u=-\log y$ of Eq.~(\ref{eq}).
\end{example}

Suppose $f_1\equiv1$, $K_1\equiv0$, and $K_2\equiv{+}1$.
Then formula~(\ref{confequiv}) describes the conformal
equivalence between the Eucludean metric $ {\delta_{ij}}$
on the plane $\BBE^2$ and the metric
$ {g_{ij}=\exp(2\cU)\,\delta_{ij}}$
on the two\/-\/dimensional sphere $ {S^2}$.
The function $\cU(x,y)$ satisfies the equation
\begin{equation}\label{eqscalplus}
\cU_{xx}+\cU_{yy}+\exp(2\cU)=0
\end{equation}
that differs from Eq.~(\ref{eq}) by the sign at the exponent (or,
equivalently, by the multiplication of the independent variables by
$\bi= {\sqrt{-1}}$). We shall say that Eq.~(\ref{eqscalplus}) is
the \emph{Liouville $\scal^{+}$-equation}.

\begin{rem}
Consider the orthogonal metric
$$\Id s^2=\alpha^2\,\Id x^2+\beta^2\,\Id y^2$$
on the sphere of raduis~$\rho$.
Then, Eq.~(\ref{eqscalplus}) is a particular (conformal)
case of the equation~(\cite{Cieslinski})
\begin{equation}\label{orth}
 {{(\alpha^{-1}\beta_x)}_x + {(\beta^{-1}\alpha_y)}_y +
\rho^{-2}\alpha\beta} = 0
\end{equation}
that describes an orthogonal coordinate net on this sphere.
Indeed, $\alpha=\beta=\rho\exp(\cU)$ is a solution of
Eq.~(\ref{orth}) if $\cU(x,y)$ is an arbitrary solution of
Eq.~(\ref{eqscalplus}). Obviously, this solution is not unique.
The pair \[\alpha=\sin\cV,\quad \beta=\rho\cV_y,\]
where $\cV$ satisfies the sine-Gordon equation \[\cV_{xy}=\sin\cV,\]
or the pair \[\alpha=\rho,\quad \beta=\rho\sin x\] are also solutions
to Eq.~(\ref{orth}).
\end{rem}

Suppose $\cE$ is a differential equation.
Throughout this article we denote the number of
dependent variables $u^j$ (the unknown functions) by $r$
and suppose that these $r$ functions depend on $n$
base coordinates (usually they are denoted by $x^1$, $\dots$, $x^n$).

In this paper, we consider the generalizations of the
two\/--\/dimensional scalar Liouville equation~\eqref{eq} to the cases
$n\geq2$ and $r\geq1$. We investigate the properties of these
generalizations by using modern methods and tools of the PDE
geometry.

We obtain a generalization of Eq.~\eqref{eq}
for $n\geq2$ by the following procedure.
Recall that the Liouville equation is the condition for
the Euclidean metric on $\BBE^n$ to be conformally equivalent to the conformal
metric on an $n$-dimensional manifold of constant scalar curvature
\begin{equation}\label{constscal}
\scal\equiv R=\const.
\end{equation}
We fix the value $R=-2$ (such that the Gaussian curvature is
$K=-1$ at $n=2$) and preserve the correlation with the two\/--\/dimensional
case of Eq.~(\ref{eq}). We also put
\begin{equation}\label{ndimmetrics}
\Id s^2=\exp(2u) {\sum_k}\Id x^k\Id x^k.
\end{equation}
Relation~(\ref{constscal}) is a nonlinear PDE for the function
$u(x^1$, $\dots$, $x^n)$.

\begin{theor}[\textup{\cite{Acta}}]
Condition~{\rm(\ref{constscal})} is
\begin{equation}\label{ndimeq}
(n-1)\Delta u+ {\tfrac{1}{2}} {(n-1)(n-2)} (\grad u)^2=\exp(2u),
\end{equation}
where $\Delta$ is the Laplace operator in the Euclidean space~$\BBE^n$
and the scalar product $({\cdot},{\cdot})$ is also induced by the
Euclidean metric.
\end{theor}

\begin{proof}
The scalar curvature $R$ of metric~(\ref{ndimmetrics})
is given by the formula
$$R=\sum\limits_{i,q}\exp(-2u)R^i_{qqi},$$
see~\cite{DNF}. Then we have
\[
R^i_{qqi}=\dd_i\,\Gamma^i_{qq}-\dd_q\,\Gamma^i_{qi}
+\Gamma^i_{pi}\Gamma^p_{qq}-\Gamma^i_{pq}\Gamma^p_{qi},
\]
where
\[
\Gamma^k_{ij}=\dd_iu\,\delta^k_j+\dd_ju\,\delta^k_i-\dd_lu\,
\delta_{ij}\delta^{kl}
\]
are the Christoffel symbols. After a straightforward calculation of the
sums in $q$, $i$, and $p=1\dots n$, we get~(\ref{ndimeq}).
\end{proof}


\subsubsection*{The Toda equations}
M.~Toda~(\cite{MToda}) considered the integrable nonlinear dynamical
system described by Eq.~\eqref{IeqMToda}, \textit{i.e.},
the one\/--\/dimensional lattice
with the exponential interaction.
Nowadays, vastest literature is devoted to the study of the properties
of system~\eqref{IeqMToda} and its various generalizations (see
\cite{DSViniti84, SokolovUMN, LeznovSaveliev, Ovch} and references
therein).
In order to obtain the two\/--\/dimensional  Toda system out
of the nonperiodic Toda lattice
\[q^i_{tt}=\frac{\dd H}{\dd q^i},\qquad i\in\BBZ,\]
where the Hamiltonian $H$ is
\[ H=- {\sum\limits_i}\exp(q^i-q^{i+1}),\] we
replace the ``acceleration'' ${\Id^2}/{\Id t^2}$ with the operator
${\dd^2}/{\dd x\dd y}$. Different stages of this replacement
process and the discovery
of the Toda equations associated with the Lie algebras (and the
Kac\/--\/Moody algebras as well, see~\cite{Bombai}) are discussed in
the papers~\cite{LSYangMills, LeznovIntegrate,  LeznovSmirnovShabat,
LeznovSaveliev, BilalGervaisIntegrate, BilalGervaisLagrange,
GervaisLetter, GervaisClassic, GervaisENS}, see
also~\cite{DSViniti84, Ovch}.

In what follows, we derive the generalizations of Eq.~\eqref{eq}
from a nontrivial viewpoint.
The resulting number $r$ of the dependent variables $u^1$, $\ldots$, $u^r$
is $r\geq1$. Namely, we use the notion
of the Laplace invariants~(\cite{SokolovUMN}).  To start with, we
simplify the calculations by making the change of coordinates such that
the Liouville equation acquires the hyperbolic form $u_{xy}=\exp(u)$.
From now on, we study the hyperbolic equations and their symmetries.
The difference between the elliptic and the hyperbolic cases vanishes
after the complexification. Still, the study of the symmetry properties
of PDE does not require the presence of a complex structure. Therefore
we assume all equations to be real in agreement with the definition on
page~\pageref{DefPDE}.

Following~\cite{SokolovUMN}, we calculate the principal Laplace
invariants $H_0$ and $H_1$ for a hyperbolic equation
\[u_{xy}=f(x, y, u, u_x, u_y);\] we have $f=\exp(u)$. Then we get
\begin{align*}
H_0& {\stackrel{\text{def}}{{}={}}}
-\bar D_y\left(\frac{\dd f}{\dd u_y}\right)
+\frac{\dd f}{\dd u_x}
\,\frac{\dd f}{\dd u_y}+\frac{\dd f}{\dd u}=\exp(u),\\
H_1& {\stackrel{\text{def}}{{}={}}}
-\bar D_x\left(\frac{\dd f}{\dd u_x}\right)
+\frac{\dd f}{\dd u_x}\,\frac{\dd f}{\dd u_y}+
\frac{\dd f}{\dd u}=\exp(u).
\end{align*}
The definition of other Laplace's invariants follows from the equations
\begin{equation}\label{eqDynamLaplace}
\bar D_{xy}(\log H_i)=-H_{i-1}+2H_i-H_{i+1},\quad i\in\BBZ.
\end{equation}
We recall that the quasilinear equation $u_{xy}=f$ is of the
\emph{Liouvillean type} if its sequence of the Laplace invariants is
finite, \textit{i.e.}, there are some $p\geq1$ and $q\geq0$
such that $H_p=H_{-q}\equiv0$.
One easily checks that the sequence $H_i$ for the Liouville equation
vanishes at once: $H_i\equiv0$ if $i$ is neither~0 nor~$1$.
This observation motivates the above
definition\footnote{We also note that the \emph{matrix} Laplace
invariants are defined for system~\eqref{eqTodaForK} of
hyperbolic differential equations, see~\cite{SokolovUMN}.%
}.
Now suppose $-q<i<p$ and make the substitution $H_i=\exp(\cU^i)$.
Now, restrict equation~\eqref{eqDynamLaplace} onto the graphs of jets of
the sections $\cU$ in the jet bundle $\BBR^r\times\BBR^2\to\BBR^2$.
Then the total derivatives $\bar D_i$ are mapped to the derivations
${\dd}/{\dd x^i}$. We finally obtain the system
\[
\cU^0_{xy}=2\exp(\cU^0)-\exp(\cU^1),\quad
\cU^1_{xy}=-\exp(\cU^0)+2\exp(\cU^1).
\]
Generally, we obtain the system of equations
\[
\cU^i_{xy}= {\sum_{j=-q+1}^{p-1}} k_{ij}\,\exp(\cU^j)
\]
such that the structure of the nondegenerate
$(p+q-1)\times(p+q-1)$-matrix $K=\|k_{ij}\|$ is
\begin{equation}\label{CartanForAr}
k_{ii}=2,\qquad k_{i,i+1}=k_{i.i-1}=-1,\qquad
k_{ij}=0\text{ if }|i-j|>1.
\end{equation}
We see that $K$ is the Cartan matrix of the Lie algebra of type
$A_{r-1}$. Now we shift the index $i$ that enumerates the variables
$\cU^i$ and make the change (\cite{LeznovSaveliev}) of the form
$\cU=K\cdot u$. Thus, we get the system
\[
u^i_{xy}=\exp(-u^{i-1}+2u^i-u^{i+1}),\quad 1\leq i\leq r,
\qquad u^0=u^{r+1}\equiv0.
\]
This is the two\/--\/dimensional Toda system associated with the type
$A_{r-1}$ Lie algebra, which is denoted by~$\gothg$.
Generally, we assign the Toda equations
\begin{equation}\label{eqTodaForK}
\bu_{xy}=\exp(K\bu)
\end{equation}
to a semisimple Lie algebra with the Cartan matrix~$K$ by following the
geometric scheme~(\cite{LeznovSaveliev}), which is discussed below.
In Chapter~4, we use the contructions supplied by this scheme and
analyse the relationship between the canonical zero\/--\/curvature
representations and B\"acklund transformations for a certain class of
differential equations.

%
First, we fix some notation.
Let $\gothg$ be a semisimple complex Lie algebra of rank $r$.
By $\{\alpha_i$, $1\leq i\leq r\}$ denote its system of
simple roots and let $K$ be the Cartan matrix of $\gothg$. We thus have
$$K=\|k_{ij}=2(\alpha_i,\alpha_j)\cdot {|\alpha_j|^{-2}},\ %
1\leq i,\ j\leq r\|.$$
By $K^{-1}=\|k^{ij}\|$ denote the
inverse matrix for $K$ and denote its elements by~$k^{ij}$.
Suppose $A$, $B\in\gothg$. Then we assume that
\begin{equation}\label{fc}
\theta=A\,\Id z+B\,\Id\zz
\end{equation}
is the flat connection form in the principal fibre bundle $G\to M$,
where $G$ is the Lie group of the Lie algebra $\gothg$.
We thus have
\begin{equation}\label{ZCR}
\Id\theta+ {\tfrac{1}{2}} [\theta,\theta]=0.
\end{equation}
In terms of the extended total derivatives we have, then,
\begin{equation}\label{ZCRCoords}
[\dd+A,\bar\dd+B]=0\quad\Leftrightarrow\quad
-\bar\dd A+\dd B+[A,B]=0.
\end{equation}

Suppose further that $H_j$ are the Cartan generators and
$E_j$, $F_j$ are the Chevalley generators of~$\gothg$, respectively,
for all $j$ such that $1\leq j\leq r=\mathrm{rank}\,\gothg$.
The commutation relations
\begin{equation}\label{Chevalley}
{}\begin{aligned}{}
{ [H_i,H_j]}&=0, & [H_i,E_j]&=k_{ji}E_j,\\
{ [H_i,F_j]}&=-k_{ji}F_j, & [E_i,F_j]&=\delta_{i,j}H_i
\end{aligned}
\end{equation}
hold.
%
We choose the following ansatz for the connection coefficients
$A$ and $B$:
\begin{equation}\label{Anzats} 
\begin{aligned}
A&=\sum_{j=1}^r\left(a_h^j(z,\zz)\cdot H_j+a_e^j(z,\zz)\cdot
   E_j\right),\\
B&=\sum_{j=1}^r\left(b_h^j(z,\zz)\cdot H_j+b_f^j(z,\zz)\cdot
   F_j\right).
\end{aligned}
\end{equation}
Then from Eq.~(\ref{ZCRCoords}) we get the relations
\begin{subequations}
\begin{gather}
\dd b_h^j-\bar\dd a_h^j+a_e^jb_f^j=0,\label{atH}\\
\bar\dd\log a_e^j=- {\sum_{i=1}^r} k_{ji}\,b_h^i,
\label{atEF}\qquad
\dd\log b_f^j= {\sum_{i=1}^r} k_{ji} \, a_h^i
\end{gather}
\end{subequations}
for the coefficients of $H_j$, $E_j$, and $F_j$, respectively,
for any $j$ such that~$1\leq j\leq r$.
From Eq.~\eqref{atEF} we obtain
$$
a_h^i= {\sum_{j=1}^r} k^{ij}\,\dd\log b_f^j ,
\qquad b_h^i=- {\sum_{j=1}^r} k^{ij}\,\bar\dd\log a_e^j,
$$
and from Eq.~\eqref{atH} we get the relation
\begin{equation}\label{TodaatH}
 {\sum_{j=1}^r}
k^{ij}\,\dd\bar\dd\log\left(a_e^j\cdot b_f^j\right) = a_e^i\cdot b_f^i
\end{equation}
for any $i$,~$1\leq i\leq r$. By definition, put
\begin{equation}\label{Subst4u}
u_i= {\sum_{j=1}^r} k^{ij}\,\log\left(a_e^j\cdot b_f^j\right).
\end{equation}
Now substitute Eq.~\eqref{Subst4u} in Eq.~\eqref{TodaatH}. We finally
get the Toda equation (see Eq.~\eqref{eqTodaForK}) associated with the
Lie algebra~$\gothg$~(\cite{LeznovSaveliev}).
In the sequel, we use the coordinates $x$ and $y$ as synonims
for the complex coordinates $z$ and $\zz$, respectively.
Also, we suggest the following convention: all symmetries, conservation
laws, and similar structures for the Toda equations
are treated up to the discrete symmetry~$x\leftrightarrow y$.

Now suppose
$K=\|k_{ij}$, $1\leq i$, $j\leq r\|$ is a nondegenerate
$(r\times r)$-matrix and $K^{-1}=\|k^{ij}\|$ is its inverse.
Assume that there is an $r$-component vector $\vec{a}$ such that
$a_i\not=0$ for any $1\leq i\leq r$
and the symmetry condition $\kappa_{ij}\equiv a_i\cdot k_{ij}=\kappa_{ji}$
holds for the matrix $\kappa=\|\kappa_{ij}\|$.
Then we say that this matrix $K$ is \emph{symmetrizable},
see also~\cite{Symmetrizator}.
By $\hat\kappa$ we denote the operator of left multiplication
by the nondegenerate matrix~$\kappa$.

The hyperbolic Toda equations associated with a nondegenerate
symmetrizable $(r\times r)$-matrix~$K$ are
\begin{equation}\label{eqToda}
\cE_\Toda=\Bigl\{F^i\equiv u^i_{xy}-
\exp\bigl(\sum_{j=1}^r k_{ij}u^j\bigr)=0,
\quad 1\leq i\leq r\Bigr\}.
\end{equation}
In particular, suppose $\gothg$ is a semisimple Lie algebra of rank
$r$, let $\{\alpha_i$, $1\leq i\leq r\}$ be the system of its
simple roots, and denote the Cartan matrix of the algebra~$\gothg$ by
$$K=\left\|k_{ij}=\frac{2(\alpha_i,\alpha_j)}{|\alpha_j|^{2}},\ 1\leq
i,j\leq r\right\|.$$
Then we set $a_i=|\alpha_i|^{-2}$; we thus have
$$\kappa_{ij}=\frac{2(\alpha_i,\alpha_j)}%
{|\alpha_i|^{2}\cdot|\alpha_j|^{2}}=\kappa_{ji}.$$


\subsubsection*{Lagrangian formalism for the Toda equation}
The Toda equation $\cE_\Toda$ is the Euler\/-\/Lagrange
equation in the following sence. Consider the action functional
\[\cL_\Toda=\int L_\Toda\,\Id x\wedge\Id y\] with the density
$$
L_\Toda=-\tfrac{1}{8}\sum_{i,j}\sum_{\mu,\nu}
g^{\mu\nu}\kappa_{ij}u^i_{;\mu}u^j_{;\nu}
+{a_i^2}\cdot\exp\Bigl(\sum_{j=1}^r k_{ij}u^j\Bigr).
$$
Here $u^j_{;\mu}\equiv D_\mu(u^i)$ and
$g^{\mu\nu}=\left(\begin{smallmatrix}0&2\\ 2&0\end{smallmatrix}\right)$
is the inverse of the metric tensor
$g_{\mu\nu}=\left(\begin{smallmatrix}0& \frac12\\
\frac12&0\end{smallmatrix}\right)$ that defines the flat metric
$\Id s^2=\Id x\,\Id y$ on the base of the jet bundle $\pi$.
In local coordinates, the density of the Lagrangian is
\begin{equation}\label{TodaLDensity}
L_\Toda=-\tfrac{1}{2}\sum_{i,j=1}^r
\kappa_{ij}u^i_{x}u^j_{y} -
\sum_{i=1}^r a_i \cdot
\exp\Bigl(\sum_{j=1}^r k_{ij}u^j\Bigr).
\end{equation}
Assign the Euler\/-\/Lagrange equations
\begin{equation}\label{ImEulerSpoilt}
\bE_u(\cL_\Toda)=
 {\Bigl|\sum_j\kappa_{ij}F^j\Bigr|}=\kappa\cdot F=0
\end{equation}
to the Lagrangian~$\cL_\Toda$. Then these equations are equivalent to
equation~\eqref{eqToda} since the matrices $K$ and $\kappa$ are
nondegenerate simultaneously due to the assumption $a_i\not=0$.

Suppose $\cE$ is a differential equation, $\vph\in\ker\bar\ell_\cE$
is its symmetry and
$\psi\in\ker\bar\ell_\cE^*$ is the generating section of its conservation law.
Now we obtain the transformation rules for
$\vph$ and $\psi$
with respect to\ the reparameterizations that preserve the equation's manifold
$\cE$ and the ideal $\cE^\infty$ of its differential consequences.
The following lemma is valid.

\begin{lemma}[\textup{\cite{TodaLawsActa}}]\label{BehaviourLemma}
Suppose $\cE=\{G^i=0$, $1\leq i\leq r\}$ be a
nonoverdetermined differential equation and let
\[     
G^i= {A^i_{\,j}}F^j
\]
be a nondegenerate transformation of the relations that define
the equation~$\cE$. Then the following two conditions hold\textup{:}
\begin{enumerate}\item
The identities
$$
\ell_G=A\cdot\ell_F,\qquad \ell_G^*=\ell_F^*\cdot {{}^t A},
$$
are fulfilled. Here $A$ is the reparameterization matrix for
the relations that describe~$\cE$.
\item
Suppose $\vph_G\in\ker\bar\ell_G$ is a symmetry of the equation
$\cE$ and $\psi_G\in\ker {\bar\ell_G^*}$ is an arbitrary solution
to Eq.~\eqref{DetEqOnPsi} for $\cE=\{G=0\}$. Now, consider the
reparameterization $G=AF$ of the relations that define the equation
$\cE\simeq\{F=0\}$\textup{.}
Then $\vph_F=\vph_G$ is a symmetry of the equation
$\cE$ again: $\vph_F\in\ker\bar\ell_F$,
while a solution $\psi_G$ of Eq.~\eqref{DetEqOnPsi} is transformed by
the rule \[\phi_G\mapsto\psi_F={}^tA\cdot\psi_G\in\ker\bar\ell_F^*.\]
\end{enumerate}
\end{lemma}

\begin{proof}
By using the definition of the linearization operator $\ell_G$
for $G=AF$, we get \[\ell_G=\ell_{AF}=A\cdot\ell_F.\] Therefore,
\[\ell_G^*= {{(A\cdot\ell_F)}^*}
=\ell_F^*\circ A^*=\ell_F^*\circ {{}^tA}.\]
If the matrix $A$ is nondegenerate, then the condition
$\bar\ell_G(\vph_G)=0$ is equivalent to
$A\cdot\bar\ell_F(\vph_G)=0$, whence we get $\vph_F=\vph_G$.
From the assumption
$\bar\ell_F^*({}^tA\cdot\psi_G)=0$ we deduce the formula
$\psi_F={}^t A\cdot\psi_G$ for the rules
of solution transformation for the equation $\bar\ell_F^*(\psi_F)=0$.
\end{proof}

Now we apply the Noether theorem
(see Theorem~\ref{InverseNoether} on page~\pageref{InverseNoether}):

\begin{cor}\label{Podkrutka}
Let the assumptions of Theorem~\ref{InverseNoether} and
Lemma~\ref{BehaviourLemma} hold.
Let $\psi\in\ker\bar\ell_F^*$ be the generating section
of a conservation law for the Euler\/-\/Lagrange equation
$\cE=\{F=0\}$.
Then there is the Noether symmetry
$\vph\in\ker\bar\ell_F$ of the equation $\cE$ such that
the relation $\psi= {{}^t A}\cdot\vph$ holds.
\end{cor}

\subsubsection*{The minimal integral for the Toda
equation}\label{SecHamToda}
Consider the Toda equation associated with a
nondegenerate symmetrizable matrix $K$. One easily  checks that
Eq.~\eqref{eqToda} admits at least one
\emph{integral}~(\cite{SokolovUMN}), that is, an expession dependent of
$u^j_\sigma$ whose total derivative $\bar D_y$ vanishes on this
equation. Namely, we have
\begin{equation}\label{EMT}
T=\tfrac{1}{2} {\sum_{i,j=1}^r} \kappa_{ij}u^i_xu^j_x
- {\sum_{i=1}^r} a_i\cdot{u^i_{xx}}\in\ker\bar D_y.
\end{equation}
Both components $T$ and $\bar T$ of the traceless energy\/--\/momentum
tensor $\Theta$ for the Euler equations~\eqref{ImEulerSpoilt} are well
known to be of form~\eqref{EMT} up to the complex conjugation:
\[\Theta=T\,\Id x+\bar T\,\Id y,\] see, \textit{e.g.},~\cite{Ovch}.
In Sec.~\ref{SecLegendre} of Chapter~2 we shall discuss some aspects
of the Hamiltonian formalism for the Toda equations
and derive~\eqref{EMT} from density~\eqref{TodaLDensity}
of the Lagrangian.
Meanwhile, we construct the differential span $\bT$ of the minimal
integral $T\in\ker\bar D_y$.
By definition, put \[T_j\equiv \bar D_x^j(T).\]
The differential consequences $T_j$ to the functional $T$ generate
the subspace $\bT\subset\ker\bar D_y$ within the kernel of the total
derivative $\bar D_y$. Indeed, any smooth function $Q$ supplies the
functional
$$Q(x,\bT)\equiv Q(x,T,T_1,\ldots,T_\mu)\in\ker\bar D_y.$$
We say that the nondegenerate symmetrizable matrix $K$ is
\emph{generic} if the integral $T$ in Eq.~\eqref{EMT} is a unique
solution to the equation $\bar D_y(T)=0$ on the corresponding Toda
equation~\eqref{eqToda}.

We emphasize that by a special choice of the matrix $K$ one can achieve
the situation such that the functional $T$ will \emph{not} be a unique
integral for the Toda equation~\eqref{eqToda}.
The criteria for the equality
$\dim\ker\bar D_y=2$ to hold at $r=2$ are found in the
paper~\cite{LeznovSmirnovShabat}.
From now on, by $\mathbf{\Omega}\subseteq\ker\bar D_y$ we denote the
subspace in $\ker\bar D_y$, which is differentially generated by
\emph{all} solutions $\Omega^i$ to the equation $\bar D_y(\Omega^i)=0$.
The total number of these solutions is denoted by $q$: $1\leq
i\leq q\leq r$. We always assume $\Omega^1\equiv T$.

\begin{example}[\textup{\cite{LeznovSmirnovShabat}}]
Consider the Cartan matrix
$K=\left(\begin{smallmatrix}
\pp2 & -1\\
-1   & \pp2
\end{smallmatrix}\right)$
of the Lie algebra $\mathfrak{sl}_3(\BBC)$.
Then the Toda equation~\eqref{eqToda} associated with $K$ admits two
integrals. The first integral $\Omega^1=T$, which is defined in
Eq.~\eqref{EMT}, acquires the form
\begin{subequations}\label{OmA2}
\begin{gather}
\Omega^1= { {(u^1_x)}^2 }-u^1_x u^2_x+
   { {(u^2_x)}^2 }-u^1_{xx}-u^2_{xx},
\label{Om1A2} \\
\intertext{while the second integral is}
\Omega^2=u^1_{xxx}+u^1_x\cdot(u^2_{xx}-2u^1_{xx})+
 { {(u^1_x)}^2 }\cdot
u^2_x -u^1_x\cdot { {(u^2_x)}^2 }.\label{Om2A2}
\end{gather}
\end{subequations}
The integral $\Omega^2$ is a solution to the equation
$\bar D_y(\Omega)=0$ but does not follow from~$\Omega^1$.
\end{example}

By~\cite{ShabatYamilov},
a necessary and sufficient condition for $r$ nontrivial independent
solutions $\Omega^i$ of the equation $\bar D_y(\Omega^i)=0$ to exist
is that $K$ be the Cartan matrix of a semisimple Lie algebra~$\gothg$.
The Toda equations associated with $\gothg$ are exactly
integrable~(\cite{LeznovIntegrate}).

\subsubsection*{The symmetry algebra of the Toda
equations}\label{SecTodaSym}
In this subsection, we assign classes of infinitesimal symmetries of
the Toda equation to the functional span $\mathbf{\Omega}$ of
differential consequences to the integrals $\Omega^i$, which were
introduced in the previous subsection.
Then, in Sec.~\ref{SecTodaNoeth} we find out which out these symmetries
are the Noether symmetries of the Lagrangian~$\cL_\Toda$.

By $\vec\Delta=|\Delta^i|$ denote the vector
of the conformal dimensions~(\cite{Strings})
\[{\Delta^i=\sum\limits_{j=1}^r k^{ij}}\]
of the Toda fields
$\exp(u)\equiv {{}^t(}\exp(u^1)$, $\ldots$, $\exp(u^r))$. This
notation is well\/--\/defined due to

\begin{state}[\textup{\cite{BilalGervaisLagrange}}]\label{FinSymState}
\begin{enumerate}\item
The transformation
\begin{equation}\label{FiniteSymDiffeo}
\begin{aligned}
x&\mapsto\cX(x),\\ y&\mapsto\cY(y),\\
u^i(x,y)&\mapsto\tilde u^i=u^i(\cX,\cY)+\Delta^i\,\log\cX'(x)\cY'(y)
\end{aligned}
\end{equation}
is a finite conformal symmetry of the Toda
equation~$\cE_\Toda$\textup{.}
\item
The Lagrangian $\cL_\Toda=\int L_\Toda\,\Id x\wedge\Id y$
is invariant with respect to\ this change\textup{.}
\item
By definition, put $\beta\equiv {\sum\limits_{i=1}^r}
 a_i\cdot\Delta^i$\textup{;}
under diffeomorphism~\eqref{FiniteSymDiffeo}, the component $T$ of the
energy\/--\/momentum tensor~$\Theta$ is transformed by the rule
\begin{equation}\label{FiniteEMTDiffeo}
T[u]\mapsto\left(\cX'(x)\right)^2\cdot T[\tilde u(\cX,\cY)]-
\beta\cdot\Bigl(\frac{\cX'''(x)}{\cX'(x)}-\frac{3}{2}
\Bigl(\frac{\cX''(x)}{\cX'(x)}\Bigr)^2\Bigr).
\end{equation}
\end{enumerate}
\end{state}

\begin{rem}\label{RemWeAdapt}
The symmetry properties, Eq.~\eqref{FiniteSymDiffeo} and
\eqref{FiniteEMTDiffeo}, of the Toda equation associated with the Lie
algebras were considered in the paper~\cite{BilalGervaisLagrange}.
The present formulation of Proposition~\ref{FinSymState} is an
extension of the cited result to the Toda equation~\eqref{eqToda}, which is now
associated with an arbitrary nondegenerate symmetrizable $r\times
r$-matrix $K$. The coefficients $\vec\Delta$ and $\beta$ are already
adapted to this general case.
\end{rem}

The infinitesimal variant of Proposition~\ref{FinSymState} is
\begin{state}\label{InfSymState}
\textup{1 (\cite{MeshkovTMPh, Ovch}).}
The infinitesimal components of conformal
symmetries~\eqref{FiniteSymDiffeo} of the Toda equation
are \[{\vph_0^f}=\square(f(x))\] up to the transformation
$x\leftrightarrow y$. Here $f$ is an arbitrary smooth function and
\begin{equation}\label{Square}
\square=\vec u_x+ {\vec\Delta}\cdot\bar D_x
\end{equation}
is a vector\/--\/valued, first\/--\/order
differential operator\textup{.}\\
\textup{2 (\cite{Vestnik2000}).}
Each point symmetry $ {\vph_0^f}$ is a Noether symmetry
of the Lagrangian $\cL_\Toda$\textup{:}
\[\cEv_{\vph_0^f}(L_\Toda\,\Id x)\in\mathrm{im}\,\Id_h.\]
\textup{3 (\cite{Ovch, BilalGervaisIntegrate}).}
The functional $T$, which was defined in Eq.~\eqref{EMT}, is a density of the
Hamiltonian for the infinitesimal conformal symmetry
$ {\vph_0^f}$\textup{:}
\begin{equation}\label{EMTGeneratesVph}
 {\vph_0^f}=A_1\cdot\bE_u\bigl(T\cdot f(x)\,\Id x\bigr),
\quad\text{где $A_1= {\hat\kappa^{-1}\cdot D_x^{-1}}$.}
\end{equation}
\end{state}

The third statement in Proposition~\ref{InfSymState} was formulated in local
coordinates in the paper~\cite{Ovch}.
In Sec.~\ref{SecLegendre} of Chapter~2 we establish the relation
between the canonical Hamiltonian formalism for the Toda equation and
the Hamiltonian operator $A_1$ above. In Theorem~\ref{CylinderTh} on
page~\ref{CylinderTh} we find out that equality~\eqref{EMTGeneratesVph}
is the root part of the infinite sequence of relations between the
hierarchy of the higher analogs of the Korteweg\/--\/de Vries
equation, see Eq.~\eqref{IKdV} on page~\pageref{IKdV},
and the commutative hierarchy within the Noether
symmetries algebra for the Toda equation (this hierarchy is constructed in
Sec.~\ref{SecAandB}).

\begin{lemma}\label{2=3}
Suppose $K$ is a symmetrizable $r\times r$-matrix
and the operator $\square$ is defined by Eq.~\eqref{Square}\textup{.}
Then the relations
\begin{equation}\label{FactorizeLin}
\begin{aligned}
\bar\ell_F\circ\square&=\bar D_x\circ\square\circ\bar D_y,\\
\bar\ell_F^*\circ\hat\kappa\circ\square&=\bar D_x\circ\hat\kappa
\circ\square\circ\bar D_y
\end{aligned}
\end{equation}
hold.
\end{lemma}

\begin{proof}
Here we express the first relation in local coordinates:
\begin{multline*}
\bar\ell_F\circ\square=\Bigl\|\delta_{ij}\,\bar D_{xy} -
k_{ij}\exp\Bigl(\sum_l k_{il}u^l\Bigr)\Bigr\|\cdot
\bigl|u^j_x+\Delta^j\cdot\bar D_x\bigr| ={}\\
\begin{aligned}
{}&=\bu_x\,\bar D_{xy}+\bu_{xx}\,\bar D_y+\bu_{xy}\,\bar D_x
+\bu_{xxy}+\vec\Delta\,\bar D_{xxy}-{}\\
{}&   
{}-\Bigl|\sum_j k_{ij}u^i_x\exp\Bigl(\sum_l k_{il}
u^l\Bigr)\Bigr| -
\Bigl|\sum_{j,p} k_{ij}k^{jp}\exp\Bigl(\sum_l k_{il}
u^l\Bigr)\Bigr|\cdot\bar D_x = {}\\
&=\bar D_x\circ {\bigl|\bu_x+\vec\Delta\,\bar D_x\bigr|}\circ\bar D_y,
\end{aligned}
\end{multline*}
Q.\,E.\,D.

The second relation is deduced from the first one by using
Lemma~\ref{BehaviourLemma} and the Helmholtz condition
\[\bar\ell_{\bE(\cL_\Toda)}= {\bar\ell^*_{\bE(\cL_\Toda)}}\]
that holds since the matrix $\kappa={}^t\kappa$ is symmetric.
\end{proof}

\begin{cor}
The vector\/--\/functions
\begin{equation}\label{symToda}
\vph=\square(\phi(x,\mathbf{\Omega}))
\end{equation}
are symmetries of the Toda equation\textup{:} $\vph\in\sym\cE_\Toda$
for any function $\phi$ that depends on an arbitrary subset
$\mathbf{\Omega}$ of the integrals $\Omega^i_j\equiv\bar
D_x^j(\Omega^i)\in\ker\bar D_y$.
\end{cor}

Formula~\eqref{symToda} gives the description of the whole symmetry
algebra $\sym\cE_\Toda$:

\begin{state}[\textup{\cite{MeshkovTMPh}}]\label{MeshkovExhaust}
\begin{enumerate}\item
Let the assumptions of Lemma~\textup{\ref{2=3}} hold\textup{;}
then any symmetry of Eq.~~\eqref{eqToda} is~\eqref{symToda}.
\item
Suppose further $K$ is such that there exist $q$ independent integrals
$\Omega^i\in\ker\bar D_y$\textup{,} where $1<q\leq r$\textup{.}
Put
\[f^i=\exp(\sum\limits_{j=1}^r k_{ij}u^j)
\text{ and }f^i_{\,j}=k_{ij}\cdot f^i.\]
Assume further that there are $r$ constant $(r\times q)$-matrices
$M_\alpha=\|(M_\alpha)^i_{\,j}\|$\textup{,}
$1\leq i\leq r$\textup{,}
$1\leq j\leq q$\textup{,}
and a constant $(r\times q)$-matrix
$\Delta=\|\Delta^i_{\,j}\|$\textup{,} $1\leq i\leq r$\textup{,}
$1\leq j\leq q$\textup{,}
such that $\mathrm{rank}\,\Delta=q$ and the equations
\[
\Delta^\beta_{\,i}\,f^\alpha_\beta=(M_\beta)^\alpha_i\,f^\beta,\qquad
f^\alpha_\beta\,(M_\gamma)^\beta_i =
  (M_\beta)^\alpha_i\,f^\beta_\gamma
\]
hold. Now, construct the first\/--\/order, $(r\times q)$-matrix
differential operator
\begin{equation}\label{SquareMatrix}
\square= {\sum_{\alpha=1}^r} M_\alpha\cdot u^\alpha_x +
   \Delta\cdot\bar D_x  
\end{equation}
and consider a vector $\phi=|\phi^i(x,\mathbf{\Omega})|$\textup{,}
$1\leq i\leq q$\textup{.}
Under these assumptions, the sections defined in Eq.~~\eqref{symToda}
exhaust all symmetries of the Toda equation~$\cE_\Toda$.
\end{enumerate}
\end{state}

Therefore, in both cases we have
$$\sym\cE_\Toda\simeq\{\vph^i=\square^i_{\,j}\phi^j(x,
\mathbf{\Omega}) \text{ mod } (x\leftrightarrow y)\},$$
where the number of columns in the operator $\square$ is equal to the
number $q$ of the independent integrals $\Omega^l$ and $\phi$ is an
arbitrary vector.

\begin{cor}\label{StArbitraryGS}
Any solution $\psi$ to the equation
$\bar\ell_F^*(\psi)$ ${}=0$ for system~\eqref{eqToda} is of the form
\begin{equation}\label{gsToda}
\psi=\hat\kappa\bigl(\square(\phi(x,\mathbf{\Omega}))\bigr).
\end{equation}
\end{cor}

We stress that the problem of finding the integrals~$\mathbf{\Omega}$
is primary for the Toda equations, and the search for the symmetries
$\vph$ as well as the selection of the Noether symmetries $\vph_\cL$
follow this problem.
We also note that each conformal symmetry~\eqref{FiniteSymDiffeo} of
the Toda equation is Noether, \textit{i.e.}, preserves the
Lagrangian~$\cL_\Toda$. Still, not each section $\vph$ of
type~\eqref{symToda} is a Noether symmetry of Eq.~\eqref{eqToda}.

\section{The Noether symmetries of the Toda
equation}\label{SecTodaNoeth}
%
%
First, we recall Example~\ref{EvolutAreNormal} on
page~\pageref{EvolutAreNormal} and obtain an important property of the Toda
equation.
This property permits the application of the generating
sections machinery in description of the conservation laws and the
Noether symmetries of the Toda equation. Namely, we have

\begin{lemma}\label{eqTodaIsLNormal}
The Toda equation $\cE_\Toda$ is $\ell$-normal.
\end{lemma}

\begin{proof}
By Example~\ref{EvolutAreNormal} on page~\pageref{EvolutAreNormal},
it is sufficient to represent the equation $\cE_\Toda$
in the evolutionary form.
Let
\[\xi=x+y,\quad \eta=x-y\] be the new independent variables.
The way to choose them is such that the equations
$$
u_{\xi\xi}^i-u_{\eta\eta}^i=\exp {\Bigl(
\sum_{j=1}^r k_{ij}u^j\Bigr)}
$$
hold. Now, put $v^i\equiv u_\eta^i$; then the equations
$\cE_{\text{ev}}\subset J^2(\BBR^2$, $\BBR^{2r})$ of the form
$$
\Bigl\{
u^i_\eta = v^i,\qquad
v^i_\eta = u^i_{\xi\xi} - \exp\Bigl(\sum_{j=1}^r
  k_{ij}u^j\Bigr)\Bigr\}
$$
are the required evolutionary representation of Eq.~\eqref{eqToda}.
\end{proof}

From relation~\eqref{ImEulerSpoilt} and Corollary~\ref{Podkrutka}
we deduce the relation $\psi=\kappa\,\vph_\cL$ between the Noether
symmetries and the generating sections of conservation
laws for the Toda equations.
This observation allows to specify a property common for all
the integrals $\Omega^i$  for Eq.~\eqref{eqToda}. Namely, suppose
$$\Id_h(\Omega^i\,\Id x)=-\tilde\nabla_i(F)\,\Id x\wedge\Id y$$
for any admissible $i$ and consider the conservation law
$[\eta]=\bigl[Q(x$, $\mathbf{\Omega})\,\Id x\bigr]$. Then we have
\[
\Id_h Q(x,\mathbf{\Omega})\,\Id x = - {\sum_{i,j}} \frac{\dd
Q}{\dd\Omega^i_j}\,D_x^j\circ\tilde\nabla_i(F)\,\Id x\wedge\Id y.
\]
By definition, the generating section $\psi_\eta$ of the conservation law
$[\eta]$ is
\begin{equation}\label{EulerWRTInt}
\psi_\eta=-\sum_{i,j}(-1)^j{\bigl(
\tilde\nabla_i\bigr)}^*\circ D_x^j
\left(\frac{\dd Q}{\dd \Omega^i_j}\right) =
-\sum_i {\bigl(\tilde\nabla_i\bigr)}^*\circ\bE_{\Omega^i}(Q).
\end{equation}
Now we compare Eq.~\eqref{gsToda} with
Eq.~\eqref{EulerWRTInt} and, by using Theorem~\ref{InverseNoether} and
Lemma~\ref{BehaviourLemma}, we obtain

\begin{theor}\label{ThNoetherStructure}
\begin{enumerate}\item
Suppose $\Omega^i$ is an integral for the Toda equation~\eqref{eqToda}
such that $D_y(\Omega^i)=\tilde\nabla_i(F)$.
Then there is the operator $\nabla_i$ such that
\[\tilde\nabla_i=\nabla_i\circ\square^*\circ\hat\kappa.\]
In particular\textup{,} consider the integral
$\Omega^1=T$ defined in Eq.~\eqref{EMT}. Then \[\nabla_1=\mathbf{1}.\]
\item
The Noether symmetries of the Toda equation are
\[
\vph_\cL=\square\circ {\sum_i}
\nabla_i^*\circ\bE_{\Omega^i}(Q(x,\mathbf{\Omega})),
\]
where $\Omega^i\in\ker\bar D_y$ are the integrals for the
equation~ $\cE_\Toda$\textup{,}
$$\bE_{\Omega^i}= {\sum\limits_{j\geq0}}(-1)^jD_x^j\cdot
\frac{\dd}{\dd\Omega^i_j}$$
is the Euler operator with respect to\ $\Omega^i$\textup{,}
$\mathbf{\Omega}$ is an arbitrary set of differential consequences
to $\Omega^i$\textup{,} and $Q$ is a smooth function.
\end{enumerate}
\end{theor}

\begin{example}
Again, consider the Toda equation~\eqref{eqToda} associated with the
root system $A_2$. Then we have $r=2$, $a_i=|\alpha_i|^{-2}=1$ for
$i=1$, $2$, and
\[
K=\begin{pmatrix}\pp2 & -1\\ -1& \pp2\end{pmatrix},\quad
K^{-1}=\frac{1}{3}\cdot\begin{pmatrix}2 & 1\\ 1& 2\end{pmatrix},\quad
\vec\Delta=\binom{1}{1}.
\]
Put $\square=\vec{u}_x+\vec\Delta\cdot\bar D_x$, see~\eqref{Square}.
The integrals $\Omega^1$ and $\Omega^2$ are defined in Eq.~\eqref{OmA2}.
One easily checks that $D_y(\Omega^1)=\square^*\circ\hat\kappa(F)$.
Therefore, $\nabla_1=\mathbf{1}$ and
\[D_y(\Omega^2)=-D_x\circ\square^*\circ\hat\kappa(F),\] thence,
$\nabla_2=-D_x$. We emphasize that the integral $\Omega^2$ is
\emph{not} equivalent to~$-D_x(\Omega^1)$.
\end{example}

\begin{rem}\label{RemOnSakovich}
In Theorem~\ref{ThNoetherStructure} we established
the relation $\nabla_1=\mathbf{1}$ to hold for the minimal
integral $T$, which is defined in Eq.~\eqref{EMT}.
Now we restrict all previous reasonings onto the subspace
$\{Q(x$, $\bT)\}\subset\ker\bar D_y$ of the kernel of the total
derivative. Thus, we consider the subspace that is generated
by the integral $T$ and its differential consequences.
Then, by Theorem~\ref{InverseNoether}, the conservation laws
$[Q\,\Id x]$ for the Toda equation $\cE_\Toda$ are in one\/-\/to\/-\/one
correspondence with the Noether symmetries \label{NoetherSymOnT}
\begin{equation}\label{EqDescribeNoether}
\vph_\cL=\square\circ\bE_T(Q(x,\bT)).
\end{equation}

In other words, the Noether symmetries $\vph_\cL$ of the Toda
equation \eqref{eqToda} associated with a generic nondegenerate
symmetrizable matrix $K$ are \eqref{EqDescribeNoether}
up to the transformation~$x \leftrightarrow y$.
This statement is an extension of the relation
(\cite{SakovichZCR})
between the Noether
symmetries and conservation laws for the scalar Liouville equation
onto $r\geq1$. We note that our scheme of
reasonings is essentially more simple than the straightforward proof
for the Liouville equation~(\textit{ibid}).

In the rest part of this section, we generalize the method of the paper
\cite{SakovichZCR}. We solve the equation \[\cEv_\vph(\cL_\Toda)=0\]
directly with respect to\ $\vph$ and thus find out which of the symmetries
$\vph=\square(\phi(x$, $\bT))\in\sym\cE_\Toda$ preserve the Lagrangian
$\cL_\Toda$ for the Toda equations. We follow the
paper~\cite{VestnikNoether}.
\end{rem}

Detailed analysis of some algebraic aspects for Eq.\ \eqref{eqToda}
shows (technically, we calculate the term
$E^{0,2}_2(\cE_\Toda)$ of the Vinogradov's $\cC$-spectral sequence,
see \cite{Vin84})
that the following two conditions are equivalent:
\[
\cEv_\vph(\cL_\Toda)=0\quad\Longleftrightarrow\quad
\bE_u(\cEv_\vph(\cL_\Toda))=0,
\]
\textit{i.e.}, a density is a total direvgence iff its variation is zero.
We note that this question was not discussed in \cite{SakovichZCR}.
Further on, we get
\begin{lemma}\label{SakovichGuessed}
The relation $\bE_{{u}}(\cEv_\vph(\cL_\Toda))=0$ is equivalent to the
condition
$$
-\bE_{{u}}\left(D_y(T)\cdot\phi(x,T,\ldots,T_m)\right)=0,
$$
where $\vph=\square(\phi)$ and the total derivative $D_y$ is \emph{not}
restricted onto $\cE_\Toda$.
\end{lemma}
\noindent%
The proof of Lemma \ref{SakovichGuessed}
is based on multiple use of the relation
$$\sum\limits_{j=1}^r\kappa_{ij}\Delta^j=a_i.$$

\begin{lemma}\label{OrderEven}
The highest order $m$ of the derivative $T_m$ in the set of $\phi$'s
arguments is even\textup{:}
$m=2\mu$, and $\phi=\phi(x,T,\ldots,T_{2\mu})$.
\end{lemma}
\begin{proof}
Put $G_i=\dfrac{\delta}{\delta u^i}
(\cEv_\vph(\cL_\Toda))$
and by $u^j_{(k,l)}$ denote the derivative $D_x^k\circ D_y^l(u^j)$.
Then we have
$$
\frac{\dd G_i}{\dd u^j_{(m+4,1)}} = -\left[(-1)^3+(-1)^{m+2}\right]
\cdot {a_i \, a_j} \cdot \frac{\dd\phi}{\dd T_m} = 0,
$$
whence $(-1)^{3}+(-1)^{m+2}=0$ and thus $m$ is even.
\end{proof}

\begin{lemma}\label{HeadNude}
The derivatives \[\frac{\dd^2\phi}{\dd T_m\dd T_l}\] vanish for all $l$
such that $\mu<l\leq m$.
\end{lemma}
\begin{proof}
It suffices to check that \[\frac{\dd G_i}{\dd u^j_{2m+4}},\quad
\frac{\dd G_i}{\dd u^j_{2m+2}},\quad \ldots,
\frac{\dd G_i}{\dd u^j_{m+6}}\] are equal to~$0$. This can be done
straighforwardly.
\end{proof}

Still, the derivative \[\frac{\dd^2\phi}{\dd T_\mu\dd T_{2\mu}}\]
is, in general, nontrivial owing to
\begin{lemma}\label{DirectEuler}
The identity
\[  
\bE_{{u}}(-D_y(T)\cdot\bE_T(P(x,T,\ldots,T_\mu)))\equiv0
\]  
holds for any function $P$.
\end{lemma}
\begin{proof}
Indeed, we have
\begin{multline*}
D_y(T)\cdot\bE_T(P)\,\Id x\wedge \Id y=\langle
D_y(T),{\ell_P^{(T)}}^*(1)\rangle=
\langle\ell_P^{(T)}(D_y(T)),1\rangle+\Id_h\gamma={}\\
\langle\cEv^{(T)}_{D_y(T)}
P(x,T,\ldots,T_\mu),1\rangle+\Id_h\gamma=
\langle D_y(P),1\rangle+\Id_h\gamma\in\ker\bE_{{u}},
\end{multline*}
where the coupling $\langle\,{}$, $\rangle$ takes values in the
horizontal $2$-forms $\omega=f\cdot\Id x\wedge\Id y$,
$\Id_h=\sum\limits_i\Id x^i\otimes D_i$ is the horizontal differential,
$\Id_h\gamma\in\ker\bE_u$ is an exact form, and both the
evolutionary derivation and the linearizations are evaluated with respect to
$T$.
\end{proof}

\begin{state}\label{ReconstructState}
A symmetry $\vph=\square(\phi(x$, $T$, $\ldots$,
$T_m))\in\ker \bar\ell_F$ is a Noether symmetry of Toda's equation
\eqref{eqToda} iff the following two conditions hold\textup{:}
\begin{enumerate}
\item $m=2\mu$, $\mu\geq0$, and
\item the function
$\phi=\bE_T(Q(x,T,\ldots,T_\mu))\in\mathrm{im}\,\bE_T$ lies in the image
of the Euler operator with respect to\ $T$, where $Q$ is an arbitrary smooth
function.
\end{enumerate}
\end{state}
\begin{proof}
We show that $\phi\in\mathrm{im}\,\bE_T$ by induction.\,Choose
$P=P(m;x,T$, $\ldots$, $T_\mu)$ such that
\begin{equation}\label{SuitableCompence}
\frac{\dd^2P(m)}{\dd T_\mu^2}=(-1)^{\mu}\cdot\frac{\dd\phi}{\dd T_m}
\end{equation}
and put
$\tilde\phi\stackrel{\mathrm{def}}{=}\phi-\bE_T(P(m))$, then we get
\[\frac{\dd\tilde\phi}{\dd T_m}\equiv0.\] By Lemma \ref{OrderEven}, the
equation
$$
\bE_{{u}}(D_y(T)\cdot\tilde\phi(x,T,\ldots,T_{m-2}))=0
$$
holds. By using Lemma \ref{HeadNude}, we choose $P(m)$
in accordance with \eqref{SuitableCompence} and
apply Lemma \ref{DirectEuler}.
Therefore, we decrease the order $m=2\mu$ to $0$ with the step $2$
inductively. Finally, we obtain
\[  
\phi=\bE_T\Bigl(\sum\limits_{i=0}^{\mu}P(2i;x,T,\ldots,T_i)\Bigr).
\]  
The proof is complete.
\end{proof}

Thus, we have obtained another description of the Noether symmetries'
class~~\eqref{EqDescribeNoether} for the Toda equation assigned to a
generic matrix~$K$.

\section{Recursion operators for the Toda
equation}\label{SecTodaRec}
In this section, we construct a continuum of the recustion operators,
which are either
local or nonlocal with respect to\ $D_x$, for the symmetry algebra of the
Toda equation. Although the structure of the symmetry algebra itself is
known, see Eq.~\eqref{symToda}, presence of the recursion operators
gives us additional information about the Toda equation and permits to
establish the relation between $\cE_\Toda$ and other mathematical
physics equations.

The explicit method by J.\,Krasil'shchik that allows construction of
the recursion operators for symmetry algebras of differential equations
was briefly described in the Introduction on
pages~\pageref{ISecRec}--\pageref{IExampleGenForms}.
The theorem below contains the result of application of this method to
the Toda equation~\eqref{eqToda} associated with a nondegenerate
symmetrizable $(r\times r)$-matrix.

\begin{theor}[\textup{\cite{TodaLawsActa}}]\label{ThContinuumTodaRec}
\begin{enumerate}\item
Equation~\eqref{eqToda} admits a continuum of
\emph{local} recursion operators
$R\colon\sym\cE_\Toda\to\sym\cE_\Toda$, which are
\[
R=\square\circ {\sum_{i,j}} f_{ij}(x,\mathbf{\Omega})\cdot
\bar D_x^j\circ \ell_{\Omega^i}.
\]
Here $f_{ij}$ are arbitrary smooth functions and the linearizations
$\ell_{\Omega^i}$ with respect to\ the integrals $\Omega^i$ for the Toda
equation are
\begin{equation}\label{LinIntegral}
\ell_{\Omega^i}=\Bigl(\ldots,\underbrace{\sum_\sigma
\frac{\dd\Omega^i}{\dd u^k_\sigma}\cdot
\bar D_\sigma}_{\text{$k$\textup{th} component}},\ldots\Bigr).
\end{equation}
\item
There is a continuum of \emph{nonlocal} recursion operators
for Eq.~\eqref{eqToda}, which are constructed in the following way.
Assign the nonlocal variables $s^i$ to the integrals $\Omega^i$ by the
compatible differentiation rules $s^i_x=\Omega^i$ and
$s^i_y=0$\textup{.} The linearizations $\ell_{s^i}$ are defined by the
formulas \[\ell_{s^i}=\bar D_x^{-1}\circ\ell_{\Omega^i},\]
where $\ell_{\Omega^i}$ are calculated by Eq.~\eqref{LinIntegral}.
Then the required recursion operators are
\[
R=\square\circ {\sum_i}
f_i(x,\mathbf{s},\mathbf{\Omega})\cdot\bar
D_x^{-1} \circ\ell_{\Omega^i},
\]
where the functions $f_i$ are arbitrary.
In general, these operators do not preserve the locality of
elements~\eqref{symToda} of the symmetry algebra~$\sym\cE_\Toda$.
\end{enumerate}
\end{theor}

\begin{proof}
First, enlarge the set of the dependent variables $u_\sigma^j$
by introducing the nonlocals $s^i$ such that their derivatives are
\begin{equation}\label{pKdVpmKdV}
s^i_x=\Omega^i,\qquad s_y=0.
\end{equation}
(In fact, any definition of $s^i_y$ is allowed if it is
compatible with the condition $s^i_{xy}=s^i_{yx}=0$).
Then, extend the total derivatives:
\[
\tilde D_x=\bar D_x+\sum\limits_i\Omega^i\,\frac{\dd}{\dd s^i},\qquad
\tilde D_y=\bar D_y,
\]
such that $[\tilde D_x,\tilde D_y]=0$.
The Cartan's flat connection is now defined on the equation
\begin{multline*}
\tilde\cE^\infty=\{\tilde D_x^{k+1}(s^i)=\bar D_x^{k}(\Omega^i),\ %
\tilde D_x^k\circ\tilde D_y(s^i)=0,\ k\geq0;\\
\bar D_\sigma(F)=0,\ |\sigma|\geq0\}.
\end{multline*}
The Cartan's generating $1$-forms
\[\omega_\Toda\in C^\infty(\tilde\cE^\infty)
\otimes\cC\Lambda^1(\tilde\cE^\infty)\]
for the recursion operators $R_\Toda$ satisfy the determining
equation \[\tilde\ell_\Toda^{[1]}(\omega_\Toda)=0,\] where
$\tilde\ell_\Toda^{[1]}$ is the restriction of the linearization
of Eq.~\eqref{eqToda} onto $H_\cC^{1,0}(\tilde\cE)$.
From the factorization
\[\bar\ell_\Toda\circ\square=\bar D_x\circ\,\square\,\circ \bar D_y\]
in Eq.~\eqref{FactorizeLin} it follows that any Cartan's $1$-form
\begin{equation}\label{gfToda}
\omega_\Toda=\square\Bigl(\sum_{i\geq0} f_i(x,
\mathbf{s},\ldots,\tilde
D_x^{k_i}(s^j))\cdot\Id_\cC\bigl(\tilde D_x^i(s^l)\bigr)\Bigr)
\end{equation}
is an element of the kernel $\ker\tilde\ell_\Toda^{[1]}$,
whence we get the statement of the theorem.
In particular, suppose that each $f_i$ does not depend on the nonlocal
variables $\mathbf{s}$. Then the resulting recursion operator is local.
\end{proof}

\begin{example}\label{OurTodaRecExample}
Consider the integral $T$ defined in Eq.~\eqref{EMT}. Its
linearization is
\begin{equation}\label{LinTWRTu}
\ell_T=\Bigl(\ldots,
\underbrace{\sum_{j=1}^r\kappa_{ij}u^j_x\cdot\bar
D_x-a_i\cdot\bar D_x^2}_{\text{$i$th component}},\ldots\Bigr) =
\square^*\circ D_x\circ\hat\kappa.
\end{equation}
Now, introduce the nonlocal variable $s$: we set $s_x=T$ and $s_y=1$.
The compatibility condition for the variable $s$ is
\begin{equation}\label{eqWaveNonlocal}
s_{xy}=0.
\end{equation}
Finally, we construct the recursion operator
\[
R_\Toda=\square\circ\bar D_x^{-1}\circ\ell_T.
\]
Apply the operator $R_\Toda$ to the translation $u_x\in\sym\cE_\Toda$.
We obtain the symmetry sequence \[\vph_k=\square(\phi_{k-1})\] that
corresponds to the sequence of functions
\begin{multline*}
\phi_{-1}=1,\quad \phi_0=s_1,\quad \phi_1=-\beta s_3+ {\tfrac{3}{2}s_1^2},\\
\phi_2=\beta^2s_5- {\tfrac{5}{2}\beta s_2^2}-5\beta s_1s_3+
 {\tfrac{5}{2}s_1^3},
\end{multline*}
\textit{etc}.
In Chapter~2 we shall investigate the properties of this recursion
operator $R_\Toda$ together with the properties of the symmetry sequence
$$\gA=\{\vph_k\equiv R_\Toda^k(\vph_0),\ \vph_0= {\vu_x}\}$$
in detailes. Meanwhile, we claim that the symmetries $\vph_k$ of the
Toda equation, which are obtained by multiple action of $R_\Toda$ to
the translation $\vph_0=\vu_x$, are local, Hamiltonian, and commute
with each other.
Also, in the next section
we shall establish the relation between the symmetry sequence $\gA$,
the Korteweg\/--\/de Vries equations~\eqref{IKdV} and
\eqref{IpKdV}, and recursion operators~\eqref{IKdVRec} for the latter
equations.
\end{example}

\newpage
\specialsection*{\textbf{Chapter~2.}
The Korteweg\/--\/de Vries hierarchies
and the Toda equations}\label{ChKdV}
In this chapter, we construct the commutative Hamiltonian hierarchy
$\gA$ of multi\/--\/component analogs for the potential modified
Korteweg\/--\/de Vries equation. This hierarchy is identified with the
commutative Lie subalgebra in $\sym\cE_\Toda$ composed by
certain Noether's symmetries. Also, we discuss some aspects of the
Hamiltonian formalism for the Toda equations themselves and establish a
link between the hierarchy $\gA$ and the higher Korteweg\/--\/de Vries
equations for Eq.~\eqref{IKdV}. The exposition follows the
papers~\cite{KdVHier, KisOvchTodaHam}.

We start our study of the relation between the Toda
equation~\eqref{eqToda} and classical equations~\eqref{IKdV}
and~\eqref{IpKdV} of mathematical physics, the Korteweg\/--\/de Vries
equations, with the following

\begin{example}\label{ExampleKdV}
Consider the hyperbolic Liouville equation
\begin{equation}\label{eqhyp}
\cE_{\mathrm{Liou}}=\{u_{xy}-\exp(2u)=0\}.
\end{equation}
The minimal integral, see Eq.~\eqref{EMT}, for the latter equation
is~(\cite{Shabat, SokolovUMN})
\begin{equation}\label{MinimalIntegrals}
T=u_1^2-u_2,\quad \bar D_y(T)=0.
\end{equation}
Introduce the nonlocal variable $s$ such that
\begin{equation}\label{pKdVKdV}
s_x=T,\qquad s_y=1,
\end{equation}
and put
\begin{equation}\label{pmKdVmKdVLiou}
\vth\equiv 2 u_1.
\end{equation}
Then, consider the symmetry
\[
\vph=(u_1+\tfrac{1}{2}\bar D_x)(T)
\]
of the Liouville equation and calculate the evolution of the variables
$u$, $\vth$, $T$, and $s$ along this symmetry. We get
\begin{align}
u_t&=-\tfrac{1}{2}u_3+u_1^3&&
\text{(potential mKdV)}\label{pmKdVLiou}\\
T_t&=-\tfrac{1}{2}T_3+3TT_1&&\text{(KdV)}\label{KdVLiou}\\
s_t&=-\tfrac{1}{2}s_3+\tfrac{3}{2}s_1^2&&
   \text{(potential KdV).}\label{pKdVLiou}
\end{align}
The Miura transformation (see~\cite{MiuraRef,RogersShadwick})
acquires the form
\begin{subequations}\label{MiuraLiou}
\begin{align}
\vth_1&=\mp 2T\mp\tfrac{1}{2}\vth^2, &
\vth_t&=\pm T_2-(\vth T_1+\vth_1 T)
\label{MiuraA} \\
T&=\mp\tfrac{1}{2}\vth_1-\tfrac{1}{4}\vth^2.&&\label{MiuraB}
\end{align}
\end{subequations}
One can treat relations~\eqref{MiuraLiou} as B\"acklund transformation
between the Korteweg\/--\/de Vries equation, see Eq.~\eqref{KdVLiou}, and
the equation
\begin{equation}\label{mKdVLiou}
\vth_t=-\tfrac{1}{2}\vth_3+\tfrac{3}{4}\vth^2\vth_1\qquad\qquad
   \text{(modified KdV).}
\end{equation}
The signs `$\pm$' and `$\mp$' in Eq.~\eqref{MiuraLiou} are induced by the
symmetry $\vth\mapsto-\vth$ of equation~\eqref{mKdVLiou}. This discrete
transformation provides B\"acklund autotransformation (\cite{ClassSym,
Wahlquist}) for Eq.~\eqref{KdVLiou}.

The recursion operator
\[
R_\Liou=D_x^2-2u_1+D_x^{-1}u_1\,D_x
\]
is common for both equations~\eqref{eqhyp} and \eqref{pmKdVLiou},
see~\cite{YS2003, Kaliappan}. This operator generates the commutative
Lie subalgebra
\[
\gA_\Liou={\mathrm{span}}_\BBR\langle\vph_k=R_\Liou^k(\vph_0),\ %
\vph_0=u_1\rangle
\]
of local higher symmetries of the potential modified Korteweg\/--\/de
Vries equation, see Eq.~\eqref{pmKdVLiou}.

Now, to the Liouville equation we assign the variable $\gv$ such that
\begin{subequations}
\begin{align}
\gv_x&=\exp(2u),\label{SubstODE}\\
\cE_\gv&=\{\gv_y=\gv^2\}\label{CovODE}
\end{align}
\end{subequations}
and therefore the compatibility condition $\gv_{xy}=\gv_{yx}$ holds.
Thence, the equation $\cE_\Liou$ is represented in the evolutionary form:
\begin{equation}\label{eqhypEvol}
u_{t_{-1}}=\vph_{-1}\equiv \gv,
\end{equation}
where the variable $t_{-1}\equiv y$ is the parameter and
$\vph_{-1}$ is the shadow of the nonlocal symmetry
\[
\tilde\cEv_{\vph_{-1}, a_{-1}}\equiv\tilde\cEv_{\vph_{-1}}+
a_{-1}\cdot\frac{\dd}{\dd\gv}
\]
such that $a_{-1}=\gv^2$.
The solution to Eq.~\eqref{CovODE} is $\gv=-{(y+\cX(x))}^{-1}$.
Apply transformation~\eqref{FiniteSymDiffeo} of the form
$\tilde y=\cY(y)$ to the ''time'' $y$. Then we obtain the potential
\[ 
\gv=-\frac{\cY'(y)}{(\cX(x)+\cY(y))}
\] 
for the genreral solution $u=\tfrac{1}{2}\log\gv_x$ of the Liouville
equation (see~\cite{Liouville, SakovichSBA}):
\begin{equation}\label{gensolhyp}
u=\tfrac{1}{2}\log\Bigl[
\frac{\cX'(x)\cdot\cY'(y)}{(\cX(x)+\cY(x))^{2}}\Bigr].
\end{equation}
We recall that functional~\eqref{MinimalIntegrals} is continuous
on formal divergences $u\to\pm\infty$ of solution~\eqref{gensolhyp},
see the paper~\cite{Acta} and references therein.
We estimate the scheme of constructing the general solution
for the Liouville equation by using the potential $\gv$ to be really
laconic and productive. A similar approach was used in~\cite{ClassSym} to
derive the Cole\/--\/Hopf transformation for the Burgers equation,
see Eq.~\eqref{ColeHopf} on page~\pageref{ColeHopf}.

Solution~\eqref{gensolhyp} of Eq.~\eqref{eqhyp} is the mapping
\[
\tau\colon\bigl\{\cX_y=0,\ \cY_x=0\bigr\}\to\cE_\Liou.
\]
Consider the evolutionary vector field $\cEv_{u_t}$ defined in
Eq.~\eqref{pmKdVLiou}. This field can be lifted onto the inverse image of
the covering $\tau$. Hence we obtain the equation $\cY_t=0$ and the
Krichever\/--\/Novikov equation
\begin{equation}\label{KN}
\cX_t=-2\cX_3+{3}{\cX_2^2}/{\cX_1}=
  -2\cX_1\cdot\{\cX,x\},
\end{equation}
where $\{\cX,x\}$ is the Schwartz derivative. Also, the evolution equation
\[
\gv_t=a_1\equiv \gv_3-\tfrac{3}{2}{\gv_2^2}{\gv_1^{-1}}
\]
holds.

The evolution equations that appear in Example~\ref{ExampleKdV}
are systemized as shown in the diagram below:
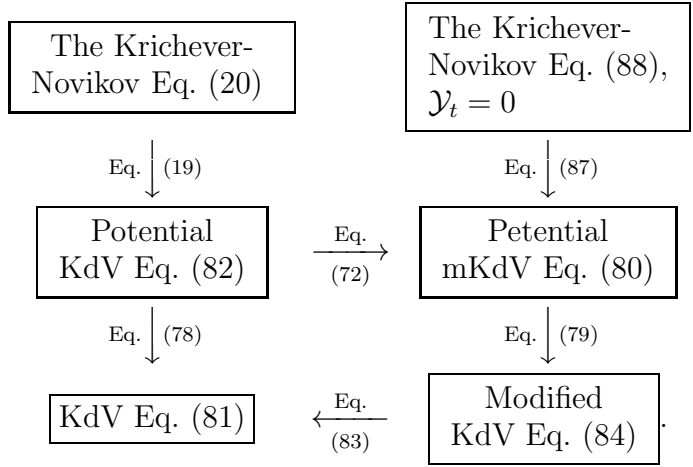
\[
  \begin{CD}
      \boxed{\begin{array}{c}\text{The Krichever-}\\
        \text{Novikov Eq.~\eqref{IKN}
            }
      \end{array}}
   @.
    \boxed{\begin{array}{l}
       \text{The Krichever-}\\
       \text{Novikov Eq.~\eqref{KN},}\\
       \cY_t=0\end{array} }
\\
   @V{\text{Eq.}}V{\eqref{IKNpKdV}}V
   @V{\text{Eq.}}V{\eqref{gensolhyp}}V
\\
      \boxed{\begin{array}{c}
        \text{Potential}\\
        \text{KdV Eq.~\eqref{pKdVLiou}}
      \end{array} }
   @>{\text{Eq.}}>{\eqref{pKdVpmKdV}}>
      \boxed{\begin{array}{c}
        \text{Petential}\\
        \text{mKdV Eq.~\eqref{pmKdVLiou}}
      \end{array} }
\\
   @V{\text{Eq.}}V{\eqref{pKdVKdV}}V
   @V{\text{Eq.}}V{\eqref{pmKdVmKdVLiou}}V
\\
      \boxed{\text{KdV Eq.~\eqref{KdVLiou}}}
      @<{\text{Eq.}}<{\text{\eqref{MiuraLiou}}}<
      \boxed{\begin{array}{c}
        \text{Modified}\\
        \text{KdV Eq.~\eqref{mKdVLiou}}
      \end{array} }.
  \end{CD}
\]
\end{example}

\section{Analogs of the potential modified Korteweg\/--\/de Vries
equation}\label{SecAandB}

\subsection{Constructing the hierarchy~$\gA$}
First, we formulate an important property of the integrals
$\Omega^i\in\ker{D_y\bigr|}_\cE$ for a Liouvillean type equation~$\cE$.
Namely, we specify the evolution of these integrals along
symmetries $\vph\in\sym\cE$ of the equation~$\cE$.

\begin{lemma}[\textup{\cite{SokolovUMN}}]\label{EvolIntByThemselves}
Consider a symmetry field $\cEv_\vph$ of a Liouvillean type equation
$\cE$. Then the evolution $\cEv_\vph(\Omega^i)$ of an arbitrary integral
$\Omega^i\in\ker\bar D_y$ for $\cE$ is an element of $\ker\bar D_y$
again.
\end{lemma}

\begin{proof}
Indeed, we have
\[
\bar D_y(\cEv_\vph(\Omega^i))=\cEv_\vph(\bar
D_y(\Omega^i))=\cEv_\vph(0)=0.
\]
\end{proof}

\begin{example}\label{EvolEMTExample}
Consider a symmetry $\vph=\square(\phi(x$, $\bT))$ of the Toda equation,
see Eq.~\eqref{symToda} on page~\pageref{symToda}. Then the relation
\begin{equation}\label{EvolEMT}
\dot T_\phi\equiv\cEv_{\square(\phi)}(T)=
 \bigl(-\beta\,\bar D_x^3+T\,\bar D_x+D_x\circ T\bigr)(\phi)
\end{equation}
holds. Here
\[\beta\equiv\sum_{i=1}^r a_i\cdot\Delta^i,\quad
\square=\bu_x+\vec\Delta\,\bar D_x,\text{ and }\Delta^i=\sum_j k^{ij}.
\]
Also, consider the nonlocal variable $s$ that was defined in
Example~\ref{OurTodaRecExample} by the equations $s_x=T$ and $s_y=1$.
Then the evolution $s_t$ of the nonlocality is described by the formula
\[\cEv_\vvph(s)=D_x^{-1}\circ\cEv_\vvph(T).\]
\end{example}

Suppose that
the sequence $\vvph_0$, $\vvph_1$, $\vvph_2$ of symmetries of the Toda
equation is assigned to the specially chosen set of the functions~$\phi$
in formula~\eqref{symToda}. We set $\phi_{-1}=1$ and obtain the symmetry
$\vph_0=\square(\phi_{-1})$. Then, we calculate the evolution $\dot
T_{\phi_{-1}}$ that corresponds to this symmetry. Now we equal the next
function $\phi_0$ to the evolution of the potential $s$. Then, we obtain
the function $\phi_1$ and the symmetries $\vph_1$, $\vph_2$ in a similar
way. The result is shown in the next diagram:
\begin{equation}\label{InitialDiagram}
\begin{diagram}
\vvph_2  &\lto^\square & \phi_1=-\beta s_3+\tfrac{3}{2}s_1^2 &
 \lto^{D_x^{-1}}&\dot T_{\phi_0}=-\beta T_3+3TT_1 & & \\
& \luto^{R_\Toda}& & \ruto^{\ell_T}& & \luto^{R_\KdV} &\\
&&\vvph_1=\square(s_1) &\lto^\square &\phi_0=s_1 & \lto^{D_x^{-1}} &
   \dot T_{\phi_{-1}}=T_1 \\
& & & \luto^{R_\Toda} & &\ruto^{\ell_T} & \\
&&&&\vvph_0=\bu_x&\lto^\square &\phi_{-1}=1 & & {}
\end{diagram}
\end{equation}
The following evolution equations are met in
diagram~\eqref{InitialDiagram}:
the Korteweg\/--\/de Vries equation is
\begin{equation}\label{KdV}
\cE_\KdV=\{T_{t}=-\beta T_3+3T\cdot T_1\}
\end{equation}
(we recall that its recursion operator is~\eqref{IRKdV} on
page~\pageref{IRKdV}); then, we also get the potential
Korteweg\/--\/de Vries equation
\begin{equation}\label{pKdV}
\cE_\pKdV=\left\{s_t=-\beta s_3+\tfrac{3}{2}s_1^2\right\},
\end{equation}
and the equation
\begin{equation}\label{pmKdV}
\cE_\pmKdV=\{\bu_{t}=\square(T(\bu_1,\bu_2)\}.
\end{equation}
Suppose $K=\|2\|$ is the Cartan matrix of the Lie algebra
$\gothg=\mathfrak{sl}_2(\BBC)$. Then the Toda equation~\eqref{eqToda}
associated with this algebra is the hyperbolic Liouville
equation~\eqref{eqhyp}, and equation~\eqref{pmKdV} is nothing else than
the scalar potential modified Korteweg\/--\/de Vries
equation~\eqref{pKdVLiou}.
Further on, suppose that the matrix~$K$ in Eq.~\eqref{eqToda} is an
arbitrary nondegenerate symmetrizable $(r\times r)$-matrix, not
necessarily the Cartan matrix of a semisimple Lie algebra $\gothg$ of rank
$r$. Then we get the $r$-component system of the third\/-\/order evolution
equations with the cubic nonlinearity. In local coordinates, this system
is
\[
u^i_t=\tfrac{1}{2}\sum\nolimits_{p,q=1}^r a_p\cdot\left\{
k_{pq}u^i_x u^p_x u^q_x + 2(\Delta^i k_{pq}-\delta_{i,q})u^p_{xx}u^q_x
- 2\Delta^i u^p_{xxx}
\right\},
\]
where $1\leq i\leq r$; we recall that
\[\Delta^i=\sum_j k^{ij}\text{ and }a_p k_{pq}=a_q k_{qp}.\]

\subsubsection*{The analogs of the potential modified
Korteweg\/--\/de Vries equation for $r=2$}
In this subsection, we consider evolution systems~\eqref{pmKdV} in case
$r=2$ and the matrix $K$ is symmetric.

\begin{example}
Let $K$ be the symmetric matrix
\[
K=\begin{pmatrix} 2 & \lambda\\ \lambda & 2\end{pmatrix},
\]
thence,
\[
\vec\Delta=\frac{1}{\lambda+2}\binom{1}{1}.
\]
Introduce new dependent variables
\[
u=u^1+u^2,\quad v=(\lambda+2)(u^1-u^2);
\]
the inverse transformation is defined by the formulas
\[
u^1=\frac{1}{2}u+\frac{1}{2(\lambda+2)}v,\qquad
u^2=\frac{1}{2}u-\frac{1}{2(\lambda+2)}v.
\]
Then, the normal form of Eq.~\eqref{pmKdV} is
\begin{equation}\label{pmKdVNormalRaw}
\begin{aligned}
u_t&=-\frac{2}{\lambda+2}u_3+\frac{2-\lambda}{(\lambda+2)^3}v_1v_2+
 \frac{\lambda+2}{4}u_1^3+\frac{2-\lambda}{4(\lambda+2)^2}u_1v_1^2,\\
v_t&=  
 -u_2v_1+
 \frac{\lambda+2}{4}u_1^2v_1+\frac{2-\lambda}{4(\lambda+2)^2}v_1^3.
\end{aligned}
\end{equation}

Further on, make the scaling transformation
\begin{equation}\label{SokHintChange}
t={(\lambda+2)}^2\cdot \tilde t,\qquad
u={(\lambda+2)}^{-1}\cdot \tilde u,\qquad
v=\tilde v,
\end{equation}
preserving the same notation $x$, $t$, $u$, and $v$.
\end{example}

\begin{prop}
In coordinates~\eqref{SokHintChange}, equations~\eqref{pmKdVNormalRaw}
become linear in $\lambda$ and acquire the form
\begin{equation}\label{pmKdVNormal}
\begin{aligned}
u_t&=-2u_3+2v_1v_2+\tfrac{1}{2}u_1^3+\tfrac{1}{2}u_1v_1^2+
\lambda\Bigl(-v_1v_2+\tfrac{1}{4}u_1^3-\tfrac{1}{4}u_1v_1^2\Bigr),\\
v_t&=-2u_2v_1+\tfrac{1}{2}u_1^2v_1+\tfrac{1}{2}v_1^3+
\lambda\Bigl(-u_2v_1+\tfrac{1}{4}u_1^2v_1-\tfrac{1}{4}v_1^3\Bigr).
\end{aligned}
\end{equation}
\end{prop}

\begin{rem}  
Consider the vector coefficient of $\lambda$ in Eq.~\eqref{pmKdVNormal}.
Then this vector\/-\/valued function  is \emph{not} a symmetry of the flow
that stands at $\lambda^0$ (and \textit{vice versa}). Namely, the
following statements hold.
\end{rem}

\begin{prop}
\begin{enumerate}
\item   
Consider the flow
\begin{align*}
u_t&=-2u_3+2v_1v_2+\tfrac{1}{2}u_1^3+\tfrac{1}{2}u_1v_1^2,\\
v_t&=-2u_2v_1+\tfrac{1}{2}u_1^2v_1+\tfrac{1}{2}v_1^3
\end{align*}
at $\lambda^0$ in Eq.~\eqref{pmKdVNormal}.
Then its symmetries
\[
\begin{pmatrix}\vph^1,\\ \vph^2\end{pmatrix}
(t,x,u,v,u_1,v_1,u_2, v_2, u_3, v_3)
\]
of order $\leq3$ are generated by
\[
\begin{pmatrix} tu_t+\tfrac{1\mathstrut}{3\mathstrut}xu_x \\
tv_t+\tfrac{1\mathstrut}{3\mathstrut}xv_x \end{pmatrix},\qquad
\begin{pmatrix} u_t \\ v_t \end{pmatrix},\qquad
\begin{pmatrix} u_x \\ v_x \end{pmatrix},\qquad
\begin{pmatrix} 1 \\ 0 \end{pmatrix},\qquad
\begin{pmatrix} 0 \\ 1 \end{pmatrix}.
\]
These generators are the scaling symmetry, two translations, and
two shifts, respectively.
The commutator with the scaling symmetry maps the translations along
$t$ and $x$ to themselves.
%
\item   
Consider the flow
\begin{align*}
u_t&=-v_1v_2+\tfrac{1}{4}u_1^3-\tfrac{1}{4}u_1v_1^2,\\
v_t&=-u_2v_1+\tfrac{1}{4}u_1^2v_1-\tfrac{1}{4}v_1^3
\end{align*}
at $\lambda^1$ in Eq.~\eqref{pmKdVNormal}.
Then the symmetries
\[
\begin{pmatrix}\vph^1\\ \vph^2\end{pmatrix}
(t,x,u,v,u_1,v_1,u_2, v_2, u_3, v_3)
\]
of order $\leq3$ of this flow are generated by the sections
\[
\begin{pmatrix} tu_t+\tfrac{1\mathstrut}{3\mathstrut}xu_x \\
tv_t+\tfrac{1\mathstrut}{3\mathstrut}xv_x \end{pmatrix},\qquad
\begin{pmatrix} u_t \\ v_t \end{pmatrix},\qquad
\begin{pmatrix} u_x \\ v_x \end{pmatrix},\qquad
\begin{pmatrix} 1 \\ 0 \end{pmatrix},\qquad
\begin{pmatrix} 0 \\ 1 \end{pmatrix},
\]
which are the scaling, the translations, and the shifts, respectively.
\end{enumerate}
\end{prop}

The conservation laws for system~\eqref{pmKdVNormal} in the normal
form are few.

\begin{prop}   
Suppose $\lambda$ is generic
\textup{(}see Remark~\textup{\ref{ExceptLambda}} below\textup{).}
Then there is a unique generating section
\[
\begin{pmatrix}\psi^1\\ \psi^2\end{pmatrix}
\bigl(t,x,u,v,u_1,v_1,u_2,v_2\bigr)=
\begin{pmatrix} u_2 \\ \frac{\lambda-2}{\lambda+2}\,v_2\end{pmatrix}
\]
of order $\leq2$ of a conservation law for the flow in
Eq.~\eqref{pmKdVNormal}\textup{;} this section corresponds to the
conserved density
\[
H=u_x^2+\frac{\lambda-2}{\lambda+2}\,v_x^2.
\]
\end{prop}

In Sec.~\ref{MagriTopo} we shall demonstrate that
equation~\eqref{pmKdVNormalRaw} is Hamiltonian with respect to\
the operator $A_1=K^{-1}\cdot\bar D_x^{-1}$ (recall
that the matrix $K$ is assumed symmetric and therefore $a_i=1$).
Now we get the Hamiltonian representation of system~\eqref{pmKdVNormal} by
using the transformation rule for Hamiltonian operators with respect to\
transformations of the dependent variables.

\begin{lemma}
Consider a Hamiltonian equation
\[u_t=A\bigl(\bE_u(\cH[u])\bigr).\]
Let $\tilde u=Q u$ be a nondegenerate transformation of the dependent
variables. Then the equation
\[\tilde u=\tilde A\bigl(\bE_{\tilde u}(\cH[\tilde u])\bigr)\]
holds for the Hamiltonian operator
\[\tilde A=Q\cdot A\cdot{}^t Q.\]
\end{lemma}

\begin{example}   
Consider Eq.~\eqref{pmKdVNormal}. Then we get the diagonal operator
\[
{\begin{pmatrix}u\\ v\end{pmatrix}}_t=2\begin{pmatrix} \lambda+2 & 0\\
0 & \frac{(\lambda+2)^2}{2-\lambda}\end{pmatrix}\cdot D_x^{-1}\bigl(
\bE_{(u,v)}(T^2\,\Id x)\bigr),
\]
where the density $h_1$ of the Hamiltonian $T^2\,\Id x$ for the
Korteweg\/--\/de Vries equation acquires the following form in the
coordinates $u$, $v$:
\[
h_1=\tfrac{1}{16}
{\bigl((\lambda-2)v_x^2-(\lambda+2)u_x^2+4(\lambda+2)u_{xx}\bigr)}^2.
\]
\end{example}

\begin{prop}   
Again, suppose $\lambda$ is generic
\textup{(}see Remark~\textup{\ref{ExceptLambda}).}
The symmetries
\[
\begin{pmatrix}\vph^1\\ \vph^2\end{pmatrix}
(t,x,u,v,u_1,v_1,u_2, v_2)
\]
of order $\leq2$ of equation~\eqref{pmKdVNormal} are generated by
the translation and the shifts, which are
\begin{equation}\label{SymGeneric}
\begin{pmatrix} u_x \\ v_x \end{pmatrix},\qquad
\begin{pmatrix} 1 \\ 0 \end{pmatrix},\qquad
\begin{pmatrix} 0 \\ 1 \end{pmatrix},
\end{equation}
respectively.
\end{prop}

\begin{rem}\label{ExceptLambda}
We emphasize that the symmetry algebra (and even the classical
symmetry algebra) of system~\eqref{pmKdVNormal}
depends on the initial matrix~$K$ essentially.
The cases $\lambda=-1$ (the matrix $K$ corresponds to the Lie algebra
$A_2$) and $\lambda=\pm2$ (the matrix $K$ is degenerate) are special.
In what follows, we analyse them separately.
\end{rem}

First, let $\lambda=-1$, that is, $K$ is the Cartan matrix for the
Lie algebra~$A_2$.
Then substitution~\eqref{SokHintChange} is
\[u=u^1+u^2,\qquad v=u^1-u^2.\]
After routine transformations, we obtain
\begin{equation}\label{pmKdVNormalA2}
\begin{aligned}
u_t&=-u_3+\tfrac{3}{2}v_1v_2+\tfrac{1}{8}u_1^3+\tfrac{3}{8}u_1v_1^2,\\
v_t&=\phantom{-u_3}-\tfrac{1}{2}u_2v_1+\tfrac{1}{8}u_1^2v_1
+\tfrac{3}{8}v_1^3.
\end{aligned}
\end{equation}

\begin{prop}   
\begin{enumerate}
\item
The symmetries
\[
\begin{pmatrix}\vph^1\\ \vph^2\end{pmatrix}
(t,x,u,v,u_1,v_1,u_2, v_2)
\]
of order $\leq2$ of flow~\eqref{pmKdVNormalA2} are
generated by the sections
\[
\begin{pmatrix} v \\
-\tfrac{1}{3}u+\tfrac{2}{3}\log v_x\end{pmatrix},\qquad
\begin{pmatrix} u_x \\ v_x \end{pmatrix},\qquad
\begin{pmatrix} 1 \\ 0 \end{pmatrix},\qquad
\begin{pmatrix} 0 \\ 1 \end{pmatrix}.
\]
The first symmetry is \emph{nonpolynomial}.
The other two are the translation and the shift, respectively.
The nonpolynomial symmetry flow can be also represented in the form
\[u_{xt}^2=\exp(u+3u_{tt}).\]
%
\item
The generating sections
\[
\begin{pmatrix}\psi^1\\ \psi^2\end{pmatrix}
(t,x,u,v,u_1,v_1,u_2, v_2)
\]
of order $\leq2$ of conservation laws for Eq.~\eqref{pmKdVNormalA2}
are
\[
\begin{pmatrix}u_2\\ 3v_2\end{pmatrix},\qquad 
\begin{pmatrix} \Psi(u,v,v_x)\\
2{\dd\Psi}/{\dd v}+2{v_2}{v_1^{-1}}\,{\dd\Psi}/{\dd v_1}-
u_1\,{\dd\Psi}/{\dd v_1} \end{pmatrix},
\]
where the function $\Psi$ satisfies the equation
\[
\frac{\dd\Psi}{\dd u}+\frac{1}{2}v_1\,\frac{\dd\Psi}{\dd v_1}-
\frac{1}{2}\Psi=0.
\]
\end{enumerate}
\end{prop}

Now, for the first time within this paper, suppose that the matrix $K$ is
degenerate. Quite untrivially, this degeneracy can occur in two distinct
ways.

First, assume that $\lambda=2$ and
$K=\left(\begin{smallmatrix}2& 2\\ 2&2\end{smallmatrix}\right)$.
Then system~\eqref{pmKdVNormalRaw} is in triangle form%
 \footnote{The classical symmetry algebra for the Toda equations
associated with this matrix is composed by the sections
 \[\vec\vph=\alpha\vec u_x+\beta(x)\,\vec\bun,\]
 where the constant $\alpha$ and the function
 $\beta(x)$ are arbitrary.}:
\begin{equation}\label{TriangleL2}
\left\{
\begin{aligned}
u_t&=-\tfrac{1}{2}u_3+u_1^3,\\
v_t&=-u_2v_1+u_1^2v_1.
\end{aligned}
\right.
\end{equation}
It consists of the potential modified Korteweg\/--\/de Vries equation
and the auxiliary dispersionless component.
We note that the variable $v$ admits an arbitrary shift.
Indeed, the section ${}^t(0$, $f(v))$ is a symmetry of
Eq.~\eqref{TriangleL2} if $f$ is arbitrary. In addition, there are two
more symmetries. They are the translation and the shift of $u$, see
Eq.~\eqref{SymGeneric}.

\begin{prop} 
If $\lambda=2$, then the generating sections
\[
\begin{pmatrix}\psi^1\\ \psi^2\end{pmatrix}
(t,x,u,v,u_1,v_1,u_2, v_2)
\]
of order $\leq2$ of conservation laws for the flow in
Eq.~\eqref{TriangleL2} are
\[
\begin{pmatrix}u_2\\ 0\end{pmatrix},\qquad
\begin{pmatrix}\exp(u)\\ 0\end{pmatrix},\qquad
\begin{pmatrix}\exp(-u)\\ 0\end{pmatrix}.
\]
The corresponding conserved densities are
\[\rho_1=-\tfrac{1}{2}u_x^2,\quad
\rho_2=\exp(u),\quad \rho_3=-\exp(-u),\]
respectively. The last two densities belong to the
\emph{nonpolynomial} conserved currents
\begin{align*}
\eta_2&=\exp(u)\,\Id x+(2u_2-u_1^2)\exp(u)\,\Id t,\\
\eta_3&=-\exp(-u)\,\Id x+(2u_2+u_1^2)\exp(-u)\,\Id t
\end{align*}
for the potential modified Korteweg\/--\/de Vries equation
\[
u_t=-2u_3+u_1^3
\]
whose right\/--\/hand side is \emph{polynomial}.
\end{prop}

\smallskip
Now, suppose $\lambda=-2$, whence
$K=\left(\begin{smallmatrix}\pp2& -2\\
 -2&\pp2\end{smallmatrix}\right)$.
Then equation~\eqref{pmKdVNormal} is
\begin{equation}\label{Normal-2}
\left\{\begin{aligned}
u_t&=-2u_3+4v_1v_2+u_1v_1^2,\\
v_t&=v_x^3.
\end{aligned}\right.
\end{equation}

\begin{prop}  
The symmetries of system~\eqref{pmKdVNormal} for
$\lambda=-2$ are described by Eq.~\eqref{SymGeneric}.
The generating sections
\[
\begin{pmatrix}\psi^1\\ \psi^2\end{pmatrix}
(t,x,u,v,u_1,v_1,u_2, v_2)
\]
of order $\leq2$ of conservation laws for the flow in right\/-\/hand side\
of Eq.~\eqref{Normal-2} are, then,
\[
\begin{pmatrix} 0 \\ \Psi(x,t,v,v_x,v_{xx}) \end{pmatrix}.
\]
Here the function $\Psi$ is subject to the equation
\[
\frac{\dd\Psi}{\dd t}-2v_1^3\,\frac{\dd\Psi}{\dd
v}-3v_1^2\,\frac{\dd\Psi}{\dd x}+v_1v_2^2\,\frac{\dd\Psi}{\dd v_2}
-6v_1v_2\,\Psi=0.
\]
\end{prop}

\subsubsection*{Factorizations of the recursion operators}
Now we return to the problem of constructing the commutative
hierarchy~$\gA$, which is associated with the initial Toda
equation~$\cE_\Toda$.

By construction, the symmetries $\vph_i$ are related by the recursion
operator
\begin{equation}\label{RecOperator}
R_\Toda=\square\circ D_x^{-1}\circ\ell_T
\end{equation}
for the Toda equation~\eqref{eqToda}. We met this operator in
Example~\ref{OurTodaRecExample} on page~\pageref{OurTodaRecExample}.
We fix\footnote{As we have already noted in Chapter~1,
we treat the Toda equations, as well as the structures on them,
up to the symmetry~$x\leftrightarrow y$.}
this notation $R_\Toda$ onwards.

In this subsection, we demonstrate that each symmetry
\[\vvph_k=\square(\phi_{k-1})\in\sym\cE_\Toda\]
defines the next function $\phi_k$ explicitly by the
relation $\phi_k=\cEv_{\vvph_k}(s)$, and therefore,
$\vvph_{k+1}=\square(\phi_k)$. We also prove that the sections
$\phi_k$ compose the hierarchy of the potential Korteweg\/--\/de Vries
equation~\eqref{pKdV}.

First, we note that the functions
$\phi_{-1}$, $\phi_0$, and $\phi_1$ are mapped successfully one to another
by the recursion operator~(\cite{JPW})
\begin{equation}\label{RpKdV}
R_\pKdV=-\beta\,D_x^2+2s_1-D_x^{-1}\circ s_2
\end{equation}
for equation~\eqref{pKdV}. Recall that this operator was obtained in
Introduction while illustrating the Cartan generating forms method
(\cite{JKKersten}) by I.~S.~Krasil'shchik.

\begin{lemma}\label{FactorRecLemma}
The following two decompositions of the recursion operators
hold\textup{:}
\begin{equation}\label{TwoSplits}
R_\Toda=\square\circ\ell_s,\qquad R_\pKdV=\ell_s\circ\square,
\end{equation}
where $\ell_s=D_x^{-1}\circ\ell_T$ by Eq.~\eqref{EqLinNonlocal}
and the linearization $\ell_T$ is defined in Eq.~\eqref{LinTWRTu}.
\end{lemma}

\begin{proof}
The first decomposition holds by construction.
To establish another one, we use the relations
\begin{multline*}
\ell_T\circ(\bu_x+\vec\Delta D_x)={}\\
\begin{aligned}{}&=
\tfrac{1}{2}\sum_{i,j}\kappa_{ij}u^i_1\,D_x\circ u^j_1+
\tfrac{1}{2}\sum_i a_i\Bigl\{\Bigl(\sum_j k_{ij}u^i_1\Bigr)D_x\circ
u^i_1 - 2D_x^2\circ u^i_1\Bigr\}+{}\\
{}&\phantom{=}\quad +
\tfrac{1}{2}\sum_{i,j}\kappa_{ij}u^i_1\Delta^jD_x^2+
\tfrac{1}{2}\sum_i a_i\Bigl\{\Bigl(\sum_j k_{ij}u^j_1\Bigr)\Delta^iD_x^2
-2\Delta^i D_x^3\Bigr\}={}\\
{}&=-\sum_i a_i\Delta^i\,D_x^3+
\tfrac{1}{2}\sum_{i,j}a_i\bigl\{k_{ij}(u^i_1u^j_2+u^i_2u^j_1)
      -2u^i_3\bigr\}+{}\\
{}&{}\quad{}+\tfrac{1}{2}\sum_{i,j}a_i\bigl\{k_{ij}(u^i_1u^j_1+u^i_1u^j_1)
-4u^i_3\bigr\}\,D_x+{}\\
{}&{}\quad{}+\tfrac{1}{2}\Bigl\{\sum_i a_i\cdot(-2u^i_1)D_x^2+
\sum_{i,j}\kappa_{ij}\Delta^ju^i_1D_x^2+
\sum_{i,j}\kappa_{ij}\Delta^iu^j_1D_x^2\Bigr\}={}
\end{aligned}\\
{}=-\beta D_x^3+T_1+2T\,D_x,
\end{multline*}
whence follows the second decomposition.
\end{proof}
\noindent%
We draw attention to the fact
that the representation of the scalar operator $R_\pKdV$
in the form of the product of the \emph{vector\/--\/valued}
operator $\square$ and the row $\ell_s$ of length $r$
seems to have been unnoticed in the literature.


Now we identify infinitesimal symmetries of differential equations with
autonomous evolution equations and, by using Lemma \ref{FactorRecLemma},
we extend diagram~\eqref{InitialDiagram} infinitely upwards. The result is
displayed in the diagram
\begin{equation}\label{BottomEvolutionDiag}
\begin{diagram}
\ldots&&\ldots\\
\uto_{R_\Toda}&\ruto^{\ell_s}&\uto_{R_\pKdV}\\
\bu_{t_2}=\vvph_2  &\lto^\square &
    s_{t_1}=\phi_1=\lefteqn{-\beta s_3+\tfrac{3}{2}s_1^2}\\
\uto_{R_\Toda}&\ruto^{\ell_s}&
\uto_{
\substack{
 R_\pKdV={}\\
     {}{}\lefteqn{\scriptstyle -\beta D_x^2+2s_1-D_x^{-1}\circ s_2}
}  
}\\
\bu_{t_1}=\vvph_1 & \lto^\square &
s_{t_0}=\phi_0=s_1\\
\uto_{R_\Toda}&\ruto^{\ell_s}&\uto_{R_\pKdV}\\
\underbrace{\bu_{t_0}=\vvph_0=\bu_x}_\gA
&\lto^\square &\underbrace{s_{t_{-1}}=\phi_{-1}=1.}_\gB
\end{diagram}
\end{equation}

The right\/-\/hand sides $\phi_k$ of the evolution equations
$s_{t_k}=\phi_k$ are generated by the recursion operator $R_\pKdV$,
therefore they are  higher symmetries of the potential
Korteweg\/--\/de Vries equation $\cE_\pKdV$.

The times $t_i$ in the equations $u_{t_k}=\vph_k$ and $s_{t_k}=\phi_k$
within diagram~\eqref{BottomEvolutionDiag} are correlated.
Namely, we have

\begin{theor}\label{ConsOfTences}
The evolution $\cEv_{\vph_k}(s)$ of the nonlocal variable $s$
along the higher symmetry $\vph_k=\square(\phi_{k-1})$
of the Toda equation~\eqref{eqToda}
coincides
with the evolution $\phi_k$ that is defined by the $k$\textup{th}
higher analog $s_{t_k}=R^{k+1}_\pKdV(1)$ of the potential
Korteweg\/--\/de Vries equation~\eqref{pKdV}\textup{:}
\[
\cEv_{\square(\phi_{k-1})}(s)=\phi_{k}=R_\pKdV(\phi_{k-1}).
\]
\end{theor}

\begin{proof}
From decompositions~\eqref{TwoSplits} it follows that
\[
R_\pKdV(\phi_k)=D_x^{-1}\left(\cEv_{\square(\phi_k)}(T)\right).\]
\end{proof}

\begin{cor}
1. Each shadow $\vvph_k$ in the covering $\tau_s$, see
Eq.~\eqref{pKdVpmKdV}, over the Toda equation can be reconstructed up to
the true nonlocal symmetry $\tilde\cEv_{\vvph_k,\phi_k}$ of the Toda
equation.\\
2. All subdiagrams
\[
\begin{CD}
\vvph_{k+2} @<{\square}<< \phi_{k+1} \\
@A{R_\Toda}AA @AA{R_\pKdV}A \\
\vvph_{k+1} @<{\square}<< \phi_k
\end{CD}
\]
in diagram~\eqref{BottomEvolutionDiag} are commutative: we have
 $R_\Toda(\vvph_{k+1})=\square(\phi_{k+1})$,
and the relations
\[
R_\Toda=\square\circ R_\pKdV\circ\square^{-1},\qquad
R_\pKdV=\square^{-1}\circ R_\Toda\circ\square
\]
hold.
\end{cor}

\begin{rem}
In the paper~\cite{Kaliappan}, the recursion operators
$R_\Toda$ and $R_\pKdV$ were noted to be conjugate to each other
in the scalar case~(\ref{eqhyp}-\ref{mKdVLiou}).
\end{rem}

\subsubsection*{Proof of the locality for the symmetry sequence $\gA$}
All previous reasonings do not imply the locality of the elements
$\phi_k$ of the sequence $\gB$ composed by higher symmetries of the
potential Korteweg\/--\/de Vries equation~\eqref{pKdV}. Indeed, these
symmetries lie in the image of the operator~$D_x^{-1}$.
Therefore, the elements~$\vph_{k+1}=\square\,(\phi_k)$ of the
sequence $\gA$ are not yet proved to be local, too.

\begin{state}[\textup{\cite{Dorfman}}]\label{local4pKdV}
Recursion operator~\eqref{RpKdV} for the potential
Korteweg\/--\/de Vries equation~\eqref{pKdV} generates the
\emph{local in $T$} sequence of the higher symmetries
\[
\phi_k=R^{k+1}_{\mathrm{pKdV}}(\phi_{-1})=\phi_k(T,
\dots, T_{2k}),
\]
where $\phi_{-1}=1$ is the shift of the dependent variable $s$
by a constant.
\end{state}

There are several ways to prove Proposition~\ref{local4pKdV}.
The method by I.~S.~Krasil'shchik (\cite{DIPS9-2002}) is based on the
weak nonlocality of the recursion operator $R_\pKdV$
(see also Eq.~\eqref{RSplitting} on page~\pageref{RSplitting}).

\begin{cor}\label{local4pmKdV}
The symmetries $\vvph_k=\square(\phi_{k-1})\in\gA$
of the Toda equation~\eqref{eqToda}
are local and depend on the derivatives
$u^j_\sigma$, $|\sigma|\geq1$ for all $k\geq0$.
\end{cor}

We also note that in the paper~\cite{YS2003} we
analysed the case $r=1$ and established the locality
of the higher symmetries $\gA\subset\sym\cE_\Liou$ directly,
leaving apart the discussion upon the properties of the
potential equation~\eqref{pKdV}.

\subsection{Commutativity of the hierarchy $\gA$}
A classical example of evolution equations that admit a commutative
symmetry subalgebra is given in the following lemma.

\begin{lemma}[\textup{\cite{IbragShabatFAP}}]\label{IsCommutative}
Suppose $\cE$ is the scalar evolution equation
\begin{equation}\label{EvolEqLinear}
u_t=u_k+f(u_{k-2},\ldots,u)
\end{equation}
such that $f$ is a polynomial. Consider the Lie subalgebra
\[\langle\vph\in\sym\cE\mid\vph=\vph(u_\sigma)\rangle
\subseteq\sym\cE\] of its symmetries that depend on the variable $u$
and its derivatives only. Then this subalgebra is commutative.
\end{lemma}

We denote by $\gB$ the minimal Lie subalgebra generated by the
symmetries $\phi_k$ of the potential Korteweg\/--\/de Vries equation;
here $k\geq-1$. From Lemma~\ref{IsCommutative} it follows that
the algebra $\gB$ is commutative:
\[
\{\phi_k, \phi_l\}=\cEv_{\phi_k}(\phi_l)-\cEv_{\phi_l}(\phi_k)=0,
\]
therefore $\gB$ coincides with the linear span
of its generators $\phi_k$:
\[\gB=\mathrm{span}_\BBR\langle\phi_k\mid k\geq-1\rangle.\]

Now we analyse the commutation
properties of the symmetries $\vph=\square(\phi(x$, $\bT))$
of the Toda equations~\eqref{eqToda}.

\begin{lemma}\label{CommutUnderSquare}
Suppose $\vph'=\square\,(\phi'(x$, $\bT))$ and
$\vph''=\square\,(\phi''(x$, $\bT))$.
Then the Jacobi bracket $\{\vph'$, $\vph''\}$, which was defined in
Theorem~\textup{\ref{ThLieSym}}, of the symmetries $\vph'$ and $\vph''$ is
\[\{\vph', \vph''\}=\square\,(\phi_{\{1,2\}}),
\]
where the bracket
\begin{equation}\label{SquareArgumentsCommut}
\phi_{\{1,2\}}=\cEv_{\vph'}(\phi'')-\cEv_{\vph''}(\phi')+\bar
D_x(\phi')\,\phi''-\phi'\,\bar D_x(\phi'')
\end{equation}
on the arguments of the operator $\square$ is induced by the
Jacobi bracket $\{\vph', \vph''\}$.
Moreover,
\[\phi_{\{1,2\}}=\phi_{\{1,2\}}(x, \bT)\]
by Eq.~\eqref{EvolEMT}.
\end{lemma}

Further on, we denote by $\gA$  the minimal Lie algebra
generated by $\vvph_k$ for all $k\geq0$.

\begin{theor}\label{AIsCommutative}
The Lie algebra $\gA\subset\sym\cE_\Toda$ is commutative\textup{:}
$[\gA,\gA]=0$, and thence
\[\gA=\mathrm{span}_\BBR\langle\vvph_k\mid
k\geq0\rangle.\]
\end{theor}

\begin{proof}
Commute two symmetries $\cEv_{\vph_{k_1}}$ and $\cEv_{\vph_{k_2}}$,
apply the resulting evolutionary vector field to the variable $s$ and
take into account the relation between $\phi_k$ and $\vph_k$. Hence we
get
\begin{multline*}
\bigl[\cEv_{\vph_{k_1}}$, $\cEv_{\vph_{k_2}}\bigr](s)=
\cEv_{\vph_{k_1}}(\phi_{k_2})-\cEv_{\vph_{k_2}}(\phi_{k_1})={}\\
\cEv^{(s)}_{\phi_{k_1}}(\phi_{k_2})-\cEv^{(s)}_{\phi_{k_2}}(\phi_{k_1})=
\{\phi_{k_1}$, $\phi_{k_2}\}=0.
\end{multline*}
Since $T=s_x$, thence
\begin{equation}\label{CommutToT}
\bigl[\cEv_{\vph_{k_1}},\cEv_{\vph_{k_2}}\bigr](T)=0.
\end{equation}

Now estimate the evolution of the integral $T$ by using
Lemma~\ref{CommutUnderSquare}. First, consider the bracket $\{\vph_{k_1}$,
$\vph_{k_2}\}$   
and then calculate $\dot T_{\phi_{\{k_1,k_2\}}}$.
Recall that $\vph_{k_1}=\square(\phi_{k_1-1})$ and
$\vph_{k_2}=\square(\phi_{k_2-1})$. Therefore,
\[
\{\vph_{k_1}, \vph_{k_2}\}=\square(\phi_{\{k_1,k_2\}}),
\]
where
\begin{multline*}
\phi_{\{k_1,k_2\}}={}\\
\cEv_{\phi_{k_1}}(\phi_{k_2-1})-
\cEv_{\phi_{k_2}}(\phi_{k_1-1})+
\bar D_x(\phi_{k_1-1})\,\phi_{k_2-1}-
\phi_{k_1-1}\,\bar D_x(\phi_{k_2-1})
\end{multline*}
owing to Eq.~\eqref{SquareArgumentsCommut}. From
Example~\ref{EvolEMTExample} it follows that
\begin{multline}
\cEv_{\{\vph_{k_1},\vph_{k_2}\}}(T)=
\cEv_{\square(\phi_{\{k_1,k_2\}})}(T)={}\\
{}=\bigl(-\beta\,\bar D_x^3+T\,\bar D_x+\bar D_x\circ
   T\bigr)(\phi_{\{k_1,k_2\}}).
   \label{BracketToT}
\end{multline}
Comparing Eq.~\eqref{CommutToT} and Eq.~\eqref{BracketToT}, we get
\begin{equation}\label{KillEvol}
\bigl(-\beta\,\bar D_x^3+T\,\bar D_x+\bar D_x\circ T\bigr)\,
\phi_{\{k_1,k_2\}}=0.
\end{equation}
In the left\/-\/hand side\ of Eq.~\eqref{KillEvol} we obtain the operator
$\hat B_2$ in total derivatives whose coefficients are elements
of $\bT$. We apply this operator to
$\phi_{\{k_1,k_2\}}(T$, $\ldots$, $T_{\mu(k_1,k_2)})$ and get $0$ in the
right\/-\/hand side Thence, $\phi_{\{k_1,k_2\}}=0$ and therefore
\[
\{\vph_{k_1},\vph_{k_2}\}=\square(0)=0.
\]
The indexes $k_1$ and $k_2$ are arbitrary,
therefore $\gA$ is commutative.
\end{proof}

\begin{state}
\label{EquivalentCommutSymmetries}%
Let $\cE_{(0)}=\{u_{t_0}=\vph_0(u_\sigma)\}$ be an evolution
equation. Suppose that the symmetry $\vph_k\in\sym\cE_{(0)}$
is assigned to each $k\geq0$ and assume that $\vph_k$ is independent
of the time $t_0$ explicitly\textup{:} $\vph_k=\vph_k(u_\sigma)$.
Then the following two statements are equivalent\textup{:}
\begin{enumerate}
\item\label{AlgIsCommutative}%
The algebra $\gA=\mathrm{span}_\BBR\langle\vph_k\mid
k\geq0\rangle$ is a commutative Lie algebra\textup{:}
$\{{\vph_k},{\vph_l}\}=0$\textup{.}
\item\label{AllAreSyms4All}%
The evolutionary vector field $\cEv_{\vph_l}$ is a symmetry
of the autonomous evolution equation
$\cE_{(k)}=\{u_{t_{k}}=\vph_{k}\}$ for each $k$, $l\geq0$.
\end{enumerate}
\end{state}

\begin{proof}
First, identify the evolutionary vector field $\cEv_{\vph_k}$ with
the autonomous evolution equation $u_{t_k}=\vph_k$. Then,
consider the equality
\[\cEv_{\vph_k}(\vph_l)=\cEv_{\vph_l}(\vph_k).\]
In its left\/-\/hand side we have
\[
\cEv_{\vph_k}(\vph_l)=\cEv_{u_{t_k}}(\vph_l)=\bigl(\bar
D_{t_k}-\tfrac{\dd}{\dd t_k}\bigr)(\vph_l)=\bar D_{t_k}(\vph_l),
\]
since $\vph_l$ does not depend on any time~$t_k$. In the right\/-\/hand
side of the equality we get
\[\cEv_{\vph_l}(\vph_k)=\ell_{\vph_k}(\vph_l).\]
Therefore, the comuutativity condition $\{\vph_k$, $\vph_l\}=0$ for the
symmetries $\vph_k$ and $\vph_l$ is equivalent to the determining
equation
\[\bigl(\bar D_{t_k}-\ell_{\vph_k}\bigr)(\vph_l)=0.\]
Thence $\vph_l$ is a symmetry of the equation $\cE_{(k)}$ for any
$k$, $l\geq0$.
\end{proof}

\begin{cor}
For any $k$, $l\geq0$, the following propositions hold:
\begin{enumerate}\item
The sections $\vvph_k\in\gA$ are not only symmetries of
the Toda equation, but also symmetries of all equations
$\cE_{(l)}=\{\bu_{t_l}=\vvph_l\}$:
\[\vvph_k\in\sym\cE_{(l)}^\infty.\]
\item
The recursion operator $R_\Toda$ is common for the whole tower of the
evolution equations~$\cE_{(l)}$:\ %
\[R_\Toda\in\mathrm{Rec}\,\cE_{(l)}.\] In particular,
\[R_\Toda=R_\pmKdV.\]
%
\end{enumerate}
\end{cor}

\begin{rem}
In the paper~\cite{Kaliappan}, the scalar potential modified
Korteweg\/--\/de Vries equation was considered in the gauge
$u_t=u_3+u_1^3$. In this case, this equation shares the recursion operator
with the sine\/-\/Gordon equation $u_{xy}=\sin u$.
\end{rem}

\begin{rem}
By using Corollary~\ref{local4pmKdV}, we deduce that the section
$\vvph'_{-1}={\const}$ is a central extension of the commutative Lie
subalgebra $\gA\subset\sym\cE_{(l)}^\infty$ of symmetries of the evolution
equation $\cE_{(l)}$ for any $l\geq0$, although
$\const\not\in\sym\cE_\Toda$.
\end{rem}

\begin{rem}
By using formula~\eqref{SquareArgumentsCommut}, we establish a curious
property of the Korteweg\/--\/de Vries equation~\eqref{KdV}.
Recall that the proof of Theorem~\ref{AIsCommutative} is based on the
equality $\phi_{\{k_1,k_2\}}=0$ that holds for all elements
$\phi_{k_1}$ and $\phi_{k_2}$ of the hierarchy~$\gB$.
Consider the first Hamiltonian structure $\hat B_1=D_x$
for Eq.~\eqref{KdV}. We know that the Hamiltonians of the higher
Korteweg\/--\/de Vries equations are in involution with respect to\ this structure.
Thence we have
\[
\bigl\langle\bE_T(\cH_i),\hat B_1\circ\bE_T(\cH_j)\bigr\rangle=0
\quad\Longleftrightarrow\quad
\phi_i\cdot D_x(\phi_j)\in\mathrm{im}\,D_x.
\]
Now we see that the last two summands in
Eq.~\eqref{SquareArgumentsCommut} are of this form.
Therefore the sum
\[
\cEv_{\phi_{k_1+1}}(\phi_{k_2})-\cEv_{\phi_{k_2+1}}(\phi_{k_1})
\in\mathrm{im}\,D_x
\]
is always a total derivative.
The feature we observe is that each of these  summands above is a total
derivative by itself, therefore generalizing the involutivity property.
Namely, by straighforward calculation we obtain $\Omega_{ij}$ such that
the relations
\[
\phi_i\cdot\frac{\dd}{\dd T}\bigl(D_x(\phi_j)\bigr)=D_x(\Omega_{ij})
\]
hold for the elements of the hierarchy~$\gB$.
Several $\Omega_{ij}$ that correspond to $i$, $j\leq2$, and the
initial elements
\begin{multline*}
\phi_{-1}=1,\quad \phi_0=s_1,\quad \phi_1=-\beta s_3+\tfrac{3}{2}s_1^2,\\
\phi_2=\beta^2s_5-\tfrac{5}{2}\beta s_2^2-5\beta s_1s_3+
\tfrac{5}{2}s_1^3
\end{multline*}
of the sequence $\gB$ are found in the table below:
\[
\boxed{
\begin{array}{ccc}
{} & {\textstyle j=1} & {\textstyle j=2} \\
{\textstyle i=-1} & 3T & -5\beta T_2+\tfrac{15}{2}T^2 \\
&& \\
{\textstyle i=0} &
\tfrac{3}{2}T^2 & 5T^3-5\beta TT_2+\tfrac{5}{2}\beta T_1^2 \\
&& \\
{\textstyle i=1}
& -\tfrac{3}{2}T_1^2+\tfrac{3}{2}T_2^2 &
  \tfrac{5}{2}\beta^2T_2^2+\tfrac{35}{8}T^4-\tfrac{15}{2}\beta T^2T_2\\
&& \\
{\textstyle i=2}
&
\begin{gathered}
\mathstrut-3\beta T_1T_3+\tfrac{3}{2}T_2^2\\
\mathstrut+\tfrac{15}{8}T^4-\tfrac{15}{2}\beta TT_1^2
\end{gathered}
&
\begin{gathered}
\mathstrut-\tfrac{5}{2}\beta^3T_3^2+\tfrac{15}{2}T^5+15\beta^2TT_1T_3\\
\mathstrut-\tfrac{25}{2}\beta T^3T_2-\tfrac{5}{2}\beta^2T_1^2T_2\\
\mathstrut+5\beta^2TT_2^2-\tfrac{75}{4}\beta T^2T_1^2.
\end{gathered}
\end{array}
}
\]
Obviously, $\Omega_{i,-1}=\Omega_{i,0}\equiv0$ for all~$i$.
\end{rem}

\section{The Hamiltonian formalism for the Euler equations%
}\label{SecLegendre}
   \label{SecHamDef} 
In this section, we consider the problem of constructing commutative
Hamiltonian hierarchies of evolution equations associated with
hyperbolic Euler systems (in particular, with the wave equation
$s_{xy}=0$ or the Toda equation~\eqref{eqToda},
see~\cite{KdVHier,LomRead2004,KisOvchTodaHam}.
Also, we interpret the canonical coordinate\/--momenta formalism,
which is widely used in mathematical physics (see~\cite{B-Sh, Ovch,
Dubrovin3} and references therein), within nonlocal Hamiltonian
operators language of the jet\/-\/bundle framework (\cite{ClassSym,
DIPS6-2002, Opava}) preserving the distinction between the coordinates
and momenta.
Next, we consider hyperbolic Euler equations (\textit{i.e.},
we specify the ansatz for the Lagrangian density) and obtain the
differential constraint between the dependent variables
$u^i$ and the momenta~$\gm_j$.
Then, we identify symmetries of these equations with autonomous
potential evolution equations; also, we investigate the relation
between Hamiltonian operators for potential and nonpotential evolution
equations that describe the evolution of coordinates and momenta,
respectively.
The aim of our reasonings is to assign new jet bundle $\pi'$ and the
jet space $J^\infty(\pi')$ to the initial Euler equation
$\cE_{E-L}\subset J^k(\pi)$, see Eq.~\eqref{ELE}, such that the latter
equation becomes an evolutionary vector field that belongs to
$\varkappa(\pi')$, while the evolutionary fields that correspond to the
momenta $\gm$ become elements of the adjoint module
$\hat\varkappa(\pi')$.
Finally, we relate commutative Lie subalgebras of the Noether
symmetries for hyperbolic Euler equations and pairs of Hamiltonian
hierarchies composed by potential and nonpotential evolution equations.

\subsection{Canonical formalism}\label{SecFormalismCoords}
Consider an abstract $2r$-dimensional dynamical system with $r$
dependent variables $u^i$, momenta $\gm_j$,
the spatial coordinates $x$, and the time $t$, defined by the Poisson
brackets~(\cite{B-Sh, Ovch})
\begin{equation}\label{PoissonDirac}
\begin{aligned}
\{u^i,u^j\}_{\bar A}&=0,\\
\{\gm_i,\gm_j\}_{\bar A}&=0,\\
\{u^i(x,t), \gm_j(x',t)\}_{\bar A}&=\bar A^i_{\,j}\,\delta(x-x');
\end{aligned}
\end{equation}
where $\bar A$ is an $(r\times r)$-matrix differential operator in total
derivatives
\[
D_\sigma={(D_{x^1})}^{\sigma_1}\cdots{(D_{x^n})}^{\sigma_n}
\]
with respect to $x$ ($\sigma=(\sigma_1$, $\ldots$, $\sigma_n)$
is a multiindex)  and the skew-symmetric
bracket acts as the derivation with respect to any of its arguments.
Assume that $\cH=[H\,\Id x]$ is a Hamiltonian with the density
\[
H(x)=H(u(x), u_{\sigma}(x); \gm(x), D_{\sigma}\gm(x)).
\]
Then the dynamics
\begin{equation}\label{HamEvol}
\begin{aligned}
\dot u&=\bigl\{u(x),H\bigl(u(x'),\gm(x')\bigr)\bigr\}_{\bar A},\\
\dot \gm&=\bigl\{\gm(x),H\bigl(u(x'),\gm(x')\bigr)\bigr\}_{\bar A}
\end{aligned}
\end{equation}
of the variables $u$ and $\gm$ is obtained in a standard way:
\[
\dot{u}=\{u(x),H(x')\}_{\bar A}{{}={}}
  \oint_{C(x)}\sum_{\sigma}\{u(x),D_{\sigma}\gm(x')\}_{\bar A}\cdot
\frac{\dd H}{\dd\left(D_{\sigma} \gm(x')\right)}\,\Id x'.
\]
Here $C(x)$ is a small contour around the point $x$.
We emphasize that this language of local coordinates and the
$\delta$-distribution is spread in many works on the Hamiltonian
formalism application in the field theory; in the sequel, we pass to
the invariant exposition soon.
Meanwhile, integrating by parts we obtain
\begin{subequations}\label{HamEqDirect}
\begin{align}
\dot{u}&=
\bar A\circ\frac{\delta H}{\delta \gm(x)}.\label{HamEqCoord}\\
\intertext{Similarly we obtain the second relation}
\dot{\gm}&=-\bar A\circ\frac{\delta H}{\delta u(x)}.\label{HamEqMomenta}
\end{align}
\end{subequations}
These reasonings motivate the following
definition of the variational bracket of
two Hamiltonians with the densities $H$ and $H'$, respectively: we set
\[
\{H,H'\}_{\bar A}=\frac{\delta H}{\delta u}\cdot\bar A\,\frac{\delta
H'}{\delta \gm}-\frac{\delta H}{\delta\gm}\cdot\bar A\,\frac{\delta
H'}{\delta u}.
\]

\begin{rem}\label{MixMomenta}
Usually, all dependent variables are treated uniformly.
In our case, the momenta can be absorbed
by the additional variables $u^{r+j}=\gm_j$ for $1\leq j\leq r$.
In other words, we enlarge the total number $m=2r$ of the
coordinates $u^j$ twice. The operators
$\bar A$ and $-\bar A$ are also united in the $(m\times m)$-matrix
\[A=\left(\begin{matrix}0&\bar A \\ \bar A^*&0\end{matrix}
\right),
\]
such that the dynamical equations take the form
\[
\Dot{\Vec{u}}=A\circ\frac{\delta\bigl(H(\bu)\bigr)}{\delta\vu},
\]
where the variation $\delta/\delta\vu$ is calculated with respect to
the new vector $\vu\equiv{}^t(u$, $\gm)$.
Due to this reason, Remark~\ref{MixMomenta} motivates definitions
\ref{DefPoissonBracket}--\ref{HamEquation},
see page~\pageref{DefHamOperator} in the Introduction.
\end{rem}

A traditional approach (\cite{ClassSym, DIPS6-2002, Opava})
to description of Hamitonian PDE dynamics does not
appeal to any distinction between the dependent variables, coordinates
$u$ and momenta $\gm$. Nevertheless, by using the double set of $m=2r$
dependent varuables $u$ and $\gm$ we describe several remarkable
properties of models of mathematical physics, for example, of the
Korteweg\/--\/de Vries equation \eqref{KdV}, the modified
Korteweg\/--\/de Vries equation \eqref{mKdVLiou}, and similar equations.
These aspects are discussed in Sec.~\ref{MagriTopo} below.

\section{Hyperbolic Euler equations}\label{SecHamLag}
Consider a first\/-\/order Lagrangian
\[
\cL=\int L(u, u_x, u_y; x, y)\,\Id x\wedge\Id y
\]
with the density
\[
L=-\tfrac{1}{2}\sum\limits_{i,j}\bar\kappa_{ij}u^i_xu^j_y+H(u;x,y),
\]
where $\bar\kappa$ is a nondegenerate, symmetric, constant
$(r\times r)$-matrix.
We note that the notation $\bar\kappa$ is correlated with
the general exposition. Suppose $r=1$ and $H=0$, then we obtain
the wave equation~\eqref{eqWaveNonlocal} (see Sec.~\ref{SecTodaRec}).
If $\bar\kappa=\kappa=\|a_ik_{ij}\|$ and the function $H$
is defined by formula~\eqref{HamTodaViaU}, then we get
the Toda equations.

Choose the independent variable $y$ for the ``time'' coordinate
(the nondegeneracy condition implies $\dd L/\dd u_y\not=0$),
leave $x$ for the spatial coordinate on the base $\BBR$
of the new jet bundle
$\pi'\colon\BBR\times\BBR \smash{\xrightarrow{u}} \BBR$
with the old fiber coordinate $u$,
and denote by $\gm_j=\dd L/\dd u^j_y$
the $j$th conjugate coordinate
(momentum) for the $j$th dependent variable
$u^j$ for any $1\leq j\leq r$:
\begin{equation}\label{RelCoordMomenta}
\gm_i = -\tfrac{1}{2}\sum\limits_{j=1}^r \bar\kappa_{ij}u^j_x.
\end{equation}
The differential constraint, Eq.~\eqref{RelCoordMomenta}, between
coordinates and momenta for the initial equation~\eqref{ELE} is our
main tool that allows to construct the Hamiltonian structures.

Consider the Legendre transform
\begin{align*}
H\,\Id x\wedge\Id y&=
\left\langle\gm,\frac{\dd L}{\dd\bu_y}\right\rangle - \cL\\
\intertext{and assign the Hamiltonian}
\cH(u, \gm)&=[H\,\Id x]
\end{align*}
to the Lagrangian $\cL$.
We decompose its density $H$ into the sum of two equal summands
and use the relation
\[u=-2\bar\kappa^{-1}\,\bar D_x^{-1}(\gm)\] in one
of the components, in agreement with~\eqref{HamEqDirect}:
\[H=\tfrac{1}{2} H[u]+\tfrac{1}{2} H[\gm].\]
The hyperbolic Euler equation
\[\cE_{E-L}=\{\bE_u(\cL)=0\}\]
is equivalent to the system
\begin{equation}\label{HamEvolEL}
u_y=\frac{\delta H}{\delta\gm},\qquad\gm_y=-\frac{\delta H}{\delta u}
\end{equation}
with respect to the \emph{canonical} Hamiltonian structure $\bar A=\bun$.
Owing to the relations
\[
\frac{1}{2}\frac{\delta}{\delta\gm}=
\underbrace{\bar\kappa^{-1}\cdot D_x^{-1}}_{A_1}
\circ\frac{\delta}{\delta u},\qquad
\frac{\delta}{\delta u}=\frac{1}{2}
\underbrace{D_x\circ\bar\kappa}_{\hat A_1}
\cdot\frac{\delta}{\delta\gm},
\]
the dynamical equations are separated:
\begin{equation}\label{eqLagHamForm}
\begin{aligned}
u_y&=A_1\circ\bE_u\bigl([H[u]\,\Id x]\bigr),\\
\gm_y&=-\tfrac{1}{2}\hat A_1\circ\bE_\gm\bigl([H[\gm]\,\Id x]\bigr).
\end{aligned}
\end{equation}

\begin{example}
Suppose that $r=1$, $\bar\kappa=\|1\|$, and the Hamiltonian is trivial:
$H\equiv0$. Then the Hamiltonian operators
\[
B_1=D_x^{-1},\qquad \hat B_1=D_x
\]
assigned to wave
equation \eqref{eqWaveNonlocal} are mutually inverse.
\end{example}

Now we apply these reasonings to the Toda equations~\eqref{eqToda},
which are assigned to Lagrangian~\eqref{TodaLDensity}.
It turns out that their Hamiltonian representation~\eqref{eqTodaHamForm}
hints the existence of the minimal integral $T$,
see Eq.~\eqref{EMT}. The conservation of $T$ is induced by the
conservation of Hamiltonian density~\eqref{HamTodaViaU} for
Eq.~\eqref{eqToda}.

We start with a general
\begin{lemma}\label{HamIsConservedLemma}
Assume that the density $H$ of a Hamiltonian $\cH=[H\,\Id x]$
for the Hamiltonian evolution equation $u_t=A(\bE_u(\cH))$
does not depend on the time $t$ explicitly. Then the density
$H$ is conserved for this equation\textup{:}
\[\bigl[\bar D_t(H)\,\Id x\bigr]=0.\]
\end{lemma}

\begin{proof}
By using the condition $\dd H/\dd t=0$, we calculate the derivative
$\bar D_t(H)$:
\begin{align*}
\bar D_t(\cH)&=\bigl\langle1,\cEv_{A\circ\bE_u(\cH)}(H)\bigr\rangle=
\sum\nolimits_\sigma \Bigl\langle\frac{\dd H}{\dd u_\sigma},
\bar D_\sigma\bigl(A\circ\bE_u(\cH)\bigr)\Bigr\rangle={} \\
\intertext{integrating by parts, we obtain}
{}&=\sum\nolimits_\sigma\bigl\langle (-1)^\sigma\,\bar D_\sigma
\Bigl(\frac{\dd H}{\dd u_\sigma}\Bigr),
A\circ\bE_u(\cH)\bigr\rangle={}\\
\intertext{since the Hamiltonian operator $A$ is
skew\/--\/symmetric, $A^*=-A$, we get}
{}&=-\bigl\langle A\circ\bE_u(\cH),\bE_u(\cH)\bigr\rangle.\\
\intertext{Again, by using the definition of the Euler operator
$\bE_u$ and integrating by parts, we have}
{}&=-\Bigl\langle\sum\nolimits_\sigma \bar
D_\sigma\bigl(A\circ\bE_u(\cH)\bigr),
\frac{\dd H}{\dd u_\sigma}\Bigr\rangle={}\\
{}&=-\bigl\langle\cEv_{A\circ\bE_u(H)}(H),1\bigr\rangle
=-\bar D_t(\cH).
\end{align*}
Moving the result to the left\/-\/hand side of the initial equality,
we finally obtain the required condition
$2\bar D_t(H)\in\mathrm{im}\,\bar D_x$.
\end{proof}

\begin{example}[\textup{\cite{LomRead2004}}]
Choose the coordinate $y$ for the
``time'', define the momenta (see~\eqref{RelCoordMomenta})
$\gm=\dd L/\dd\bu_y$:
\begin{equation}\label{Momenta}
\gm_i=\tfrac{1}{2}\sum\limits_{j=1}^{r}\kappa_{ij}u^j,
\end{equation}
and obtain the density $H_\Toda$ of the Hamiltonian
$\cH_\Toda$:
\[
H_\Toda(u,\gm)=-\tfrac{1}{2}\sum\limits_{i=1}^r a_i\,
\exp\left(\frac{2}{a_i} \,D_x^{-1}(\gm_i)\right)
-\tfrac{1}{2}\sum\limits_{i=1}^r a_i\,
\exp\Bigl(\sum\limits_{j=1}^r k_{ij}u^j\Bigr).
\]
We have already decomposed the density $H$ into the sum of two components
that depend on $\gm_i$ and $u^j$, respectively.
Then, the canonical Hamiltonian representation of the Toda equation
$\cE_\Toda$ is
\[
\left\{
\begin{aligned} \dot u^i&=\frac{\delta
H_\Toda}{\delta \gm_i}= D_x^{-1}\Bigl(\exp\Bigl(\sum\limits_{j=1}^r
k_{ij}u^j\Bigr)\Bigr),\\
\dot \gm_i&=-\frac{\delta H_\Toda}{\delta u^i}=
-\frac{1}{2}\sum\limits_{j=1}^r\kappa_{ij}\,\exp\Bigl(
\sum\limits_{l=1}^r k_{jl}u^l\Bigr).
\end{aligned}
\right.
\]
Therefore, in terms of the dependent variables $\bu$, we have
\begin{equation}\label{HamTodaViaU}
H_\Toda(\bu)=\sum\limits_{i=1}^r a_i\,
\exp\Bigl(\sum\limits_{j=1}^r k_{ij}u^j\Bigr)
\end{equation}
and
\begin{equation}\label{eqTodaHamForm}
\Dot{\bu}=A_1\circ\bE_{\bu}\bigl(\cH_\Toda(\bu)\bigr),
\end{equation}
where $A_1=\hat\kappa^{-1}\cdot D_x^{-1}$.
Recall that an evolution representation for the Liouville
equation~\eqref{eqhyp} was obtained earlier in Eq.~\eqref{eqhypEvol}.

In order to correlate the resulting expressions with~\eqref{EMT},
we apply the transformation $x\leftrightarrow y$ to the
Hamiltonian equation~\eqref{eqTodaHamForm}, consider the
Hamiltonian density~\eqref{HamTodaViaU}, which is independent of $y$,
and use Lemma~\ref{HamIsConservedLemma}. Then we obtain
\begin{equation}\label{TodaHamDensityIsConserved}
\bar D_x(H_\Toda) =
\bar D_y\Bigl(\sum\limits_{i=1}^r a_iu^i_{xx}\Bigr) =
\bar D_y\Bigl(\tfrac{1}{2}\sum\limits_{i,j=1}^r
     \kappa_{ij}u^i_xu^j_x\Bigr),
\end{equation}
whence follows expression~\eqref{EMT}.
\end{example}

By using the Hamiltonian representation~\eqref{eqTodaHamForm},
we describe the negative elements $\vph_k$ of the sequence $\gA$
with $k<0$:

\begin{state}
The right-hand side of the Hamiltonian evolution equation
\begin{equation}\label{eqTodaHam'}\tag{\ref{eqTodaHamForm}${}'$}
u_y=\hat\kappa^{-1}\circ D_x^{-1}\circ\bE_u(\cH_\Toda)
\end{equation}
is the inverse image of the translation $u_{t_0}=\vph_0$
with respect to\ the mapping $R_\Toda$\textup{;} the translation
$u_y$ is the element $\vph_{-1}$ of the sequence~$\gA$.
\end{state}

\begin{proof}
By Lemma~\ref{FactorRecLemma}, we have
\[
R_\Toda(\vvph_{-1})=\square
\left(\cEv_{\bu_y}(s)\right)=\square(\bar D_y(s))=\square(1)=u_x.
\]
\end{proof}

Therefore, representations \eqref{eqTodaHamForm} and \eqref{eqTodaHam'}
of the Toda equation provide the translations
\begin{align*}
u_y&=\vph_{-1}=\hat\kappa^{-1}\circ\bar
D_x^{-1}\circ\bE_u(H_\Toda)\\
\intertext{and}
u_x&=\vph_{0}=\hat\kappa^{-1}\circ\bar
D_y^{-1}\circ\bE_u(H_\Toda)
\end{align*}
that are related by the recursion operator $R_\Toda$:
\[
\begin{diagram}
\vph_{-2}=\bar\square(\bar T) &
\lto_{R'_\Toda} & u_y=\vph_{-1} &
\pile{\rto^{R_\Toda\,{}}\\ \lto_{R'_\Toda}} &
\vph_0=u_x & \rto^{R_\Toda\,{}} & \vph_1=\square(T)
\end{diagram}
\]  
The operator $R'_\Toda$ obtained from $R_\Toda$
by using the discrete symmetry $x\leftrightarrow y$
generates the symmetries $\vph_k\in\gA$ of the Toda equation
with $k\leq-1$.

Now we return to the discussion on the Hamiltonian formalism for the
Euler equations $\cE_{E-L}=\{\bE_u(\cL)=0\}$.
We identify the symmetries $\vph$ of these equations
with the evolution equations $u_t=\vph$.
Consider a symmetry $\vph(u_x$, $u_{xx}$, $\ldots)$ of Eq.~\eqref{ELE},
\textit{i.e.}, the \emph{potential} evolution equation
\begin{subequations}\label{SymEq}
\begin{align}
u_t&=\vph(u_x,u_{xx}\ldots),\label{SymEqCoord}\\
\intertext{then the induced evolution $\gm_t$ of the momenta
is described by the \emph{non}\-po\-ten\-tial equation}
\gm_t&=-\tfrac{1}{2} \bar\kappa\cdot
D_x\bigl({}^t\vph(\gm,\gm_x,\ldots)\bigr).\label{SymEqMom}
\end{align}
\end{subequations}
In addition, assume that the evolution $\vph$ is Hamiltonian:
\[
\dot u=\frac{1}{2}\frac{\delta H}{\delta\gm},\qquad
\dot\gm=-\frac{1}{2}\frac{\delta H}{\delta u}.
\]
Then we have
\begin{equation}\label{HamSymEqCoords}
\begin{aligned}
\dot u&= \underbrace{\bar\kappa^{-1}
   \cdot D_x^{-1} }_{A_1}\,\frac{\delta H}{\delta u},\\
\dot\gm&=-\tfrac{1}{2}\underbrace{D_x\cdot\bar\kappa}_{\hat A_1}\,
   \frac{\delta H}{\delta\gm},
\end{aligned}
\end{equation}
\textit{i.e.}, \emph{both} equations~\eqref{SymEq} are Hamitonian
simultaneously and their Hamiltonian structures $A_1$ and $\hat A_1$
are mutually inverse.

We note two classical examples of the pairs of evolution
equations that admit mutually inverse Hamiltonian operators.

\begin{example}[\textup{\cite{LomRead2004}}]
The potential KdV equation \eqref{pKdV} is Hamiltonian with respect to\ the
operator $B_1=D_x^{-1}$, while $\hat B_1=D_x$ is the first Hamiltonian
operator (see \eqref{IKdVHam}) for the KdV equation~\eqref{KdV}.
One can easily verify that Eq.~\eqref{pKdV} is compatible with the wave
equation \eqref{eqWaveNonlocal}, \textit{i.e.}, the right\/-\/hand side\ $\phi_1$ in
the potential KdV equation $s_t=\phi_1$ is a symmetry of $s_{xy}=0$,
while the nonpotential equation~\eqref{KdV} describes the evolution of
the momentum $T=s_x$ (up to the constant factor $-1/2$).
\end{example}

The potential and nonpotential modified Korteweg\/--\/de Vries
equations, Eq.~\eqref{pmKdV} and \eqref{mKdV}, respectively,
supply another example.
In fact, the succeeding subsections~\ref{SecpKdVKdV} and \ref{SecMiura}
contain the detailed analysis of these two examples.
We demonstrate that the first pair defines the symmetry
hierarchy $\gB$ for wave equation~\eqref{eqWaveNonlocal}
and the second pair is related with the hierarchy~$\gA$ of symmetries of
the Toda equation~\eqref{eqToda}.

Now we return to Eq.~\eqref{HamSymEqCoords} and
note an important property of the Hamiltonian equations
\begin{align*}
\dot u&=A_1\circ\bE_u(\cH)\\
\intertext{and}
\dot\gm&=-\tfrac{1}{2}\hat
A_1\circ\bE_\gm(\cH)=-\tfrac{1}{2}\bE_u(\cH).
\end{align*}
It turns out that we have already met the expression in the
right\/-\/hand side of the latter equation in Lemma~\ref{OpavaGenf}
on page~\pageref{OpavaGenf}. Lemma~\ref{OpavaGenf} assigns
the generating sections $\psi_\eta=\bE_u(\eta_0)$ to the densities
$\eta_0$ of conservation laws $[\eta]$ for evolution equations.
Formulas~\eqref{SymEq} give an interpretation of Lemma~\ref{OpavaGenf}:
the generating sections $\psi$ describe (up to the sign) the evolution
$\dot\gm$ of the momenta $\gm$ along the Hamiltonian symmetries $\vph$
of the initial equation~$\cE$.

Moreover, we relate two pairs of mappings of different types.
Assume that there are recursion operators $R_u$ and $R_\gm$
for \emph{different} evolution equations \eqref{SymEqCoord} and
\eqref{SymEqMom}, respectively, and consider the mappings
\[
R_u\colon\varkappa\to\varkappa,\qquad
\cT_u\colon\hat\varkappa\to\hat\varkappa
\]
which produce symmetries and generating sections of conservation
laws for the same equation~$\cE_u$, respectively.
The relation~(\cite{DIPS6-2002})
\begin{equation}\label{RelateRec}
\cT_u=R_u^*
\end{equation}
holds for evolution equations; see, for example,
diagram~\eqref{NLSDiag} on page~\pageref{NLSDiag}.
In terms of the \eqref{SymEq},
Eq.~\eqref{RelateRec} means that the skew\/--\/symmetric recursion
operator
\begin{equation}\label{HamViaRec}
A_R=\begin{pmatrix} 0 & R_u \\ -R_u^* & 0 \end{pmatrix}
\end{equation}
is defined for the Hamiltonian representation~\eqref{eqLagHamForm} of
the initial Euler equation $\cE=\{\bE_u(\cL)=0\}$.
This situation is realized for the above\/--\/mentioned pairs of
equations, see~\eqref{TwoDecompRpKdV} on page~\pageref{TwoDecompRpKdV}.
We also note that
the condition $R_v=R_u^*$ is valid in this case.

Now we study the problem of constructing a bi-Hamiltonian hierarchy
(to be more precise, a pair of hierarchies with respect to\ $u$ and $\gm$)
by using the recursion operator~\eqref{HamViaRec}.
Namely, we find the conditions for the operator $A_R$ to
induce the \emph{second} structure $\{u$, ${\gm\}}_{A_R}$
on the Hamiltonians $\cH_i$ in the diagram
\[
\begin{CD}
\cH_0 @. \cH_1 @. \cH_2 @. {} \\
@V{\frac{\delta}{\delta\gm}}VV @V{\frac{\delta}{\delta\gm}}VV
@V{\frac{\delta}{\delta\gm}}VV @.\\
\vph_0 @>>{R}> \vph_1 @>>{R}> \vph_2 @>>{R}> {\ldots}
\end{CD}
\]
such that the Jacobi identity~\eqref{Jacobi} holds.
The commutativity condition $[\vph_i$, $\vph_j]=0$
is sufficient for the Jacobi identity to hold for bracket
\eqref{PBracket}, supplied by the operator $A_R$; the sections
$\gm_{t_i}=\psi_i$ also commute in this case.  We denote
by $\mathfrak{U}$ the minimal Lie algebra generated by the sections
$u_{t_i}=\vph_i$.
We emphasize that, in general, we require
these operators \eqref{HamViaRec} to provide the Lie algebra structure
\eqref{Jacobi} on the Hamiltonians $\cH_i\in\bar H^n(\pi)$ but not on
the whole horizontal cohomology group $\bar H^n(\pi)$. Therefore, our
concept extends Definition \ref{DefHamOperator} of a Hamiltonian
operator since we treat the particular case of a hyperbolic Euler
equation $\cE$ and a sequence of its Hamiltonian symmetries $\vph_i$
whose Hamiltonians are $\cH_i$, respectively.
Further on, suppose $\mathfrak{U}$ is a Lie subalgebra of the
\emph{Noether} symmetries of the Lagrangian $\cL$. This assumption is
sufficient for the existence of the
Hamiltonians $\cH_i$ such that
\begin{equation}\label{MagriRect}
\vph_i=R\,\frac{\delta\cH_{i-1}}{\delta\gm}=\bun\cdot
\frac{\delta\cH_{i}}{\delta\gm},\qquad
\psi_i=-R^*\,\frac{\delta\cH_{i-1}}{\delta u}=-\bun\cdot
\frac{\delta\cH_{i}}{\delta u}.
\end{equation}
Indeed, the existence of the conserved densities is prescribed
by the Noether theorem (see Theorem~\ref{InverseNoether} on
page~\pageref{InverseNoether}).

We obtain a more usual description of the pair of the Magri
schemes (see~\cite{Magri}) by using the splitting~\eqref{HamSymEqCoords}:
\begin{equation}\label{MagriElem}
\begin{diagram}
&&{\psi_{i+1}}&&{\ldots}&& \psi_{i+1}&&{\ldots}&&{}\\
&&&\rdto^{A_1}&\uto_{R}&& &\luto^{\hat A_2} & \uto_{R} &&\\
&&\uto^{R^*}&&\vph_{i+1}&&
  \uto^{R^*}&&\vph_{i} & \lto^{\bE_\gm} & \cH_{i}\\
&&&\ruto^{A_2}&&& &\ldto^{\hat A_1}&&&\\
\cH_i&\rto^{\bE_u}&\psi_i&&\uto_{R,}&& \psi_{i}&&\uto_{R}&&\\
&&&\rdto^{A_1}&&& &\luto^{\hat A_2}&&&\\
&&\uto^{R^*}&&\vph_i&&
   \uto^{R^*}&&\vph_{i-1}&\lto^{\bE_\gm} & \cH_{i-1}.\\
&&&\ruto^{A_2}&\uto_{R}&& & \ldto^{\hat A_1} & \uto_{R}&&\\
&&{\psi_{i-1}}&&{\ldots}&& \psi_{i-1}&&{\ldots}&&
\end{diagram}
\end{equation}
Here the operators $A_1$ and $\hat A_1$ are defined by
constraint \eqref{RelCoordMomenta} between $u$ and $\gm$, while the
second operators $A_2$ and $\hat A_2$ originate from the relations
\[
R=A_2\circ A_1^{-1},\quad R^*=\hat A_2\circ\hat A_1^{-1}
\]
respectively.
In the sequel, we preserve the notation and mark
the operators $\hat A_{1,2}$ and $\hat B_{1,2}$ for the
{non}potential equations~\eqref{mKdV} and \eqref{KdV},
respectively, with the `hat' sign.

The notion of the $\ell^*$-\emph{covering} in the category of
differential equations was introduced in \cite{DIPS6-2002}, and the
unifying approach towards the recursion operators
$R\colon\varkappa\to\varkappa$, the conjugate recursion operators
$\cT\colon\hat\varkappa\to\hat\varkappa$, the Hamiltonian structures
$A\colon\hat\varkappa\to\varkappa$, and the symplectic structures $\hat
A\colon\varkappa\to\hat\varkappa$ was elaborated for the Magri scheme
\eqref{MagriElem} technique.
The return from diagram~\eqref{MagriElem} to~\eqref{MagriRect}
provides new interpretation of these structures and their
interrelations.

\begin{rem}
The correlation between the generating sections of conservation laws
for the initial Euler equation
$\cE_{E-L}=\{F\equiv\bar\kappa\cdot\bE_u(\cL)=0\}$
\textup{(}these sections are denoted by $\psi_\cL$\textup{)}
and for the same equation in its evolution
representation~\eqref{eqLagHamForm}
\textup{(}we denote these sections by $\psi$\textup{)}
is given by the diagram
\begin{equation}\label{GSChainDiag}
\begin{CD}
\psi @>{-D_x^{-1}}>> \psi_\cL @>{\bar\kappa^{-1}}>> \vph,
\end{CD}
\end{equation}
where the first arrow follows from the definition of a generating
section\textup{:}
\[
\Id_h\eta=\nabla(F)\,\Id\bx=\nabla\circ D_x(D_x^{-1}(F))\,\Id\bx,
\]
and, therefore,
\[
\psi_\cL=\nabla^*(1),\quad \psi=-D_x\circ\nabla^*(1),
\]
while the second arrow is induced by Lemma~\ref{BehaviourLemma}.
\end{rem}

\section{Properties of the Korteweg\/--\/de Vries
hierarchies}\label{MagriTopo}
In this section, we illustrate the concept
(\cite{Ovch, LomRead2004, KisOvchTodaHam}) formulated
in the preceding section.
In what follows, we consider the bi\/--\/Hamiltonian equations
\eqref{KdV},  \eqref{pKdV}, \eqref{pmKdV},  and the
$r$-component analogs
\begin{equation}\label{mKdV}
\cE_{\mKdV}=\{\vth_{t_1}=D_x\circ\hat\kappa\circ\square(T)\}
\end{equation}
of the modified Korteweg\/--\/de Vries equation~\eqref{mKdVLiou}.
We also introduce the new variable
\begin{equation}\label{pmKdVmKdV}
\vth=\kappa\cdot u_x
\end{equation}
that is subject to the latter equation.
This dependent variable differs from the momenta
\[\gm_\Toda=-\tfrac{1}{2}(\kappa\,u_x)^*\]
by the constant factor $(-2)$. This gauge simplifies the
calculations and resulting expressions.

Now we study the relations between the equations
$\cE_\pKdV$ and $\cE_\KdV$,
and between $\cE_\pmKdV$ and $\cE_\mKdV$, respectively.
We claim that the potentials $u$ and $s$ for $\vth$ and $T$
are such that the linearizations
\[
\ell_\vth^{(u)}=D_x\circ\hat\kappa, \quad
\ell_T^{(s)}=D_x
\]
are equal to the first Hamiltonian structures $\hat
A_1$ and $\hat B_1$ for the equations $\cE_{\mKdV}$ and $\cE_{\KdV}$,
respectively. The first Hamiltonian structures $A_1$ and $B_1$ for
$\cE_{\text{p(m)KdV}}$ are inverse with respect to
$\hat A_1$ and $\hat B_1$.
The Hamiltonians for the equations $\cE_{\text{(p)KdV}}$ are
well\/-\/known. The Hamiltonians $\cH_k$ for the equations
$\cE_{\text{(p)mKdV}}$ are described
in final Theorem~\ref{CylinderTh}
on page~\pageref{CylinderTh}.

\subsection{The Korteweg\/--\/de Vries equation}\label{SecpKdVKdV}
Let $\cL_s$ be the Lagrangian
\[\cL_s=-\bigl[\tfrac{1}{2}s_xs_y\,\Id x\wedge\Id y\bigr].\]
Assign the wave equation
\[
\cE_s=\{s_{xy}=0\},
\]
see Eq.~\eqref{eqWaveNonlocal} on page~\pageref{eqWaveNonlocal},
to $\cL_s$ and consider the commutative Lie subalgebra
\[
\gB={\mathrm{span}}_\BBR\langle R_\pKdV^{k+1}(\phi_{-1}),\ %
\phi_{-1}=1,\ k\geq0\rangle\subset\sym\cE_s
\]
of its symmetries.
We identify this subalgebra with the hierarchy of the higher
potential Korteweg\/--\/de Vries evolution equations,
see Eq.~\eqref{pKdV}.
We continue using the variable $T=s_x$, whose evolution is described by
the elements of the \emph{non}potential Korteweg\/--\/de Vries
hierarchy for Eq.~\eqref{KdV}, instead of the canonical momentum
$\gm_s=-\tfrac{1}{2}s_x$ for the wave equation.
By~\eqref{HamSymEqCoords}, these two hierarchies
admit the pair of mutually inverse Hamiltonian operators
$\smash{\hat B_1}=B_1^{-1}=D_x$ for
the equations $\cE_\KdV$ and $\cE_\pKdV$, respectively.
The coinciding elements within the initial part of the
Magri schemes (see Eq.~\eqref{MagriElem})
for $\cE_\KdV$ and $\cE_\pKdV$ are underlined in the diagram below:

\begin{multline}
\begin{diagram}
&&&&\ldots&&\\
&&&\ruto_{B_2}&&&\\
h_1[s] 
&\rto^{\bE_s}&\underline{\underline{ -\beta
  s_4+3s_1s_2}}&&\uto_{R_\pKdV}&&\ldots\\
&\rdto^{-\id{}}&&\rdto^{B_1=D_x^{-1}}&&\ruto_{\hat B_2}&\\
&&h_1[T]=\bigl[\tfrac{1}{2}(\beta T_1^2+T^3)\,\Id x\bigr]&\rto^{\bE_T}&
  -\beta s_3+\tfrac{3}{2}s_1^2&& \uto^{R_\KdV}\\
&&&\ruto_{B_2}&&\rdto^{\hat B_1=D_x}&\\
h_0[s] 
&\rto^{\bE_s}&
  \phantom{MM}\underline{s_2}&&\uto^{R_\pKdV}&&
  \underline{\underline{-\beta T_3+3T\cdot T_1}}\\
&\rdto^{-\id{}}&&\rdto^{B_1}&&\ruto_{\hat B_2}&\\
&&h_0[T]=\bigl[\tfrac{1}{2}T^2\,\Id x\bigr]
  &\rto^{\bE_T}&s_1&&\uto^{R_\KdV}\\
&&&&&\rdto^{\hat B_1}&\\
&&&&\uto^{R_\pKdV=R_\KdV^*} && \underline{T_1}\\
&&&&&\ruto_{\hat B_2=
\lefteqn{\scriptstyle-\beta\,D_x^3+D_x\circ T+T\cdot D_x}}&\\
&&h_{-1}[T]=[T\,\Id x]&\rto^{\bE_T}& 1 &&
\end{diagram}\\
{}\label{IdentifyKdVBottomDiag}
\end{multline}
The initial terms of this diagram are given by the Hamiltonians whose
densities are known from the vastest literature:
\[
h_{-1}=[T\,\Id x],\qquad
h_0=[\tfrac{1}{2}T^2\,\Id x],\qquad
h_1=[\tfrac{1}{2}(\beta T_x^2+T^3)\,\Id x],\quad
\text{etc.}
\]

Now we consider an example to the relation~\eqref{RelateRec}.
Namely, the identity
\[
R_\pKdV=R_\KdV^*
\]
is satisfied by the recursion operator
$R_\pKdV$  
for the potential Korteweg\/--\/de Vries
equation~\eqref{pKdV} and the recursion operator
\[
R_\KdV=   
   -\beta D_x^2 + 2T + T_1\cdot D_x^{-1}
\]
for the Korteweg\/--\/de Vries equation~\eqref{KdV}.
The recursion $R_\pKdV$ is factorized by the Hamiltonian structures as
follows:
\begin{multline}\label{TwoDecompRpKdV}
\overbrace{(-\beta\,D_x+s_1\cdot D_x^{-1}+D_x^{-1}\circ s_1)}^{B_2}
\circ D_x = {}\\
  D_x^{-1}\circ\underbrace{(-\beta\,D_x^3+D_x\circ T+
      T\cdot D_x)}_{\hat B_2}.
\end{multline}
The second Hamiltonian structure $\hat B_2$ for Eq.~\eqref{KdV}
is a factor in the right\/-\/hand side of Eq.~\eqref{TwoDecompRpKdV}.
Recall that this structure appeared also in Remark~\ref{EvolEMTExample}
on page~\pageref{EvolEMTExample}.
Another important property of this structure is the correlation between
the equation $\cE_\KdV$, the Toda equation $\cE_\Toda$,
and the Virasoro algebra (\cite{Bombai}). We analyse the latter
property in Remark~\ref{RemVir}, see the end of this subsection.

The Hamiltonian operators $B_1$ and $\hat B_1$ are inverse,
therefore the Magri sche\-mes \eqref{MagriElem} for
Eq.~\eqref{pKdV} and~\eqref{KdV} are correlated.
\begin{equation}\label{IdentifyKdVDiag}
\begin{diagram}
&&&&\ldots&&\\
&&&\ruto_{B_2}&&&\\
h_k[s]&\rto^{\bE_s}&\underline{\underline{%
  \psi_k}}&&\uto_{\lefteqn{{\scriptstyle R_\pKdV=R_\KdV^*}}}&&\ldots\\
&\rdto^{-\id{}}&&\rdto^{B_1=D_x^{-1}}&&\ruto_{\hat B_2}&\\
&&h_k[T]&\rto^{\bE_T}&s_{t_k}=\phi_k=\Psi_k&&
  \uto_{R_\KdV=R_\pKdV^*}\\
&&&\ruto_{B_2}&&\rdto^{\hat B_1=B_1^{-1}}&\\
h_{k-1}[s]&\rto^{\bE_s}&\underline{\psi_{k-1}}&&\uto^{R_\pKdV}&&
  \underline{\underline{T_{t_k}=\Phi_k=\psi_k}}\\
&\rdto^{-\id{}}&&\rdto^{B_1}&&\ruto_{\hat B_2}&\\
&&h_{k-1}[T] &\rto^{\bE_T}&s_{t_{k-1}}=\phi_{k-1}=\Psi_{k-1}
     &&\uto_{R_\KdV}\\
&&&&&\rdto^{\hat B_1}&\\
&&&&\uto^{R_\pKdV=R_\KdV^*} && \underline{T_{t_{k-1}}=\Phi_{k-1}
   =\psi_{k-1} }.\\
&&&&&\ruto_{\hat B_2}&\uto_{R_\KdV}\\
&&&& \ldots &&\ldots
\end{diagram}
\end{equation}
Here $\phi_k$ and $\Phi_k$ are symmetries and
$\psi_k$ and $\Psi_k$ are the generating functions of conservation
laws for the potential Korteweg\/--\/de Vries equation
\begin{align*}
s_{t_1}&=\phi_1\\
\intertext{and the Korteweg\/--\/de Vries equation}
T_{t_1}&=\Phi_1,
\end{align*}
respectively.
We conclude that
the opposite sides of diagram~\eqref{IdentifyKdVDiag} are
identified with the one\/-\/step vertical shift.

\begin{cor}
We conclude
from diagram~\eqref{IdentifyKdVDiag} that the symmetries $\Phi_k$ of
the bi-Hamiltonian hierarchy for the Korteweg\/--\/de Vries
equation are the gradients $\psi_k$ of the
Hamiltonians $h_k[s]$ for the
potential Korteweg\/--\/de Vries equation~\eqref{pKdV},
and \textit{vice versa}.
\end{cor}

The equations $\cE_{\KdV}$ and $\cE_{\pKdV}$ share the same set of the
Hamiltonians with the densities $h_k[T]$ and $h_k[s]$, respectively.
Thence we obtain the following property of
symmetry subalgebra $\gA$ of the Toda equation:

\begin{theor}\label{NoetherSubalgebra}
The generators $\vph_k$ of the commutative Lie algebra $\gA$
are the Noether symmetries of the Toda equation\textup{:}
\[\vph_k\in\sym\cL_\Toda.\]
\end{theor}

\begin{proof}
Indeed, we have
\[
\gA\ni\vph_k=\square(\phi_{k-1})=\square\circ\bE_T(h_k[T])\in
\sym\cL_\Toda\subset\sym\cE_\Toda
\]
by Remark~\ref{RemOnSakovich} on page~\pageref{RemOnSakovich}.
\end{proof}

\begin{rem}\label{LawsForARem}
The conservation laws $[\eta_k]$ associated with the generating
sections \[\psi_{k+1}^\Toda=\hat\kappa\cdot\square\circ\bE_T(h_k)\]
(see Eq.~\eqref{EqDescribeNoether} on page~\pageref{EqDescribeNoether}
and diagram~\eqref{GSChainDiag}) are exactly the Hamiltonians
$h_k\,\Id x$ of the higher Korteweg\/--\/de Vries equations
$s_{t_k}=\phi_k$.
\end{rem}

This fact, $[h_k\,\Id x]\in\bar H^1(\cE_\Toda)$, was noted in a
similar situation in~\cite[\S 10]{DSViniti84} but required a nontrivial
proof there.

By Lemma~\ref{HamIsConservedLemma},
the densities $h_k$ of the Hamiltonians for both Korteweg\/--\/de Vries
equations~\eqref{KdV} and \eqref{pKdV}
are conserved on the corresponding higher analogs
$T_{t_k}=D_x(\phi_k)$ and $s_{t_k}=\phi_k$
of these equations. We have
\[
\bar D_{t_k}(h_k)=\bar D_x(\Omega_k^\KdV).
\]
We claim that the hierarchy $\gB$ is composed by conserved densities for
the hierarchy $\gA$ of potential modified equation~\eqref{pmKdV}.
We need the following useful lemma to prove this claim.

\begin{lemma}[\textup{\cite{Opava}}]\label{EvThroughEuler}
The relation
\[
\bE(\cEv_\vph(\cL))=\cEv_\vph(\bE(\cL))+\ell_\vph^*(\bE(\cL))
\]
holds for any $\vph\in\varkappa$ and $\cL\in\bar\Lambda^n(\pi)$.
\end{lemma}

\begin{proof}
Suppose $\Delta\in\CDiff(\varkappa$, $\bar\Lambda^n(\pi))$.
By multiple integrating by parts, we transform
the expression $\ell_{\Delta(\vph)}$ to
\[\ell_{\Delta_0(\vph)}+D_x\circ\Delta'(\vph),\] where
the order of $\Delta_0$ is zero and
$\Delta'(\vph)\in\CDiff(\varkappa$, $\bar\Lambda^n(\pi))$;
then by using Eq.~\eqref{PreDetEq} we obtain
\[
\bE(\Delta(\vph))=\ell^*_\vph(\Delta^*(1))+\ell^*_{\Delta^*(1)}(\vph)
\]
for any section $\vph\in\varkappa$.
Now, let $\cL\in\bar\Lambda^n(\pi)$ be a form and put
$\Delta=\ell_\cL\colon\varkappa\to\bar\Lambda^n(\pi)$.
The linearization $\ell_{\mathstrut\bE(\cL)}=\ell^*_{\bE(\cL)}$
of the image of the Euler operator is self\/--\/adjoint, hence we
obtain the equality
\[
\bE(\ell_\cL(\vph))=
  \ell^*_\vph(\bE(\cL))+\ell_{\bE(\cL)}(\vph),
\]
whence follows Lemma~\ref{EvThroughEuler}.
\end{proof}

\begin{rem}
In particular, from this lemma it follows that any Noether symmetry
$\vph_\cL$ of a Lagrangian $\cL$ (we recall that the condition
$\cEv_\vph(\cL)=0$ holds on $J^\infty(\pi)$ in this case)
is also a symmetry of the Euler equation~\eqref{ELE} that is assigned
to~$\cL$, \textit{i.e.},
\[\sym\cL\subseteq\sym\cE.\]
Lemma~\ref{EvThroughEuler} gives an explanation why the inverse statement
is wrong.
\end{rem}

\begin{state}[\textup{\cite{KdVHier}}]\label{CylinderLaws}
Suppose $k\geq0$ is arbitrary.
Then the $k$\textup{th} term $\phi_k=\bE_T(h_k\,\Id x)$
of the hierarchy $\gB$ is a conserved density for the
$k$\textup{th} higher potential modified
Korteweg\/--\/de Vries equation.
\end{state}

\begin{proof}
First, we recall Theorem~\ref{ConsOfTences} that describes the
correlation of the times within the hierarchies $\gA$ and $\gB$.
Second, we apply Lemma~\ref{EvThroughEuler} that provides the rule to
commutate the Euler operator and an evolutionary derivation. Third,
we recall that the higher symmetries $\phi_k$ are independent of the jet
variable~$s$. Thence we obtain
\begin{multline*}
\bar D_{t_k}(\phi_k)=
\cEv_{\phi_{k}}(\bE_T(h_k))=
\underbrace{\bE_T(\overbrace{\cEv_{\phi_{k}}(h_k)}^{%
\bar D_{t_k}(h_k)\lefteqn{%
{\scriptstyle{}=\bar D_x(\Omega_k^\KdV)}}} %
) }_{\equiv0} - \ell^*_{\phi_{k}}(\bE_T(h_k))={}\\
{}=
-\sum_{j>0}(-1)^j \bar D_x^j\circ\left(\frac{\dd\phi_{k}}{\dd s_j}
\cdot\bE_T(h_k)\right)\in\mathrm{im}\,\bar D_x.
\end{multline*}
Therefore the density $\phi_k$ is conserved on the equation
$\cE_{(k)}=\{u_{t_k}=\vph_k\}$.
\end{proof}

\begin{rem}\label{RemVir}
The second Hamiltonian structure
\[
\hat B_2=-\beta\,\bar D_x^3+\bar D_x\circ T+T\cdot\bar D_x,
\]
for the equation $\cE_{\KdV}$ endowes the Fourier coefficients
$\gt_k$ of the energy\/--\/momentum component (see Eq.~\eqref{EMT})
\[
T=\sum_{k\in\BBZ}\frac{\gt_k}{x^{k+2}}
\]
for the Toda equation~\eqref{eqToda} with the structure of the
Virasoro algebra (\cite{Bombai}) with the central charge
$c=-\tfrac{3}{4}\beta$. We have
\begin{equation}\label{VirRel}
\begin{aligned}
2\pi\bi\,[\gt_n,\gt_m]&=2(n-m)\,\gt_{n+m}-\beta\cdot(n^3-n)\,
\delta_{n+m,0},\\
[\gt_k,\beta]&=0.
\end{aligned}
\end{equation}
We emphasize that
the known property \cite{BilalGervaisLagrange, OvchWarsaw}
of the integral $T$ is adapted to the case of the Toda
equation~\eqref{eqToda} associated with a nondegenerate symmetrizable
matrix~$K$ that is not necessarily the Cartan matrix of a semisimple Lie
algebra. The central charge $c$ is  expressed by using the constant
$\beta=\sum_i a_i\Delta^i$ that depends on the symmetrizator $\vec a$.
See Remark~\ref{RemWeAdapt} on page~\pageref{RemWeAdapt} for
a similar situation.

In the paper~\cite{ForKac}, we considered a class of generalizations of
the Virasoro algebra, see Eq.~\eqref{VirRel}, such that the determining
relations for those generalizations are given by an $N$-ary
skew\/--\/symmetric bracket and~$N\geq2$.
\end{rem}

\subsection{The analogs of the modified Korteweg\/--\/de Vries
equation}\label{SecMiura}
In this subsection, we investigate the relation
between the potential and nonpotential modified
Korteweg\/--\/de Vries equations that correspond to the
symmetry subalgebra $\gA\subset\sym\cL_\Toda\subset\sym\cE_\Toda$
of the Euler type Toda equations~\eqref{eqToda}.
We analyse the properties of the Miura transformation
$T=T(\vth$, $\vth_x)$ that maps solutions of the modified
Korteweg\/--\/de Vries equation~\eqref{mKdV} to solutions
of Eq.~~\eqref{KdV} and thence we establish the invariant nature of the
operator $\square$, see Eq.~\eqref{Square} on page~\pageref{Square},
that appeared in description of the structure of symmetries for the Toda
equations. Also, we construct the Hamiltonian structures for the
hierarchy $\gA$ and prove that modified equations~\eqref{pmKdV} and
\eqref{mKdV} share the Hamiltonians $[h_k\,\Id x]$ with the
Korteweg\/--\/de Vries equations~\eqref{KdV} and~\eqref{pKdV}.

We recall that  the dependent variable $\vth$ introduced in
Eq.~\eqref{pmKdVmKdV} satisfies the modified Korteweg\/--\/de
Vries equation~\eqref{mKdV} and its higher analogs~$\cE_{\mKdV(k)}$.
By Corollary~\ref{local4pmKdV} on page~\pageref{local4pmKdV},
all right\/-\/hand sides in the equations $\cE_{\mKdV(k)}$ are local
with respect to~$\vth$.

Similarly to Sec.~\ref{SecpKdVKdV}, we establish
identifications between the higher symmetries of the potential equation
$\cE_\pmKdV$ and the generating sections of conservation laws
for the equation~$\cE_\mKdV$, and \textit{vice versa}:
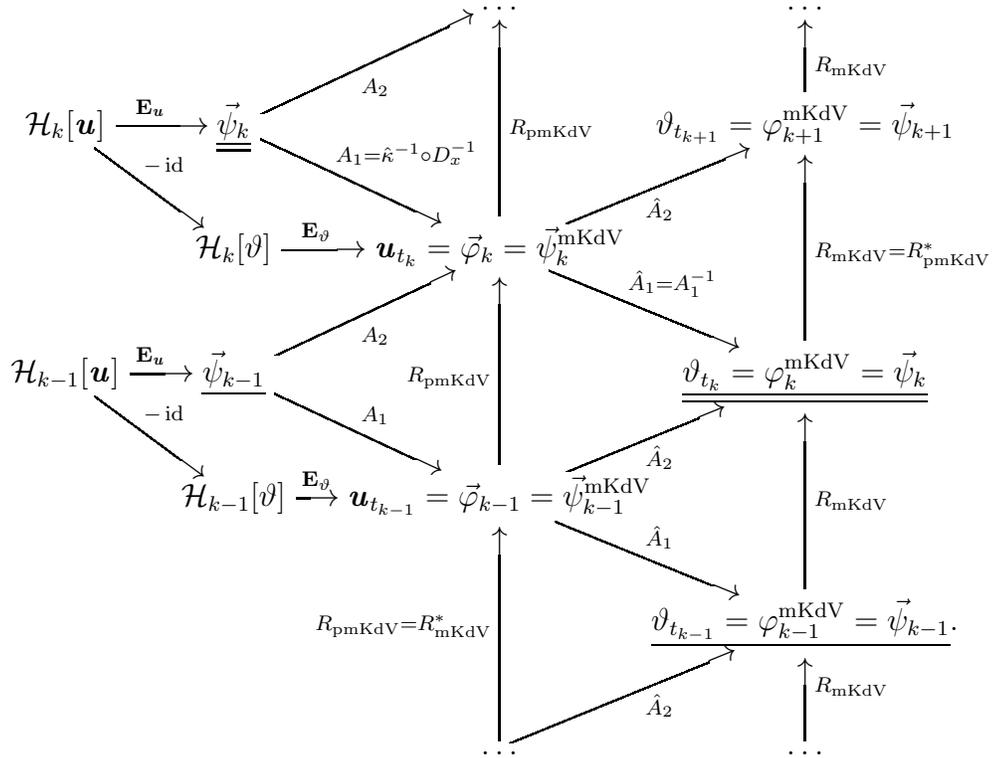
\[  
\begin{diagram}
&&&&\ldots&&\ldots\\
&&&\ruto_{A_2}&&&\uto_{R_\mKdV}\\
\cH_k[\bu]&\rto^{\bE_{\bu}}&\underline{\underline{%
  \vpsi_k}}&&
  \uto_{R_\pmKdV}&&\vth_{t_{k+1}}=\vph^{\mKdV}_{k+1}=\vpsi_{k+1}\\
&\rdto^{-\id{}}&&\rdto^{A_1=\hat\kappa^{-1}\circ
   D_x^{-1}}&&\ruto_{\hat A_2}&\\
&&\cH_k[\vth]&\rto^{\bE_\vth}&\bu_{t_k}=\vvph_k=\vpsi_k^\mKdV&&
  \uto_{R_\mKdV=R_\pmKdV^*}\\
&&&\ruto_{A_2}&&\rdto^{\hat A_1=A_1^{-1}}&\\
\cH_{k-1}[\bu]&\rto^{\bE_{\bu}}
   &\underline{\vpsi_{k-1}}&&\uto^{R_\pmKdV}&&
  \underline{\underline{\vth_{t_k}=\vph^{\mKdV}_{k}=\vpsi_k}}\\
&\rdto^{-\id{}}&&\rdto^{A_1}&&\ruto_{\hat A_2}&\\
&&\cH_{k-1}[\vth] &\rto^{\bE_\vth}&
     \bu_{t_{k-1}}=\vvph_{k-1}=\vpsi_{k-1}^\mKdV
     &&\uto_{R_\mKdV}\\
&&&&&\rdto^{\hat A_1}&\\
&&&&\uto^{R_\pmKdV=R_\mKdV^*} &&
   \underline{\vth_{t_{k-1}}=\vph^\mKdV_{k-1}=\vpsi_{k-1} }. \\
&&&&&\ruto_{\hat A_2}&\uto_{R_\mKdV}\\
&&&& \ldots && \ldots
\end{diagram}
\]  

The Lie algebras $\gA$ and $\gB$ are commutative,
see Lemma~\ref{IsCommutative} and Theorem~\ref{AIsCommutative}.
Therefore, Jacobi's identities~\eqref{Jacobi} are valid for the
operators $A_{1,2}$ and $B_{1,2}$ that are either induced by
relations~\eqref{HamSymEqCoords} for the subscript $1$ or obtained
from the recursions~\eqref{HamViaRec} by using the Magri
scheme~\eqref{MagriElem} for the subscript $2$.
By construction, the recursion operators $R_{\text{p(m)KdV}}$
admit the standard decompositions
\begin{align*}
R_\pmKdV&=A_2\circ A_1^{-1}=\square\circ D_x^{-1}\circ\ell_T,\\
R_\pKdV&=B_2\circ B_1^{-1},
\end{align*}
where the skew\/--\/symmetric factors $A_{1,2}$ and $B_{1,2}$ are
\[
\left\{
  \begin{aligned}
    A_1&=\hat\kappa^{-1}\circ D_x^{-1},\\
    A_2&=\square\circ D_x^{-1}\circ\square^*,
  \end{aligned}
\right.\qquad\left\{
  \begin{aligned}
    B_1&=D_x^{-1},\\
    B_2&=-\beta\,D_x+s_1\cdot D_x^{-1}+D_x^{-1}\circ s_1.
  \end{aligned}
\right.
\]

We denote by $\ell_T^{u}$ the linearization of
functional~\eqref{EMT} with respect to\
the dependent variables $u$ and by
$\ell_T^\vth$ its linearization with respect to\ $\vth$.
The operator $\ell_T^u$ is given in Eq.~\eqref{LinTWRTu} on
page~\pageref{LinTWRTu}, and we have
\[
\ell_T^\vth=\bigl(\ldots, \sum_\sigma \frac{\dd
T}{\dd \vth^i_\sigma}\,D_\sigma, \ldots\bigr).
\]

\begin{lemma}\label{AdjointIsLin}
The relation $\square^*=\ell_T^\vth$ holds\textup{.} Moreover,
\[
\ell_T^{\bu}=\square^*\circ\ell_\vth^{\bu}.\label{RemChainRule}
\]
\end{lemma}

\begin{proof}
The verification of the first statement is straightforward.
Expressing $u_x$ via $\vth$, $u_x=\kappa^{-1}\vth$, we write down
functional \eqref{EMT} in terms of~$\vth$: we have
\[
T=\tfrac{1}{2}\sum\limits_{l,m}\kappa^{lm}\vth^l\vth^m-
\sum\nolimits_l\Delta^l\cdot\vth^l_x,
\]
where $\kappa^{-1}=\|\kappa^{lm}\|$ and the identity
\[
\sum\limits_{i=1}^r a_i\cdot\kappa^{ij}=\Delta^j
\]
is used. Consequently, we obtain
\[
\square^*={}^t\bigl(\kappa^{-1}\cdot\vth-\vec\Delta\cdot\bar
D_x\bigr).
\]
The second statement follows from the definition of the linearization
on page~\pageref{UnivLinear}.
\end{proof}

Several decompositions follow from this lemma.
For example, we obtain
\[
\ell_T=\square^*\circ A_1^{-1}=\square^*\circ\hat A_1 =
\square^*\circ D_x\circ\hat\kappa=\square^*\circ\ell_\vth^{\bu}.
\]
The recursion operator $R_\pKdV$ is factorized to the product
\begin{multline*}
R_\pKdV(\phi_{k-1}) =
\cEv_{\square(\phi_{k-1})}(s)=\ell_s(\square(\phi_{k-1}))={}\\
 {} = D_x^{-1}\circ\ell_T\circ\square(\phi_{k-1})=
B_1\circ\square^*\circ\hat A_1\circ\square(\phi_{k-1})
\end{multline*}
of the Hamiltonian operators glued together by the operators
$\square$ and $\square^*$.

\begin{state}[\textup{\cite{KdVHier,
      KisOvchTodaHam}}]\label{NoethAreHamTh}
Each Noether symmetry
\[\vph_\cL=\square\,\circ\,\bE_T(Q(x,\bT))\in\sym\cL_\Toda\]
of the Toda equation
associated with a nondegenerate symmetrizable matrix $K$
is Hamiltonian with respect to\ the Hamiltonian structure
$A_1=\kappa^{-1}\cdot D_x^{-1}$ and the Hamiltonian
$\cH=[Q(x,\bT)]$\textup{:}
\[\vph_\cL=A_1\circ\bE_u(\cH).\]
\end{state}

The proof of Proposition~\ref{NoethAreHamTh} follows from the
definition of the Euler operator, $\bE(H\,\Id x)=\ell_H^*(1)$,
and Lemma~\ref{AdjointIsLin}.


\begin{rem}\label{RemWhatAreHamiltonians}
The propetry of the Noether point symmetries
\[\vph_0^f=\square\circ\bE_T(T\cdot
f(x)\,\Id x)\]
of the Toda equation to be Hamiltonian,
\[\vph_0^f=A_1\circ\bE_u(T\cdot f(x)\,\Id x),\]
was established in~\cite{Ovch}. One easily verifies that
equation~\eqref{pmKdV} is also Hamiltonian
with respect to\ the cohomology class
$h_0$, see diagram~\eqref{IdentifyKdVDiag} on
page~\pageref{IdentifyKdVDiag}:
\[\vph_1=A_1\circ\bE_u(h_0\,\Id x).\]
\end{rem}

Now we extend Remark~\ref{RemWhatAreHamiltonians} onto the whole
hierarchy $\gA$. Here we formulate the most remarkable relation between
the hierarchy $\gA$ for the potential modified Korteweg\/--\/de Vries
equation~\eqref{pmKdV} and the hierarchy~$\gB$ for scalar
equation~\eqref{KdV}.

\begin{theor}[\textup{\cite{KdVHier,
      KisOvchTodaHam}}]\label{CylinderTh}
The Hamiltonian $[h_k\,\Id x]$ for the $k$\textup{th}
higher Korteweg\/--\/de Vries equation is the Hamiltonian for the
Noether symmetry $\vph_k\in\gA$ for any integer $k\geq0$.
\end{theor}

This theorem follows from Proposition~\ref{NoethAreHamTh}
and Remark~\ref{LawsForARem} on page~\pageref{LawsForARem}
that assign the Noether symmetries
$\vph_k\in\gA$ to the conserved densities~$h_k$.

\begin{rem}
The Lagrangian representation of the Hamiltonian equations
$\cE_{(k)}$ within the hierarchy $\gA$ is
\[
\cE_{(k)}=\Bigl\{A_1\circ\bE_u\bigl(
\bigl[-\tfrac{1}{2}\sum_{i,j}\kappa_{ij}u^i_xu^j_{t_k}-h_{k-1}\bigr]\,
\Id x\bigr)=0\Bigr\},\quad k\geq0.
\]
\end{rem}


\noindent%
Here we complete the description of the Magri scheme for the
equation~\eqref{pmKdV}. The Toda equation~\eqref{eqTodaHamForm} is the
first nonlocal term of the symmetry tower below:

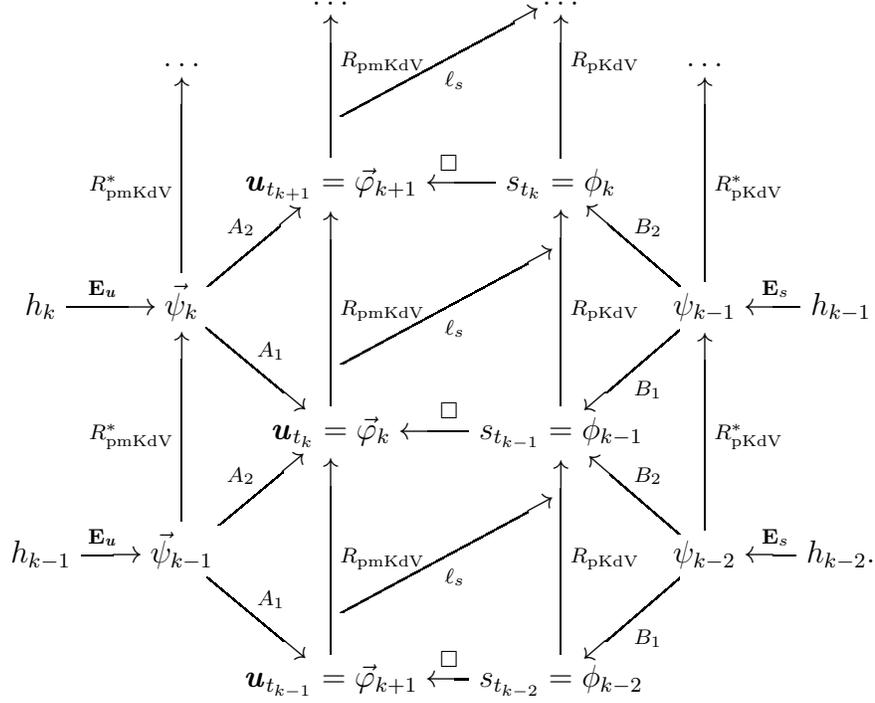
\[  
\begin{diagram}
&&&&\ldots&&\ldots&&&&\\
&&\ldots&&\uto_{R_\pmKdV}&\ruto_{\ell_s}&
  \uto_{R_\pKdV}&&\ldots&&\\
&&&&&&&&&&\\
&&\uto^{R_\pmKdV^*}&&\bu_{t_{k+1}}=\vvph_{k+1}&\lto^\square&
  s_{t_k}=\phi_k && \uto_{R_\pKdV^*}&&\\
 &&&\ruto^{A_2}&&&&\luto^{B_2}&&&\\
h_k&\rto^{\bE_{\bu}}&\vpsi_k&&
 \uto_{R_\pmKdV}&\ruto_{\ell_s}&\uto_{R_\pKdV} &&
 \psi_{k-1}&\lto^{\bE_s}&h_{k-1} \\
&&&\rdto^{A_1}&&&&\ldto_{B_1}&&&\\
&&\uto^{R_\pmKdV^*}&&\bu_{t_{k}}=\vvph_{k}&\lto^\square&
  s_{t_{k-1}}=\phi_{k-1} && \uto_{R_\pKdV^*}&&\\
 &&&\ruto^{A_2}&&&&\luto^{B_2}&&&\\
h_{k-1}&\rto^{\bE_{\bu}}&\vpsi_{k-1}&&
 \uto_{R_\pmKdV}&\ruto_{\ell_s}&\uto_{R_\pKdV} &&
 \psi_{k-2}&\lto^{\bE_s}&h_{k-2}. \\
&&&\rdto^{A_1}&&&&\ldto_{B_1}&&&\\
&&&&\bu_{t_{k-1}}=\vvph_{k+1}&\lto^\square&
  s_{t_{k-2}}=\phi_{k-2} && &&
\end{diagram}
\]  


Finally, we have the following assertion
\begin{state}[\textup{\cite{KdVHier,KisOvchTodaHam}}]
\begin{enumerate}\item
The substitution
\begin{equation}\label{Miura}
T=T(\vth, \vth_x) \colon
\cE_\mKdV\xrightarrow{T(\vth,\vth_x)}\cE_\KdV
\end{equation}
is the Miura transformation between the higher equations
\begin{align*}
\cE_{\mKdV(k)}&=\{\vth_{t_k}=D_x\cdot\hat\kappa(\vph_k)\}\\
\intertext{and}
\cE_{\KdV(k)}&=\{T_{t_k}=D_x(\phi_k)\}.
\end{align*}
\item
The operator
\[\square^*=\ell_T^\vth\colon\vph\mapsto\cEv_\vph(T)\]
maps
the Lie algebra $\sym\cE_\mKdV\ni\vph$ of symmetries of the equation
$\cE_\mKdV$ into the symmetry Lie algebra $\sym\cE_\KdV$ for the
Korteweg\/--\/de Vries equation~\eqref{KdV},
see Example~\textup{\ref{EvolEMTExample}}.
\item
The operator $\square=\square^{**}$ maps the generating sections of
conservation laws which are dual to the symmetries,
\[\square\colon\phi_k=\bE_T(h_k\,\Id x)\mapsto\vph_k\in\gA,\]
in the opposite direction.
\end{enumerate}
\end{state}


\begin{rem}
From diagrams~\eqref{BottomEvolutionDiag}
and~\eqref{IdentifyKdVDiag} we see that the successive
application of the mappings $\square^*=\ell_T^\vth$ and
$\square=\square^{**}$ does {not} preserve the index of a higher
symmetry:
\begin{equation}\label{Pentagon}
\begin{diagram}
\vpsi_k=\vph^\mKdV_{k}&\rto^{\square^*}&
  \psi_k=\Phi_{k}&\rto^{B_1}&\phi_k\\
\uto^{\hat A_1}&&&&\uto_{R_\pKdV}\\
\vvph_k=\vpsi_k^\mKdV&& \lto^{\square=\square^{**}}&&\phi_{k-1}
\end{diagram}
%
\end{equation}
and the recursion operator $R_\pKdV$ measures the difference in their
action.
\end{rem}

\newpage
\specialsection*{{
\textbf{Part~II.}}
Group\/--\/theoretic properties of the mathematical physics equations:
methods and applications}
Within Part~II of the present paper, we illustrate the cohomological
concepts and algorithms in the geometry of differential equations
(\cite{ClassSym, JKKersten, Opava, Vin84}). Namely, we apply these
methods to the study of the properties of the dispersionless Toda
equation, the multi\/--\/component nonlinear Schr\"odinger equation,
the Liouville equation, and related systems.
Invariant solutions, Noether's symmetries, local and nonlocal
conservation laws, weakly nonlocal recursion operators, parametric
families of B\"acklund (auto)transformations, and zero\/--\/curvature
representations are the structures, which are obtained for these
equations of mathematical physics.

\specialsection*{\textbf{Chapter~3.}
Symmetries, solutions, and conservation laws for
differential equations}\label{ChHeavNLS}
In this chapter, we consider two examples of applications of the
PDE geometry methods; namely, we analyse the geometric properties of
the dispersionless Toda equation (\cite{HeavLaws, SolHeav})
and obtain the structures (\cite{YS2004}) for
the multi\/-\/component analog of the
nonlinear Schr\"odinger equation, which is related with the former
one (see~\cite{Dubrovin3}).

\section{Nonlinear Schr\"odinger equation}\label{SecNLS}
In this section, we study the properties of the $m$-component analogs
of the nonlinear Schr\"odinger equation~(\cite{NwavesNLS, Chaos,
Sukhorukov})
\begin{equation}\label{NLS}
\Psi_t=\bi\Psi_{xx}+\bi f(|\Psi|)\,\Psi,
\end{equation}
where $\Psi$ is an $m$-component vector ($m\geq1$),
$\bi=\smash{\sqrt{-1}}$, and $f\in C^1(\mathbb{R})$ is a smooth
function.

The scalar ($m=1$) nonlinear Schr\"odinger equation describes the
propagation of light pulses or wave packets in the
linear dissipation and nonlinear self\/-\/focusing media, \emph{e.g.}, in
optical fibers, nonlinear cristals and gases, multicomponent
Bose\/--\/Einstein condensates at zero temperature, \textit{etc}.
The collective evolution of a complex of space\/--\/incoherent solitons
in a nonlinear media with a Kerr\/--\/like nonlinearity is described by
system~\eqref{NLS} of the nonlinear Schr\"odinger equations, where
$\Psi={}^t(\Psi^1$, $\ldots$, $\Psi^m)$ are the amplitudes of the light
wave, $$\mathcal{I}=\sum\limits_{i=1}^m|\Psi^i|^2$$ is the density of
energy, $t$ is the coordinate along the direction of the wave
propagation, and $x$ is the coordinate along the front of the wave.

Suppose $f=\id$, then this equation is known (\cite{NwavesNLS}) to
admit a commutative bi\/-\/Hamiltonian hierarchy of higher symmetries
and an infinite number of conserved densities in involution. Still, for
an arbitrary $f$ this is not so. In the sequel, we obtain the symmetry
algebra of the latter equation in the physically actual case of a
homogeneous function $f$ of weight $\Delta$; also, we desribe the set
of $m^2$ conserved currents,
which generalize the previously known $m$ separate conservation laws of
energy of an $i$th mode and the conservation law of total momentum of
the system. Those conservation laws correspond to two types of
Hamiltonian symmetries of the Schr\"odinger equation, the scaling
transformation and the translation.

So, consider the $m$-component analog of the nonlinear Schr\"odinger
equation, the multi\/--\/soliton complex~(\cite{Chaos})
\begin{equation}\label{NLS2Parts}
\begin{aligned}
F^k&\equiv\Psi^k_t-\bi\Psi^k_{xx}-\bi f(\cI)\cdot\Psi^k=0,\\
\bar F^k&\equiv\bar\Psi^k_t+\bi
\bar\Psi^k_{xx}+\bi f(\cI)\cdot\bar\Psi^k=0,\quad
1\leq k\leq m,\:\cI\equiv\Psi\bar\Psi.
\end{aligned}
\end{equation}
This equation is Hamiltonian for an arbitrary function $f$:
\begin{equation}\label{NLS1stHam}
{\begin{pmatrix}\Psi\\ \bar\Psi\end{pmatrix}}_t =
\begin{pmatrix}\pp0&1\\ -1&0\end{pmatrix}
\begin{pmatrix} \delta\cH_\NLS/\delta\Psi\\
\delta\cH_\NLS/\delta\bar\Psi\end{pmatrix},
\end{equation}
where the density $H_\NLS$ of the Hamiltonian
$\cH_\NLS=[H_\NLS\,\Id x]$ is
\[
H_\NLS=-\bi\Psi_x\bar\Psi_x+\bi\int\nolimits^\cI f(\cI)\,\Id\cI.
\]
We also note that $H_\NLS$ is a conserved density for
Eq.~\eqref{NLS2Parts}, quite similarly to Eq.~\eqref{EMTHeav} for the
dispersionless Toda equation or to Eq.~\eqref{TodaHamDensityIsConserved}
on page~\pageref{TodaHamDensityIsConserved} for the Toda
equation~\eqref{eqToda}.
From the Hamiltonian representation~\eqref{NLS1stHam}
we deduce that the complex conjugate variables $\bar\Psi$ are the
canonically conjugate variables (the momenta) for the dynamical coordinates
$\Psi$; in Sec.~\ref{SecLegendre} we investigated a generalization
of this situation.

Equation~\eqref{NLS} is known to possess the following property
(see~\cite{Chaos} and references therein): in addition to the
conservation of the total energy $\Psi\bar\Psi$, there are $m$ separately
conserved densities
\begin{equation}\label{EnergyOfModes}
Q_i=\Psi^i\cdot\bar\Psi^i,\qquad \bar D_t(Q_i)=0.
\end{equation}
Physically speaking, there is no energy transfer between the modes
$\Psi^i$. Still, this observation about the properties of the nonlinear
Schr\"odinger equation is uncomplete, since these $m$ integrals of
motion~\eqref{EnergyOfModes} are only particular instances within the
set of $m^2$ currents (\cite{YS2004})
\begin{equation}\label{CLNLS}
\eta_{ij}=\Psi^i\bar\Psi^j\,\Id x +
  \bi\left(\Psi^i_x\,\bar\Psi^j - \Psi^i\,\bar\Psi^j_x\right)\,\Id t,
\qquad\bar\Id_h\eta_{ij}=0.
\end{equation}
These currents are conserved on Eq.~\eqref{NLS} with an arbitrary
nonlinearity $f(\cI)$. They reflect the correlation between the
intensities of the light modes with \emph{different} indexes $i$ and
$j$, $1\leq i$, $j\leq m$. The set of conservation
laws~\eqref{CLNLS} remained unnoticed in the papers~\cite{Chaos,
Sukhorukov}, and the arising conservation restrictions seem to have not
been taken into consideration by the authors when making computer
experiments.

The generating sections of the conservation laws $[\eta_{ij}]$
defined in Eq.~\eqref{CLNLS} are
$$\vpsi_{(ij)}={}^t(\psi_{(ij)},\bar\psi_{(ij)}),$$
where
$$\psi^i_{(ij)}=\bar\Psi^j,\quad \bar\psi_{(ij)}^j=\Psi^i,$$
and
$$\psi^{i'}_{(ij)}=\bar\psi_{(ij)}^{j'}=0\text{ for }i'\not=i,\ j'\not=j.$$
The canonical Hamiltonian structure
$$\Gamma_1=\left(\begin{matrix}\pp0&1\\
-1&0\end{matrix}\right)$$ assigns the point symmetries
$$
\vvph_{(ij)}={}^t(\vph_{(ij)},\bar\vph_{(ij)}):\ %
\vph_{(ij)}^i=\Psi^j,\ \bar\vph_{(ij)}^j=-\bar\Psi^i
$$
to these sections. Suppose $i=j$, then we get the gauge
symmetry~(\cite{NwavesNLS})
\begin{equation}\label{NLSScaling}
A_\lambda(\vvph_{(ii)})\colon\Psi^i\mapsto\exp(\lambda)\Psi^i,
\:\bar\Psi^i\mapsto\exp(-\lambda)\bar\Psi^i.
\end{equation}
The Hamiltonians of the symmetries $\vvph_{(ii)}$ are the conserved
densities $Q_i$ in Eq.~\eqref{EnergyOfModes}.

In the paper~\cite{NwavesNLS}, the case $f(\cI)=\cI$ of cubic
nonlinearity in Eq.~\eqref{NLS} was analysed. Then, the cubic
$m$-component equation~\eqref{NLS} admits the recursion operator
%
\begin{multline*}
\label{RecNLS}
R_\NLS=\begin{pmatrix} -D_x & 0\\ 0 & D_x\end{pmatrix}
+\frac{1}{2}\begin{pmatrix} -\Psi\\ \pp\bar\Psi\end{pmatrix}\cdot
D_x^{-1}\circ\bigl(\bar\Psi,\Psi\bigr)-\\
\frac{1}{2}\sum_{i,j}\vvph_{(ij)}\cdot D_x^{-1}\circ
\vpsi_{(ij)}.
\end{multline*}
This operator generates an infinite sequence of the
commuting local higher symmetries of Eq.~~\eqref{NLS}, starting with
scaling symmetry~\eqref{NLSScaling}:
\begin{equation}\label{NLSDiag}
\begin{CD}
 Q_i\,\Id x @. @. \cH_\NLS @. @. \cH_4 @. {}\\
@VV{\bE_{\vec\Psi}}V @. @VV{\bE_{\vec\Psi}}V @.
     @VV{\bE_{\vec\Psi}}V @. \\
\vpsi_{(ii)} @>{R_\NLS^*}>> \vpsi_1 @>{R_\NLS^*}>> \vpsi_2
   @>{R_\NLS^*}>> \vpsi_3 @>{R_\NLS^*}>> \vpsi_4
   @>{R_\NLS^*}>> {\ldots} \\
@VV{\Gamma_1}V @VV{\Gamma_1}V @VV{\Gamma_1}V @VV{\Gamma_1}V
   @VV{\Gamma_1}V @. \\
\vvph_{(ii)} @>{R_\NLS}>> \vec\Psi_x @>{R_\NLS}>> \vec\Psi_t
   @>{R_\NLS}>> \vvph_3 @>{R_\NLS}>> \vvph_4
   @>{R_\NLS}>> {\ldots\mathbf{.}}
\end{CD}
\end{equation}
This diagram is an example of a sequence of Hamiltonian symmetries such
that there are no Hamiltonians for a half of these symmetries,
$\vvph_{2k+1}$: we have
$\vpsi_{2k+1}\not\in\mathrm{im}\,\bE_{\vec\Psi}$.

\begin{rem}
We have obtained the formulae for the generating sections
$\vpsi_{(ij)}$ of conservation laws~\eqref{CLNLS} and the
Hamiltonian symmetries~$\vvph_{(ij)}$. Now we conclude that
the recursion operator $R_\NLS$ is \emph{weakly nonlocal}
(\cite{DIPS9-2002}), \emph{i.e.}, this operator is decomposable to the form
\begin{equation}\label{RSplitting}
R=\text{local part}
+\sum\nolimits_\alpha\vph_\alpha\circ D_x^{-1}\circ\psi_\alpha,
\end{equation}
where $\vph_\alpha$ are symmetries and
$\psi_\alpha$ are generating sections of conservation laws.
We recall that in the preceding chapter we used the recursion
operators' decompositions similar to \eqref{RSplitting}, and then we generated
the commuting symmetry sequences for the Toda equations
and established nontrivial properties of these sequences.
\end{rem}

We finally note that the symmetries $\vvph$ that are independent on
$x$ and $t$ (only such symmetries were considered in~\cite{NwavesNLS})
do not exhaust the whole algebra of classical symmetries for the
nonlinear Schr\"odinger equation~\eqref{NLS}. Moreover, the resulting
symmetry algebra of this equation is noncommutative.

\begin{example}
Suppose there are no restrictions for the nonlinearity function $f(\cI)$
in Eq.~\eqref{NLS2Parts}. Then the point symmetry algebra of the
latter equation is generated by the fields whose sections are
\[
\vvph_{(ij)},\ \vec\Psi_x,\ \vec\Psi_t,\ %
\begin{pmatrix}\vph^i\\ \bar\vph^i\end{pmatrix} =
\begin{pmatrix}2t\,\Psi^i_x-\bi x\,\Psi^i\\
2t\,\bar\Psi^i_x+\bi x\,\Psi^i\end{pmatrix}.
\]
Further on, suppose additionally that $f$ is homogeneous,
$$f(\lambda\cI)=\lambda^\Delta\cdot f(\cI).$$ Then
equation~\eqref{NLS} acquires the scaling symmetry
\[
\begin{pmatrix} \vph^i\\ \bar\vph^i \end{pmatrix} =
\begin{pmatrix}
2\Psi^i+\Delta\,x\,\Psi^i_x
+2\Delta\,t\,\Psi^i_t\\
2\bar\Psi^i+\Delta\,x\,\bar\Psi^i_x
+2\Delta\,t\,\bar\Psi^i_t
\end{pmatrix}.
\]
\end{example}

\section{The dispersionless Toda equation}\label{SecHeav}
The dispersionless Toda equation is an analog of Eq.~\eqref{eqToda}
with the continuous variation of the index $j$ that enumerates the
dependent variables $u^j$. In this section, the classical symmetry
algebra of the dispersionless Toda equation is computed
and $5$ classes of conservation laws for the latter equation are
reconstructed; also, some questions of the
Lagrangian formalism with higher-order derivatives are discussed.

Consider the hyperbolic Toda equations
\begin{equation}\label{GeneqToda}
\bu_{xy}=\exp(K\bu),\qquad\bu'_{xy} = K\cdot\exp(\bu'),
\end{equation}
associated with the type $A_{r-1}$ Lie algebras with the Cartan
matrices $K$, introduce an additional independent variable $z\in\BBR$,
and extend the values of the discrete index $j\in[1$, $r]$, which
enumerates the dependent variables $u^j$, onto the whole line $\BBR$.
Also, let the rank $r$ tend to infinity: $r\to\infty$. Then for any
section $\bu$ of the bundle $\pi$ we set
$$u^j={\bu(x,y,z)\bigr|}_{z=j\varepsilon},$$ where
$\varepsilon$ is the cell step.

As $r\to\infty$ and $\varepsilon\to+0$, the continuous limit equation,
which is refered as either the dispersionless Toda equation, or the
heavenly equation, or the Boyer\/--\/Finley equation
(\cite{BoyerFinley}), also appeares in many models, \emph{e.g.}, in gravity
(\cite{SavelievTMPh, BoyerFinley}) as a reduction of the antiselfdual
vacuum Einstein equation \mbox{(ASDVEE)}.
The heavenly equation $\cE_\heav$ exists due to the special structure
(see Eq.~\eqref{CartanForAr})
%
of the Cartan matrices $K=\|k_{ij}\|$
of the type $A_r$ algebras. As a matter of fact,
instead of the Cartan $(r\times
r)$-matrices $K$ one should consider (\cite{SavelievTMPh,
HeavenlyExists}) the Cartan operator $\hat K=-D_z^2$.
Then the scalar equations, which originate from Eq.~\eqref{GeneqToda},
are~(\cite{Alfinito, BoyerFinley})
\begin{subequations}\label{BothHeavenly}
\begin{align}
\hat\cE=\{\hat F&\equiv u_{xy}-\exp(-u_{zz})=0\} \label{OurHeavenly}\\
u'_{xy}&=-D_z^2\circ\exp(u'); \label{HeavenlyAsEL}
\end{align}
\end{subequations}
the Cartan operator $\hat K$ defines the sign ``${-}$''
in the one\/--\/dimensional equations
\begin{equation}\label{eqHeav1D}
u_{\tau\tau}=\exp(-u_{zz}),\qquad
u'_{\tau\tau}=-D_z^2\circ\exp(u')
\end{equation}
that appeared in the Introduction.

Following the papers~\cite{SavelievTMPh, Dubrovin3}, we consider the limit
\[
\lim_{\varepsilon\to+0}\lim_{r\to\infty}
L_\Toda=D_z^{-1}\left(\tfrac{1}{2}u_x u_{yzz}-\exp(-u_{zz})\right)
\]
of the Lagrangian density $\cL_\Toda$, defined in
Eq.~\eqref{TodaLDensity}, as $r\to\infty$ and $\varepsilon\to+0$.
Then the Lagrangian $\cL_\Toda$ itself is mapped to the functional
\[\hat\cL=\iint\Id x\Id y\int\text{\L}\,\Id z\] with the density
\[\text{\L}=-\tfrac{1}{2}\,u_{xz} u_{yz}-\exp(-u_{zz})\] that depends on
the second derivatives of the sections $u=u(x,y,z)$.
Applying the Euler operator $\bE_u$ to $\hat\cL$, we obtain the
equation
\begin{equation}\label{ImEulerHeav}
\cE_{\heav}=\left\{F_\heav\equiv
u_{xyzz}-D_z^2\circ\exp(-u_{zz})=0\right\};
\end{equation}
we see that double integrating in $z$ maps Eq.~\eqref{ImEulerHeav}
to Eq.~\eqref{OurHeavenly}, while the substitution
$u'=-u_{zz}$ maps it to Eq.~\eqref{HeavenlyAsEL}.

\subsection{Symmetries and exact solutions}
%
Computing the symmetries $\vph\in\Sym\hat\cE$ of
Eq.~\eqref{OurHeavenly} by using the analytic transformations software
\textsf{Jet} (\cite{Jet97}), we get the following solution:
\begin{state}\label{SymPointHeavState}
The point symmetries \[\vph(x,y,z,u,u_x,u_y,u_z)\]
of Eq.~\eqref{OurHeavenly}, which solve the determining equation
\[\bar D_{xy}(\vph)+\exp(-u_{zz})\cdot\bar D_z^2(\vph)=0,\]
are
\begin{subequations}\label{SymPointHeav}
\begin{align}
\vph_1[f]&=\Bigl(u_x-\tfrac{1}{2}z^2\,\bar
D_x\Bigr)\,f(x), &
\bar\vph_1[g]&=\Bigl(u_y-\tfrac{1}{2}z^2\,\bar
D_y\Bigr)\,g(y),\label{SymHeavConf}\\
\vph_2&=-\tfrac{1}{2}{z}u_z+u-\tfrac{1}{2}{z^2}, & &
  \label{SymHeavScale}\\
\vph_3&=u_z, & & \label{SymHeavTransl}\\
\vph_4[q]&=q(x)\,z, & \bar\vph_4[\bar q]&=\bar q(y)\,z,
  \label{SymHeavPoly1}\\
\vph_5[r]&=r(x), & \bar\vph_5[\bar r]&=\bar r(y),\label{SymHeavPoly0}
\end{align}
\end{subequations}
where $f$, $q$, and $r$ are arbitrary smooth functions of $x$
and $g$, $\bar q$, and $\bar r$ are functions of~$y$.
%
The commutation rules for symmetries~\eqref{SymPointHeav} are given in
the skew\/--\/symmetric table below\textup{:}
{\normalsize
{\renewcommand{\arraystretch}{0}%
\footnotesize
\[
\text{%
\begin{tabular}{|c||c|c|c|c|c|c|c|c|}
\hline
\strut & $\vph_1[f]$ & $\bar\vph_1[g]$ &
  $\vph_2$ & $\vph_3$ & $\vph_4[q]$ &
  $\bar\vph_4[\bar q]$ & $\vph_5[r]$ & $\bar\vph_5[\bar r]$ \\
\hline
\rule{0pt}{2pt} &&&&&&&&\\
\hline
\strut$\vph_1[f]$ & $0$ & $0$ & $0$ & $\vph_4[{-}f']$ &
   $\vph_4[{-}fq']$ & $0$ & $\vph_5[{-}fr']$ & $0$ \\
\hline
\strut$\bar\vph_1[g]$ && $0$ & $0$ & $\bar\vph_4[{-}g']$
   & $0$ & $\bar\vph_4[{-}g\bar q]$ & $0$ & $\bar\vph_5[{-}g\bar r']$ \\
\hline
\strut$\vph_2$ &&& $0$ & $\vph_3{+}\vph_4[2]$ & $\vph_4[q]$
   & $\bar\vph_4[\bar q]$ & $\vph_5[2r]$ & $\bar\vph_5[2\bar r]$ \\
\hline
\strut$\vph_3$ &&&& $0$ & $\vph_5[{-}q]$ & $\bar\vph_5[{-}\bar q]$
   & $0$ & $0$ \\
\hline
\strut$\vph_4[q]$ &&&&& $0$ & $0$ & $0$ & $0$ \\
\hline
\strut$\bar\vph_4[\bar q]$ &&&&&& $0$ & $0$ & $0$ \\
\hline
\strut$\vph_5[r]$ &&&&&&& $0$ & $0$ \\
\hline
\strut$\bar\vph_5[\bar r]$ &&&&&&&& $0$ \\
\hline
\end{tabular}
}
\]
}
} 
\end{state}

\begin{rem}
1. The operator $\hat\square=u_x-\tfrac{1}{2}z^2\,\bar D_x$ in
Eq.~\eqref{SymHeavConf} is an analog of the operator
$\square$, which was introduced in Eq.~\eqref{Square} on
page~\pageref{Square}.\\
2. The symmetries of Eq.~\eqref{HeavenlyAsEL} that correspond to
\eqref{SymHeavConf}--\eqref{SymHeavTransl} were obtained in
\cite{Alfinito}; the symmetries $\vph_4$, $\vph_5\in\ker D_z^2$
of the equation $\hat\cE$ demonstrate the (inessential)
distinction between the geometries of Eq.~\eqref{OurHeavenly} and
Eq.~\eqref{HeavenlyAsEL}, this distinction is provided by the substitution
$u'=-u_{zz}$.
\end{rem}

\begin{rem}
In fact, equations~\eqref{BothHeavenly} and \eqref{ImEulerHeav}
are huge inspite of an apparent compactness of the notation.
Indeed, one can estimate the number of $k$th order
internal coordinates for these equations or the total number
of nontrivial relations that define the $k$th prolongation
$\cE_\heav^{(k)}$.
Therefore the calculation of the higher symmetry algebra
$\sym\cE^\infty_\heav$ for Eq.~\eqref{ImEulerHeav} is really difficult.

The passage from difference\/--\/differential equations
\eqref{eqTodaForK} associated with Cartan's matrix~\eqref{CartanForAr}
to their \emph{dispersionless} limit~\eqref{OurHeavenly}
provides the additional constraint $$u_{zzz}=0.$$
The latter equation must be taken into consideration when describing
the Lie algebra $\sym\cE^\infty_\heav$ composed by \emph{all}
symmetries of the limit dispersionless Toda equation.
\end{rem}

\subsubsection{Constructing exact solutions to the dispersionless Toda
equation}
By using the structure of the point symmetry Lie algebra $\Sym{\cE}$ for
Eq.~\eqref{OurHeavenly} and applying various geometric methods
(\cite{ClassSym}), the problem of constructing exact solutions to the
heavenly equation was considered in~\cite{Alfinito}
and~\cite{Martina}.
In the former paper, invariant solutions of the equation
$u_{xy}=\pm{(\exp(u))}_{zz}$ were obtained; they correspond to
Eqs.~\eqref{Instanton} and~\eqref{ConfLiouSol} of the present article.
A class of solutions that are
non-invariant with respect to\ the whole Lie algebra $\Sym\cE$ was pointed out
in~\cite{Martina}.
The approach of this paper is wider than in~\cite{Alfinito} in the
following sence: any time an equation under consideration being reduced to
an auxiliary equation for a function that depends on fewer arguments
than the initial one, we construct the point symmetry Lie algebra for
the new equation and obtain its invariant solutions that are
parameterized by arbitrary functions.
Therefore, what we get is much more than a
list of particular solutions. Also, we note that the method of solving
Eq.~\eqref{OurHeavenly} together with the constraint $\varphi_i=0$
is similar to the scheme
(\cite{Martina}) used to construct \emph{non}-invariant solutions of
differential equations. Indeed, we treat the variable $z$ in the
constraints $\varphi_i=0$ as a formal parameter in order to write down
these additional equations $\varphi_i=0$ in total differentials.

\textsl{The symmetry $\vph_1\in\Sym\hat\cE$.}
Consider the generator
\[
\vph_1=\vph_1[f]+\bar\vph_1[g]=u_x\,f(x)-\tfrac{1}{2}z^2\,f'(x)+
   u_y\,g(y)-\tfrac{1}{2}z^2\,g'(y)
\]
of an infinitesimal conformal symmetry of Eq.~\eqref{OurHeavenly}; this
generator depends on two arbitrary smooth functions $f$ and $g$.
Now we represent the invariance condition $\vph_1=0$ in the
characteristic form and obtain the first integral
\[
t=\int^x{\Id x}/{f(x)}-\int^y{\Id y}/{g(y)}
\]
that is independent on the variable~$u$.
In order to construct another integral $C_2$ we treat the
base coordinate $z$ as a parameter:
$$\tfrac{1}{2}z^2\log f(x)+\tfrac{1}{2}z^2\log g(y)-u=C_2(z).$$
Then, a solution $u(x,y,z)$ to the equation $\varphi_1=0$ is defined by
the condition $$\Pi(t, C_2(z))=0,$$ where $\Pi$ is an arbitrary (for a
while) function of two ``integrals'', one of which depends on~$z$.
Solving this relation with respect to\ $u$, we get
\begin{equation}\label{eqHeavSolClass}
u=\tfrac{1}{2}z^2\log f(x)g(y) + \Phi(t,z).
\end{equation}
Moreover, if $u$ is a solution to Eq.~\eqref{OurHeavenly}, then
it is necessary that the function $\Phi$ satisfies the
one\/--\/dimensional equation
\begin{equation}\tag{\ref{eqHeav1D}${}'$}\label{OurHeav1D}
\Phi_{tt}+\exp(-\Phi_{zz})=0.
\end{equation}
The reduced equation is nothing else than the one\/--\/dimensional
analog of Eq.~\eqref{eqHeav1D}. In a similar situation, the hyperbolic
Liouville equation is reduced to solving its one\/--\/dimensional
analog when one constructs solutions that are invariant with respect to\ the
conformal symmetries of the initial Liouville equation.
We emphasize that the second ``integral'' $C_2(z)$ of the equation
$\varphi_1=0$  was obtained by using a nontrivial interpretation of the
coordinate $z$ as a parameter which is auxiliary with respect to\ the
independent variables $x$ and $y$. This approach seems to have not been
used in~\cite{Alfinito}; in the sequel, we apply it to
Eq.~\eqref{OurHeav1D} again.
We use the \textsf{Jet} (\cite{Jet97}) software and compute the
generators of the point symmetry Lie algebra of Eq.~\eqref{OurHeav1D};
then, we obtain the invariant solutions of this equation. The result is
given in

\begin{lemma}
The Lie algebra of classical symmetries of Eq.~\eqref{OurHeav1D} is
generated by eight evolutionary vector fields whose
generating functions are
\begin{align*}
\phi_1&=t\Phi_t-z^2,& \phi_2&=\Phi_t,& \phi_3&=z\Phi_z+z^2-2\Phi,&
\phi_4&=\Phi_z,\\
\phi_5&=zt,& \phi_6&=t,& \phi_7&=z,& \phi_8&=1.
\end{align*}
\end{lemma}

Further on, we point out the families of solutions to
Eq.~\eqref{OurHeav1D} and assign the
solution class~\eqref{eqHeavSolClass} for the dispersionless Toda
equation (see Eq.~\eqref{OurHeavenly}) to each of them.

\textsl{The symmetry $\phi_1$.}
Consider the equation $\phi_1=0$; again, we require the variable~$z$ to
be a formal parameter, then we have $z^2\log|t|-\Phi=C(z)$. Hence we
express the solution $\Phi$, substitute it into Eq.~\eqref{OurHeav1D},
and solve the latter with respect to\ $C$; we finally get
\[
\Phi(t,z)=z^2\log|t|-\tfrac{1}{4}z^2\log|z|-\tfrac{1}{8}z^2+C_1z+C_2,
\]
where $C_1$ and $C_2$ are arbitrary constants.

\textsl{The symmetry $\phi_2$.}
One easily checks that the real $\phi_2$-invariant solutions of
Eq.~\eqref{OurHeav1D} do not exist at all, since the equation
$\exp(-\Phi_{zz})=0$ is insoluble over $\BBR$.
Still, we consider a more general case: namely, we seek solutions to
Eq.~\eqref{OurHeav1D} that are invariant with respect to\ the linear
combination $$\phi_{(a:b)}=\phi_2+(a:b)\phi_4$$ of the symmetries, where
$(a:b)\in\mathbb{RP}^1$, \emph{i.e.}, we seek the propagating wave solutions.
Substituting the function $\Phi(z-(a:b)\,t)$ into
Eq.~\eqref{OurHeav1D}, we get the equation
\begin{equation}\label{Eq4Bifur}
(a:b)^2\,\Psi=-\exp(-\Psi).
\end{equation}
Here by $\Psi$ we denote the second derivative $\Phi''$ of $\Phi$
with respect to\ its argument $$w=z-(a:b)\,t.$$
Suppose the
condition $$-\sqrt{\rme}<(a:b)<\sqrt{\rme}$$ is satisfied,
then the obstacle for the
low\/--\/velocity $\phi_{(a:b)}$-invariant solutions to
exist is clear from
Eq.~\eqref{Eq4Bifur}. Indeed, there is the critical (minimal) velocity
$|a:b|=\sqrt{\rme}$  and the wave solution
$$\Phi=-\tfrac{1}{2}{\bigl(z\pm\sqrt{\rme}\,t\bigr)}^2+
\alpha\bigl(z\pm\sqrt{\rme}\,t\bigr)+\beta,$$
where $\alpha$, $\beta\in\BBR$.
Quite remarkably, if the velocity
$\sqrt{\rme}<|a:b|<\infty$ is greater then the minimum velocity,
then there are distinct roots $\Psi_{1,2}$ of
Eq.~\eqref{Eq4Bifur} and
Eq.~\eqref{OurHeav1D} admits \emph{two} solutions
\begin{equation}\label{SolPoly1DHeav}
\Phi=\tfrac{1}{2}\Psi_{1,2}\cdot w^2+\alpha w+\beta,\qquad
w=z-(a:b)\,t,\qquad \alpha,\beta\in\BBR,
\end{equation}
similtaneously.
This pair of roots $\Psi_{1,2}$ appeares after the obvious
bifurcation: no solutions of Eq.~\eqref{Eq4Bifur} and~\eqref{OurHeav1D}
are assigned to a low inclination $(a:b)^2$ of the straight line
$y=(a:b)^2\,\Psi$ on the $0\Psi y$-plane; then, at the inclination
$(a:b)^2=\rme$ this straight line is tangent to the graph of exponent
$y=-\exp(-\Psi)$ at the point $(-1$, $-\rme)$, while at a greater
incline $\rme<(a:b)^2<\infty$ the point of tangency is split to the
pair of distinct intersection points
$(\Psi_{1,2}$, $(a:b)^2\Psi_{1,2})$,
where $-\infty<\Psi_1<-1<\Psi_2<0$.
Each admissible value of $\Psi$ determines the second derivative of a
polynomial solution~\eqref{SolPoly1DHeav} of Eq.~\eqref{OurHeav1D}.
A unique root $\Psi=0$ and the $\phi_4$-invariant solution
of Eq.~\eqref{OurHeav1D} (see below) is assigned to the point
$a:b=\infty$.

\textsl{The symmetry $\phi_3$.}
First, we solve the auxiliary ordinary differential equation:
\begin{equation}\label{AuxODE}
xy'(x)-2y=-x^2\quad\Longrightarrow\quad
y(x)=\bigl(\gamma-\tfrac{1}{2}\log x^2\bigr)\cdot x^2.
\end{equation}
We shall use formulae~\eqref{AuxODE} twice, in order to construct the
solutions of Eq.~\eqref{OurHeav1D} that are invariant with respect to\ the
symmetry $\phi_3$ and to get the $\vph_2$-invariant solutions of the
initial equation~\eqref{OurHeavenly}.
In these two cases, the function $\gamma$ depends on two distinct sets
of variables: the first case is $\gamma=\gamma(t)$ and the second case is
$\gamma=\gamma(x,y)$.

Substituting the expression $$\Phi=(\gamma(t)-\tfrac{1}{2}\log z^2)\cdot
z^2$$ that appeared in Eq.~\eqref{AuxODE} into Eq.~\eqref{OurHeav1D}
and paying attention to the sign of the constant of integration, we get
the following expressions for the function $\gamma(t)$:
\begin{align*}
\gamma_1&=\tfrac{3}{2}-\log\Bigl[\sqrt{\epsilon}\sinh2\,\mathrm{artanh}\,\exp
\bigl(\pm\sqrt{\epsilon}(t-t_0)\bigr)\Bigr],\\
\gamma_2&=\tfrac{3}{2}+\log\bigl[\pm t-t_0\bigr],\\
\gamma_3&=\tfrac{3}{2}-\log\Bigl[\sqrt{\epsilon}\cosh\log\tan\bigl(
\sqrt{\epsilon}(\pm t-t_0)\bigr)\Bigr],
\end{align*}
where $\epsilon>0$ and $t_0\in\BBR$.
Now we return to formula~\eqref{eqHeavSolClass} and assign the solution
class
\begin{equation}\label{Instanton}
u=\tfrac{1}{2}z^2\log f(x)g(y)+
(\gamma_i(t)-\tfrac{1}{2}\log z^2)\cdot z^2,\quad i=1,2,3,
\end{equation}
of the dispersionless Toda equation~\eqref{OurHeavenly}
to each of the functions $\Phi(t$,~$z)$.

\textsl{The symmetry $\phi_4$.}
Consider the constraint $\Phi_z=0$; then, a solution to the equation
$\Phi_{tt}=1$ is a polynomial
$$\Phi(t)=\tfrac{1}{2}t^2+C_1t+C_2,$$ where $C_1$ and $C_2$ are
constants of integration; the corresponding solutions of the
dispersionless Toda equation~\eqref{OurHeavenly} are again defined in
Eq.~\eqref{eqHeavSolClass}.

\smallskip\textsl{The symmetry $\vph_2\in\Sym\hat\cE$.}
Now we construct the $\vph_2$-invariant solutions of
Eq.~\eqref{OurHeavenly}. First, we consider the auxiliary ordinary
differential equation in Eq.~\eqref{AuxODE}; then, we obtain the
hyperbolic $\scal^{+}$-Liouville equation
(see Eq.~\eqref{eqscalplus} on page~\pageref{eqscalplus}).
Now, subsitute the expression
$(\gamma(x$, $y)-\tfrac{1}{2}\log z^2)\cdot z^2$ for $u$ in the equation
$u_{xy}=\exp(-u_{zz})$ and make the transformation
$$\cX=x\,\exp(3/2),\quad \cY=y\,\exp(3/2).$$ Hence we get the Liouville
equation $$\gamma_{\cX\cY}=\exp(-2\gamma)$$ whose solutions
(\cite{Liouville}) are easily obtained from formula~\eqref{gensolhyp}
on page~\pageref{gensolhyp} by a trivial coordinates change; we have
\[
\gamma(\cX,\cY)=-\tfrac{1}{2}\log\Bigl[{-f'(\cX)g'(\cY)}%
{{\Bigl\{\mathbf{Q}\bigl( {[f(\cX)+g(\cY)]}^2 \bigr)\Bigr\}
}^{-2}}\Bigr],
\]
where the mapping $\mathbf{Q}$ is one of $\sin$, $\id$, or $\sinh$,
whence we  finally obtain
\begin{equation}\label{ConfLiouSol}
u(x,y,z)=\frac{z^2}{2}\log\frac{
{\Bigl\{\mathbf{Q}\bigl({[f(\rme^{3/2}x)+
     g(\rme^{3/2}y)]}^2\bigr)\Bigr\} }^2          }%
{-z^2 f'(\rme^{3/2}x)g'(\rme^{3/2}y)}.
\end{equation}
This class of solutions of Eq.~\eqref{OurHeavenly} is in fact present
in paper~\cite{Alfinito}, and their physical interpretation is known
(\emph{ibid.}): these expressions provide the instanton solutions
(\cite{BoyerFinley}) of the antiselfdual vacuum Einstein equations
(ASDVEE).

\smallskip\textsl{The symmetry $\vph_3\in\Sym\hat\cE$.}
In order to obtain the $\vph_3$-invariant solution of
Eq.~\eqref{OurHeavenly}, we solve the equations
$u_z=0$ and $u_{xy}=\exp(-u_{zz})$ together. The solution is
\[
u(x,y,z)=xy+f(x)+g(y),
\]
where $f$ and $g$ are arbitrary functions.

The symmetries $\varphi_4$ and $\varphi_5$ of Eq.~\eqref{OurHeavenly}
depend neither on the unknow function $u$ nor its derivatives.
Therefore,
if the equation $u_{xy}=\exp(-u_{zz})$ is overdetermined by the
condition $\varphi_4=0$ or $\varphi_5=0$, then the problem of
search for exact solutions to the dispersionless Toda equation is not
simplified.

\subsection{The Noether symmetries and conservation laws}
In this subsection, we consider the problem of constructing the
conservation laws for the dispersionless Toda
equations~\eqref{BothHeavenly} and \eqref{ImEulerHeav}.
First, we discuss the general correlation between an Euler
equation~\eqref{ELE} and the fixed set of conservation laws which
reflect the conservation properties of the energy\/--\/momentum tensor.

\subsubsection{On the Lagrangian formalism involving
higher\/--\/order derivatives}
Suppose that $\cL=\int L\,\Id\bx$ is a Lagrangian with the
second order derivatives and $g_{\mu\nu}$ is the corresponding
metric; then we can construct a conservative analog of the
energy\/--\/momentum tensor $T^{\nu\mu}$ that satisfies the equation
\begin{equation}\label{EMTConserv}
\sum\limits_\mu\bar D_\mu(T^{\nu\mu})=0,
   \qquad \bar D_\mu\equiv{D_\mu\bigr|}_{\cE}.
\end{equation}
The equations of motion $\cE$ are
\begin{equation}\label{eqMotionHigher}
\sum\limits_{\mu,\nu}
D_\mu D_\nu\left(\frac{\dd L}{\dd u^i_{;\mu;\nu}}\right)=
\sum\limits_{\nu}
D_\nu\left(\frac{\dd L}{\dd u^i_{;\nu}}\right)-\frac{\dd L}{\dd u^i}
\end{equation}
and the tensor $T^{\nu\mu}$ coincides with its classical definition
(\cite{B-Sh}) if the higher\/--\/order term
\[\left\|\frac{\dd L}{\dd u^a_{;\mu;\nu}}\right\|\]
is trivial.
Again, we use the notation $u^i_{;\mu}\equiv D_{x^\mu}(u^i)$
and $u^i_{;\mu;\nu}\equiv D_{x^\mu}\cdot D_{x^\nu}(u^i)$.
We raise and lower the indexes $;\mu$, $;\nu$ by using
the metric $g_{\mu\nu}$.
By a straighforward substitution into Eq.~\eqref{EMTConserv} one checks
that the tensor
$$
T^{\nu\mu}=-g^{\nu\mu}L+
\sum\limits_i\frac{\dd L}{\dd u^i_{;\mu}}\,u^{i;\nu}+
\sum\limits_{i,\lambda}\Bigl[
 \frac{\dd L}{\dd u^i_{;\mu;\lambda}}
 \cdot D_\lambda\left(u^{i;\nu}\right) -
D_\lambda\,\Bigl(\frac{\dd L}{\dd u^i_{;\mu;\lambda}}\Bigr)\cdot
u^{i;\nu}\Bigr]
$$
satisfies the required restrictions (\cite{HeavLaws}), and therefore
condition~\eqref{EMTConserv} provides conservation laws for
system~ \eqref{eqMotionHigher}. Still, we note an important feature of
Eq.~\eqref{ImEulerHeav}: as $r\to\infty$ and $\varepsilon\to+0$ in the
Toda equations, the metric $g_{\mu\nu}$ becomes degenerate
and the Lagrangian $\hat\cL$ looses its covariance. Indeed, it is
impossible to find a nondegenerate metric $\hat g_{\mu\nu}$ such that
\[\text{\L}\sim\tfrac{1}{8}{u_{;\mu}}^{;z}{u^{;\mu}}_{;z}+
\exp(-u_{;z}^{;z}).\]
For the same reason we cannot define the energy\/--\/momentum tensor
$T^{\nu\mu}$ for the equation $\cE_\heav$ as the variation of the
Lagrangian $\hat\cL_\heav$ with respect to\ the metric $\hat g_{\nu\mu}$
since the nondegenerate metric does not exist.

\subsubsection{Constructing the conservation laws}
From the preceding section it is clear that the search of conservation
laws for Eq.~\eqref{ImEulerHeav} is less trivial than finding the
symmetry Lie fields by their generating functions and requires
additional reasonings.
In the sequel, we find the symmetries $\sym\cE_\heav$ of the
dispersionless Toda equation, select its
Noether symmetries $\sym\hat\cL_\heav$ in a fixed coordinate system,
and reconstruct the conservation laws by using the homotopy method,
which is described in~\cite{ClassSym, Vin84} and~\cite{Vestnik2000}.

\begin{lemma}[\textup{\cite{HeavLaws}}]\label{WhichAreNoether}
The generating functions~\eqref{SymHeavConf},
\eqref{SymHeavTransl}\/--\/\eqref{SymHeavPoly0}
are the \emph{Noether} symmetries of the equation $\cE_\heav$\textup{,}
while~\eqref{SymHeavScale} is not.
\end{lemma}

\begin{state}[\textup{\cite{HeavLaws}}]\label{HeavLawsState}
The conserved currents $\eta_i\in\bar\Lambda^2(\cE_\heav)$,
assigned to the Noether symmetries $\vph_i$ of
Eq.~\eqref{ImEulerHeav}, are
\begin{align*}
\eta_1&
=\Bigl\{\frac{3}{8}f(x)\,uu_{xyzz}+\frac{1}{12}f(x)\,u_zu_{xyz}-
\frac{1}{24}f(x)\,u_{xy}u_{zz}+\frac{1}{24}f(x)\,u_yu_{xzz}\\
&
-\frac{1}{12}f'(x)\,u_y
-\frac{1}{12}\,f(x)u_{xz}u_{yz}+\frac{z}{6}f'(x)\,u_{yz}
+\frac{1}{8}f(x)u_xu_{yzz}
\\&
-\frac{z^2}{8}f'(x)\,u_{yzz}
-f(x)\,\int\nolimits_0^1 t^2uu_{zzz}^2\,\exp(-tu_{zz})\,\Id t
\\&\strut\qquad
+f(x)\,\int\nolimits_0^1 tuu_{zzzz}\,\exp(-tu_{zz})\,\Id t
\Bigr\}\,\omx
\\&
+\Bigl\{-\frac{1}{8}f'(x)\,uu_{xzz}-\frac{1}{8}f(x)\,uu_{xxzz}+
\frac{1}{4}f''(x)\,u+\frac{1}{12}f'(x)\,u_zu_{xz}
\\&
+\frac{1}{12}f(x)\,u_zu_{xxz}
-\frac{z}{6}f''(z)\,u_z
-\frac{1}{24}f'(x)\,u_xu_{zz}-\frac{1}{24}f(x)\,u_{xx}u_{zz}
\\&
+\frac{z^2}{24}f''(x)\,u_{zz}+\frac{1}{6}f(x)\,u_xu_{xzz}
-\frac{1}{12}f'(x)\,u_x-\frac{1}{12}f(x)\,u_{xz}^2
\\&\strut\qquad
+\frac{z}{6}f'(x)\,u_{xz}-\frac{z^2}{8}f'(x)\,u_{xzz}
\Bigr\}\,\omy
\\&
+\Bigl\{-\frac{1}{6}f(x)\,u_{xy}u_{xz}+\frac{1}{3}f(x)\,u_xu_{xyz}
-\frac{1}{12}f'(x)\,u_xu_{yz}-\frac{1}{4}f'(x)\,uu_{xyz}
\\&
-\frac{z}{6}f''(x)\,u_y-\frac{z^2}{4}f'(x)\,u_{xyz}
+\frac{1}{12}f(x)\,u_yu_{xxz}-\frac{1}{4}f(x)\,uu_{xxyz}
\\&
+\frac{1}{12}f(x)\,u_zu_{xxy}
-\frac{1}{12}f(x)\,u_{xx}u_{yz}
+\frac{z}{6}f'(x)\,u_{xy}+\frac{1}{12}f'(x)\,u_yu_{xz}
\\&
{+}\frac{1}{12}f'(x)\,u_zu_{xy}{+}\frac{z^2}{12}f''(x)\,u_{yz}
{+}f(x)\cdot\!\!\int\nolimits_0^1 \bigl[ -tuu_{xzzz}
{-}tu_{xz}u_{zz}{+}tu_xu_{zzz}
\\&
{+}tu_zu_{xzz}{+}t^2uu_{xzz}u_{zzz}
{-}t^2u_xu_{zz}u_{zzz} \bigr] \exp(-tu_{zz})\,\Id t
+f'(x)\cdot\!\!\int\nolimits_0^1 \bigl[ -u_z
\\&\strut\qquad
{+}zu_{zz}
{-}\frac{z^2}{2}u_{zzz}
{-}tuu_{zzz}
{+}\frac{z^2}{2}tu_{zz}u_{zzz}\bigr] \exp(-tu_{zz})\,\Id t
\Bigr\}\,\omz,
\end{align*}
\begin{align*}
\eta_3&=\left\{\frac{5}{24}u_zu_{yzz}-\frac{1}{8}u_{yz}u_{zz}+
\frac{1}{24}u_yu_{zzz}-\frac{1}{8}uu_{yzzz}\right\}\,\omx\\
{}&+
\left\{\frac{5}{24}u_zu_{xzz}-\frac{1}{8}u_{xz}u_{zz}+
\frac{1}{24}u_xu_{zzz}-\frac{1}{8}uu_{xzzz}\right\}\,\omy\\
{}&+
\left\{\frac{1}{4}uu_{xyzz}{+}\frac{1}{3}u_zu_{xyz}{-}\frac{1}{12}u_{xy}u_{zz}{-}%
\frac{1}{6}u_{xz}u_{yz}{+}\frac{1}{12}u_xu_{yzz}{+}%
\frac{1}{12}u_yu_{xzz}\right.\\
{}&{}\qquad\left.{}
+u_{zz}\exp(-u_{zz})+\exp(-u_{zz})+u_zu_{zzz}\exp(-u_{zz})
\lefteqn{\phantom{\frac{1}{2}}}\right\}\,\omz,
\end{align*}
\begin{align*}
\eta_4&=\left\{-\frac{1}{6}q(x)\,u_{yz}+
\frac{z}{4}q(x)\,u_{yzz}\right\}\,\omx \\
{}&+
\left\{\frac{1}{6}q'(x)\,u_z-\frac{z}{12}q'(x)\,u_{zz}-
\frac{1}{6}q(x)\,u_{xz}+\frac{z}{4}q(x)\,u_{xzz}\right\}\,\omy \\
{}&+
\left\{\frac{z}{2}q(x)\,u_{xyz}-\frac{1}{6}q(x)\,u_{xy}-
\frac{z}{6}q'(x)\,u_{yz}
+\frac{1}{6}q'(x)\,u_y\right.\\
  {}&{}\qquad\left.{}-zq(x)\,u_{zzz}\exp(-u_{zz})+q(x)\,\exp(-u_{zz})
\lefteqn{\phantom{\frac{1}{2}}}\right\}\,\omz,
\end{align*}
\begin{align*}
\eta_5&=\frac{r(x)}{4}\,u_{yzz}\, \omx
+ \left\{-\frac{r'(x)}{12}\,u_{zz}+
   \frac{r(x)}{4}\,u_{xzz}\right\}\,\omy \\
{}&+
\left\{-\frac{r'(x)}{6}\,u_{yz}+\frac{r(x)}{2}\,u_{xyz}+
r(x)\,u_{zzz}\exp(-u_{zz})\right\}\,\omz.
\end{align*}
Any divergence $\bar\Id_h(\eta_i)$ equals $0$ on the equation
$\cE_\heav$\textup{;} calculating the divergences does not require
precise evaluation of the integrals in the
homotopy parameter $t$ in rational functions,
because differentiating with respect to\ $x$ or $z$
under the integral sign is allowed.
\end{state}

Proving Lemma~\ref{WhichAreNoether} and Proposition~\ref{HeavLawsState}
is based on Lemma~\ref{EvThroughEuler} and involves the general scheme of
reconstruction of conservation laws by their generating sections.

\subsubsection*{Method of reconstruction of conserved currents}
Now we describe a method (\cite{Vin84}) for constructing the
differential $(n-1)$-forms that are exact on the equation
$\cE=\{F=0\}$; we follow the paper~\cite{Vestnik2000}.

Consider the function $f=u/\tau$ and the evolutionary vector field
$\cEv_f$ whose flow is
$$A_\tau\colon(x^i, u^j_\sigma)\mapsto(x^i, \tau u^j_\sigma),$$
then we have
\begin{equation}\label{diffpart}
\frac{dA_\tau^*(\omega)}{d\tau}=
A_\tau^*\left(\cEv_f(\omega)\right) = A_\tau^*\left(\ell_\omega(f)\right)
\end{equation}
for any differential form~$\omega$.
Let this form $\omega$ be $$\omega=\langle F,\psi\rangle=F\psi\,dx^1
\wedge\ldots\wedge dx^n$$
and suppose that $\eta$ is the desired current which is assigned to
the generating function $\psi$ on the equation \mbox{$\{F=0\}$:}
$\Id_h(\eta)=\nabla(F)$ and $\psi=\nabla^*(1)$.
We note that the right\/-\/hand side\ in Eq.~(\ref{diffpart}) contains the term
$$\ell_\omega(f)=\langle\ell_\omega^*(1),f\rangle+
\Id_h G(\ell_\omega\circ f),$$
where the first summand is cohomological to $0$ since the image of the
Euler operator is trivial if the Lagrangian density is a total
divergence,
\[
\ell_\omega^*(1)=\ell_{\langle F,\psi\rangle}^*(1)=
\bE(\langle F,\psi\rangle)=\bE(\Id_h\eta)=0.
\]
Next, the mapping $G\colon\cC{\rm Diff}(\mathcal{F},
\bar{\Lambda}^n)\to\bar{\Lambda}^{n-1}$, which is restricted onto the
equation $\cE$, is defined by the rule
$$
G\Bigl(\sum_\sigma a_\sigma D_\sigma\Bigr)=
\sum_{|\sigma|>0}\sum_{j\in\sigma}\frac{(-1)^{|\sigma|-1}}{|\sigma|}
D_{\sigma-1_j}(a_\sigma)
\omega_{(-j)},
$$
where $$\omega_{(-j)}=(-1)^{j+1}\,\Id x^1\wedge
\ldots\widehat{\Id x^j}\ldots\wedge\Id x^n$$ and
$\sigma-1_j$ is the result of excluding an index $j$ from a multiindex
$\sigma=(\sigma_1$, $\ldots$, $\sigma_n)$.

Now we integrate Eq.~(\ref{diffpart}) in $\tau$ from $0$ to $1$ and get
\begin{multline*} 
\int\nolimits_{0}^{1}\frac{\displaystyle d}{\displaystyle d\tau}
A_\tau^*(\omega)\,d\tau= A_1^*(\omega)
- A_{0}^*(\omega) =
\langle F,\psi\rangle- A_{0}^*\left(\langle F,\psi\rangle
\right) = {} \\
{}= \int\nolimits_{0}^1 A_\tau^*\left(\ell_\omega(f)\right)d\tau
= \Id_h\int\nolimits_{0}^1 A_\tau^*\left(G(\ell_\omega\circ f)
\right)d\tau = \Id_h\eta,
\end{multline*}
whence we obtain the required formula for the current~$\eta$.

\begin{proof}[Proof of Lemma~\textup{\ref{WhichAreNoether}} and
Proposition~\textup{\ref{HeavLawsState}}]
First we note that symmetries \eqref{SymPointHeav} of
Eq.~\eqref{OurHeavenly} are also symmetries of the Euler
equation~\eqref{ImEulerHeav}:\linebreak
$\Sym\hat\cE\subset\sym\cE_\heav$, because
the relation $\ell_{F_\heav}=D_z^2\circ\ell_{\hat F}$ holds due to
Lemma~\ref{LinWRTDLemma} on page~\pageref{LinWRTDLemma}.
By using Lemma~\ref{EvThroughEuler}, we check that symmetries
\eqref{SymHeavConf} and \eqref{SymHeavTransl}\/--\/\eqref{SymHeavPoly0}
of Eq.~\eqref{ImEulerHeav} are Noether, while for symmetry
\eqref{SymHeavScale} we have
\[
\ell_{F_\heav}(\vph_2)+\ell^*_{\vph_2}(F_\heav)=2u_{xyzz}-
\tfrac{3}{2}D_z^2(\exp(-u_{zz}))\not=0.
\]

From Theorem~\ref{InverseNoether} it follows that the Noether
symmetries of Eq.~\eqref{ImEulerHeav} are in one-to-one correspondence
with the generating sections of its conservation laws.
By Corollary~\ref{Podkrutka} on page~\pageref{Podkrutka}, this
correspondence
for the equation $\cE_\heav$
is provided by the identity mapping.
%
Following the reasonings scheme above, to each of four Noether
symmetries $\vph_i$ we assign the conserved current $\eta_i$;
the arising expressions are rather large and therefore omitted.
Nevertheless, the result is correct, for the conditions
$\bar\Id_h(\eta_i)=0$ may be checked straightforwardly.
\end{proof}

\begin{rem}
Suppose that the arbitrary function $f$ in the conserved current
$\eta_1$ is $f\equiv1$. Then $\eta_1$ is a countinuous in
$z$ analog of component~\eqref{EMT} of the energy\/--\/momentum
tensor $\Theta=T\,\Id x+\bar T\,\Id y$ for the Toda
equations~\eqref{eqToda}. We recall that the conserved current $T\,\Id
x$ for Eq.~\eqref{eqToda} corresponds to the Noether symmetry
$\vph_0^1=\square(\bE_T(T\,\Id x))$ $\in\sym\cL_\Toda$.
Another (nonlocal) conserved current
\[
\eta=u_{xz}\exp(-u_{zz})\,\Id x\wedge\Id y
+ \bigl(\tfrac{1}{2}u_{xz}^2-u_{xx}\bigr)\,\Id x\wedge\Id z,
\]
for Eq.~\eqref{ImEulerHeav} (see Eq.~\eqref{EMTHeav}
on page~\pageref{EMTHeav}) could be also treated as an analog of the
integral~\eqref{EMT} for Eq.~\eqref{eqToda} due to the following
reason. Consider the Hamiltonian representation
\begin{equation}\label{HeavHamForm}
u_y=D_x^{-1}\circ D_z^{-2}\circ\bE_u\bigl( H_\heav\,\Id x
\wedge\Id z\bigr),
\end{equation}
of Eq.~\eqref{ImEulerHeav}, here $$H_\heav=-\exp(-u_{zz})$$ is the
continuous limit of the Hamiltonian, see Eq.~\eqref{HamTodaViaU}, for
the Toda equations. By Lemma~\ref{HamIsConservedLemma},
the density $H_\heav$ is conserved on the related
Hamiltonian equation~\eqref{HeavHamForm}, therefore the current $\eta$
is also conserved.
\end{rem}


\newpage
\specialsection*{\textbf{Chapter~4.}
B\"acklund transformations and zero\/--\/curvature
representations}\label{ChZCR}
In this chapter, we investigate the relationship
between B\"acklund transformations and zero\/--\/curvature
representations for the hyperbolic Liouville equation, the wave
equation, and the Liouville $\scal^{+}$-equation. They are
\begin{equation}\label{eqList}
u_{xy}=\exp(2u),\qquad v_{xy}=0,\qquad \Upsilon_{xy}=\exp(-2\Upsilon),
\end{equation}
respectively. In Sec.~\ref{SecTodaEq} on
page~\pageref{ConfEquivExample} we we discussed a natural geometric
scheme that provides Eqs.~\eqref{eqList} and gives their interpretation.
In what follows, we construct one\/--\/parametric
families of B\"acklund transformations and zero\/--\/curvature
representations and consider exmaples of integrating
B\"acklund transformations and zero\/--\/curvature
in nonlocal variables.

\section{B\"acklund transformations and their deformations}%
\label{SecDeform}
In this section, we study the properties of one\/--\/parametric
deformations of B\"acklund transformations for Eq.~\eqref{eqList}.
Consider the covering
$\tau_t\colon\tilde\cE_t\to\cE_\Liou$
defined by the extended total derivatives
\begin{equation}\label{CoveringStructure}
\tilde D_x=\bar D_x+\tilde{u}_x\,\frac{\dd}{\dd \tilde{u}},\quad
\tilde D_y=\bar D_y+\tilde{u}_y\,\frac{\dd}{\dd
\tilde{u}},\qquad[\tilde D_x, \tilde D_y]=0.  \end{equation} We assume
that the partial derivatives of the nonlocal variable $\tilde{u}$
with respect to\ $x$ and $y$ are %
\begin{subequations}\label{ruletoderiv}
\begin{align} \tilde{u}_{x}&=u_{x}+\exp(-t)\cdot\exp(\tilde{u}+u),\\
\tilde{u}_{y}&=-u_{y}+2\exp(t)\cdot\sinh(\tilde{u}-u),
\end{align}
\end{subequations}
respectively.
Recalling Remark~\ref{BacklundBySymmetryRem} on
page~\pageref{BacklundBySymmetryRem}, let
the diffeomorphism $\mu$ be the swapping
$u\leftrightarrow \tilde{u}$ of the
fiber variable $u$ and the nonlocal variable $\tilde{u}$ combined with
the mapping $x\mapsto-x$ and $y\mapsto-y$.
Then diagram~(\ref{AlternativeAuto}) supplies
B\"acklund autotransformation
$\cB(\tilde\cE_t$, $\tau_t$, $\tau_t\circ\mu$,
$\cE_\Liou)$ for Eq.~(\ref{eqhyp2}), and
for the Liouville equation.
The equations $\tilde\cE_t$ of this B\"acklund autotransformation
(\cite{DoddBacklund}) are, obviously,
\begin{subequations}\label{BacklundAuto}
\begin{align}
(\tilde{u}-u)_{x}&=\exp(-t)\cdot\exp(\tilde{u}+u),\label{BacklundAuto1}\\
(\tilde{u}+u)_{y}&=2\exp(t)\cdot\sinh(\tilde{u}-u).\label{BacklundAuto2}
\end{align}
\end{subequations}

Put $$u_k\equiv\frac{\dd^ku}{\dd x^k}\text{ and }
\dy{u}{k}\equiv\frac{\dd^ku}{\dd y^k}$$ for any $k\in\BBN$.  Consider
the scaling symmetry
\begin{align}
X^0&=-{x}\,\frac{\dd}{\dd{x}}+{y}\,\frac{\dd}{\dd{y}}\notag\\
\intertext{of the
Liouville equation and extend $X^0$ onto entire $\cE_\Liou^\infty$:}
\hat X&=-{x}\,\frac{\dd}{\dd{x}}+{y}\,\frac{\dd}{\dd{y}}
+\sum\limits_{k\geq1}ku_k\,\frac{\dd}{\dd u_k}
-\sum\limits_{k\geq1}k\dy{u}{k}\,\frac{\dd}{\dd\dy{u}{k}}.\label{ShadowLift}
\end{align}

\begin{state}\label{CannotLiftState}
The symmetry $\hat X$ cannot be extended to a symmetry of the covering
equation $\tilde\cE_t$.
\end{state}

\begin{proof}
Assume the converse. By $\tilde\cEv_\vph$ denote the evolutionary
vector field
$$\sum\limits_\sigma\tilde{D}_\sigma(\vph)\cdot\frac{\dd}{\dd u_\sigma}$$
on $\tilde\cE_t$, where $\vph\in C^\infty(\tilde{\cE}_t)$. By
definition, set $$\tilde\ell_F(\vph)=\tilde\cEv_\vph(F).$$
Suppose there is a smooth function $a\in C^\infty(\tilde\cE_t)$ such
that the linearized system %
\begin{equation}\label{eqonnonlocalsym}
\tilde\ell_F(\vph)=0,\qquad \tilde
D_{x^i}(a)=\tilde\cEv_{\vph,a}(\tilde{u}_{x^i})
\equiv\left(\tilde\cEv_{\vph}+a\,\frac{\dd}{\dd \tilde{u}}\right)
(\tilde{u}_{x^i})
\end{equation}
holds. This means that the field
$\tilde\cEv_{\vph,a}$ is a local
symmetry of the covering equation $\tilde\cE_t$ and $\hat X$ is
extended onto $\tilde\cE_t$ constructively.
Nevertheless, system (\ref{eqonnonlocalsym}) is
not compatible since
$$
\tilde D_{x}\circ\tilde D_{y}(a)\neq\tilde D_{y}\circ\tilde D_{x}(a).
$$
Indeed, $\tilde D_{x}\circ\tilde D_{y}(a)-\tilde D_{y}\circ\tilde
D_{x}(a)$ does not depend on $a$ at all and equals
\begin{gather*}
  x u_x^2  {\rme}^{t + u - \tilde{u}} + u_x y u_y {\rme}^{t + \tilde{u} - u} -
x u_x {\rme}^{2 \tilde{u}} - u_x y u_y {\rme}^{t + u - \tilde{u}}\\ - 2 y u_y
       {\rme}^{2 t + \tilde{u} + u} + 2 x u_x {\rme}^{2 t + \tilde{u} + u} -
       x u_x^2  {\rme}^{t + \tilde{u} - u} + x {\rme}^{2 u} u_x\\ - y {\rme}^{2
       u} u_y + 2 {\rme}^t x u_x^2 + y u_y {\rme}^{2 \tilde{u}} - 2 {\rme}^t
       y u_y u_x \neq0.
\end{gather*}
This contradiction concludes the proof.
\end{proof}

Therefore, the scaling symmetry $\hat X$ is a $\tau_t$-\emph{shadow} only
(\textit{i.e.}, a solution to the equation $\tilde\ell_F(\vph)=0$) and
generates the family of the covering equations $\tilde\cE_t$ over
$\cE_\Liou$; these $\tilde\cE_t$ are parametrized by $t\in\BBR$.

Consider the Cartan distribution $\tilde\cC_t$ on the equation
$\tilde\cE_t$ and by $U_t$ denote the Cartan's connection form
on~$\tilde\cE_t$. In local coordinates, the form $U_t$ is
\begin{multline}
U_t=\sum\limits_\sigma \Id_\cC(u_\sigma)\otimes\frac{\dd}{\dd
u_\sigma} +\Bigl(\Id\tilde{u}-
\bigl(u_{x}+\exp(\tilde{u}+u-t)\bigl)\,\Id{x}+{}\\
{}+\bigl(u_{y}-2\exp(t)\sinh(\tilde{u}-u)\bigl)\,\Id{y}\Bigl)
\otimes\frac{\dd}{\dd \tilde{u}},
\label{StructElem}
\end{multline}
where $\Id_\cC$ is the Cartan differential. In coordinates, we have
$$\Id_\cC(u_\sigma)=\Id u_\sigma-
\sum\limits_i\bar D_i(u_{\sigma})\,\Id x^i.$$

Let $\Omega\in\ID(\Lambda^\mu(\tilde\cE))$, then
$\mu$ is the \emph{degree} of the derivation~$\Omega$.
By $\fn{{\cdot}}{{\cdot}}$ we denote the
\emph{Fr\"olicher\/--\/Nijenhuis bracket} (\cite{JKIgonin, JKKersten}):
\begin{equation}\label{DefFNBracket}
\fn{\Omega}{\Theta}(f)=\IL_\Omega(\Theta(f))-(-1)^{\mu\nu}
\cdot\IL_\Theta
(\Omega(f)),
\end{equation}
where $\Omega,\Theta\in\ID(\Lambda^\ast(\cE))$,
$f\in C^\infty(\cE)$, and the degrees $\mu$, $\nu$
are $\mu=\deg\Omega$, $\nu=\deg\Theta$, respectively.
We also assume that
$$\IL_\Omega=[i_\Omega,\Id]\colon
\Lambda^k(\cE)\to\Lambda^{k+\deg\Omega}(\cE)$$
is the Lie derivative and
$$i_\Omega\colon\Lambda^k(\cE)\to\Lambda^{k+\deg\Omega-1}(\cE)$$ is the
inner product.
The Fr\"olicher\/--\/Nijenhuis bracket is a natural geometric structure
in differential calculus. The de Rham differential~$\Id$ and the
Richardson\/--\/Nijenhuis bracket
$[\![{\cdot},{\cdot}]\!{]}^{\mathrm{RN}}$
(see~\cite{ForKac} for its
applications) are other examples of natural structures.

\begin{theor}[\cite{JKIgonin}]
Let $\tau\colon \tilde\cE\to\cE$ be a covering and $A_t\colon
\tilde\cE\to \tilde\cE$ be a smooth family of diffeomorphisms such
that $A_0=\id$ and $\tau_t=\tau\circ A_t\colon \tilde\cE\to \cE$ is the
covering for any $t\in\BBR$.  Then the evolution of the Cartan
connection form $U_{\tau_t}$ is
\begin{equation}\label{DeformStructElem}
\frac{dU_{\tau_t}}{dt}=\fn{\hat X_t}{U_{\tau_t}},
\end{equation}
where $\hat X_t$ is a $\tau_t$-shadow for any $t\in\BBR$.
\end{theor}

Suppose $\tilde\cE$ is a
finite\/-\/dimensional manifold, then there is the isomorphism
$$\ID(\Lambda^*(\tilde\cE))\simeq\Lambda^*(\tilde\cE)\otimes\ID(\tilde\cE).$$
Thence, any
derivation $\Omega\in\ID(\Lambda^*(\tilde\cE))$ is
decomposable to the finite
sum such that the summands are $\Omega=\omega\otimes X$, where
$\omega\in\Lambda^*(\tilde\cE)$ and $X\in\ID(\tilde\cE)$.
The Fr\"olicher\/--\/Nijenhuis bracket of such elements is
\begin{multline}
\fn{\omega\otimes X}{\theta\otimes Y}=\omega\wedge\theta\otimes[X,Y]
+\omega\wedge\IL_X(\theta)\otimes(Y)+ {}\\
{}+(-1)^i\,\Id\omega\wedge(X\inner\theta) \otimes Y-
(-1)^{ij}\,\theta\wedge\IL_Y(\omega)\otimes X- {}\\
{}-(-1)^{(i+1)j}\,\Id\theta
\wedge(Y\inner\omega)\otimes X,\label{FNinCoords}
\end{multline}
where $X$, $Y\in\ID(\tilde\cE)$, $\omega\in\Lambda^i(\tilde\cE)$ and
$\theta\in\Lambda^j(\tilde\cE)$.
If the dimension of $\tilde\cE$ is not necessarily finite,
then there is the embedding
$$\Lambda^\ast(\tilde\cE)\otimes\ID(\tilde\cE)\subset\ID(\Lambda^\ast
(\tilde\cE))$$ defined by the rule
$$(\omega\otimes X)(f)=X(f)\omega$$ for any function
$f\in C^\infty(\tilde\cE)$.

Consider the coverings $\tau_t\colon\tilde\cE_t\to\cE$
defined in Eq.~\eqref{CoveringStructure}. Then we have
\begin{equation}\label{RHSDeform}
\frac{dU_t}{dt}=\left(\exp(\tilde{u}+u-t)\,\Id{x}
-2\exp(t)\sinh(\tilde{u}-u)\,\Id{y}\right)\otimes\frac{\dd}{\dd
\tilde{u}}.
\end{equation}

We claim that the scaling symmetry $\hat X$ is the
$\tau_t$-shadow such that the evolution of the
connection form $U_t$ (\ref{StructElem}) in the covering
$\tau_t$, see Eq.~\eqref{CoveringStructure}, is given by
(\ref{RHSDeform}) in virtue of equation~(\ref{DeformStructElem}).
We need Lemmas~\ref{NonlocalVar}-\ref{HigherVars} to prove this.

\begin{lemma}\label{NonlocalVar}
$\fn{\hat
X}{U_t}\inner\Id\tilde{u}=\bigl({dU_t}/{dt}\bigr)\inner\Id\tilde{u}$.
\end{lemma}

\begin{lemma}\label{LowerVars}
$\fn{\hat X}{U_t}\inner\Id{x}=\fn{\hat X}{U_t}\inner\Id{y}=
\fn{\hat X}{U_t}\inner\Id u=0$.
\end{lemma}

\begin{proof}
The proof of Lemmas~\ref{NonlocalVar} and~\ref{LowerVars}
is based on successive application of formula~(\ref{FNinCoords}).
The coefficient of ${\dd}/{\dd x}$ is
$$-\sum\limits_\sigma
\Id_\cC u_\sigma\wedge\IL_{{\dd}/{\dd u_\sigma}}(-x) -
\Id_\cC\tilde{u}\wedge\IL_{{\dd}/{\dd\tilde{u}}}(-x)=0;$$
the coefficient of ${\dd}/{\dd y}$ is calculated analogously.
Now consider Eq.~\eqref{ShadowLift} and
decompose $\hat X$ in the coordinates $\langle x$, $y$, $u_k$,
$\dy{u}{k}\rangle$. We get
$\hat{X}=\sum\limits_\alpha \omega_\alpha\otimes X_\alpha$,
where $\omega_\alpha$ is a $0$-form and
$X_\alpha$ is a derivation for any $\alpha$. We have
$i=0$ and $j=1$ in Eq.~\eqref{FNinCoords}. Thence we get
$$\sum\limits_\alpha\bigl( \omega_\alpha\wedge\IL_{X_\alpha}(\Id_\cC u)
+ \Id\omega_\alpha\wedge(X_\alpha\inner\Id_\cC
u)\bigr)\otimes\frac{\dd}{\dd u},$$
where the first summand is
$$\sum\limits_\alpha \omega_\alpha\wedge\Id(X_\alpha\inner\Id_\cC
u)+\sum\limits_\alpha\omega_\alpha\wedge\bigl(X_\alpha\inner\Id(\Id_\cC
u)\bigr)=-u_x\,\Id x+u_y\,\Id y,$$
and the second summand equals
$u_x\,\Id x-u_y\,\Id y$, whence their sum is also trivial.
Finally, we calculate the coefficient
$$\sum\limits_\alpha\bigl(\omega_\alpha\wedge\IL_{X_\alpha}(\Id_\cC\tilde{u}) +
\Id\omega_\alpha\wedge(X_\alpha\inner\Id_\cC\tilde{u})\bigr)$$
of
${\dd}/{\dd\tilde{u}}$ by using the explicit formula for
$\Id_\cC\tilde{u}$. The first summand is equal to
$-u_x\,\Id x-u_y\,\Id y$, and the second summand is
$$\bigl(u_x+\rme^{-t}\exp(\tilde{u}+u)\bigr)\,\Id x + \bigl(u_y-2\rme^t
\sinh(\tilde{u}-u)\bigr)\,\Id y.$$
Consequently, we obtain the
required expression
$$\bigl(\rme^{-t}\exp(\tilde{u}+u)\,\Id x-2\rme^t\sinh(\tilde{u}-u)\,\Id
y\bigr)\otimes\frac{\dd}{\dd\tilde{u}}.$$
This concludes the proof.
\end{proof}

However, the calculation of coefficients of
${\dd}/{\dd u_k}$ or ${\dd}/{\dd\dy{u}{k}}$
in $\fnh{\hat X}{U_t}$ is untrivial for $k\geq1$.
First, we have

\begin{lemma}[\textup{\cite{DeformLiou}}]\label{StrangeRelation}
Let $u(x)$ and $f(u)$ be smooth functions and
$D_x$ be the total derivative
with respect to~$x$. Denote $u_k\equiv D^k_x(u(x))$, $k\geq0$,
$u_0\equiv u$. Then the relation
\begin{equation}\label{DerivExpRelation}
n\cdot D_x^n(f(u))=\sum\limits_{m=1}^nm\,u_m\,\frac{\dd}{\dd u_m}
  D_x^n(f(u))
\end{equation}
holds for any integer $n\geq1$.
\end{lemma}

The proof of Lemma~\ref{StrangeRelation} is based on
Lemma~\ref{RemMarvanManual} and Corollary~\ref{CorMarvanManual}
below.

\begin{lemma}\label{RemMarvanManual}
Let the assumptions of Lemma~\ref{StrangeRelation} hold.
Take an integer $n>0$ and a positive integer $l\leq n-1$.
Then the relation
\begin{equation}\label{RelMarvan}
D_x\left(\frac{\dd}{\dd u_l}\,D^{n-1}_x(f(u))\right)=
 \frac{\dd}{\dd u_l}\, D^n_x(f(u))-
 \frac{\dd}{\dd u_{l-1}}\, D^{n-1}_x(f(u))
\end{equation}
is valid.
\end{lemma}

\begin{cor}\label{CorMarvanManual}
Let the assumptions of Lemma~\textup{\ref{RemMarvanManual}} hold.
Then we also have
\begin{multline}
(n+1)u_{n+1}\,\frac{\dd}{\dd u_{n+1}}D_x^{n+1}(f(u))
 =(n+1)u_{n+1}\,\frac{\dd}{\dd u_{n}}D_x^{n}(f(u))={}\\
{}=(n+1)u_{n+1}\cdot f'(u).\label{CorMarvan}
\end{multline}
\end{cor}

\begin{proof}[Proof of Lemma~\textup{\ref{StrangeRelation}
   (\cite{DeformLiou})}]
We prove (\ref{DerivExpRelation}) by induction on $n$. For ${n=1}$,
relation (\ref{DerivExpRelation}) holds. For $n\geq1$, we have
\begin{align*}
&(n+1)\,D_x^{n+1}(f(u))=D_x(n\,D_x^n(f(u))+D_x^n(f(u)))={}\\
\intertext{by the inductive assumption,}
&\quad  {}=D_x\left(\sum\limits_{m=1}^nmu_m\,\frac{\dd}{\dd
u_m}\,D_x^n(f(u))+D_x^n(f(u))\right)={}\\
\intertext{by the Leibnitz rule,}
&\quad  {}=\sum\limits_{m=1}^nmu_{m+1}\,\frac{\dd}{\dd
u_m}\,D_x^n(f(u))+{}\\
&\quad {} \quad {}+\sum\limits_{m=1}^nmu_m\,D_x\frac{\dd}{\dd
u_m}\,D^n_x(f(u))+D_x\,D_x^n(f(u))={}\\
\intertext{by Eq.~(\ref{RelMarvan}) applied to the second sum,}
&\quad  {}=\sum\limits_{m=1}^nmu_m\,\frac{\dd}{\dd
u_m}\,D_x^{n+1}(f(u))+\sum\limits_{m=1}^nmu_{m+1}\,\frac{\dd}{\dd
u_m}\,D_x^n(f(u))-{}\\
&\quad  \qquad{}-\sum\limits_{m=1}^nmu_m\,\frac{\dd}{\dd
u_{m-1}}\,D_x^n(f(u))+D_xD_x^n(f(u))={}\\
\intertext{by the definition of $D_x$ and
the subscript shift in the latter sum,}
&\quad  {}=\sum\limits_{m=1}^nmu_m\,\frac{\dd}{\dd
u_m}\,D_x^{n+1}(f(u))
+\sum\limits_{m=0}^n(m+1)u_{m+1}\,\frac{\dd}{\dd u_m}\,D_x^n(f(u))-{}\\
&\quad  \qquad{}-\sum\limits_{m=0}^{n-1}(m+1)u_{m+1}\,\frac{\dd}{\dd
u_m}\,D_x^n(f(u))={}\\
\intertext{since almost all summands in the latter two sums coincide,}
&\quad  {}=\sum\limits_{m=1}^nmu_m\,\frac{\dd}{\dd
u_m}\,D_x^{n+1}(f(u)) +(n+1)u_{n+1}\,\frac{\dd}{\dd
u_n}\,D_x^n(f(u))={}\\
\intertext{by (\ref{CorMarvan}),}
&\quad
{}=\sum\limits_{m=1}^nmu_m\,\frac{\dd}{\dd u_m}\,D_x^{n+1}(f(u))
+(n+1)u_{n+1}\,\frac{\dd}{\dd u_{n+1}}\,D_x^{n+1}(f(u))={}\\
&\quad  {}=\sum\limits_{m=1}^{n+1}mu_m\,\frac{\dd}{\dd
u_m}\,D_x^{n+1}(f(u)).  
\end{align*}
The proof is complete.
\end{proof}

Here we offer another proof of Lemma~\ref{StrangeRelation}.
This proof is based on the weights technique.
The idea of this proof was communicated by V.\,V.\,Trushkov.

\begin{proof}[Proof of
Lemma~\textup{\ref{StrangeRelation} (\cite{NonAbel})}]
By definition, put the \emph{weight}
$\wt(u_k)=k$, $\wt(u_k\cdot u_l)=k+l$,
and $\wt(u_{k_1}+u_{k_2})=k_1$ if $k_1=k_2$. We have
\begin{equation}\label{WRTDerOfSmooth}
D^n_x(f(u))=\sum\limits_{m=1}^n P_{n,m}\cdot f^{(m)}(u),
\end{equation}
where $P_{n,m}=\sum\limits_{\vec\jmath}\const(n,m)\cdot u_{j_1}\cdot\ldots\cdot
u_{j_{l(n,m)}}$. We claim that $P_{n,m}$ is a differential
polynomial such that
\begin{equation}\label{NormForJ}
j_1+\ldots+j_{l(n,m)}=n\qquad \forall\vec\jmath,\quad\forall\ n,\ m.
\end{equation}
We prove this fact by induction on $n$. Indeed, if $\wt(P_{n,m})=n$,
then $\wt(D_x(P_{n,m})))=n+1$ due to the Leibnitz rule. Besides,
$$
D^{n+1}_x(f(u))=\sum\limits_{m=1}^n\left( D_x(P_{n,m})\cdot f^{(m)}(u)
+ P_{n,m}\cdot u_1\cdot f^{(m+1)}(u) \right)
$$
and therefore the weight $\wt(D_x^{n+1}(f(u)))$ is well defined and equals
$n+1$.

Now, consider the \emph{weight counting operator}
\begin{equation}\label{WCO}
\mathcal{W}\equiv\sum\limits_{m\geq1}m\cdot u_m\,\frac{\dd}{\dd u_m}
\end{equation}
that acts on the right\/-\/hand side\ in Eq.~\eqref{WRTDerOfSmooth}:
\begin{multline*}
\sum\limits_{m\geq1}m\cdot u_m\,\frac{\dd}{\dd u_m} \circ
 \sum\limits_{k=1}^n \sum\limits_{\vec\jmath} \const(n,k)\cdot
 u_{j_1}\cdot\ldots\cdot u_{j_{l(n,k)}}\cdot f^{(k)}(u)={}\\
{}=\sum\limits_{k=1}^n \sum\limits_{\vec\jmath} \const(n,k)\cdot n\cdot
 u_{j_1}\cdot\ldots\cdot u_{j_{l(n,k)}}\cdot f^{(k)}(u)=
n\cdot D_x^n(f(u)),
\end{multline*}
since condition~\eqref{NormForJ} holds for all multiindexes
$\vec\jmath$. We finally conclude that the functions $D_x^n(f(u))$
are eigenfunctions for operator \eqref{WCO} and the integers $n\in\BBN$
are the eigenvalues.  Therefore, relation \eqref{DerivExpRelation} is a
solution to the problem $\lambda\cdot\vph=\mathcal{W}(\vph)$. The proof
is complete.  \end{proof}

\begin{lemma}\label{HigherVars}
$\fn{\hat X}{U_t}\inner\Id u_k=\fn{\hat X}{U_t}\inner\Id\dy{u}{k}=0$
for any $k\geq1$.
\end{lemma}

\begin{proof}
Let $k\in\BBN$. Taking into account Eq.~(\ref{FNinCoords}),
consider the $1$-form
$$
\fn{\hat X}{U_t}\inner\Id u_k=\Bigl((k-1)\bar
D_yu_k-\sum\limits_{l=1}^{k-1}lu_l\,\frac{\dd}{\dd u_l}\bar D_x^{k-1}
(\exp(2u))\Bigr)\cdot\Id y.
$$
We see that all coefficients of $\Id x$, $\Id u$, $\Id u_l$,
$\dy{\Id u}{l}$ vanish for all $l\geq1$. Finally, note that $\bar
D_yu_k=\bar D_x^{k-1}(\exp(2u))$. By Lemma \ref{StrangeRelation}, the
coefficient of $\Id y$ is trivial.  Arguing as above, we conclude that
$\fnh{\hat X}{U_t}\inner d\dy{u}{k}=0$.  This completes the proof.
\end{proof}

\begin{theor}\label{FinalDeformTh}
The $\tau_t$-shadow in Eq.~{\rm(\ref{ShadowLift})} satisfies
the relation $$ \fnh{\hat X}{U_t}=\left(
\exp(\tilde{u}+u-t)\,d{x}-2\exp(t)\sinh(\tilde{u}-u)\,d{y}
		 \right)\otimes\frac{\dd}{\dd \tilde{u}},
$$
\textit{i.e.}, the abelian group of diffeomorfisms $A_t=\exp(t\hat X)$ provides
smooth one\/-\/paramet\-ric family {\rm(\ref{ruletoderiv})} of
one\/-\/dimensional coverings over
Liouville's equation $\cE_\Liou$.
These coverings correspond to B\"acklund
autotransformations~\eqref{BacklundAuto} for the
Liouville equation, see diagram~\eqref{AlternativeAuto}.  Evolution of
the connection form is given by~{\rm(\ref{RHSDeform})}.  \end{theor}

\begin{proof}
Decompose the bracket $\fn{\hat X}{U_t}$ with respect to\ the basis
$\langle\frac{\dd}{\dd x}$, $\frac{\dd}{\dd y}$, $\frac{\dd}{\dd u_k}$,
$\frac{\dd}{\dd u_{\bar k}}$, $\frac{\dd}{\dd\tilde u}\rangle$.  By
Lemmas~\ref{LowerVars} and~\ref{HigherVars}, all coefficients of the
derivations $\frac{\dd}{\dd x}$, $\frac{\dd}{\dd y}$, $\frac{\dd}{\dd
u_k}$, $\frac{\dd}{\dd u_{\bar k}}$ equal $0$ for $k\geq0$. Thence,
Lemma~\ref{NonlocalVar} supplies the required expression, which is
given in Eq.~\eqref{RHSDeform}.  \end{proof}

\begin{rem}
Consider B\"acklund transformation~(\cite{DoddBacklund})
\begin{equation}\label{BtsLiouWave}
(v-u)_x=\rme^{-t}\exp(u+v),\; (v+u)_y=-\rme^{t}\exp(u-v),
\end{equation}
$t\in\BBR$, between the Liouville equation~(\ref{eqhyp}) and the wave
equation $v_{xy}=0$, as well as B\"acklund
transformation~(\cite{DoddBacklund})
\begin{equation}\label{BtsLiouScalPlus}
(\Upsilon-u)_x=2\rme^{-t}\cosh(\Upsilon+u),\;
(\Upsilon+u)_y=-\rme^{t}\exp(u-\Upsilon),
\end{equation}
$t\in\BBR$, between $\cE_\Liou$ and the Liouville $\scal^{+}$-equation
$\Upsilon_{xy}=\exp(-2\Upsilon)$.
Then these transformations possess the property similar to one
described in Theorem~\ref{FinalDeformTh}.
Namely, scaling symmetry~(\ref{ShadowLift}) of $\cE_\Liou$
is the required $\tau_t$-shadow in both cases.
The proof of these theorems is quite analogous to the proof of
Theorem~\ref{FinalDeformTh} above and appeals to useful
identity~\eqref{DerivExpRelation} in total derivatives.
\end{rem}

\section{Integrating B\"acklund\,transformations
in nonlocal\,variables}\label{SecIntegrating}
In this section, we study nonlocal
aspects of integrating B\"acklund transformations between PDE.
The aim of this section
is to illustrate a natural scheme that provides nonlocal variables
associated with a certain PDE and to obtain nonlocal structures
for the hyperbolic Liouville equation.
We reconstruct the coverings $\tau_j$ from Eq.~\eqref{BacklundAuto} and
\eqref{BtsLiouWave}--\eqref{BtsLiouScalPlus} and demonstrate the
nonlocal variables to be potentials for the fiber variables $u$, $v$,
and $\Upsilon$. In what follows, we use the notation $\cE_u$ as a
synonim of $\cE_\Liou$, meaning that the Liouville equation
$u_{xy}=\exp(2u)$
is imposed on the fiber variable~$u$.

Now we construct the one-dimensional non-abelian coverings such that we
can integrate B\"acklund transformations \eqref{BacklundAuto},
\eqref{BtsLiouWave}--\eqref{BtsLiouScalPlus} in the corresponding
nonlocal variables.
We fix an arbitrary $t\in\BBR$ and
consider the extended total derivatives
\begin{align}
\tilde D^{\cE_u}_x &= \bar D^{\cE_u}_x -\rme^{2u}\frac{\dd}{\dd\Xi_t},
    &
\tilde D^{\cE_u}_y &= \bar D^{\cE_u}_y +
(\Xi_t^2+2u_y\Xi_t-\rme^{2t})\frac{\dd}{\dd\Xi_t},\notag\\
%
\tilde D^{\cE_u}_x &= \bar
   D^{\cE_u}_x-\rme^{2u}\frac{\dd}{\dd\Xi_\infty},
  &
\tilde D^{\cE_u}_y &= \bar D^{\cE_u}_y +
(\Xi_\infty^2+2u_y\Xi_\infty)\frac{\dd}{\dd\Xi_\infty},\notag\\
%
\tilde D^{\cE_u}_y &= \bar D^{\cE_u}_y-\rme^{2u}\frac{\dd}{\dd\Xi_t'},
 &
\tilde D^{\cE_u}_x &=
  \bar D^{\cE_u}_x+({\Xi_t'}^2+2u_x\Xi_t'+\rme^{-2t})
  \frac{\dd}{\dd\Xi_t'},\notag\\
\tilde D^{\cE_v}_x &= \bar D^{\cE_v}_x
  +\rme^{2v}\frac{\dd}{\dd\Xi^v_t},
  &
\tilde D^{\cE_v}_y &= \bar D^{\cE_v}_y + (2v_y\Xi^v_t+\rme^{2t})
  \frac{\dd}{\dd\Xi^v_t},\notag\\
\tilde D^{\cE_\Upsilon}_y &= \bar D^{\cE_\Upsilon}_y +
     \rme^{-2\Upsilon} \frac{\dd}{\dd\Xi_t^\Upsilon},
 &
\tilde D^{\cE_\Upsilon}_x &= \bar D^{\cE_\Upsilon}_x +
 ((\Xi_t^\Upsilon)^2-2\Upsilon_x\Xi_t^\Upsilon
     +\rme^{-2t}) \frac{\dd}{\dd\Xi_t^\Upsilon}.
\label{Nonlocals}
\end{align}
We see that the extended derivatives commute
in all cases: $$[\tilde D_x, \tilde D_y]=0.$$
Therefore the coverings
\begin{multline}\label{Coverings}
\tau_t\colon\tilde\cE_t\to\cE_u^\infty,\quad
\tau_\infty\colon\tilde\cE_\infty\to\cE_u^\infty,\quad
\tau'_t\colon\tilde\cE'_t\to\cE_u^\infty,\\
\tau^v_t\colon\tilde\cE^v_t\to\cE_v^\infty,\quad
\tau^\Upsilon_t\colon\tilde\cE^\Upsilon_t\to\cE_\Upsilon^\infty
\end{multline}
are well defined. The explicit form of the covering equations
$\tilde\cE_t$, $\tilde\cE_\infty$, $\tilde\cE'_t$, $\tilde\cE^v_t$,
and $\tilde\cE^\Upsilon_t$ is discussed in Remark~\ref{ExplicitCovEq}.

\begin{rem}
Coverings \eqref{Coverings} with nonlocal variables
\eqref{Nonlocals} cannot be reduced to local
conservation laws for the underlying equations $\cE_u$, $\cE_v$, and
$\cE_\Upsilon$; these coverings are called \emph{non\/--\/abe\-li\-an}.
We also claim that the $t$-parameterized coverings
(\textit{e.g.} $\tau_t$ at the points $t_1$ and $t_2$) are equivalent
($\tau_{t_1}\simeq\tau_{t_2}$), \textit{i.e.}, there is a functional dependence
between the nonlocal variables ($\Xi_{t_1}$ and $\Xi_{t_2}$ in our
case). We have
$$
\Xi_{t_1}+y\cdot\exp(2t_1)=\Xi_{t_2}+y\cdot\exp(2t_2)=
\Xi_{t=-\infty}\qquad\forall t_1,t_2\in\BBR;
$$
similar relations hold for other coverings~\eqref{Coverings}.
\end{rem}

\begin{rem}\label{ExplicitCovEq}
The covering equations can be obtained explicitly since
each nonlocal variable in \eqref{Nonlocals} is a
potential for at least one of the dependent variables $u$, $v$,
and $\Upsilon$. For example,
$$u=\frac{1}{2}\log\Bigl(-\frac{\dd\Xi_\infty} {\dd x}\Bigr).$$ Now we
derive the covering equations imposed on the fiber variable $\Xi_t$ and
its limit $\Xi_\infty$ at the point $t=-\infty$:  %
\begin{equation}\label{CovEqns}
\tilde\cE_t=\left\{
\frac{\dd\Xi_t}{\dd y}
 =\Xi_t^2+\frac{\Xi_t\cdot \frac{\dd^2\Xi_t}{\dd x\dd y}}
   {\frac{\dd\Xi_t}{\dd x}}-\exp(2t)
\right\},
\end{equation}
where $t\in\BBR\cup\{-\infty\}$. The equations $\tilde\cE_t'$,
$\tilde\cE^v_t$, and $\tilde\cE^\Upsilon_t$ are obtained analogously.
\end{rem}

\subsection{Integrating in nonlocal variables}
We emphasize that transformations~\eqref{BacklundAuto},
\eqref{BtsLiouWave}, and~\eqref{BtsLiouScalPlus}
cannot be integrated in local variables. Therefore we consider
one\/-\/dimensional non\/-\/abelian coverings \eqref{Coverings}
and extend the sets of
(local) variables with new nonlocal ones, see~\eqref{Nonlocals},
and then we integrate the transformations successfully.
The results can be summarized as follows.

\begin{theor}[\textup{\cite{NonAbel}}]\label{IntegrBacklundInNonlocal}
Consider B\"acklund \textup{(}auto\textup{)}\-trans\-for\-ma\-ti\-ons
\eqref{BacklundAuto}, \eqref{BtsLiouWave}--\eqref{BtsLiouScalPlus}
for the equations $\cE_u$, $\cE_v$, and $\cE_\Upsilon$.
These transformations
are integrated in nonlocal variables explicitly\textup{:}
\begin{enumerate}
\item B\"acklund autotransformation \eqref{BacklundAuto} for
$\cE_u$\textup{:}  
$$
\tilde{u}=u+t-\log\Xi_t \text{ \textup{и} }
u=t+\tilde{u}-\log\Xi_t[\tilde{u}](-x,-y), $$ \textit{i.e.}, to inverse
the transformation and obtain $u[\tilde u]$, the inversion $x\mapsto-x$
and $y\mapsto-y$ is required.  \item B\"acklund transformation
\eqref{BtsLiouWave} between
$\cE_u$
and the wave equation
$v_{xy}=0$\textup{:} $$ v=u+t-\log\Xi_\infty \text{ and, conversely, }
u=v+t-\log\Xi^v_t.
$$
\item B\"acklund transformation \eqref{BtsLiouScalPlus} between
$\cE_u$
and the $\scal^{+}$\/--\/Li\-o\-u\-vi\-lle equation
$\Upsilon_{xy}=\exp(-2\Upsilon)$\textup{:}
$$
\Upsilon=-u+t+\log\Xi'_t \text{ and, conversely, }
u=-\Upsilon-t-\log\Xi^\Upsilon_t.
$$
\end{enumerate}
\end{theor}
\begin{proof}
We consider the case $\tilde{u}[u](x,y)$ within B\"acklund
autotransformation \eqref{BacklundAuto}. By definition, put
$\cU=\exp(\tilde{u})$ and $\cT=\exp(-\tilde{u})$.
From Eq.~\eqref{BacklundAuto} we obtain the Bernoulli equation
$$
\cU_x=u_x\cdot\cU+\exp(u-t)\,\cU^2,
$$
whence $\cU^{-1}=\cT=\exp(-u-t)\cdot\Xi$, where the nonlocal
variable $\Xi$ is such that $\tilde D_x(\Xi)=-\exp(2u)$,
and also get the Riccati equation
\begin{equation}\label{RiccatiInAuto}
\cT_y=u_y\cdot\cT+\exp(u+t)\,\cT^2-\exp(t-u).
\end{equation}
Substituting $\exp(-u-t)\cdot\Xi$ for $\cT$ in
\eqref{RiccatiInAuto}, we get $\tilde
D_y(\Xi)=\Xi^2+2u_y\Xi-\exp(2t)$.
Now we refer~\eqref{Nonlocals} and
compare the result with the derivatives $\tilde D_x(\Xi_t)$ and $\tilde
D_y(\Xi_t)$.

The proof of other five cases is quite analogous. Assume
$f(x$, $y)\in\{u$, $\tilde{u}$, $v$, $\Upsilon\}$
is a known solution to the PDE $\cE_f$. Then we obtain
either two Bernoulli equations for Eq.~\eqref{BtsLiouWave} or one
Bernoulli equation and one Riccati equation for Eq.~\eqref{BacklundAuto}
and Eq.~\eqref{BtsLiouScalPlus} after a proper change of variables.
Solving these ordinary differential
equations for the solution
$g(x$, $y)\in\{u$, $\tilde{u}$, $v$, $\Upsilon\}$ to the
PDE $\cE_g$ that is related with $\cE_f$ by one of B\"acklund
(auto)transformations \eqref{BtsLiouWave}--\eqref{BtsLiouScalPlus}, we
finally obtain the rules to differentiate the nonlocal variable,
which appears in one of the coverings~\eqref{Coverings}.
\end{proof}


Consider diagram~\eqref{BacklundDiag} that
arises in the definition of a
B\"acklund transformation and apply Eq.~\eqref{AlternativeAuto}
to Theorem~\ref{IntegrBacklundInNonlocal}.
Consider the coverings $\tau_1$ and $\tau_2$ in Eq.~\eqref{Nonlocals}.
We see that in all cases one of these two mappings is a first order
differential operator that depends on the nonlocal variable only,
while another projector is a zero order morphism. Thence we have

\begin{theor}
Consider equations~\eqref{CovEqns}. Then the following relations hold:
\begin{align*}
u&=\tfrac{1}{2}\log{(-\Xi_t)}_x, &
\tilde u&= t+\log\bigl[{\Xi_t^{-1}} { \sqrt{{(-\Xi_t)}_x} }\bigr],\\
u&=\tfrac{1}{2}\log{(-\Xi_\infty)}_x, &
v&= t+\log\bigl[{\Xi_\infty^{-1}}{ \sqrt{{(-\Xi_\infty)}_x} }\bigr],\\
u&=\tfrac{1}{2}\log{(-\Xi'_t)}_y, &
\Upsilon&= t+\log \bigl[\frac{\Xi'_t}{ \sqrt{{(-\Xi'_t)}_y} }\bigr],\\
%
%
v&=\tfrac{1}{2}\log{(\Xi^v_t)}_x, &
u&= t+\log\bigl[ {(\Xi^v_t)}^{-1}{ \sqrt{{(\Xi^v_t)}_x} }\bigr],\\
\Upsilon&=-\tfrac{1}{2}\log{(\Xi^\Upsilon_t)}_y, &
u&=
 -t+\log\bigl[{(\Xi^\Upsilon_t)}^{-1}{\sqrt{{(\Xi^\Upsilon_t)}_y}
}\bigr].
\end{align*}
\end{theor}

In other words, the nonlocal variables that satisfy Eq.~\eqref{CovEqns}
are potentials for \emph{both}
solutions $f,\ g$
to the equations $\cE_f,\ \cE_g$ within the list
$\cE_u$, $\cE_v$, and $\cE_\Upsilon$.
The property of all the coverings in \eqref{Coverings} to be
nonlinear differential operators of order not greater than $1$ is a
special feature of these equations.

\subsection{On nonlocal symmetries}
Let $\cE$ be an equation and $\tau\colon\tilde\cE\to\cE$ be a
covering. By definition, a $\tau$-\emph{shadow} $\vph$ is a solution
of the linearized equation $\tilde\ell_{\cE}(\vph)=0$.  The shadow
fields
$\tilde\cEv_\vph\in{\bar{\mathrm{D}}}_\cC(\tilde\cE)$ are not true
nonlocal symmetries since they do not describe the evolution of the
nonlocal variable and, generally, not all of them can be extended up
to nonlocal symmetries.

We have proved that B\"acklund
transformations~\eqref{BacklundAuto}, \eqref{BtsLiouWave},
and~\eqref{BtsLiouScalPlus} themselves do contain certain information
about nonlocal variables such that these transformations can be
integrated successfully. The corresponding non-abelian coverings in
\eqref{Coverings} provide nonlocal conservation laws for the
underlying differential equations. Still, the structures on the
covering equations are ``too close'' to the initial ones, so that the
point symmetries of the initial equations and the
classical symmetries of the covering equations
are in one\/-\/to\/-\/one correspondence such that we get no
nonlocal symmetries except the liftings of local transformations.

Now we introduce new nonlocal variables such that the symmetries we are
in search of depend on them.
Let $\rSi_t=\Xi_t+u_y$ be the new nonlocal variable such that
\begin{align*}
\tilde D^{\cE_u}_x(\rSi_t)&=0,\\
\tilde D^{\cE_u}_y{(\rSi_t)}&=
{\left(\rSi_t\right)}^2+u_{yy}-u_y^2-\exp(2t).
\end{align*}
Consider the limit of
$\rSi_t$ as $t\to-\infty$. We see that at the point $t=-\infty$
there appeares the automodel variable
$$\Sigma_\infty=u_x+\frac{\exp(2u)}{\Xi_\infty}$$ such that $\tilde
D^{\cE_u}_y(\Sigma_\infty)=0$. We claim that
$$\rSi_\infty=\lim_{t\to-\infty} \rSi_t$$ and $\Sigma_\infty$ differ by
the discrete symmetry $x\leftrightarrow y$. Indeed, consider
the derivatives $\tilde D_x$ and $\tilde D_y$ of $\Sigma_\infty$ and
$\rSi_\infty$. We get $$\tilde
D_x(\Sigma_\infty)=\Sigma_\infty^2+u_{xx}-u_x^2,\quad \tilde
D_y(\Sigma_\infty)=0.$$ Also, we have
$$\tilde D_x(\rSi_\infty)=0,\quad \tilde
D_y(\rSi_t)=(\rSi_\infty)^2+u_{yy}-u_y^2.$$ Therefore we use the
nonlocal variable $\Sigma_\infty$ only and treat all relations up to
the symmetry transformation $x\leftrightarrow y$ for the Liouville
equation. By definition, put $\Sigma_t=(x\leftrightarrow
y)\cdot(\rSi_t)$: we  have $$\tilde
D_x(\Sigma_t)=\Sigma_t^2+u_{xx}-u_x^2-\exp(2t)\text{ and }\tilde
D_y(\Sigma_t)=0.$$
%
%
Nonlocal variables enable us to find shadows of nonlocal symmetries
for Eq.~\eqref{eqhyp} and to extend  these shadows up to true
nonlocal symmetries of the Liouville equation.

\begin{state}
\label{StateLiouShadows}
\begin{enumerate}\item
Let $f(t,x,\Sigma_t)$ be a smooth function.
Then the generating function
\begin{equation}\label{ShadowWithT}
\vph=\frac{1}{2}
({\Sigma_t}^2+u_{xx}-u_x^2-\exp(2t))\cdot\frac{\dd f}{\dd
\Sigma_t}+\frac{1}{2}\frac{\dd f}{\dd x} + u_x\cdot f
\end{equation}
is a $\tau_t$-shadow of a nonlocal symmetry of the Liouville
equation.
\item 
Let $f(x,\Sigma_\infty)$ be a smooth function.
Then the second order $\tau$-shadow
$\vph(x,\varSigma_\infty,u,u_x,u_{xx})$
for the Liouville equation is
\begin{equation}\label{SecondOrdNonlocal}
\vph=\frac{1}{2}
   (\Sigma_\infty^2+u_{xx}-u_x^2)\cdot\frac{\dd f}{\dd \Sigma_\infty}
    +\frac{1}{2}\frac{\dd f}{\dd x} + u_x f
     =\tilde\square\bigl(f(x,\Sigma_\infty)\bigr).
\end{equation}
\end{enumerate}
Nonlocal shadows~\eqref{ShadowWithT} and \eqref{SecondOrdNonlocal}
belong to the class~\textup{(}this class was considered in the
paper~\textup{\cite{Shabat})} of
solutions~\eqref{symToda} to the equation
$\tilde\ell_F(\vph)=0$. This class is now provided by the operator
$$\tilde\square=u_x+\tfrac{1}{2}\tilde D_x$$ that contains the
extended total derivative~$\tilde D_x$.
\end{state}


\subsubsection*{Recontruction of nonlocal symmetries}
In order to extend the $\tau_t$-shadows $\tilde\cEv_\vph$ up to
true nonlocal symmetries
$$\tilde\cEv_{\vph,a}=\tilde\cEv_\vph+a\cdot\frac{\dd}{\dd\Sigma_t},$$
where $a\in C^\infty(\tilde\cE)$ and $t\in\BBR\cup\{-\infty\}$, we
solve the equations $$ \tilde D_x(a)=\tilde\cEv_{\vph,a}(\tilde
D_x(\Sigma_t)),\quad \tilde D_y(a)=\tilde\cEv_{\vph,a}(\tilde
D_y(\Sigma_t)) $$ for the function $a$.

%

\begin{state}
\label{ReconstructShadows}
\begin{enumerate}
\item 
Let $f(t)$ be a smooth function and
the functions $\vph$ and $a(t,\Sigma_t,u_x,u_{xx})$ be defined by the
relations
\begin{equation}\label{ReconstructWithT}
\vph= u_x\cdot f(t),\quad
a=(\Sigma_t^2+u_{xx}-u_x^2-\exp(2t))\cdot f(t).
\end{equation}
Then the field $\tilde\cEv_\vph+a\cdot{\dd}/{\dd\Sigma_t}$
is a true nonlocal symmetry of Eq.~\eqref{eqhyp}.
\item 
Let $f(x)$ be a smooth function and
the functions $\vph$ and  $a(\Sigma_\infty$, $u_x$, $u_{xx})$ be given in
\begin{equation}\label{ReconstructWithoutT}
\vph=u_x\,f(x) +\frac{1}{2}\frac{df}{dx},\:
a=(\Sigma_\infty^2+u_{xx}-u_x^2)\,
f(x) + \Sigma_\infty\,\frac{df}{dx}+\frac{1}{2}\frac{d^2f}{dx^2}.
\end{equation}
Then the field $\tilde\cEv_\vph+a\cdot{\dd}/{\dd\Sigma_\infty}$
is a true nonlocal symmetry of Eq.~\eqref{eqhyp}.
\end{enumerate}
\end{state}

The proof of Propositions~\ref{StateLiouShadows}
and~\ref{ReconstructShadows} is very extensive and cannot be completed
without application of the \textsf{Jet} (\cite{Jet97}) software for
analytic transformations. The \textsf{Jet} environment allows
to set the determining equations in a dialogue mode, select the
simplest differential consequences to these equations, and then specify
the expressions for the unknown symmetries.

Nonlocal symmetry \eqref{ReconstructWithT}
is defined up to elements of $\mathrm{CD}(\tilde\cE)\ni g\cdot\tilde
D_x$, $g\in C^\infty(\tilde\cE)$.
Thence, the nonlocal symmetry class~\eqref{ReconstructWithT} is
$[\tilde\cEv_{\vph,a}]=[-f(t)\cdot{\dd}/{\dd x}]$, where
$f(t)\cdot{\dd}/{\dd x}$ is the translation.
Symmetry~\eqref{ReconstructWithoutT}
is the lifting of a classical point symmetry $\vph_0^f$, see
Proposition~\ref{FinSymState} on page~\pageref{FinSymState}.
As usual, a class of nonlocal symmetries $\tilde\cEv_{\vph,a}$ can be
obtained from Eq.~\eqref{ReconstructWithT} and
\eqref{ReconstructWithoutT} by using the discrete
transformation~$x\leftrightarrow y$.

\subsection{On permutability of B\"acklund transformations}
Now we illustrate the permutability property of B\"acklund
(auto)transformations \eqref{BacklundAuto}, \eqref{BtsLiouWave},
and~\eqref{BtsLiouScalPlus}.
\begin{state}
\label{PermutBaecklundState}
\begin{enumerate}
\item
Let $u^j$, $j=\mathrm{i,ii}$, be solutions to Eq.~\textup{(\ref{eqhyp})}
such that $\cB_u(u,u^j;t_j)=0$, $t_j\in\BBR$.
Then there is a unique solution $u'''(x,y)$ to the system
\begin{equation}\label{commutesys}
\left\{\begin{array}{ccc}\cB_u(u',u''';t_2)&\!=\!&0,\\
\cB_u(u'',u''';t_1)&\!=\!&0.
\end{array}\right.
\end{equation}
Namely, the solution $u'''$ satisfies the relation
\begin{equation}\label{u3}
\exp(u''')=\exp(u)\cdot\frac{k_2\exp(u')-k_1\exp(u'')}
{k_2\exp(u'')-k_1\exp(u')},
\end{equation}
where~$k_j\equiv\exp(t_j)$.
\item\label{Case2Backl}
Let $j=\mathrm{i,ii}$ and $t_j\in\BBR$.
Suppose both $v^j$ are solutions to the wave equation $v_{xy}=0$ such
that $\cB_{uv}(u,v^j;t_j)=0$. Also, suppose that $u^j$ are solutions to
the Liouville equation such that $\cB_{uv}(u^j,v;t_j)=0$.  Then there
are unique solutions $u'''$ and $v'''$ to the systems $$
\left\{\begin{array}{ccc}\cB_{uv}(u''',v';t_2)&\!=\!&0\\
\cB_{uv}(u''',v'';t_1)&\!=\!&0
\end{array}\right.
\qquad\text{ and }\qquad
\left\{\begin{array}{ccc}\cB_{uv}(u',v''';t_2)&\!=\!&0\\
\cB_{uv}(u'',v''';t_1)&\!=\!&0\lefteqn{,}
\end{array}\right.
$$
respectively. Denote $k_j\equiv\exp(t_j)$, then the following relations
hold:
\begin{align*}
\exp(u''')&=\exp(u)\cdot\frac{k_2\exp(v')-k_1\exp(v'')}
{k_2\exp(v'')-k_1\exp(v')},\\
\exp(v''')&=\exp(v)\cdot\frac{k_1\exp(u'')-k_2\exp(u')}
{k_2\exp(u'')-k_1\exp(u')}.
\end{align*}
\item\label{Case3Backl}
Let $j=\mathrm{i,ii}$ and $t_j\in\BBR$.
Assume that $\Upsilon^j$ are solutions to the
$\scal^{+}$-equation $\cE_\Upsilon$ such that
$\cB_{u\Upsilon}(u,\Upsilon^j;t_j)=0$, and suppose that
$u^j$ are solutions to the Liouville equation such that
$\cB_{u\Upsilon}(u^j,\Upsilon;t_j)=0$.
Then there are unique solutions $u'''$ and $\Upsilon'''$ to the systems
$$
\left\{\begin{array}{ccc}\cB_{u\Upsilon}(u''',\Upsilon';t_2)&\!=\!&0\\
\cB_{u\Upsilon}(u''',\Upsilon'';t_1)&\!=\!&0
\end{array}\right.
\qquad\text{ and }\qquad
\left\{\begin{array}{ccc}\cB_{u\Upsilon}(u',\Upsilon''';t_2)&\!=\!&0\\
\cB_{u\Upsilon}(u'',\Upsilon''';t_1)&\!=\!&0\lefteqn{,}
\end{array}\right.
$$
respectively. We also have
\begin{align*}
\exp(u''')&=\exp(u)\cdot\frac{k_2\exp(\Upsilon')-k_1\exp(\Upsilon'')}
{k_2\exp(\Upsilon'')-k_1\exp(\Upsilon')},\\
\exp(\Upsilon''')&=\exp(\Upsilon)\cdot\frac{k_1\exp(u'')-k_2\exp(u')}
{k_2\exp(u'')-k_1\exp(u')},
\end{align*}
where $k_j\equiv\exp(t_j)$.
\end{enumerate}
\end{state}

\begin{proof}
We consider the case of B\"acklund
autotransformation~\eqref{BacklundAuto} only; cases~\ref{Case2Backl}
and~\ref{Case3Backl} are treated analogously.
Consider the subsystem in Eq.~(\ref{commutesys}) that consists of
relations~(\ref{BacklundAuto}) which contain the derivatives with respect to\
$x$ only. Then the solution $u'''$ defined in Eq.~(\ref{u3}) follows
from the linear dependence of the left\/-\/hand side\ within Eq.~\eqref{commutesys}
and is unique. Consider another subsystem composed by the relations in
Eq.~\eqref{BacklundAuto} that contain the derivatives with respect to~$y$. Then
there are two solutions to this subsystem; we denote them by $u'''$ and
$\bar u'''$. The solution $u'''$ is defined in Eq.~(\ref{u3})
and $\bar u'''$ is obtained from the relation
$$
\exp(\bar u''')=\exp(-u)\cdot\frac{k_1\exp(u')-k_2\exp(u'')}
{k_2\exp(-u')-k_1\exp(-u'')}.
$$
The solution $\bar u'''$ is irrelevant. Therefore, the function $u'''$ is a
unique solution to the whole system~(\ref{commutesys}).
\end{proof}

\begin{rem}
Proposition~\ref{PermutBaecklundState} means that the diagrams
\begin{gather*}
\begin{CD}
  u @>{t_1}>> u' \\
@V{t_2}VV @VV{t_2}V \\
u'' @>>{t_1}> u'''
\end{CD}
\quad,\qquad
\begin{CD}
  u @>{t_1}>> v'     @>{t_3}>> u'' \\
@V{t_2}VV @VV{t_2}V       @VV{t_2}V \\
v'' @>>{t_1}> u'     @>>{t_3}> v'''
\end{CD}
\quad,\\
\intertext{and the diagram}
\begin{CD}
  u @>{t_1}>> \Upsilon'     @>{t_3}>> u'' \\
@V{t_2}VV @VV{t_2}V       @VV{t_2}V \\
\Upsilon'' @>>{t_1}> u'     @>>{t_3}> \Upsilon'''
\end{CD}
\end{gather*}
are commutative for any $t_1$, $t_2$, $t_3\in\BBR$.
\end{rem}

\section{Zero\/--\/curvature representations}\label{SecBuildZCR}
In this section, we illustrate the relationship between the parametric
families of zero\/--\/curvature representations and B\"acklund
transformations for Eq.~\eqref{eqList}. In Sec.~\ref{SecTodaEq} we
assigned the Liouville equation to the Lie algebra
$\gothg=\mathfrak{sl}_2(\BBC)$, and now we benefit from the use of
different representations of this algebra.

%
There is a natural equivalence (\cite{Brandt}) between
$\gothg$-valued zero\/--\/curvature representations of a PDE $\cE$ and
the special type coverings over the equation $\cE$. Further on, we
study the case 
$\gothg\simeq\mathfrak{sl}_2(\BBC)$. It is
essential that there is a
representation of $\mathfrak{sl}_2(\BBC)$ in vector
fields. We use it in order to construct the required
coverings over the hyperbolic Liouville equation
\begin{equation}\label{eqhyp2}
\cE_{\mathrm{Liou}}=\{u_{xy}=\exp(2u)\}.
\end{equation}
Here `$2$' is the Cartan $1\times1$-matrix of the Lie algebra $A_1$
while $x$ and $y$ are the coordinates in the standard
two\/--\/dimensional extension $(z$, $\bar
z)\hookrightarrow\BBC^2\ni(x$, $y)$.
By $\langle e$, $h$, $f\rangle$ we denote the
canonical basis such that 
\begin{equation}\tag{\ref{Chevalley}${}'$}
[h,e]=2e,\qquad [h,f]=-2f,\qquad [e,f]=h.
\end{equation}
Consider the representation
$\rho\colon\mathfrak{sl}_2(\BBC)\to\ID(\BBC_2[\Xi])$
of the Lie algebra $\gothg$ in the space of
polynomial\/-\/valued derivations:
\begin{subequations}
\begin{align}
\rho(e)&=1\cdot\frac{\dd}{\dd\Xi},&
\rho(h)&=-2\Xi\cdot\frac{\dd}{\dd\Xi},&
\rho(f)&=-\Xi^2\cdot\frac{\dd}{\dd\Xi} \label{sl2asFields}\\
\intertext{such that the Lie bracket is the commutator of vector
fields: $[A$, $B]=A\circ B-B\circ A$.
This representation of the Lie algebra $\mathfrak{sl}_2(\BBC)$
was used in the paper~\cite{ForKac}, where a class of
$N$-ary analogs for the Lie algebraswas constructed.
Also, consider the matrix representation}
\varrho(e)&=\begin{pmatrix}0&1\\0&0\end{pmatrix},&
\varrho(h)&=\begin{pmatrix}1&0\\0&-1\end{pmatrix},&
\varrho(f)&=\begin{pmatrix}0&0\\1&0\end{pmatrix}
\end{align}
\end{subequations}
such that the Lie bracket is the matrix commutator:
$[A,B]=A\cdot B-B\cdot A$.

Given an equation $\cE$, consider the flat connection form
\eqref{fc} in the bundle
$C^\infty(\cE^\infty)\otimes G\to\BBC^2$, where
$G$ is the Lie group of $\gothg$. Suppose
$A$, $B\in C^\infty(\cE^\infty)\otimes\gothg$. The zero-curvature
condition \eqref{ZCR}, which is equivalent to the relation
$$[\bar D_x+A,\bar D_y+B]=0,$$
is satisfied on the differential equation $\cE$.
Finally, we obtain the matrix equation
\begin{equation}\label{ZCRCoordsOnEq}
\bar D_yA-\bar D_xB-[A,B]=0.
\end{equation}
Now, decompose the matrices $A$ and $B$ with respect to\ the basis in the
representation $\varrho\colon\mathfrak{g}\to\{M\in\mathrm{Mat}(2,2)\mid
\tr M=0\}$:
$$
A=a_e\otimes\varrho(e)+a_h\otimes\varrho(h)+a_f\otimes\varrho(f),\quad
B=b_e\otimes\varrho(e)+b_h\otimes\varrho(h)+b_f\otimes\varrho(f),
$$
where $a_\mu$, $b_\nu\in C^\infty(\cE^\infty)$, and construct the
one-dimensional covering $\tau$ over $\cE$ such that
$\Xi$ is the nonlocal variable,
the extended total derivatives $\tilde D_x$ and $\tilde D_y$ are
\begin{align*}
\tilde D_x&=\bar D_x+
a_e\otimes\rho(e)+a_h\otimes\rho(h)+a_f\otimes\rho(f),\\
\tilde D_y&=\bar D_y+
b_e\otimes\rho(e)+b_h\otimes\rho(h)+b_f\otimes\rho(f),
\end{align*}
respectively, and the rules to differentiate the variable $\Xi$ are
\begin{equation}\label{Rules2Deriv}
\begin{aligned}
\tilde D_x(\Xi)&=\Id x\inner
(a_e\otimes\rho(e)+a_h\otimes\rho(h)+a_f\otimes\rho(f)),\\
\tilde D_y(\Xi)&=\Id y\inner
(b_e\otimes\rho(e)+b_h\otimes\rho(h)+b_f\otimes\rho(f)).
\end{aligned}
\end{equation}
We see that the Maurer\/--\/Cartan condition~\eqref{ZCR},
which is satisfied on $\cE$, is
equivalent to the compatibility condition
$[\tilde D_{\mathstrut x},\tilde D_{\mathstrut y}]=0$
that holds in virtue of the equation $\cE^\infty$.

\begin{example}
First, we obtain B\"acklund transformation between the Liouville
and the wave equations.
Consider Eq.~\eqref{TodaatH} on page~\pageref{TodaatH}
and choose the gauge
$$a_e\equiv a_e^1=\exp(\kappa u),\ %
b_f\equiv b_f^1=\exp\bigl((2-\kappa)u\bigr),$$
for an arbitrary constant $\kappa$. Then the covering equation
$\tilde\cE$ is
\begin{equation}\label{BtsLiouWaveGauged}
\left\{\begin{aligned}
v_x&=(\kappa-2)u_x+\exp(\kappa u-v)\\
v_y&=\kappa u_y-\exp((2-\kappa)u+v)
\end{aligned}\right\}.
\end{equation}
Here the variable $v=\log\Xi$ is a transformation of the
nonlocal variable $\Xi$, see Eq.~\eqref{Rules2Deriv}. The
compatibility condition for Eq.~\eqref{BtsLiouWaveGauged} is
$$
v_{xy}=(\kappa-1)\exp(2u).
$$
Finally, suppose $\kappa=1$, then  Eq.~\eqref{BtsLiouWaveGauged} is
the B\"acklund transformation (\cite{DoddBacklund})
\begin{equation}\tag{\ref{BtsLiouWaveGauged}${}_1$}
\begin{aligned}
(v+u)_x&=\exp(u-v),\\ (v-u)_y&=-\exp(u+v)
\end{aligned}
\end{equation}
between the Liouville equation Eq.~\eqref{eqhyp2} and the wave
equation
\begin{equation}\label{eqwave}
v_{xy}=0,
\end{equation}
while the coordinate $\Xi$ is exactly the one that allows integrating
equation~(\ref{BtsLiouWaveGauged}${}_1$) in nonlocal variables, see
Sec.~\ref{SecIntegrating}.
\end{example}

\begin{rem}
Transformation~(\ref{BtsLiouWaveGauged}${}_1$) is the particular
case $t=0$, $k\equiv\exp(t)=1$ within the family of
B\"acklund transformations~\eqref{BtsLiouWave}
between Eq.~\eqref{eqhyp2} and Eq.~\eqref{eqwave}.
We note that the mapping $k\mapsto-k$ is
the replacing of representation \eqref{sl2asFields} with
the representation $\bar
\rho\colon\mathfrak{sl}_2(\BBC)\to\ID(\BBC_2[\Xi])$ such that
\begin{align*}
\bar\rho(e)&=-1, & \bar\rho(h)&=-2\Xi, & \bar\rho(f)&=\Xi^2.
\end{align*}
\end{rem}

\subsubsection*{Zero-curvature representations constructed by using
B\"acklund transformations}
Consider B\"acklund autotransformation~\eqref{BacklundAuto}
%
%
for the Liouville equation~\eqref{eqhyp2}
and B\"acklund transformation~\eqref{BtsLiouScalPlus}
between the Liouville equation and the Liouville $\scal^{+}$-equation
\[  
\cE_\Upsilon=\{\Upsilon_{xy}=\exp(-2\Upsilon)\}
\]   
(we recall that the latter equation is the hyperbolic representation of
the Gauss equation for the conformal metric of constant curvature
$+1$, see Example~\ref{ConfEquivExample} on
page~\pageref{ConfEquivExample}).
Next, consider the coverings over the Liouville equation that provide
these transformations. Then, these coverings \emph{exceed}
ansatz~\eqref{Anzats}.

Namely, the flat connection form
for B\"acklund autotransformation \eqref{BacklundAuto} is
\begin{align*}
\theta   
&=\begin{pmatrix} -\frac{1}{2}u_x & 0\\ -\exp(u-t)&\frac{1}{2}u_x
       \end{pmatrix}\,\Id x+
       \begin{pmatrix} \frac{1}{2}u_y & -\exp(t+u)\\
       -\exp(t-u)&-\frac{1}{2}u_y
       \end{pmatrix}\,\Id y.\\
\intertext{The $\mathfrak{sl}_2$-valued flat connection form $\theta$
for B\"acklund transformation~\eqref{BtsLiouScalPlus} is}
%
\theta   
&=\begin{pmatrix} -\frac{1}{2}u_x & \exp(-t-u)\\
         -\exp(u-t)&\frac{1}{2}u_x
       \end{pmatrix}\,\Id x+
       \begin{pmatrix} \frac{1}{2}u_y & -\exp(t+u)\\
         0&-\frac{1}{2}u_y
       \end{pmatrix}\,\Id y.
\end{align*}
By construction, these $\mathfrak{sl}_2$-valued forms
are zero\/--\/curvature representations for the Liouville equation.


\subsubsection*{B\"acklund transformations constructed by using
zero\/--\/curvature representations}
The problem of constructing multi\/--\/parametric families of B\"acklund
transformations by using known zero\/--\/curvature representations for
Eq.~\eqref{eqhyp2} was discussed in the paper~\cite{SakovichZCR};
later, this problem was studied thoroughly by V.~Golovko
in~\cite{GolovkoYS2003}.

Now the exposition follows~\cite{SakovichZCR} and~\cite{GolovkoYS2003}.
Let us find three $\mathfrak{sl}_2$\/-\/valued classes of
zero\/--\/curvature representations
$\theta\in\bar\Lambda^1(\cE_{\mathrm{Liou}}^\infty)\otimes
\mathfrak{sl}_2(\BBC)$
for the hyperbolic Liouvile equation \eqref{eqhyp2}.
Assume that $A=A(u_x)$, $B=B(u)$, and suppose $[A,B]\not=0$.
Then Eq.~(\ref{ZCRCoords}${}'$)
on page~\pageref{ZCRCoordsOnEq} is reduced to
$$
u_x^{-1}\frac{\partial{A}}{\partial{u_x}} -\exp(-2u)\frac{\partial{B}}
{\partial{u}}- [u_x^{-1} A,\exp(-2u) B]=0.  $$ Applying
${\partial^2}/{\partial{u}\,\partial{u_x}}$ to this identity, we
get the equation $[M$, $N]=0$, where
$$M=\frac{\partial{(A/u_x)}}{\partial{u_x}},\qquad
N=\frac{\partial{(B/\exp(2u))}}{\partial{u}}.$$  There are three
possible cases:
\begin{enumerate}
\item $M=0$, \item $N=0$, and \item $M=r(u_x)\cdot C$,
$N=s(u)\cdot C$, where $C\not=0$~ is a constant
$\mathfrak{sl}_2(\BBR)$-valued matrix аnd $r$, $s\in
C^{\infty}(\BBR)$~are smooth functions.
\end{enumerate}
Hence we get three gauge
non-equivalent clases of zero\/--\/curvature representations:  %
\begin{case}[$M=0$]\label{c0}
There are no nontrivial solutions to Eq.~(\ref{ZCRCoords}${}'$).
\end{case}
\begin{case}[$N=0$]
There are two classes of zero\/--\/curvature representations.
They are
\begin{subequations}\label{Case2}
\begin{align}
A&=\lefteqn{\begin{pmatrix} 2\alpha u_x+2\beta & 2\alpha\\
      u^2_x(1-2\alpha)-4\beta u_x+2\gamma &
      -2\alpha u_x-2\beta
  \end{pmatrix},}&&\notag\\
&& B&=\begin{pmatrix} 0 & 0\\ \exp(2u)& 0
  \end{pmatrix},\label{c2a} \\
A&=\lefteqn{\begin{pmatrix}
    \alpha u^2_x+2\beta & 2\delta\exp(-2\alpha u_x)\\
    2\gamma\exp(2\alpha u_x) & -2\alpha u^2_x-2\beta
  \end{pmatrix},}&&\notag\\
&& B&=\begin{pmatrix} \alpha\exp(2u) & 0\\ 0 & -\alpha\exp(2u)
  \end{pmatrix},\label{c2b}
\end{align}
\end{subequations}
\end{case}
\begin{case}[$M\not=0$, $N\not=0$]
Another solution to Eq.~(\ref{ZCRCoords}${}'$) is
\begin{equation}\label{c3}
A=\begin{pmatrix} u_x & 1\\ 0 & -u_x
  \end{pmatrix},\,\,
B= \begin{pmatrix} 0 & \alpha\exp(-2u) \\ \exp(2u) & 0
\end{pmatrix},
\end{equation}
where $\alpha$, $\beta$, $\gamma$, and $\delta$ are arbitrary constants.
\end{case}

By \cite{Brandt}, any $\mathfrak{sl}_2$-valued
zero\/--\/curvature representation for $\cE$ provides
some special type covering over the equation $\cE$.
Consider the representation
$\rho'\colon\mathfrak{sl}_2(\BBC)$
$\to\ID(\BBC[[v]])$:
$$
\rho'(e)=\exp(-v)\frac{\partial}{\partial v},\qquad
\rho'(f)=-\exp(v)\frac{\partial}{\partial v},\qquad
\rho'(h)=-2\frac{\partial}{\partial v},
$$
of the Lie algebra $\mathfrak{sl}_2(\BBC)$ in the space of differential
operators on $\BBC$, \textit{i.e.}, $v\in\BBC$ is the nonlocal variable
and
$\rho'\colon\mathfrak{sl}_2(\BBC)$
$\to\ID(\BBC[[v]])$. Then
we construct one\/-\/dimensional coverings
over the Liouvile equation $\cE_\Liou$ that
provide B\"ac\-k\-lund transformations
between $\cE_\Liou$
and some differential equations that depend on the initial
zero\/--\/curvature representation.

\begin{state}[\textup{\cite{GolovkoYS2003}}]
Representations~\eqref{c2a}, \eqref{c2b}, and~\eqref{c3} correspond to
B\"acklund transformations between $\cE_\Liou$ and the equations
\begin{align*}
v'_x&=
  \begin{aligned}[t]
{}&\tfrac{1}{4}(2\alpha-1)\exp(v')\left(\frac{v'_{xy}}{v'_y}-v'_x
\right)^2+{}\\
{}&2(\beta \exp(v')-\alpha)\left(\frac{v'_{xy}}{v'_y}-v'_x\right)
+2\alpha \exp(-v')-4\beta-2\gamma \exp(v'),
  \end{aligned}
\\
v''_x&=-\frac{\alpha {v''_{xy}}^2}{2{v''_y}^2}-4\beta+2\delta
 \exp\left(-
\frac{\alpha v''_{xy}}{v''_y}-v''\right)-2\gamma \exp\left(\frac{\alpha
v''_{xy}}{v''_y}+v''\right),
\\
{v'''_{xy}}^2&=\exp(-2v''')({v'''_y}^2+4\alpha).
\end{align*}
If $\alpha=0$ in representations~\eqref{c3}, then we get
B\"acklund transformation
between $\cE_\Liou$ and the wave equation Eq.~\eqref{eqwave}.
\end{state}

Finally, we analyse the removability of the parameters $\alpha$,
$\beta$, $\gamma$, and $\delta$ in
zero\/--\/curvature representations~\eqref{Case2}--\eqref{c3}
with respect to\ the gauge transformations
\[
A\mapsto SAS^{-1}-(D_xS)S^{-1},\qquad
B\mapsto SBS^{-1}-(D_yS)S^{-1}.
\]
The result is described in

\begin{rem}[\textup{\cite{GolovkoYS2003}}]
The parameter $\beta$ in Eq.~\eqref{c2a} is removable by the gauge
transformation
$$S=a\cdot\left(\begin{matrix} 1 & 0\\ {\beta}/{\alpha} & 1
  \end{matrix}\right)$$
that depends on an arbitrary constant $a\in\BBC$\textup{;}
under this transformation
$\gamma\mapsto\gamma+{\beta^2}/{\alpha}$.  All other parameters
$\alpha$, $\beta$, $\gamma$, and $\delta$ in zero\/--\/curvature
representations~\eqref{Case2}--\eqref{c3} are non-removable.  \end{rem}

\newpage
\appendix
\section{Geometric methods of solving boundary\/--\/value problems}%
\label{SecAppendix}
In this appendix, we consider several methods for solving
boundary\/--\/value
problems (mainly, the Dirichlet problem) for nonlinear equations of
mathematical physics. Then we compare and analyse the results of a
computer experiment in applying the described algorithms, see
also~\cite{Dirichlet}. Again, the methods which we study are based on the
geometry of the jet spaces (\cite{ClassSym, Opava, Pommaret}) and treating
differential equations $\cE$ (and their prolongations $\cE^\infty$ as
well) as submanifolds  $\cE\subset J^k(\pi)$  of the jet space
of order $k$ for a certain fibre bundle $\pi$ (respectively, we have
$\cE^\infty\subset J^\infty(\pi)$). By using this approach, we pay
attention to the analysis of the following objects:
they are
\begin{itemize}
\item
differential equations
and sections $s$ such that the jet $j_k(s)\subset \cE$ defines a solution
of $\cE$,
\item
boundary\/--\/value
problems $\cP=(\cE,\cD,f={s|}_{\partial \cD})$ in a domain $\cD$,
\item
and their deformations~$\dot \cP$.
\end{itemize}
We emphasize that the geometrically motivated technology of solving the
boundary\/--\/value
problems is discussed within this appendix. We do not stop on
particular properties of some solutions in practical situations.
Basic definitions and concepts were formulated in the Introduction, see
also~\cite{ClassSym, Opava, Pao, Pommaret}.
We use the elliptic Liouville equation
$$\cE_{\text{Liou}}=\{u_{z\bar z}=\exp(2u)\}$$
as a basic example.

The appendix is organized as follows.
In Sec.~\ref{S1}, the method of monotonous iterations for the solutions
$u^t$ of the boundary\/--\/value problem $\cP$ is described. Here
$t\in\BBN$. This method is based of the theory of differential
inequalities~(\cite{Pao}). In Sec.~\ref{S2}, we construct a method for
solving the boundary\/--\/value problems that involves the
evolution representation of the equation $\cE$ under study.
We shall see that this interpretation is admissible in a sufficiently
general situation (\cite{ClassSym, Pommaret}).
Then, in Sec.~\ref{S3} we consider various methods based on the
deformations $\dot f$ of the boundary conditions $f$ and simultaneous
invariance of the equation $\cE$:
\[
\dot \cP=(\cE,\cD,\dot f).
\]
In Sec.~\ref{S4} we describe the relaxation method based on the
substitution
\[
\cP\mapsto \cP'=(\cE',\cD\times\BBR_{+}, f\otimes\id_t)
\]
of the boundary\/--\/value problem such that solutions $u$ of the problem
$\cP$ are stable stationary solutions for the new problem $\cP'$ with
respect to the evolution equation~$\cE'$.
The deformations
\[\dot \cP=(\dot \cE,\cD,f),
\]
which are considered in Sec.~\ref{S5}, are in some sence antipodal to ones
described in Sec.~\ref{S3}. Now, the boundary condition $f$ is invariant
and the equation $\cE(t)$ moves. The ``planting'' of a nonlinearity could
be an example. In the final section of this appendix, we discuss the
results of a computer experiment in practical application of all these
methods.

\begin{rem}
In what follows, we assume that there is a unique classical solution to
the boundary\/--\/value problem at hand (see~\cite{Gilbarg}).
Also, we recall that the transformation laws of solutions with respect to
transformations of the independent variables are known for some equations,
\textit{e.g.}, the conformally invariant Toda equations, see
Proposition~\ref{FinSymState} on page~\pageref{FinSymState}). Therefore,
the domain $\cD$, where the problem $\cP$ is solved, can be chosen
relatively regular (we set $\cD\sim B_0^1$ for Eq.~\eqref{eqToda}).
\end{rem}

\begin{rem}
First, we point out a practically useful way to construct a set of
solutions (that can be large enough) of a boundary\/--\/value problem
$\cP=(\cE$, $\cD$, $f)$.
Assume that the equation $\cE$ is invariant with respect to
a vector field $X$. Denote its generating section by $\varphi_X$ and
suppose that $A_t$ is the flow of the field $X$ such that
\[
A_t(\partial \cD)=\partial\cD\text{ and  }A_t(j_k(f))=j_k(f),
\]
that is, the boundary condition is unvariant with respect to the symmetry
$X$ of the equation $\cE$ at hand. Then problem $\cP$ can be reduced to
finding solutions of the boundary\/--\/value problem
\[\cP'=(\cE\cap\{\varphi_X=0\},\cD,f)\]
that are invariant with respect to $X$ at any point in~$\cD$.
If the problem $\cP'$ has a solution, then, in general, this solution may
be not unique.
\end{rem}

\begin{example}
Consider the Dirichlet boundary\/--\/value problem
\begin{equation}\label{DirichletCirc}
\left\{
u_{xx}+u_{yy}=\exp(2u),\qquad
{u|}_{r=1}=f,\quad f\in C^{0,\lambda}
\right\}
\end{equation}
in the unit disc $B_0^1$ and assume that the boundary condition is
homogeneous and trivial, $f\equiv0$.
The solutions for the Liouville equation $\cE_{\text{Liou}}$ that are
invariant with respect to its point symmetries are
\begin{subequations}\label{InvLiou}
\begin{align}
u&=\frac{1}{2}\log\frac{v_x^2+v_y^2}{{\text{sinh}}^2v},\label{InvLiouA}\\
u&=\frac{1}{2}\log\frac{v_x^2+v_y^2}{v^2},\label{InvLiouB}\\
u&=\frac{1}{2}\log\frac{v_x^2+v_y^2}{\sin^2 v},\label{InvLiouC}
\end{align}
\end{subequations}
where $v$ is a harmonic function,
$$\Delta v=0.$$
Consider the radial symmetry
$$X=-y\,\frac{\partial}{\partial x}+x\,\frac{\partial}{\partial y}.$$
Then the solutions $u(r)=u(\sqrt{x^2+y^2})$  of
problem~\eqref{DirichletCirc} are the following:
\begin{subequations}\label{SolvCirc}
\begin{align}
u_1(r)&=\log\frac{2(1+\sqrt{2})}{(1+\sqrt{2})^2-r^2},\label{SolvCircA}\\
u_2(r)&=-\log r-\log(1-\log r),\label{SolvCircB}\\
u_3(r)&=\frac{1}{2}\log\frac{\alpha^2r^{-2}}{\sin^2(\alpha\beta-\alpha\log
r)},\qquad
\beta=\frac{\pm 1}{\alpha}\arcsin\alpha,\ \alpha\neq 0.\label{SolvCircC}
\end{align}
\end{subequations}
A unique classical solution $u_1$ has no singularity at the point $0$.
The solution $u_2$ has an integrable singularity at $r=0$, and its
gradient on the border vanishes.
The solution $u_3$ has the cardinal set of logarithmic singularities along
the radius $r$. These singularities accumulate as~$r\to0$.
\end{example}

\begin{rem}
Suppose that the equation $\cE$ at hand is Euler.
Then, obviously, one can search solutions of the Dirichlet
boundary\/--\/value problem by using the direct minimization of the action
functional. We emphasize that the projective methods can be used in
addition to the lattice discretization methods.
\end{rem}

\subsection{The monotonous iterations method}\label{S1}
The monotonous iterations method is a useful instrument in the theory of
differential inequalities (\cite{Pao}). This method allows to contruct
solutions of the Dirichlet boundary\/--\/value problems
\begin{equation}\label{Problem}
\left\{\Delta u=h(u,x),\quad {u|}_{\partial \cD}=f,\qquad
\partial \cD\in C^{1+\varepsilon},\quad f,h\in C^{0,\lambda}
\right\}
\end{equation}
and then check the local uniqueness of these solutions.

\begin{define}
A function  $\alpha\in C(\overline{\cD})\cap C^2(\cD)$
(resp, $\beta$) is a \emph{lower} (\emph{upper}) solution of
problem~\eqref{Problem} is the following two conditions hold:
\begin{enumerate}
\item $\Delta\alpha-h(\alpha$, $x)\geq0$ in $\cD$ and
\item $f(x)\geq\alpha(x)$ on $\partial \cD$
\end{enumerate}
(respectively, ${\leq}$).
Suppose further $\alpha(x)\leq\beta(x)$ for each
$x\in\overline{\cD}$.
Then we set $\alpha\preceq\beta$.
\end{define}

\begin{state}[\textup{\cite{Pao}}]
Assume that a lower and an upper solutions
of problem~\eqref{Problem} are ordered, $\alpha\preceq\beta$.
Suppose there is a nonnegative constant $C\in\BBR$ such that
$$h(u_1,x)-h(u_2,x)\leq C\cdot(u_1-u_2)$$
for any $x\in\overline{\cD}$, $u_1$, and
$u_2$ provided that $\alpha\leq u_2\leq u_1\leq\beta$.
Consider two sequences
$\underline{U}=\{\underline{u}^k\}$ and $\overline{U}=\{\overline{u}^k\}$
of solutions of the problem
\begin{equation}\label{Iterate}
\Delta u^k-C\,u^k=h(u^{k-1},x)-C\,u^{k-1},\quad {u^k|}_{\partial
\cD}=f, \qquad x\in \cD,\;k\in\BBN.
\end{equation}
Here the inital values are
$\underline{u}^0=\alpha$ and $\overline{u}^0=\beta$, respectively.
Then the monotonously nondecreasing (nonincreasing)
sequence $\underline{U}$ ($\overline{U}$) converges to a solution
$\underline{u}$ (resp., $\overline{u}$) for problem~\eqref{Problem}.
Moreover, we have
$\underline{u}\leq\overline{u}$ in $\overline{\cD}$
and the bound $\underline{u}\leq u_*\leq\overline{u}$ holds for
any solution $u_*\in[\alpha$, $\beta]$.

Suppose further that $h$ is monotonous with respect to $u$\textup{:}
$h(u_1,x)-h(u_2,x)\geq0$ provided that
$\alpha\preceq u_2\preceq u_1\preceq\beta$.
Then the solution for problem~\eqref{Problem} is unique on
$[\alpha,\beta]$: $\underline{u}=\overline{u}$.
\end{state}

The notions of a lower and upper solution, which were introduced
in Sec.~\ref{S1}, and the method of proving the local uniqueness for
solutions of problem~\eqref{Problem} will be used in Sec.~\ref{S4}.
The relaxation method will be described there. In that case, the
evolution of the superscript $k$ is continuous unlike in
problems~\eqref{Iterate}.

\subsection{Evolutionary representation
of differential equations}\label{S2}
The following remarkable result was in fact obtained in the papers
concerning the formal theory of differential equations
(see~\cite{ClassSym, Pommaret} and references therein):
any equation that satisfies the assumtions of the `$2$-line theorem'
for the Vinogradov's ${\cC}$-spectral sequence (\cite{ClassSym, Opava})
admits a representation in the form of an evolution equation. An
illustration is given in
Example~\ref{EvolutAreNormal} on page~\pageref{EvolutAreNormal}.
In practice, this means that any equation that does not have gauge
symmetries (unlike the Maxwell, the Yang\/--\/Mills, and the Einstein
equations and similar systems) admits a set of coordinate transformations
and differential substitutions that map it to an evolution equation
(possibly, the new equation is imposed on a larger set of dependent
variables). The class of equations subject to the assumptions of the
`$2$-line theorem' is really wide. Of course,
fixed\/-\/precision integrating of evolution equations is simpler that
integrating of arbitrarily chosen equations of mathematical physics.
Therefore, solving the initial boundary\/--\/value problem is divided to
two stages. First, we find an evolution representation of the equation
$\cE$. Then we reconstruct the boundary conditions for the auxiliary
dependent variables if there appeares a necessity to introduce them.

\begin{example}
Consider the boundary\/--\/value problem
\begin{equation}\label{LiouShoot}
\begin{gathered}
u_{xx}\pm u_{yy}=\exp(2u),\qquad \cD=\{(x,y),|x|<1,|y|<1\},\\
u(1,y)=f_1(y),\; u(x,1)=f_2(x),\; u(-1,y)=f_3(y),\;
u(x,-1)=f_4(x)
\end{gathered}
\end{equation}
for the elliptic (resp., hyperbolic) Liouville equation
$\cE_{\text{Liou}}$.
Introduce the additional dependent variable $v=u_y$.
Then the equation $\cE_{\text{Liou}}$ has the form
$$
\frac{\partial}{\partial y}
\left(\begin{matrix} u\\ v \end{matrix}\right)
= \left(\begin{matrix} v \\ \mp u_{xx}\pm\exp(2u)\end{matrix}\right)
$$
and the conditions on the bound $\partial \cD$ are split to
the initial and the boundary conditions:
\begin{align*}
u(x,-1)&=f_4(x), &  u(1,y)&=f_1(y),  & u(-1,y)&=f_3(y)\\
v(x,-1)&=v[f_2], & v(1,y)&=df_1(y)/dy, & v(-1,y)&=df_3(y)/dy.
\end{align*}
Thence, solving boundary\/--\/value problem~\eqref{LiouShoot}
is reduced to reconstruction of the initial section $v(x,-1)$
by the terminal section $u(x,1)=f_2(x)$.
In practice, this can be done by using the conjugated gradient method.
\end{example}

\subsection{Evolution of boundary conditions}\label{S3}
Recall that any symmetry $\varphi\in\text{sym}\,\cE^\infty$ of an equation
$\cE=\{\vec F=0\}$ is an element of the kernel $\ker\bar\ell_F$
of the linearization
$$
\ell_F=\left\|\sum_\sigma\frac{\partial F_i}{\partial u^j_\sigma}
D_\sigma\cdot \bun_{ij}\right\|
$$
restricted onto $\cE$. The evolutionary vector field
$$\cEv_\varphi=\sum_{j,\sigma}\bar
D_\sigma(\varphi^j)\,\frac{\partial}{\partial u^j_\sigma}
$$
commutes with the total derivatives $D_i$ and is tangent to the infinite
prolongation
$$\cE^\infty=\{D_\sigma(F)=0,\ |\sigma|\geq0\}\subset
J^\infty(\pi).$$
The latter is the projective limit of smooth epimorphisms by definition.
The manifold $\cE^\infty$ that is defined by the
closed algebra of smooth functions is in fact infinite\/--\/dimensional,
therefore, formally, the field $\cEv_\varphi$ has no flow. Indeed, there
is no Cauchy theorem for the initial\/--\/value problem
\begin{equation}\label{Cauchy}
\dot u_\sigma=\bar D_\sigma(\varphi),|\sigma|\geq0,\qquad
u(t=0)=u
\end{equation}
composed by the cardinal set of equations.
The integral trajectories for problem~\eqref{Cauchy} exist provided that
$\varphi(x,u,D_i(u^j))$ is a contact symmetry. In particular, the point
summetries that are linear with respect to the derivatives suit well.
From Eq.~\eqref{Cauchy} it follows that the functions $\varphi$ describe
the correlated evolution of the dependent variables and their derivatives.
This evolution preserves solutions of the equation $\cE$ if
$\varphi\in\ker\bar\ell_F$.

\subsubsection{}    
Suppose a solution $u_0$ of some boundary\/--\/value problem
$\cP_0=(\cE$, $\cD$, $f_0)$ for an equation $\cE$ is known.
Then, deform the problem $\cP=(\cE,\cD,f)$ such that
$f(t=0)=f_0$ and $f(t=1)=f$.
Here by $\dot f(t)$ we denote the deformation velocity
of the boundary condition on $\partial \cD$ and by
$\varphi(t)$ the corresponding deformation of the solution.
One easily checks that $\varphi\in\ker\bar\ell_F$.
Suppose further that the problem
\begin{equation}\label{Lin4vph}
\varphi\in\ker\bar\ell_F,\qquad {\varphi|}_{\partial \cD}=\dot f
\end{equation}
is soluble at each $t\in[0,1)$. Then,
$$u=u_0+\int_0^1\varphi(t)\,dt$$
is the required solution for the problem~$\cP$.

The condition $\varphi\in\ker\bar\ell_F$ is the decomposition of the
deformation $\dot u=\varphi$ to a power series in $t$.
The analysis of the deformation equations that are coefficiets of the
higher powers of $t$ is interesting by itself, see~\cite{Kumei}.
For example, insolubility of the equation at $t^2$ implies
impossibility to solve
the problem $\cP$ by using this deformation method.

\begin{rem}
The Liouville equation $\cE_{\text{Liou}}$ has the following peculiar
feature. Its general solution, see Eq.~\eqref{gensolhyp}, that
depends on an arbitrary holomorphic function $v(z)$ is always invariant
with respect to a point symmetry $X$ whose coefficients are related with
$v(z)$ by the Abel transformation. The curve $X(t)$ in the space
$\text{sym}\,\cE_{\text{Liou}}$ is assigned to the deformation $\dot f$ of
the boundary condition. This curve is such that the solution $u(t')$ is
invariant with respect to $X(t')$ for any $t'\in[0,1]$.
Therefore, the solution of boundary\/--\/value
problem~\eqref{DirichletCirc} can be reduced to analysis of the equations
that define the curve~$X(t)$.
\end{rem}

\subsubsection{}   
Consider a particular covering strucuture $\tilde \cE^\infty\to \cE^\infty$
over the equation $\cE$ at hand,
that is, suppose that the substitution $u=u[v]$ maps
solutions $v\in\text{Sol}\,\tilde \cE$ to solutions
$u\in \text{Sol}\,\cE$.
Several examples are well known: the Cole\/--\/Hopf substitution
\begin{equation}\label{ColeHopf}
u=\frac{v_x}{v},
\end{equation}
the Miura transformation $u=v_x-v^2$, and the formula
\begin{equation}\label{gensolell}
u=\frac{1}{2}\log[4\partial v\cdot\bar\partial\bar v/(1-v\bar v)^2]
\end{equation}
by Liouville (\cite{Liouville})
that relate the Burgers and the heat equations,
the Korteweg\/--\/de Vries equation and the modified
Korteweg\/--\/de Vries equation, and the Liouville and the
Cauchy\/--\/Riemann equations, respectively.
We recall that all three expressions~\eqref{InvLiou} are transformed to
Eq.~\eqref{gensolell} by an appropriate change of variables.
Also, suppose the condition $f$ is fixed. Then there can be several
substitutions $u[v]$ that solve the boundary\/--\/value problem~$\cP$.
Therefore, the reconstruction problem for the function $v$ in the whole
domain $\cD$ is incorrect by Hadamars.
Still, assume that a class of the substitutions $v\mapsto u$ is fixed.
Then one easily obtains the coordinate expressions for the equations
$$\psi\equiv\dot v\in\ker\tilde\ell_{\tilde \cE},$$
which are analogous to Eq.~\eqref{Lin4vph}, plus the condition
$$\tilde \cEv_\psi(u[v])=\dot f$$
defined by the boundary functions $f$ and $f_0$,
and the quadrature
$$v(1)=v(0)+\int_0^1\psi(t)\,dt.$$
Thence we construct the solution $u$ of the initial boundary\/--\/value
problem $\cP$ in the whole domain $\cD$ by using the exact formula
$u=u[v]$ and the solution $v(1)$ of the covering equation~$\tilde \cE$.
Still, we see that the condition $\tilde\cEv_\psi(u[v])=\dot f$
may not even be defined by an operator with directional derivative
if the function $u$ depends on the gradient $\text{grad}\,\,v$
explicitly.

Nawadays, there exist regular algorithmic methods (\cite{ClassSym}, see
also Chapter~4) that allow obtaining and classification of the coverings
over equations of mathematical physics. These algoriths are already
available as environments (\cite{Jet97}) for the symbolic transformations
software. We also recall that the concept of coverings over differential
equations is closely related with the theory of B\"acklund transformations
and zero\/--\/curvature representations (see the preceding section) for
PDE. The latter structures also provide some classes of the substitutions
$u[v]$ for a given equation~$\cE$.

\subsubsection{}   
Now we consider in more details
solving the Dirichlet boundary\/--\/value problem,
see Eq.~\eqref{DirichletCirc},  for the Liouville equation
$\cE_{\text{Liou}}$ within the class of substitutions~\eqref{InvLiouB} by
using the methods described above.
Recall that the harmonic functions $v(t)$ admit the representations
$P[g]$ via the Poisson kernel $P$ (see \cite{Gilbarg}) by their
boundary value $g\equiv {v|}_{\partial \cD}$.
Therefore, the initial boundary\/--\/value problem $\cP$ which is solved by
using the homotopy $\dot f$ of the solution $u$ such that $f(t)$ are its
values on $\partial \cD$ is reduced to the integral equation
$$
\frac{d}{dx}P[g]\cdot \frac{d}{dx}P[\dot g]
+\frac{d}{dy}P[g]\cdot \frac{d}{dy}P[\dot g]=
(g\dot g+g^2\cdot(f-f_0))\cdot\exp(2f)
$$
with respect to the deformation $\dot g={\psi|}_{\partial \cD}$
of the bounder value for the harmonic function $v$.
Suppose the boundary value for $v$ is obtained, then an approximation of $v$
on a lattice in $\cD$ can be obtained by using multiple averaging of the
values of $v$ in the neighbouring interpolation points
(see also Sec.~\ref{S4} below).

In Remark~\ref{Rem6.2} on page~\pageref{Rem6.2},
the results of a computer experiment in application of this method for
solving boundary\/--\/value problem~\eqref{DirichletCirc} are discussed.

\subsubsection{}\label{Sec3.4}
Finally, we note that solutions of some other equation $\cE'$ (in general,
distinct from the equation $\cE$ at hand) can be used for construction of
the initial sections $u_0$ that correspond to the boundary values
$f_0$ on $\partial \cD$. Here we assume that the solutions $u(t)$ satisfy
a degenerate equation $\cE'$ as~$t\to0$.

\begin{example}
Suppose $f\rightrightarrows-\infty$ such that $\max f-\min
f\leq \text{const}<\infty$.
Then a solution $u$ of boundary\/--\/value problem~\eqref{DirichletCirc}
in the disc $B_0^1$ approximated with respect to the norm
$\|\cdot\|_{C^0(\bar B_0^1)}$ by the harmonic function $P[f]$
as closely as desired.
Let
$$
\Delta u(t)=\exp(2u(t)),\qquad f(t)=f+\log t,\quad 0<t\leq1,
$$
then in the same notation we have
$$
\Delta\varphi(t)-2\exp(2u(t))\varphi(t)=0,\quad
\dot f(t)=t^{-1},\quad
u(t)=u_{t_{\min{}}}+\int_{t_{\min{}}}^t\!\!\!\varphi(\tau)d\tau,
$$
where $t_{\min{}}\to+0$ and $u_{t_{\min{}}}=P[f+\log t_{\min{}}]$
is the initial approximation.
Now suppose $t$ is small. We expand the solution $u(t)$ by using the
Hadamars lemma and obtain $u(t)=\log t+P[f]+t^2\cdot U$. If $t\to+0$, then
the function $U$ satisfies the homogeneous problem
$$
\Delta U=\exp(2P[f]),\qquad {U|}_{\partial B_0^1}=0
$$
for the Poisson equation. Its solution is
$$U(p)=\int_{B_0^1}\exp(2P[f])(q)\cdot G(p,q)\,dq,$$
where $G$ is the Green function (\cite{Gilbarg})
for the Laplace operator in the unit disc.
\end{example}

In fact, the method described in Sec.~\ref{Sec3.4} is based on the
simultaneous deformation $\dot \cP$ of the boundary condition $f$ and the
equation $\cE$ itself. Indeed, we have $\cE(0)=\cE'$ and
$\cE(t)=\cE$ for $t\in(0,1]$ by construction. In Sec.~\ref{S4}, we consider
an opposite situation: an auxiliary equation $\cE'$ is used at any
value of the parameter $t$. Here we assume that solutions of $\cE'$ tend to
solutions of the problem $\cP$ as $t$ increases.

\subsection{The relaxation method}\label{S4}
Complement the mixed boundary\/--\/value problem, which is a generalization
of problem \eqref{Problem}, by the relaxation term $\partial u/\partial t$.
Assume that $\alpha$ and $\beta$ are the lower and the upper
solutions of the initial stationary problem, respectively.
Then, extend the boundary value $f$ onto
$\partial \cD\times\BBR_{+}$ and fix an initial approximation $u_0$ such
that $\alpha\preceq u_0\preceq\beta$: hence we obtain
\begin{equation}\label{Relax}
\frac{\partial u}{\partial t}-\Delta u=-h(u,x),\quad
a\,\frac{\partial u}{\partial\vec{n}}+b\,u=f,\quad
u(x,0)=u_0(x).
\end{equation}

\begin{state}[\textup{\cite{Pao}}]
Let the above assumptions hold.
Assume further that $h\in C^1_u[\alpha,\beta]$.
Then the following two statements are equivalent\textup{:}
\begin{enumerate}
\item the stationary solution $u_s$ of problem~\eqref{Relax}
is unique in $[\alpha,\beta]$\textup{;}
\item the solution $u_s\in[\alpha,\beta]$ is asymptotically stable
such that the stability domain is $[\alpha,\beta]$.
\end{enumerate}
\end{state}

Here we discussed a method based on the replacement of the elliptic equation
$\cE$ in problem~\eqref{Problem} by the evolution equation~$\cE'$.
This method is an alternative to the evolutionary representation method for
the initial equation $\cE$ (see Sec.~\ref{S2}).

\subsection{Deformation of the equation}\label{S5}
Now we consider the deformations $\dot \cP$ of the boundary\/--\/value
problem $\cP(t)$ that are induced by the evolution $\dot \cE$ of the
equation $\cE(t)=\{F(t)=0\}$ at hand.
Here we assume that the boundary value $f$ remains invariant.
Namely. consider the homotopy
$$\cE(t)=(1-r(t))\cdot \cE_0+r(t)\cdot \cE,$$
where $r(0)=0$ and $r(1)=1$.
Then the velocity $\varphi$ of deformation of the solution $u(t)$ is subject
to the boundary\/--\/value problem
$$
\bar\ell_{F(t)}(\varphi)+r'(t)\cdot(F-F_0)=0,\qquad
{\varphi|}_{\partial \cD}=0.
$$
Therefore, the solution of the problem $\cP$
is defined by the quadrature $u=u_0+\int_0^1\varphi(t)\,dt$.
We emphasize that this method provides reliable solution
approximations in computations.

\subsection{Discussion and practical hints}\label{S6}
\subsubsection{}  
Consider the deformation $(1-r(t))\cdot \cP_0+r(t)\cdot \cP$
of the boundary values (see Sec.~\ref{S3})
or of the equation itself (see Sec.~\ref{S5}).
Then the smooth step\/-\/like homotopy function
$$r(t)=\exp(-\text{ctg}^2(\pi t/2)),\qquad 0<t<1,$$
is more preferable than the linear motion $\dot \cP=\cP-\cP_0$.
Then, the final stage of solving the boundary\/--\/value problem
$\cP$ is in fact equivalent to the Newton method that starts with the
approximation obtained in the initial computations.

\subsubsection{}\label{Rem6.2}
The comparative analysis of practical computations by using the algorithms
described in this appendix was carried out in 2002 in the dimploma papers
by A.~V.~Punina (Chair of Higher Mathematics, ISPU;
Master thesis ``Comparative analysis of the methods for solving the
elliptic Liouville equation by using the homotopies'')
and N.~P.~Cheluhoeva (Chair of Higher Mathematics, ISPU;
Master thesis ``Integral equations on the border method studied for
solving the elliptic Liouville equation'').
The following assertion is in order.

\begin{itemize}
\item
The method of deforming the equation $\cE$ (see Sec.\ref{S5})
is twice more precise with respect to the absolute deviation
from known exact solutions than the method based on the deformation of
the boundary function, see Sec.~\ref{S3}, although the latter method
converges $3$-$5$ times faster.
\item
The relaxation method is the simplest among the algorithms such that the
evolutions of $t$ is continuous. The interative method is also
preferable for solving the Laplace equation that appeared in
Sec.~\ref{S3} since the use of the Poisson kernel requires greater time
and the Gauss method cannot be speeded up owing to the strongly sparse
matrix of the discrete Laplace operator~$\Delta$.
In Sec.~\ref{S3}, we described the reduction of the Dirichlet
border\/--\/value problem to the one\/--\/dimensional problem and the
integral equation. This method is very sensitive with respect to
smoothness of the boundary conditions $f(t)$ since the Poisson kernel
has a singularity on $\partial B_0^1$ and the harmonic functions are
less smooth on $\partial B_0^1$ than in $\cD$ (where they are infinitely
smooth).
We also see that a solution of the Liouville equation, when moving along
$t$, can be ``attracted'' to a nonclassical (of type~\eqref{SolvCircB}
or~\eqref{SolvCircC}) solution which satisfies the same boundary
conditions. This can happen if the time step is suffucuently large.
\end{itemize}

Summarizing, we conclude that all these methods demonstrated comparable
precisions. Therefore, the choice of an algorithm should be based on the
actual properties of the problem at hand.

\subsubsection{} 
Conservation laws for the analysed equation~$\cE$ serve an important
instrument for the precision control in computations.
Nowadays, there exist regular methods (\cite{ClassSym, Opava, Vin84}) of
reconstruction of the exhaustive set of conservation laws for the
equations subject to the `$2$-line theorem'. These methods are realized
in the form of the computer software~(\cite{Jet97}) for systems of
analytic transformations.

\subsubsection{} 
Application of the methods of soving boundary\/--\/value problems
described in this appendix does neither neglect nor underesteeme
the standard check of continuity of the resulting solutions, their
H\"older or the Sobolev space properties an so on.
Meanwhile, we hope that these practical ideas will compliment the
everyday set of instruments for numerical analysis of nonlinear
equations of mathematical physics.

\newpage
\section*{\textbf{Final remarks}}
\subsubsection*{1.}
Recently, Demskoi and Startsev (\cite{StarHere}) analysed the
correlation between the integrals $\mathbf{\Omega}$
and symmetries $\vph$ for the Liouvillean
hyperbolic systems. They assigned the operators $\bar\square$
that factor symmetries of these systems to the product
\[
\vph=\bar\square(\phi(x,\mathbf{\Omega}))
\]
to the linearizations $\ell_{\Omega^i}$ of the integrals $\Omega$.
These results are reported in the note~\cite{StarHere}, which is found
in this issue. In fact, they generalize the statements of
Lemma~\ref{2=3} on page~\pageref{2=3} and Lemma~\ref{AdjointIsLin}
on page~\pageref{AdjointIsLin} to the case of arbitrary integrals
$\Omega^i$, $i\geq1$ (we recall that $\Omega^1\equiv T$).

\subsubsection*{2.}
The commutative Hamiltonian hierarchy
$\gA$ of the local Noether symmetries
$\vph_k\in\sym\cL_\Toda$, where $k\geq0$, was constructed in Part~I.
We identified this hierarchy with the sequence of higher
$r$-component analogs for the potential modified Korteweg\/--\/de
Vries equation and related $\gA$ with the bi\/--\/Hamiltonian
hierarchy $\gB$ for the scalar potential Korteweg\/--\/de Vries
equation, see Eq.~\eqref{pmKdV} on page~\pageref{pmKdV}. The
degeneracy of the matrix coefficients of the higher-order derivtives
is an immanent feature of these evolution equations.

A comment is in order. In the
fundamental paper~\cite{DSViniti84}, the integrable
bi\/--\/Hamiltonian hierarchy $\gC$ of the Drinfel'd\/--\/Sokolov
equations was assigned to the hyperbolic Toda system associated with a
semisimple Lie algebra $\gothg$ or a Kac\/--\/Moody algebra
$\hat{\gothg}$.  The hierarchy $\gC$ was related with the
sequence $\gD$ of the multi\/--\/component
Korteweg\/de Vries equation's analogs.
Unlike in the case of the hierarchy $\gA$, the symbols of the
Drinfel'd\/--\/Sokolov equations are always nondegenerate.  Therefore,
it is quite reasonable to check whether there is an interrelation
between these two systems $\gA$ and $\gC$, as well as $\gB$ and $\gD$.
Also, the necessary and sufficient conditions for $\gA$ to be
bi\/--\/Hamiltonian remain unclear.

Suppose the hierarchy $\gA$ is assigned to the Toda
equation~\eqref{eqToda} which is associated with a nondegenerate
symmetrizable matrix $K$.  We conjecture that the hierarchy $\gA$ is
bi\/-\/Hamiltonian if and only if $K$ is the Cartan matrix of a
semisimple Lie algebra. We also conjecture that the pair $(A_1$,
$A_2)$ of the operators is Hamiltonian in the sence of
Definition~\ref{DefHamOperator} on page~\pageref{DefHamOperator}
under the same assumptions.

\subsubsection*{3.}
Chapter~3 contains the exposition of the geometric structures for
the scalar heavenly equation, which is a continuous dispersionless
limit of the $r$-component Toda systems as $r$ tends to infinity.  We
observed that local structures for the limit equation are relatively
few, and therefore a nonlocal setting must be introduced.  Meanwhile,
local Noether's symmetries, conservation laws, and recursion operators
for the Toda equations themselves were considered in Part~I.
Therefore it is logical to describe the permutability properties of the
diagram
\[
\begin{CD}
\cE_\Toda @>>> \text{local structures for $\cE_\Toda$}\\
@V{\substack{r\to\infty,\\
   \varepsilon\to+0,\\
       u_{zzz}=0}%
       }VV @VV{\text{\textbf{?}}}V\\
\cE_\heav @>>> \text{nonlocal structures for $\cE_\heav$},
\end{CD}
\]
that links together the local geometry of $\cE_\Toda$ and (yet
undiscovered) nonlocal geometry of $\cE_\heav$.

\subsubsection*{4.}
In Chapter~4, we applied the cohomological schemes developed by
I.~S.~Krasil'shchik and constructed one\/--\/parametric families of
B\"acklund transformations for Eq.~\eqref{eqList} only, leaving apart
the general case~(\cite{Andreev}) of B\"acklund transformations for
the Toda equations associated with the semisimple Lie algebras.
Still, from the resulting expressions in~\cite{Andreev} it is clear
that the scaling symmetry is the required generator of deformations
for any $r\geq1$ and any $\gothg$ of rank~$r$.

\medskip
\centerline{{}\hspace{4cm}{}\hrulefill{}\hspace{4cm}{}}

\medskip
The author hopes that the reasonings of the present paper demonstrate
the profits one obtains by using the invariant
coordinate\/--\/free approach towards the mathematical physics
equations. The reader will enjoy an excursion to the world of the
PDE\/-\/related algebraic structures in the recent paper~\cite{ForKac},
where a natural class of $N$-ary generalizations for the
Lie\/--\/algebra structures (in particular, of the symmetry algebra
$\sym\cE_\Toda$ for the Toda equations) was considered.

\newpage
\subsection*{Acknowledgements}
The author thanks I.~S.~Krasil'shchik for fruitful discussions and
contructive criticism, and also thanks A.~V.~Ovchinnikov,
V.~V.~Sokolov, and A.~M.~Verbovetsky, for their remarks and advice.
The author is grateful to V.~M.~Buchstaber, E.~V.~Ferapontov,
V.~A.~Golovko, P.~Kersten,
B.~G.~Konopel'chenko, V.~G.~Marikhin, A.~K.~Pogrebkov, A.~V.~Samokhin,
A.~B.~Shabat, R.~Vitolo, V.~A.~Yumaguzhin, and to the participants of
the research seminar in algebra and geometry of differential equations
(Independent University of Moscow) for useful discussions.
The author thanks M.~Marvan, who created the \textsf{Jet}
(\cite{Jet97}) analytic transformations software, for a version of the
program and practical hints.

The major part of this research was done at the Moscow State
University. Also, the author is grateful to the universities of Twente,
Lecce, and Salerno, where a part of the work was completed, for warm
hospitality.

The research was partially supported by the scholarship
of the Government of the Russian Federation, the INTAS grant
YS~2001/2-33, and the Lecce University grant n.~650~CP/D.

\bigskip
\rightline{Translated by \textsc{the Author}.}

\end{document}